\documentclass[preprint,12pt]{elsarticle}

\usepackage{amsmath,amssymb}
\usepackage{color}

\usepackage{bm}

\journal{Materials Today Quantum}

\begin{document}

\begin{frontmatter}

\title{Stabilization mechanisms of magnetic skyrmion crystal and multiple-$Q$ states based on momentum-resolved spin interactions}

\author{Satoru Hayami$^a$ and Ryota Yambe$^b$}

\affiliation{Graduate School of Science, Hokkaido University, Sapporo 060-0810, Japan}
\affiliation{Department of Applied Physics, The University of Tokyo, Tokyo 113-8656, Japan }

\begin{abstract}
Multiple-$Q$ states as represented by a magnetic skyrmion crystal and hedgehog crystal have been extensively studied in recent years owing to their unconventional physical properties. 
The materials hosting multiple-$Q$ states have been so far observed in a variety of lattice structures and chemical compositions, which indicates rich stabilization mechanisms inducing the multiple-$Q$ states. 
We review recent developments in the research of the stabilization mechanisms of such multiple-$Q$ states with an emphasis on the microscopic spin interactions in momentum space. 
We show that an effective momentum-resolved spin model is a canonical model for not only understanding the microscopic origin of various multiple-$Q$ states but also exploring further exotic multiple-$Q$ states with topological properties. We introduce several key ingredients to realize the magnetic skyrmion crystal with the skyrmion numbers of one and two, hedgehog crystal, meron-antimeron crystal, bubble crystal, and other multiple-$Q$ states. 
We also review that the effective spin model can be used to reproduce the magnetic phase diagram in experiments efficiently. 
\end{abstract}




\end{frontmatter}


\section{Introduction}

Magnetism is one of the main subjects in condensed matter physics. 
According to real-space spin alignments in crystals, magnetic states are classified into collinear and noncollinear ones. 
In collinear magnetic states, all the spins are parallel or antiparallel to each other, as found in the ferromagnetic (FM) states, staggered antiferromagnetic (AFM) states, and spin density waves with sinusoidal modulations. 
In noncollinear magnetic states, a pair of spins is neither parallel nor antiparallel in the spin alignment, which induces a nonzero vector spin chirality between two spins, i.e., $\bm{S}_i \times \bm{S}_j \neq \bm{0}$. 
Such noncollinear magnetic states are further classified into coplanar and noncoplanar ones by a scalar spin chirality defined as $\bm{S}_i \cdot (\bm{S}_j \times \bm{S}_k)$; the coplanar (noncoplanar) magnetic states are characterized by zero (nonzero) scalar spin chirality. 
The types of magnetic states are related to the emergence of physical phenomena; a collinear FM state induces the anomalous Hall effect~\cite{Karplus_PhysRev.95.1154,smit1958spontaneous, Nagaosa_RevModPhys.82.1539}, a noncollinear magnetic state with a coplanar spiral structure induces a spin-dependent electric polarization~\cite{kimura2003magnetic, Katsura_PhysRevLett.95.057205, Mostovoy_PhysRevLett.96.067601, SergienkoPhysRevB.73.094434, Harris_PhysRevB.73.184433,tokura2014multiferroics,cardias2020first}, and a noncoplanar magnetic state with breakings of both spatial inversion and time-reversal symmetries induces nonlinear nonreciprocal transports~\cite{Hayami_PhysRevB.106.014420}. 

Among the above magnetic states, noncoplanar magnetic states can exhibit nontrivial topological properties characterized by a nonzero integer topological number once the uniform component of the scalar spin chirality is present, which is in contrast to collinear and coplanar magnetic states~\cite{batista2016frustration}. 
A typical example is a magnetic skyrmion with a swirling spin texture, where the topological (skyrmion) number corresponds to the number of times the constituent spins wrap the unit sphere~\cite{skyrme1962unified, Bogdanov89, Bogdanov94, rossler2006spontaneous, nagaosa2013topological}. 
Other examples are the three-sublattice umbrella-type magnetic state in the kagome system~\cite{Ohgushi_PhysRevB.62.R6065} and the four-sublattice tetrahedral magnetic state in the triangular-lattice system~\cite{Kurz_PhysRevLett.86.1106, Martin_PhysRevLett.101.156402, Akagi_JPSJ.79.083711, Kumar_PhysRevLett.105.216405}. 
These magnetic states with noncoplanar spin textures in real space lead to nontrivial topology in momentum space through the spin Berry phase~\cite{berry1984quantal, Loss_PhysRevB.45.13544, Ye_PhysRevLett.83.3737, Ohgushi_PhysRevB.62.R6065, Shindou_PhysRevLett.87.116801, Xiao_RevModPhys.82.1959}, which become the origin of the topological Hall and Nernst effects~\cite{Ohgushi_PhysRevB.62.R6065,taguchi2001spin,tatara2002chirality,Neubauer_PhysRevLett.102.186602,Shiomi_PhysRevB.88.064409, Hamamoto_PhysRevB.92.115417,nakazawa2018topological}. 
In addition, since the magnetic skyrmion is regarded as a topologically protected particle, intriguing current-induced motions~\cite{Jonietz_skyrmion, yu2012skyrmion, jiang2017direct, yu2020motion}, microwave and laser-induced spin excitations~\cite{okamura2013microwave, Mochizuki_PhysRevLett.108.017601, Onose_PhysRevLett.109.037603}, and nucleation by current and electric field pulses~\cite{jiang2015blowing, buttner2017field, hrabec2017current}, have been clarified. 
These features indicate that the skyrmion-hosting materials are promising for future spintronics applications, such as high-density information bit~\cite{fert2013skyrmions, fert2017magnetic, zhang2020skyrmion} and logic devices~\cite{zhang2015magnetic, luo2018reconfigurable, Chauwin_PhysRevApplied.12.064053}.

The magnetic skyrmions have been experimentally found in various materials with different lattice structures and chemical compositions~\cite{Tokura_doi:10.1021/acs.chemrev.0c00297}, where the magnetic skyrmions form a periodic array, i.e., skyrmion crystal (SkX). 
The SkXs were first observed in the chiral magnet MnSi~\cite{Muhlbauer_2009skyrmion, Neubauer_PhysRevLett.102.186602, Jonietz_skyrmion, Adams_PhysRevLett.107.217206, Bauer_PhysRevB.85.214418, Bauer_PhysRevLett.110.177207, Chacon_PhysRevLett.115.267202, muhlbauer2016kinetic, reiner2016positron} and other B20 compounds~\cite{yu2010real,Munzer_PhysRevB.81.041203, adams2010skyrmion, yu2011near, Gallagher_PhysRevLett.118.027201, Turgut_PhysRevMaterials.2.074404, Spencer_PhysRevB.97.214406, Balasubramanian_PhysRevLett.124.057201, Borisov_PhysRevMaterials.6.084401}, where the spatial inversion symmetry is absent so that the Dzyaloshinskii-Moriya (DM) interaction~\cite{dzyaloshinsky1958thermodynamic,moriya1960anisotropic} is 
present~\cite{rossler2006spontaneous, Yi_PhysRevB.80.054416, Butenko_PhysRevB.82.052403, Wilson_PhysRevB.89.094411, Mochizuki_PhysRevLett.108.017601, Banerjee_PhysRevX.4.031045,Gungordu_PhysRevB.93.064428, Rowland_PhysRevB.93.020404,Leonov_PhysRevB.96.014423}. 
Simultaneously, the SkXs have been observed in other noncentrosymmetric materials, such as intermetallic compounds~\cite{tokunaga2015new,karube2016robust, Li_PhysRevB.93.060409,karube2018disordered, Karube_PhysRevB.98.155120}, oxides~\cite{seki2012observation, Adams2012, Seki_PhysRevB.85.220406, Kurumaji_PhysRevLett.119.237201}, sulfides~\cite{kezsmarki_neel-type_2015}, monolayers~\cite{heinze2011spontaneous,romming2013writing}, Heusler compounds~\cite{nayak2017discovery,peng2020controlled}, Co-Zn-Mn alloys~\cite{karube2018disordered}, and $f$-electron compounds like EuPtSi~\cite{kakihana2018giant,kaneko2019unique,tabata2019magnetic, kakihana2019unique, Mishra_PhysRevB.100.125113, takeuchi2019magnetic,Matsumura_PhysRevB.109.174437} and EuNiGe$_3$~\cite{Goetsch_PhysRevB.87.064406, Fabreges_PhysRevB.93.214414, singh2023transition, matsumura2023distorted}. 
Subsequently, it was revealed that the SkXs also appear in centrosymmetric materials, such as Gd$_2$PdSi$_3$~\cite{Saha_PhysRevB.60.12162,kurumaji2019skyrmion, Hirschberger_PhysRevLett.125.076602, Nomoto_PhysRevLett.125.117204, Kumar_PhysRevB.101.144440,Spachmann_PhysRevB.103.184424}, Gd$_3$Ru$_4$Al$_{12}$~\cite{Nakamura_PhysRevB.98.054410, hirschberger2019skyrmion, Hirschberger_10.1088/1367-2630/abdef9}, GdRu$_2$Si$_2$~\cite{khanh2020nanometric, Yasui2020imaging, khanh2022zoology, Matsuyama_PhysRevB.107.104421, Wood_PhysRevB.107.L180402, hayami2023widely, eremeev2023insight, nomoto2023ab, Spethmann_PhysRevMaterials.8.064404}, EuAl$_4$~\cite{Shang_PhysRevB.103.L020405, kaneko2021charge, takagi2022square, Zhu_PhysRevB.105.014423, Gen_PhysRevB.107.L020410,Moya_PhysRevB.108.064436, Yang_PhysRevB.109.L041113, Miao_PhysRevX.14.011053}, and GdRu$_2$Ge$_2$~\cite{yoshimochi2024multi}, where the DM interaction is absent. 
Furthermore, a plethora of topological spin textures except for the SkX have been identified~\cite{gobel2021beyond}, such as the antiferro (AF) SkX in MnSc$_2$S$_4$~\cite{Gao2016Spiral,gao2020fractional, Rosales_PhysRevB.105.224402, takeda2024magnon}, the hedgehog crystal in MnSi$_{1-x}$Ge$_{x}$~\cite{tanigaki2015real,kanazawa2017noncentrosymmetric,fujishiro2019topological, Kanazawa_PhysRevLett.125.137202} and SrFeO$_3$~\cite{Ishiwata_PhysRevB.84.054427, Ishiwata_PhysRevB.101.134406,Rogge_PhysRevMaterials.3.084404, Onose_PhysRevMaterials.4.114420}, meron crystals in a magnetic alloy~\cite{yu2018transformation}, skyrmionium~\cite{Zhang_PhysRevB.94.094420, Pylypovskyi_PhysRevApplied.10.064057, zhang2018real, barts2021magnetic} in a ferromagnet-magnetic topological insulator heterostructure~\cite{zhang2018real}, and hopfion~\cite{Liu_PhysRevB.98.174437, Sutcliffe_PhysRevLett.118.247203,sutcliffe2018hopfions, Tai_PhysRevLett.121.187201, Gobel_PhysRevResearch.2.013315, Liu_PhysRevLett.124.127204, Raftrey_PhysRevLett.127.257201} in Ir/Co/Pt multilayers~\cite{kent2021creation}. 
In addition, more exotic topological spin states like the CP$^2$ skyrmion~\cite{Garaud_PhysRevB.87.014507, Akagi_PhysRevD.103.065008, Amari_PhysRevB.106.L100406, zhang2023cp2} have been theoretically proposed, whose emergence has been suggested in candidate materials with $ABX_3$, $BX_2$, and $AB$O$_2$, where $A$, $B$, and $X$ represent an alkali metal, transition metal, and halogen atom, respectively~\cite{zhang2023cp2}.  

In this way, the exploration of the SkXs and other topological spin states has been still a central issue. 
From the theoretical point of view, it is important to clarify the microscopic conditions of the lattice structures, magnetic interactions, anisotropy, and so on to realize these magnetic states. 
On the other hand, such theoretical analyses are limited to specific situations, since numerical simulations need tremendous computational cost owing to large length-scale spin textures, as found in the SkXs. 
In addition, there is no comprehensive guide on which magnetic interactions and anisotropy play an important role in stabilizing topological spin states. 
Thus, it is desired to establish a theoretical strategy in order to efficiently and comprehensively investigate the instability toward the SkXs and other multiple-$Q$ states.

An effective spin model has been recently proposed to understand the origin of multiple-$Q$ states~\cite{hayami2021topological}; the multiple-$Q$ states are represented by a superposition of multiple spin density waves and include periodic arrays of topological spin textures, such as the SkXs. 
This model is characterized by a momentum-resolved spin interaction rather than a conventional real-space spin interaction, which enables us to investigate the instability toward the ground-state spin configuration without incurring computational costs.
Its low-computational cost enables us to perform a number of simulations while changing the lattice structures, magnetic interactions, anisotropy, and so on. 
The effective spin model has so far uncovered new stabilization mechanisms of the SkXs: biquadratic interaction~\cite{Hayami_PhysRevB.95.224424}, symmetric anisotropic interaction~\cite{Hayami_PhysRevB.103.024439, Hayami_PhysRevB.103.054422, yambe2021skyrmion}, staggered DM interaction~\cite{Hayami_PhysRevB.105.014408}, high-harmonic wave-vector interaction~\cite{Hayami_PhysRevB.105.174437, hayami2022multiple}, and so on. 
Along the line, the realization of the SkXs has given a deep understanding of the experimental phase diagrams in SkX-hosting materials, such as Gd$_3$Ru$_4$Al$_{12}$~\cite{Hirschberger_10.1088/1367-2630/abdef9}, GdRu$_2$Si$_2$~\cite{khanh2022zoology}, EuAl$_4$~\cite{takagi2022square}, EuPtSi~\cite{hayami2021field}, EuNiGe$_3$~\cite{singh2023transition}, and GdRu$_2$Ge$_2$~\cite{yoshimochi2024multi}. 
In addition to the SkXs, the effective spin model has described the stabilization mechanisms of other multiple-$Q$ states, such as the hedgehog crystal~\cite{Okumura_PhysRevB.101.144416}, meron-antimeron crystal (MAX)~\cite{Hayami_PhysRevB.104.094425}, bimeron~\cite{Hayami_PhysRevB.103.224418}, bubble crystal~\cite{Hayami_PhysRevB.108.024426}, AF skyrmion crystal~\cite{Yambe_PhysRevB.107.014417, Hayami_doi:10.7566/JPSJ.92.084702}, and unconventional vortex crystals~\cite{hayami2021phase}. 
The result for the unconventional vortex crystals has also accounted for the experimental phase diagrams in Y$_3$Co$_8$Sn$_4$~\cite{takagi2018multiple} and CeAuSb$_2$~\cite{seo2021spin}. 

In this article, we give an overview of the instability toward the SkXs and other multiple-$Q$ states, which have been obtained through the analyses based on the effective spin model. 
We summarize their various stabilization mechanisms by focusing on the momentum-resolved spin interactions under different crystal point groups including the noncentrosymmetric point groups like $C_3$, $C_{\rm 4v}$, $C_{\rm 6v}$, $T_{\rm d}$, $O$, and $T$ and centrosymmetric point groups like $D_{\rm 2h}$, $C_{\rm 4h}$, $D_{\rm 4h}$, $D_{\rm 3d}$, $D_{6d}$, $T_{\rm h}$, and $O_{\rm h}$. 
We show that the topological spin states including the SkXs with the skyrmion number of one and two, hedgehog crystal, MAX, and vortex crystal are realized by the competition among several magnetic interactions. 
The present effective spin model ubiquitously describes the emergence of the SkXs and multiple-$Q$ states irrespective of the lattice structures and detailed model parameters, which can be used to explain experimental results, explore new materials hosting the multiple-$Q$ states, and suggest exotic multiple-$Q$ states in a systematic way.

The organization of this paper is as follows. 
In Sec.~\ref{sec: Skyrmion crystal and multiple-$Q$ states under crystallographic point groups}, we introduce the effective spin model with the momentum-resolved spin interaction. 
We also present key spin interactions leading to multiple-$Q$ instability, which are summarized in two tables (Tables~\ref{tab:noncentro} and \ref{tab:centro}). 
Then, we review the stability of the SkXs and other multiple-$Q$ states by taking several systems in noncentrosymmetric point groups in Sec.~\ref{sec: Stabilization mechanisms in noncentrosymmetric magnets} and centrosymmetric point groups in Sec.~\ref{sec: Stabilization mechanisms in centrosymmetric magnets}. 
In noncentrosymmetric systems, we discuss the stabilization mechanisms of the SkX, hedgehog crystal, and MAX by focusing on the role of the DM interaction. 
In centrosymmetric systems, we discuss the stabilization mechanisms of the SkX, hedgehog crystal, AF SkX, and bubble crystal by focusing on the differences from noncentrosymmetric systems. 
Section~\ref{sec: Summary and perspective} is devoted to the summary and future perspective.

\section{Skyrmion crystal and multiple-$Q$ states under crystallographic point groups}
\label{sec: Skyrmion crystal and multiple-$Q$ states under crystallographic point groups}

In this section, we review the microscopic ingredients to induce the multiple-$Q$ states including the SkX in various lattice structures based on the effective spin model. 
First, we introduce the effective spin model in momentum space in Sec.~\ref{sec: Effective spin model in momentum space}. 
Next, we discuss the microscopic origin of the single-$Q$ spiral state in the cases of centrosymmetric and noncentrosymmetric systems in Sec.~\ref{sec: Microscopic origin of single-$Q$ spiral state}. 
Then, we show the microscopic spin interactions included in the effective spin model in Sec.~\ref{sec: Microscopic interactions leading to multiple-Q instability}, which can bring about the multiple-$Q$ instabilities. 
We mainly focus on the role of the external magnetic field (Sec.~\ref{sec: External magnetic field}), the biquadratic interaction (Sec.~\ref{sec: Biquadratic interaction}), the anisotropic interaction (Sec.~\ref{sec: Anisotropic interaction}), the sublattice-dependent interaction (Sec.~\ref{sec: Sublattice-dependent interaction}), and the high-harmonic wave-vector interaction (Sec.~\ref{sec: High-harmonic wave-vector interaction}) on the stabilization tendency of the multiple-$Q$ states. 
We also discuss other ingredients leading to the multiple-$Q$ states in Sec.~\ref{sec: Other mechanisms}.

\subsection{Effective spin model in momentum space}
\label{sec: Effective spin model in momentum space}

\begin{figure}[tp!]
\begin{center}
\includegraphics[width=1.0\hsize]{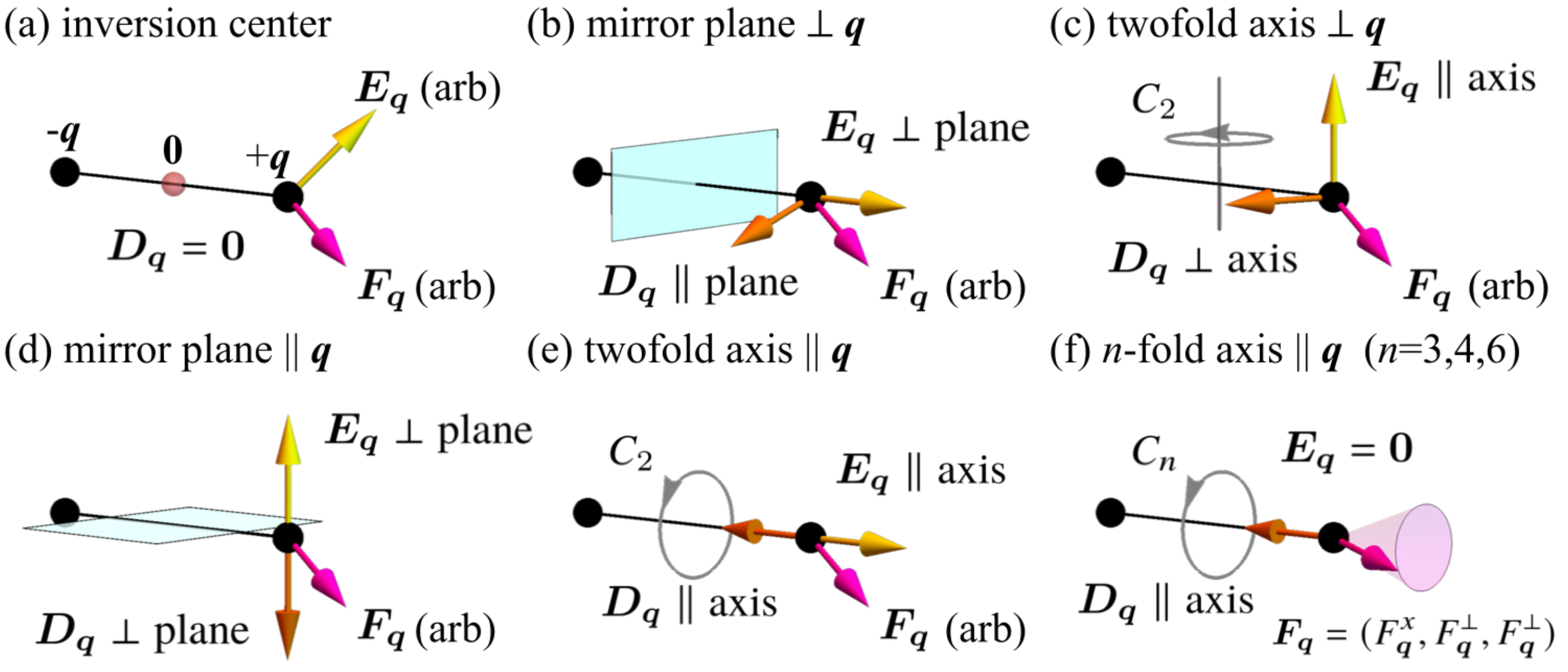} 
\caption{
\label{fig: interaction}
Six symmetry rules for the momentum-resolved spin interactions, $\bm{D}_{\bm{q}}$, $\bm{E}_{\bm{q}}$, and $\bm{F}_{\bm{q}}$, with the wave vector $\bm{q}$ in momentum space: (a) spatial inversion at $\bm{q}=\bm{0}$, (b) mirror perpendicular to $\bm{q}$, (c) twofold rotation perpendicular to $\bm{q}$, (d) mirror parallel to $\bm{q}$, (e) twofold rotation around $\bm{q}$, and (f) $n$-fold ($n=3,4,6$) rotation around $\bm{q}$. 
Reprinted figure with permission from~\cite{Yambe_PhysRevB.106.174437}, Copyright (2022) by the American Physical Society.
}
\end{center}
\end{figure}

Let us introduce the effective spin model incorporating the momentum-resolved bilinear exchange interaction~\cite{Hayami_PhysRevB.95.224424, Yambe_PhysRevB.106.174437, Yambe_PhysRevB.107.174408}. 
We start by considering the spin model without the sublattice degree of freedom, which is given by 
\begin{align}
\label{eq:model_bilinear_single}
\mathcal{H}^{\rm (2)} =- \sum_{\bm{q},\alpha_\mathrm{s}, \beta_\mathrm{s}} S^{\alpha_\mathrm{s}}_{\bm{q}} X^{\alpha_\mathrm{s}\beta_\mathrm{s}}_{\bm{q}} S^{\beta_\mathrm{s}}_{-\bm{q}},
\end{align}
where $S^{\alpha_\mathrm{s}}_{\bm{q}}$ is the Fourier transform of the classical or quantum spin $S^{\alpha_\mathrm{s}}_i$ in real space and $(x_\mathrm{s}, y_\mathrm{s}, z_\mathrm{s})$ are cartesian spin coordinates. 
$X^{\alpha_\mathrm{s}\beta_\mathrm{s}}_{\bm{q}}$ represents the general form of the bilinear exchange interaction, which is expressed by the $3\times 3$ matrix as
\begin{align}
\label{eq:qspace_Xq}
X_{\bm{q}}=
\begin{pmatrix}
F_{\bm{q}}^{x_\mathrm{s}} & E_{\bm{q}}^{z_\mathrm{s}}+iD_{\bm{q}}^{z_\mathrm{s}} & E_{\bm{q}}^{y_\mathrm{s}}-iD_{\bm{q}}^{y_\mathrm{s}} \\
E_{\bm{q}}^{z_\mathrm{s}}-iD_{\bm{q}}^{z_\mathrm{s}} & F_{\bm{q}}^{y_\mathrm{s}}  & E_{\bm{q}}^{x_\mathrm{s}}+iD_{\bm{q}}^{x_\mathrm{s}} \\
E_{\bm{q}}^{y_\mathrm{s}}+iD_{\bm{q}}^{y_\mathrm{s}} & E_{\bm{q}}^{x_\mathrm{s}}-iD_{\bm{q}}^{x_\mathrm{s}}  & F_{\bm{q}}^{z_\mathrm{s}}  
\end{pmatrix}, 
\end{align}
where $\bm{D}_{\bm{q}}=(D^{x_\mathrm{s}}_{\bm{q}}, D^{y_\mathrm{s}}_{\bm{q}}, D^{z_\mathrm{s}}_{\bm{q}})$, $\bm{E}_{\bm{q}}=(E^{x_\mathrm{s}}_{\bm{q}}, E^{y_\mathrm{s}}_{\bm{q}}, E^{z_\mathrm{s}}_{\bm{q}})$, and $\bm{F}_{\bm{q}}=(F^{x_\mathrm{s}}_{\bm{q}}, F^{y_\mathrm{s}}_{\bm{q}}, F^{z_\mathrm{s}}_{\bm{q}})$ are real coupling constants. 
$\bm{D}_{\bm{q}}$ corresponds to the antisymmetric off-diagonal interaction (DM interaction) for the interchange of the spin components $(\alpha_\mathrm{s} \leftrightarrow \beta_\mathrm{s})$ in Eq.~(\ref{eq:model_bilinear_single}), while $\bm{E}_{\bm{q}}$ and $\bm{F}_{\bm{q}}$ correspond to the symmetric off-diagonal interaction and symmetric diagonal interaction, respectively.
Owing to the antisymmetric (symmetric) property, the relations of $\bm{D}_{\bm{q}}=-\bm{D}_{-\bm{q}}$, $\bm{E}_{\bm{q}}=\bm{E}_{-\bm{q}}$, and $\bm{F}_{\bm{q}}=\bm{F}_{-\bm{q}}$ hold. 
The real-space counterparts of the $\bm{q}$-resolved bilinear exchange interactions $(\bm{D}_{\bm{q}}, \bm{E}_{\bm{q}}, \bm{F}_{\bm{q}})$ are the short-range/long-range anisotropic exchange interactions $(\bm{D}_{ij}, \bm{E}_{ij}, \bm{F}_{ij})$, where $\bm{D}_{ij}$, $\bm{E}_{ij}$, and $\bm{F}_{ij}$ are the real coupling constants for $S^{\alpha_{\rm s}}_{i}S^{\beta_{\rm s}}_{j}-S^{\alpha_{\rm s}}_{j}S^{\beta_{\rm s}}_{i}$, $S^{\alpha_{\rm s}}_{i}S^{\beta_{\rm s}}_{j}+S^{\alpha_{\rm s}}_{j}S^{\beta_{\rm s}}_{i}$, and $S^{\alpha_{\rm s}}_{i}S^{\alpha_{\rm s}}_{j}$ on the $(ij)$ bond in real space, respectively. 
Especially, $\bm{E}_{\bm{q}}$ and $\bm{F}_{\bm{q}}$ (or $\bm{E}_{ij}$ and $\bm{F}_{ij}$) include the $\Gamma$-type interaction in the Kitaev model for $\alpha_s\neq \beta_s$~\cite{Rau_PhysRevLett.112.077204} and the Ising-type interaction for $\alpha_s = \beta_s$.
The minus sign in Eq.~(\ref{eq:model_bilinear_single}) means that $\bm{q}$ channels giving the largest positive eigenvalues of $X^{\alpha_\mathrm{s}\beta_\mathrm{s}}_{\bm{q}}$ are important in investigating the ground states.

Nonzero components in $X_{\bm{q}}$ in Eq.~(\ref{eq:qspace_Xq}) are determined by the symmetry of crystals as well as the wave vectors. 
Similarly to Moriya's rule for real-space spin interactions~\cite{moriya1960anisotropic}, the symmetry conditions inducing nonzero $\bm{D}_{\bm{q}}$, $\bm{E}_{\bm{q}}$, and $\bm{F}_{\bm{q}}$ are summarized as the following six rules~\cite{Yambe_PhysRevB.106.174437}: 
\begin{itemize}
\item[(a)] The spatial inversion symmetry imposes ``$\bm{D}_{\bm{q}}=\bm{0}$", while there is no constraint on $\bm{E}_{\bm{q}}$ and $\bm{F}_{\bm{q}}$, as shown in Fig.~\ref{fig: interaction}(a).
\item[(b)] The mirror symmetry to the plane perpendicular to $\bm{q}$ imposes ``$\bm{D}_{\bm{q}}\parallel$ plane" and ``$\bm{E}_{\bm{q}}\perp$ plane", while there is no constraint on $\bm{F}_{\bm{q}}$, as shown in Fig.~\ref{fig: interaction}(b).
\item[(c)] The twofold rotational symmetry around the axis perpendicular to $\bm{q}$ imposes ``$\bm{D}_{\bm{q}}\perp$ axis" and ``$\bm{E}_{\bm{q}}\parallel$ axis", while there is no constraint on $\bm{F}_{\bm{q}}$, as shown in Fig.~\ref{fig: interaction}(c).
\item[(d)] The mirror symmetry to the plane parallel to $\bm{q}$ imposes ``$\bm{D}_{\bm{q}}\perp$ plane" and ``$\bm{E}_{\bm{q}}\perp$ plane", while there is no constraint on $\bm{F}_{\bm{q}}$, as shown in Fig.~\ref{fig: interaction}(d).
\item[(e)] The twofold rotational symmetry around the axis parallel to $\bm{q}$ imposes ``$\bm{D}_{\bm{q}}\parallel$ axis" and ``$\bm{E}_{\bm{q}}\parallel$ axis", while there is no constraint on $\bm{F}_{\bm{q}}$, as shown in Fig.~\ref{fig: interaction}(e).
\item[(f)] The $n$-fold ($n=3,4,6$) rotational symmetries around the axis parallel to $\bm{q}$ imposes ``$\bm{D}_{\bm{q}}\parallel$ axis", ``$\bm{E}_{\bm{q}}=\bm{0}$", and ``$\bm{F}_{\bm{q}} =(F_{\bm{q}}^{x_\mathrm{s}},F_{\bm{q}}^\perp,F_{\bm{q}}^\perp)$", as shown in Fig.~\ref{fig: interaction}(f).
\end{itemize}
Here, $x_\mathrm{s}$ is set along the $\bm{q}$ direction and each symmetry operation leaves the origin $\bm{q}=(0,0,0)$ invariant.

\begin{figure}[tp!]
\begin{center}
\includegraphics[width=1.0\hsize]{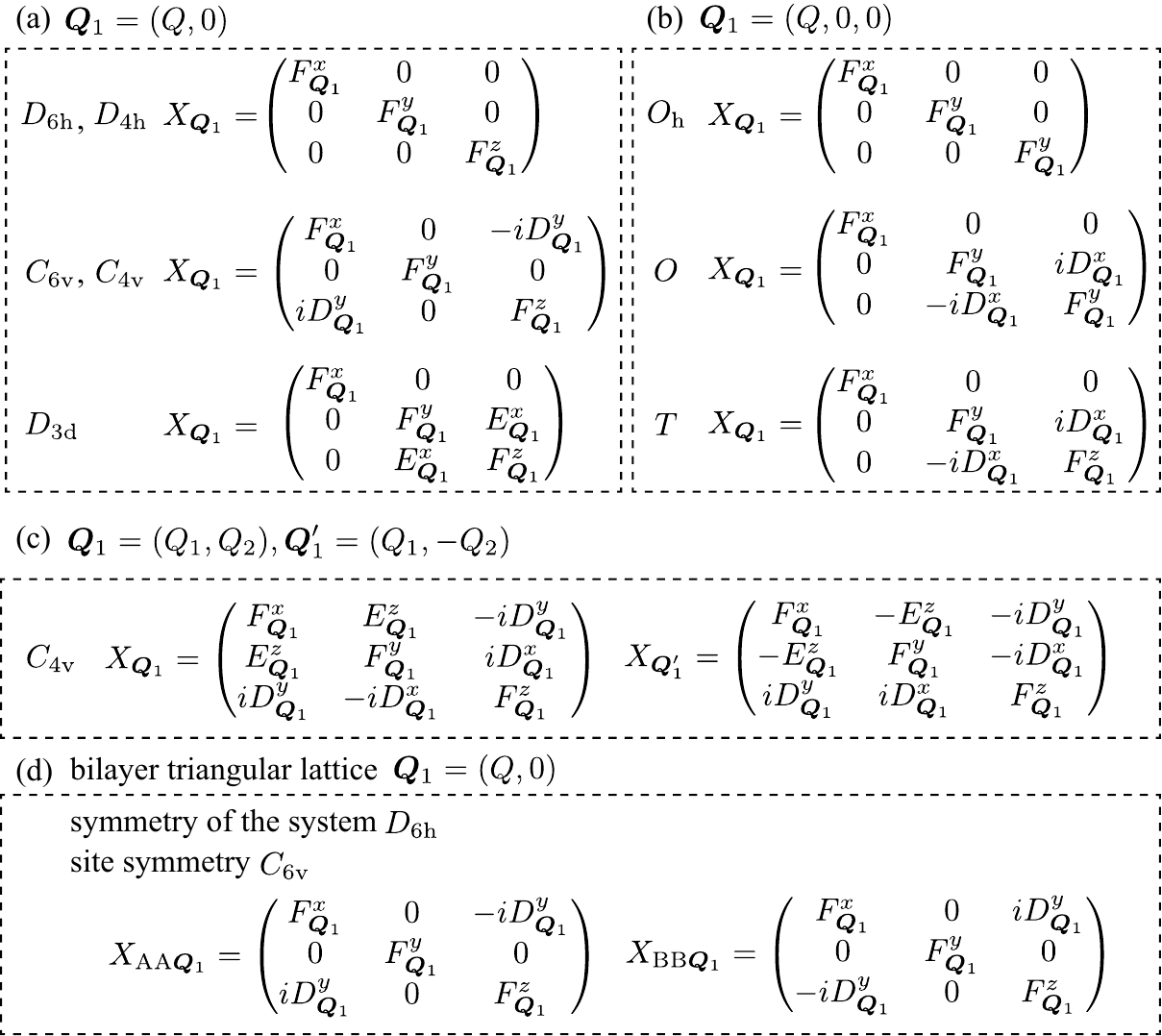} 
\caption{
\label{fig: int_matrix}
Examples of interaction matrices $X_{\bm{Q}_1}$ under several point groups~\cite{Yambe_PhysRevB.106.174437, Yambe_PhysRevB.107.174408}. 
(a) $X_{\bm{Q}_1}$ for the high-symmetric wave vector $\bm{Q}_1=(Q, 0)$ under (top) $D_{\rm 6h}$ and $D_{\rm 4h}$, (middle) $C_{\rm 6v}$ and $C_{\rm 4v}$, and (bottom) $D_{\rm 3d}$. 
(b) $X_{\bm{Q}_1}$ for the high-symmetric wave vector $\bm{Q}_1=(Q, 0, 0)$ under (top) $O_{\rm h}$, (middle) $O$, and (bottom) $T$. 
(c) $X_{\bm{Q}_1}$ and $X_{\bm{Q}'_1}$ for the low-symmetric wave vectors $\bm{Q}_1=(Q_1, Q_2)$ and $\bm{Q}'_1=(Q_1, -Q_2)$ under $C_{\rm 4v}$. 
(d) $X_{{\rm AA}\bm{Q}_1}$ and $X_{{\rm BB}\bm{Q}_1}$ for the high-symmetric wave vector $\bm{Q}_1=(Q, 0)$ under $D_{\rm 6h}$ with the site symmetry $C_{\rm 6v}$. 
} 
\end{center}
\end{figure}

These rules indicate that nonzero components of $\bm{D}_{\bm{q}}$, $\bm{E}_{\bm{q}}$, and $\bm{F}_{\bm{q}}$ largely depend on the point group symmetry and wave-vector direction.
Figure~\ref{fig: int_matrix} shows several examples of the interaction matrix $X_{\bm{q}}$ for different point groups and wave vectors $\bm{Q}_1$. 
In the case of the centrosymmetric point groups $D_{\rm 6h}$ and $D_{\rm 4h}$ with the two-dimensional wave vector $\bm{Q}_1=(Q, 0)$, $F^x_{\bm{Q}_1}$, $F^y_{\bm{Q}_1}$, and $F^z_{\bm{Q}_1}$ become nonzero and are independent from each other, while $\bm{D}_{\bm{Q}_1}$ and $\bm{E}_{\bm{Q}_1}$ vanish, as shown in Fig.~\ref{fig: int_matrix}(a). 
Meanwhile, when the polar-type noncentrosymmetric point groups $C_{\rm 6v}$ and $C_{\rm 4v}$ are considered so that the symmetry in terms of the horizontal mirror plane parallel to $\bm{Q}_1$, as well as the spatial inversion, is lost, the DM interaction $D^y_{\bm{Q}_1}$ is additionally induced owing to the rules in Figs.~\ref{fig: interaction}(b) and \ref{fig: interaction}(c). 
Furthermore, when the symmetry lowering from $D_{\rm 6h}$ to $D_{\rm 3d}$ happens by breaking the symmetries with respect to the twofold rotation around the $z$ axis and horizontal mirror perpendicular to the $z$ axis while the mirror symmetry with respect to the $yz$ plane is kept, $E^x_{\bm{Q}_1}$ is additionally induced. 
In this way, the number of independent anisotropic interactions increases as the crystal symmetry is lowered. 

Such a situation also happens in the case of the cubic point groups with the three-dimensional wave vector $\bm{Q}_1=(Q, 0, 0)$~\cite{Yambe_PhysRevB.107.174408}. 
The number of independent anisotropic interactions is two, i.e., $F^x_{\bm{Q}_1}$ and $F^y_{\bm{Q}_1}$, for the high-symmetric $O_{\rm h}$ point group, as shown in Fig.~\ref{fig: int_matrix}(b). 
When the symmetry is lowered to the noncentrosymmetric point group $O$ where the inversion symmetry is lost, the DM interaction $D^x_{\bm{Q}_1}$ is additionally induced. 
In addition, the symmetric anisotropic interaction $F^z_{\bm{Q}_1} \neq F^y_{\bm{Q}_1}$ is induced when the symmetry is further lowered to the point group $T$ by breaking the fourfold rotational symmetry.  

The number of independent anisotropic interactions also depends on the symmetry of the wave vectors in momentum space.  
For example, when the anisotropic interactions at the wave vector $\bm{Q}_1=(Q_1, Q_2)$ with $Q_1 \neq Q_2$, $Q_1 \neq \pi$, and $Q_2 \neq \pi$ under the $C_{\rm 4v}$ symmetry are considered, 6 independent anisotropic interactions are defined, as shown in the left panel of Fig.~\ref{fig: int_matrix}(c). 
Compared to the case of $\bm{Q}_1=(Q, 0)$ in Fig.~\ref{fig: int_matrix}(a), $D^y_{\bm{Q}_1}$ and $E^z_{\bm{Q}_1}$ are additionally induced. 
We also show the anisotropic interaction matrix at symmetry-related wave vector $\bm{Q}'_1=(Q_1, -Q_2)$ in the right panel of Fig.~\ref{fig: int_matrix}(c), where each tensor component in $X_{\bm{Q}'_1}$ is related to that in $X_{\bm{Q}_1}$ by the mirror symmetry in the $xz$ plane.
These results indicate that the effective spin model has multiple anisotropic interactions depending on the crystals and wave vectors, which can be a source of complicated magnetic states including the multiple-$Q$ state, as discussed in Sec.~\ref{sec: Anisotropic interaction}.
It is noted that the anisotropic interactions can appear even in the high-symmetric point groups such as $O_{\rm h}$ and $D_{\rm 6h}$ according to the discrete rotational symmetry. 
  
The extension of the anisotropic interaction in Eq.~(\ref{eq:model_bilinear_single}) to the multi-sublattice case is straightforwardly performed once the permutation symmetry between sublattices is appropriately taken into account~\cite{Yambe_PhysRevB.106.174437}. 
The effective spin Hamiltonian in the multi-sublattice system is generally given by
\begin{align}
\label{eq:qspace_multisub}
\mathcal{H}^{(2)}_m = -\sum_{\bm{q},\alpha_\mathrm{s},\beta_\mathrm{s},\eta,\eta'}S^{\alpha_\mathrm{s}}_{\eta\bm{q}} \mathcal{X}^{\alpha_\mathrm{s}\beta_\mathrm{s}}_{\eta\eta'\bm{q}}  S^{\beta_\mathrm{s}}_{\eta'-\bm{q}}, 
\end{align}
where $\eta, \eta'$ represent the indices for the number of sublattices $N_{\rm atom}$. 
In contrast to $X_{\bm{q}}$ in Eq.~(\ref{eq:qspace_Xq}), the interaction matrix $\mathcal{X}_{\eta\eta'\bm{q}}$ with $\mathcal{X}_{\eta\eta'\bm{q}}=\mathcal{X}_{\eta\eta'\bm{q}}^\dagger$ is characterized by 18 independent degrees of freedom, which is expressed as 
\begin{align}
\label{eq:qspace_Xq_sub}
\mathcal{X}_{\eta\eta'\bm{q}}=
\begin{pmatrix}
\mathcal{F}_{\eta\eta'\bm{q}}^{x_\mathrm{s}} & \mathcal{E}_{\eta\eta'\bm{q}}^{z_\mathrm{s}}+\mathcal{D}_{\eta\eta'\bm{q}}^{z_\mathrm{s}} & \mathcal{E}_{\eta\eta'\bm{q}}^{y_\mathrm{s}}-\mathcal{D}_{\eta\eta'\bm{q}}^{y_\mathrm{s}} \\
\mathcal{E}_{\eta\eta'\bm{q}}^{z_\mathrm{s}}-\mathcal{D}_{\eta\eta'\bm{q}}^{z_\mathrm{s}} & \mathcal{F}_{\eta\eta'\bm{q}}^{y_\mathrm{s}}  & \mathcal{E}_{\eta\eta'\bm{q}}^{x_\mathrm{s}}+\mathcal{D}_{\eta\eta'\bm{q}}^{x_\mathrm{s}} \\
\mathcal{E}_{\eta\eta'\bm{q}}^{y_\mathrm{s}}+\mathcal{D}_{\eta\eta'\bm{q}}^{y_\mathrm{s}} & \mathcal{E}_{\eta\eta'\bm{q}}^{x_\mathrm{s}}-\mathcal{D}_{\eta\eta'\bm{q}}^{x_\mathrm{s}}  & \mathcal{F}_{\eta\eta'\bm{q}}^{z_\mathrm{s}} 
\end{pmatrix}, 
\end{align}
where $\bm{\mathcal{D}}_{\eta\eta'\bm{q}}=(\mathcal{D}^{x_\mathrm{s}}_{\eta\eta'\bm{q}}, \mathcal{D}^{y_\mathrm{s}}_{\eta\eta'\bm{q}}, \mathcal{D}^{z_\mathrm{s}}_{\eta\eta'\bm{q}})$, $\bm{\mathcal{E}}_{\eta\eta'\bm{q}}=(\mathcal{E}^{x_\mathrm{s}}_{\eta\eta'\bm{q}}, \mathcal{E}^{y_\mathrm{s}}_{\eta\eta'\bm{q}}, \mathcal{E}^{z_\mathrm{s}}_{\eta\eta'\bm{q}})$, and $\bm{\mathcal{F}}_{\eta\eta'\bm{q}}=(\mathcal{F}^{x_\mathrm{s}}_{\eta\eta'\bm{q}}, \mathcal{F}^{y_\mathrm{s}}_{\eta\eta'\bm{q}}, \mathcal{F}^{z_\mathrm{s}}_{\eta\eta'\bm{q}})$ are the complex coupling constants; $\mathcal{X}_{\eta\eta'\bm{q}}$ represents the intrasublattice (intersublattice) interaction for $\eta=\eta'$ ($\eta \neq \eta'$). 
It is noted that $\mathrm{Re}(\bm{\mathcal{D}}_{\eta\eta\bm{q}})=\mathrm{Im}(\bm{\mathcal{E}}_{\eta\eta\bm{q}})=\mathrm{Im}(\bm{\mathcal{F}}_{\eta\eta\bm{q}})=\bm{0}$, which is consistent with the single-sublattice ($N_{\rm atom}$=1) result. 
$\mathrm{Re}(\bm{\mathcal{D}}_{\eta\eta'\bm{q}})$,  $\mathrm{Im}(\bm{\mathcal{E}}_{\eta\eta'\bm{q}})$, and  $\mathrm{Im}(\bm{\mathcal{F}}_{\eta\eta'\bm{q}})$ [$\mathrm{Im}(\bm{\mathcal{D}}_{\eta\eta'\bm{q}})$,  $\mathrm{Re}(\bm{\mathcal{E}}_{\eta\eta'\bm{q}})$, and  $\mathrm{Re}(\bm{\mathcal{F}}_{\eta\eta'\bm{q}})$] are antisymmetric (symmetric) for $\eta \leftrightarrow \eta'$. 
$\mathrm{Re}(\bm{\mathcal{D}}_{\eta\eta'\bm{q}})$,  $\mathrm{Re}(\bm{\mathcal{E}}_{\eta\eta'\bm{q}})$, and $\mathrm{Re}(\bm{\mathcal{F}}_{\eta\eta'\bm{q}})$ [$\mathrm{Im}(\bm{\mathcal{D}}_{\eta\eta'\bm{q}})$,  $\mathrm{Im}(\bm{\mathcal{E}}_{\eta\eta'\bm{q}})$, and  $\mathrm{Im}(\bm{\mathcal{F}}_{\eta\eta'\bm{q}})$] are symmetric (antisymmetric) for $\bm{q} \leftrightarrow -\bm{q}$.  
As in the case of $N_{\rm atom}=1$, the nonzero components of $\mathcal{X}_{\eta\eta'\bm{q}}$ are determined by the symmetries in terms of the wave vector $\bm{q}$ and the crystal symmetry.
The different point from the $N_{\rm atom}=1$ case is to consider the permutation among the sublattices in each symmetry operation.

The sublattice degree of freedom can bring about additional anisotropic interactions that do not appear in the single-sublattice case. 
As an example, we show the interaction matrices $X_{\rm AA\bm{Q}_1}$ and $X_{\rm BB\bm{Q}_1}$ for $\bm{Q}_1=(Q,0)$ in the centrosymmetric bilayer triangular-lattice system under the $D_{6\rm h}$ symmetry in Fig.~\ref{fig: int_matrix}(d), where we suppose that the site symmetry for two sublattices (layers) A and B is given by $C_{\rm 6v}$ and two sublattices are connected by the spatial inversion symmetry. 
Compared to the interaction matrix under the $D_{6\rm h}$ symmetry in Fig.~\ref{fig: int_matrix}(a), one finds that the DM interaction $D^y_{\bm{Q}_1}$ is additionally induced as a result of the $C_{\rm 6v}$ site symmetry in each sublattice. 
Thus, the DM interaction has the sublattice-dependent form, although their signs are opposite for sublattices A and B owing to the presence of the global spatial inversion symmetry under $D_{\rm 6h}$~\cite{hayami2016emergent}. 
Since the DM interaction often leads to instability toward the SkXs, this sublattice-dependent DM interaction is also expected to induce the SkX even in the centrosymmetric lattice structure, as detailed in Secs.~\ref{sec: Sublattice-dependent interaction} and \ref{sec: Effect of sublattice-dependent interaction}.

Similarly to the bilinear exchange interaction in Eq.~(\ref{eq:model_bilinear_single}), we can define higher-order multi-spin interactions. 
In the following sections, we mainly focus on the four-spin interactions, whose general form is given by 
\begin{align}
\label{eq:qspace_4th}
\mathcal{H}^{(4)} = \sum_{\bm{q}_1,\bm{q}_2,\bm{q}_3,\bm{q}_4,\alpha_\mathrm{s},\beta_\mathrm{s},\alpha'_\mathrm{s},\beta'_\mathrm{s}} X^{\alpha_\mathrm{s}\beta_\mathrm{s}\alpha'_\mathrm{s}\beta'_\mathrm{s}}_{\bm{q}_1\bm{q}_2\bm{q}_3\bm{q}_4} S^{\alpha_\mathrm{s}}_{\bm{q}_1}   S^{\beta_\mathrm{s}}_{\bm{q}_2}S^{\alpha'_\mathrm{s}}_{\bm{q}_3}   S^{\beta'_\mathrm{s}}_{-\bm{q}_4} \delta_{\bm{q}_1+\bm{q}_2+\bm{q}_3+\bm{q}_4, l\bm{G}}, 
\end{align}
where the single-sublattice case is considered for simplicity; $\delta$ is the Kronecker delta and $\bm{G}$ is the reciprocal lattice vector ($l$ is an integer). 
The real-space counterpart of the four-spin interaction is given by $S^{\alpha_\mathrm{s}}_{i}   S^{\beta_\mathrm{s}}_{j}S^{\alpha'_\mathrm{s}}_{k}   S^{\beta'_\mathrm{s}}_{l}$. 
The nonzero components of $X^{\alpha_\mathrm{s}\beta_\mathrm{s}\alpha'_\mathrm{s}\beta'_\mathrm{s}}_{\bm{q}_1\bm{q}_2\bm{q}_3\bm{q}_4} $ are determined by the symmetry in terms of the wave vector and point group. 
It is noted that the odd-order terms with respect to the spin can appear only when the time-reversal symmetry is broken~\cite{Hayami_PhysRevB.95.224424, Nikolic_PhysRevB.103.155151}.

\begin{figure}[tp!]
\begin{center}
\includegraphics[width=0.75\hsize]{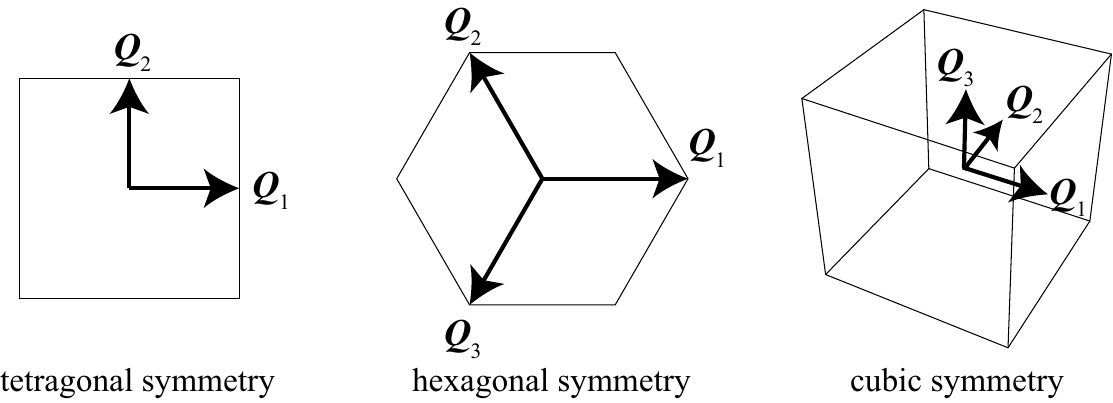} 
\caption{
\label{fig: qvector}
Symmetry-related ordering wave vectors under (a) the tetragonal symmetry, (b) hexagonal symmetry, and (c) cubic symmetry; $\bm{Q}_1$ lies along the $x$ direction. 
The interactions at $-\bm{Q}_1$, $\pm \bm{Q}_2$, and $\pm \bm{Q}_3$ give the same magnitude of that at $\bm{Q}_1$. 
} 
\end{center}
\end{figure}

To summarize, the effective spin model with the momentum-resolved interaction in the case of the single-sublattice system is expressed as 
\begin{align}
\label{eq: Ham}
\mathcal{H}= \mathcal{H}^{(2)} + \mathcal{H}^{(4)} + \mathcal{H}^{\rm ex}, 
\end{align}
where $\mathcal{H}^{\rm ex}$ is added to represent the effect of external fields, such as magnetic and electric fields. 
In the following analyses, we regard the spins as classical ones with a fixed length $|\bm{S}_i|=1$ for simplicity. 
Hereafter, we neglect the spin index ${\rm s}$ for the cartesian spin coordinates $x_{\rm s}$, $y_{\rm s}$, and $z_{\rm s}$ by taking along the $x$, $y$, and $z$ directions in real space, respectively, for simplicity.

The effective spin model in Eq.~(\ref{eq: Ham}) can be used to reproduce the phase diagram in experimental materials. 
There are two important experimental inputs: One is the ordering wave vectors and the other is the symmetry of the lattice structure. 
In terms of the ordering wave vectors, small-angle neutron and/or resonant elastic X-ray scattering experiments are powerful tools. 
Once one obtains these two pieces of information, one can construct an effective spin model with symmetry-allowed anisotropic exchange interactions. 
One may also introduce other factors, such as multi-spin interactions and high-harmonic wave-vector interactions, depending on the observed magnetic phases.
In order to help the construction of the spin model, we describe the dominant mechanisms invoking the multiple-$Q$ states in the following subsections.

\subsection{Microscopic origin of single-$Q$ spiral state}
\label{sec: Microscopic origin of single-$Q$ spiral state}

Before discussing the instability toward the multiple-$Q$ states in the model in Eq.~(\ref{eq: Ham}), we show the microscopic origin of the single-$Q$ spiral states. 
In other words, we discuss when and how finite-$q$ states have lower energy than the FM and collinear AFM states. 
For that purpose, we consider the isotropic bilinear spin model, which is given by 
\begin{align}
\label{eq: Ham_iso}
\mathcal{H}^{\rm (2)}_{\rm iso} =-  \sum_{\bm{q}}J_{\bm{q}}  \bm{S}_{\bm{q}} \cdot \bm{S}_{-\bm{q}},
\end{align}
where $J_{\bm{q}}$ is the isotropic coupling.
This isotropic model corresponds to the effective spin model in Eq.~(\ref{eq:model_bilinear_single}) with $\bm{D}_{\bm{q}}=\bm{0}$, $\bm{E}_{\bm{q}}=\bm{0}$, and $\bm{F}_{\bm{q}}=(J_{\bm{q}},J_{\bm{q}},J_{\bm{q}})$, which is obtained by neglecting the spin-orbit coupling of the system.
The ground state is obtained by the Luttinger-Tisza method, where its ordering wave vector gives a maximum value of $J_{\bm{q}}$. Thus, to obtain the instability toward a finite-$Q$ ordering, the relation of $J_{\bm{Q}}\geq J_{\bm{q}}$ must be satisfied for arbitrary $\bm{q}$ in the Brillouin zone; $\bm{Q}$ is different from FM ordering wave vector $\bm{q}=\bm{0}$ and staggered AFM ordering wave vector corresponding to time-reversal invariant momenta.

There are mainly three microscopic situations that favor the finite-$q$ ordering. 
The first is the frustrated exchange interaction in real space. 
For example, the ordering wave vector for the ground-state spin configuration of the triangular-lattice Heisenberg model with the nearest-neighbor FM interaction $J_1<0$ and the third-nearest-neighbor AFM interaction $J_3>0$ is given by 
\begin{align}
Q=2 \cos^{-1}\left[ \frac{1}{4}\left(1+\sqrt{1-\frac{2J_1}{J_3}} \right)\right],
\end{align}
where $\bm{Q}=(Q, 0)$. 
Thus, the single-$Q$ spiral state, whose spin configuration is given by $\bm{S}_i = (\cos \bm{Q} \cdot \bm{r}_i, \sin \bm{Q} \cdot \bm{r}_i,0)$ ($\bm{r}_i$ is the position vector at site $i$), becomes the ground state when $J_3 > -J_1/4$. 
The second is the Ruderman-Kittel-Kasuya-Yosida (RKKY) interaction in itinerant electron systems like the Kondo lattice model consisting of localized spins and itinerant electrons, where the exchange interaction between localized spins is mediated by the kinetic motion of itinerant electrons~\cite{Ruderman, Kasuya, Yosida1957}. 
In this case, the nesting property of the Fermi surface determines the ordering wave vector $\bm{Q}$, since the RKKY interaction depends on the bare susceptibility of the itinerant electrons.  
When the bare susceptibility becomes the largest at $\bm{Q}$, the single-$Q$ spiral state is stabilized in the ground state.  
For the above two cases, the spiral plane is arbitrary owing to the spin rotational symmetry of the spin Hamiltonian.
The third is the additional introduction of the DM interaction in the form of $i\bm{D}_{\bm{q}}\cdot (\bm{S}_{\bm{q}} \times \bm{S}_{-\bm{q}})$, which arises from the spin-orbit coupling in noncentrosymmetric lattice structures. 
In this case, even when the isotropic exchange interaction favors the FM spin configuration, the single-$Q$ spiral state can be realized in the presence of $\bm{D}_{\bm{q}}$, since the DM interaction tends to twist the spins; the spiral plane is fixed depending on the direction of the DM vector $\bm{D}_{\bm{q}}$.

Owing to the rotational symmetry around the principle axis, the above single-$Q$ spiral state is degenerate in terms of the direction of the wave vectors. 
For example, there are four (six) equivalent wave vectors in the Brillouin zone under the tetragonal (hexagonal and cubic) symmetry when one of the dominant ordering wave vectors is characterized by $\bm{Q}_1=(Q, 0, 0)$, as shown in Fig.~\ref{fig: qvector}: $\bm{Q}_2=(0, Q, 0)$ for the tetragonal symmetry,  $\bm{Q}_2=(-Q/2, \sqrt{3}Q/2, 0)$ and $\bm{Q}_3=(-Q/2, -\sqrt{3}Q/2, 0)$ for the hexagonal (trigonal) symmetry, and $\bm{Q}_2=(0, Q, 0)$ and $\bm{Q}_3=(0, 0, Q)$ for the cubic symmetry. 
In this sense, it is enough to take into account the interactions at a few symmetry-related ordering wave vectors in evaluating the internal energy; the interactions at other wave vectors give almost no contributions. 
Then, the effective spin model in Eq.~(\ref{eq: Ham}) can be simplified by extracting the dominant interactions at a particular set of ordering wave vectors. 
Specifically, the model in Eq.~(\ref{eq: Ham_iso}) can be rewritten as 
\begin{align}
\label{eq: Ham_iso_q}
\mathcal{H}^{\rm (2)}_{\rm iso} =- 2J \sum_{\nu}  \bm{S}_{\bm{Q}_\nu} \cdot \bm{S}_{-\bm{Q}_\nu},
\end{align}
where $\bm{Q}_\nu$ corresponds to the ordering wave vector that gives the maximum of $J_{\bm{q}}$ in Eq.~(\ref{eq: Ham_iso}); $\nu$ is the index for the symmetry-related ordering wave vectors, and prefactor 2 means the contribution from $-\bm{Q}_\nu$. 
Since the magnitude of the interaction at $\bm{Q}_\nu$ is the same as each other, we set $J \equiv J_{\bm{Q}_\nu} = 1$ as the energy unit of the model. 
Although such a simplification enables us to investigate the multiple-$Q$ instability in an efficient way, the lack of interactions at other wave vectors makes the analyses about the impurity effect and isolated topological defects difficult.

As the multiple-$Q$ spin configuration often consists of spiral states with symmetry-related wave vectors, the single-$Q$ spiral state can be replaced by the multiple-$Q$ states by additionally considering the effect of the anisotropic interactions, multi-spin interactions, external fields, and so on. 
An efficient search for the multiple-$Q$ states while varying such additional model parameters can be performed based on the effective spin model in Eq.~(\ref{eq: Ham_iso_q}). 
We discuss the multiple-$Q$ instabilities based on the effective spin model in Eq.~(\ref{eq: Ham_iso_q}) in noncentrosymmetric lattice structures in Sec.~\ref{sec: Stabilization mechanisms in noncentrosymmetric magnets} and centrosymmetric lattice structures in Sec.~\ref{sec: Stabilization mechanisms in centrosymmetric magnets}.

\subsection{Microscopic interactions leading to multiple-$Q$ instability}
\label{sec: Microscopic interactions leading to multiple-Q instability}

\begin{table}[tb!]
\begin{center}
\caption{
Classification of topological spin crystals in noncentrosymmetric point groups (PGs), which is mainly obtained by the effective spin model in Eq.~(\ref{eq: Ham}). 
In the ``System" column, TL, SL, and CL represent the triangular lattice, square lattice, and cubic lattice, respectively. 
In the ``Interaction" column, DM, SA, BQ, EA, and EP stand for the Dzyaloshinskii-Moriya interaction, symmetric anisotropic interaction, biquadratic interaction, easy-axis-type interaction, and easy-plane-type interaction, respectively. 
In the ``Spin texture" column, the prefixes S and T stand for square and triangular, respectively. 
SkX and MAX stand for the skyrmion crystal and meron-antimeron crystal, respectively. 
Some of the spin textures are presented in Fig.~\ref{fig: ponti}. 
}
\label{tab:noncentro}
\begingroup
\renewcommand{\arraystretch}{1.1}
\scalebox{0.8}{
 \begin{tabular}{ccccc}
 \multicolumn{4}{l}{\fbox{noncentrosymmetric system}} \\
 \hline \hline
PG & System & Interaction & Spin texture & Relevant materials \\ \hline
$C_3$ & TL (screw) & DM & T-SkX~\cite{Hayami_PhysRevB.105.224411} &  \\
$C_3$ & CL & DM \& SA & hedgehog~\cite{Kato_PhysRevB.104.224405} &  \\
$D_4$, $D_{\rm 2d}$, $C_{\rm 4v}$ & SL & DM \& SA & S-vortex~\cite{Kato_PhysRevB.104.224405} &  \\
$D_{\rm 2d}$ & CL & DM & S-SkX~\cite{Hayami_PhysRevB.107.174435} &  \\
$C_{\rm 4v}$ & SL & DM & hybrid S-SkX~\cite{Hayami_PhysRevB.109.054422} & EuNiGe$_3$~\cite{singh2023transition} \\ 
$C_{\rm 4v}$ & SL & DM \& SA & hybrid S-SkX~\cite{Hayami_PhysRevLett.121.137202} & \\ 
$C_{\rm 4v}$ & SL & DM \& SA & S-MAX~\cite{Hayami_PhysRevLett.121.137202} & CeAlGe~\cite{Puphal_PhysRevLett.124.017202}\\ 
$D_{\rm 3h}$ & TL & DM & T-bimeron~\cite{Hayami_PhysRevB.105.224423} & \\ 
$C_{\rm 6v}$ & TL (bilayer) & EA & T-SkX~\cite{Okigami_doi:10.7566/JPSJ.91.103701} &  \\
$C_{\rm 6v}$ & TL & EP \& BQ & T-vortex~\cite{hayami2020multiple} & Y$_3$Co$_8$Sn$_4$~\cite{takagi2018multiple} \\
$C_{\rm 6v}$ & TL & EP \& BQ & T-MAX~\cite{Hayami_PhysRevB.104.094425} &  \\
$O$, $T$ & CL & DM \& BQ & T-SkX~\cite{hayami2021field} & EuPtSi \\
$O$, $T$ & CL & DM & 6$Q$ SkX~\cite{Hayami_PhysRevB.107.174435} &  \\
$O$, $T$ & CL & DM \& BQ & hedgehog~\cite{Okumura_PhysRevB.101.144416} & MnSi$_{1-x}$Ge$_x$ \\
$T$ & CL & DM \& SA & hedgehog~\cite{Kato_PhysRevB.104.224405} &  \\
 \hline\hline
\end{tabular}
}
\endgroup
\end{center}
\end{table}

\begin{table}[tb!]
\begin{center}
\caption{
Classification of topological spin crystals in centrosymmetric point groups (PGs), which is mainly obtained by the effective spin model in Eq.~(\ref{eq: Ham}). 
In the ``System" column, TL, SL, and CL represent the triangular lattice, square lattice, and cubic lattice, respectively. 
In the ``Interaction" column, DM, SA, BQ, HH, EA, EP, and SIA stand for the Dzyaloshinskii-Moriya interaction, symmetric anisotropic interaction, biquadratic interaction, high-harmonic wave-vector interaction, easy-axis-type interaction, easy-plane-type interaction, and single-ion anisotropy, respectively. 
In the ``Spin texture" column, the prefixes S, T, and R stand for square, triangular, and rectangular, respectively. 
SkX, MAX, and TVX stand for the skyrmion crystal, meron-antimeron crystal, and tetra-axial vortex crystal, respectively. 
Some of the spin textures are presented in Fig.~\ref{fig: ponti}. 
}
\label{tab:centro}
\begingroup
\renewcommand{\arraystretch}{1}
\scalebox{0.72}{
 \begin{tabular}{ccccc}
\multicolumn{4}{l}{\fbox{centrosymmetric system}} \\
 \hline \hline
PG & System & Interaction & Spin texture & Relevant materials \\ \hline
 \hline\hline
$D_{\rm 2h}$ & TL & SA  & T-SkX~\cite{Hayami_doi:10.7566/JPSJ.91.093701} &  \\
$D_{\rm 2h}$ & SL & EA \& HH  & S-SkX~\cite{hayami2023orthorhombic} &  \\
$D_{\rm 4h}$ & SL & SA \& BQ  & S-SkX~\cite{Hayami_PhysRevB.103.024439, hayami2023widely} & GdRu$_2$Si$_{2}$~\cite{khanh2022zoology} \\
$D_{\rm 4h}$ & SL & EA \& HH  & R-SkX~\cite{Hayami_PhysRevB.105.174437, Hayami_PhysRevB.108.094416} & EuAl$_4$~\cite{takagi2022square}, GdRu$_2$Ge$_{2}$~\cite{yoshimochi2024multi} \\
$D_{\rm 4h}$ & SL & EA \& HH  & S-SkX~\cite{Hayami_PhysRevB.105.174437, Hayami_PhysRevB.108.094416} & EuAl$_4$~\cite{takagi2022square}, GdRu$_2$Ge$_{2}$~\cite{yoshimochi2024multi} \\
$D_{\rm 4h}$ & SL & EP & S-bimeron~\cite{hayami2023anisotropic} &  \\
$D_{\rm 4h}$ & SL & EA \& BQ  & S-bubble~\cite{Hayami_PhysRevB.108.024426, Hayami_PhysRevB.108.094415} & CeAuSb$_2$~\cite{seo2021spin} \\
$D_{\rm 4h}$ & SL (bilayer) & DM  & S-SkX~\cite{hayami2022square} & 
  \\
$D_{\rm 4h}$ & SL & SA \& BQ  & $n_{\rm sk}=2$ S-SkX~\cite{hayami2022multiple} &  \\
$C_{\rm 4h}$ & SL & SA \& BQ  & S-SkX~\cite{Hayami_PhysRevB.105.104428} &  \\
$C_{\rm 3i}$, $C_{\rm 6h}$ & TL & SIA & T-SkX~\cite{hayami2022skyrmion} & \\
$D_{\rm 3d}$, $D_{\rm 6h}$ & TL & SIA  & T-SkX~\cite{hayami2022skyrmion} & \\
$D_{\rm 3d}$ & TL & SA  & T-SkX~\cite{yambe2021skyrmion} &  \\
$D_{\rm 3d}$ & TL & SA  & $n_{\rm sk}=2$ T-SkX~\cite{yambe2021skyrmion} &  \\
$D_{\rm 6h}$ & TL & thermal & T-SkX~\cite{Okubo_PhysRevLett.108.017206, Mitsumoto_PhysRevB.104.184432, Mitsumoto_PhysRevB.105.094427} &  \\
$D_{\rm 6h}$ & TL & EA & T-SkX~\cite{leonov2015multiply, Lin_PhysRevB.93.064430, Hayami_PhysRevB.93.184413} &  \\
$D_{\rm 6h}$ & TL & EP & T-bimeron~\cite{Hayami_PhysRevB.103.224418} &  \\
$D_{\rm 6h}$ & TL & impurity & T-SkX~\cite{Hayami_PhysRevB.94.174420} &  \\
$D_{\rm 6h}$ & TL & SA  & T-SkX~\cite{Hayami_PhysRevB.103.054422} & GdRu$_3$Al$_{12}$~\cite{Hirschberger_10.1088/1367-2630/abdef9}
  \\
$D_{\rm 6h}$ & TL (bilayer) & DM  & T-SkX~\cite{Hayami_PhysRevB.105.014408} & 
  \\
$D_{\rm 6h}$ & TL (trilayer) & DM  & T-SkX~\cite{Hayami_PhysRevB.105.184426} & 
  \\
$D_{\rm 6h}$ & TL & SA  & $n_{\rm sk}=2$ T-SkX~\cite{Hayami_PhysRevB.103.054422} & 
  \\
$D_{\rm 6h}$ & TL & BQ  & $n_{\rm sk}=2$ T-SkX~\cite{Ozawa_PhysRevLett.118.147205, Hayami_PhysRevB.99.094420} &  \\
$D_{\rm 6h}$ & honeycomb & BQ  & AF-SkX~\cite{Yambe_PhysRevB.107.014417} &  \\
$D_{\rm 6h}$ & TL (bilayer) & EP \& BQ  & AF-SkX~\cite{Hayami_doi:10.7566/JPSJ.92.084702} &  \\
$D_{\rm 6h}$ & TL (trilayer) & BQ  & AF-SkX~\cite{Hayami_PhysRevB.109.014415} & 
  \\
$D_{\rm 6h}$ & honeycomb & BQ  & $n_{\rm sk}=2$ AF-SkX~\cite{Yambe_PhysRevB.107.014417} &  \\
$D_{\rm 6h}$ & TL & thermal & TVX~\cite{hayami2021phase} &  \\
$D_{\rm 6h}$ & TL & six-spin int. & TVX~\cite{hayami2021phase} &  \\
$D_{\rm 6h}$ & TL & EA \& BQ & T-bubble~\cite{Hayami_10.1088/1367-2630/ac3683, Utesov_PhysRevB.105.054435} & \\
$T_{\rm h}$ & CL & SA & hedgehog~\cite{Yambe_PhysRevB.107.174408}  & \\
$O_{\rm h}$ & CL & BQ & hedgehog~\cite{Okumura_doi:10.7566/JPSJ.91.093702}  & SrFeO$_3$\\

 \hline\hline
\end{tabular}
}
\endgroup
\end{center}
\end{table}

\begin{figure}[tp!]
\begin{center}
\includegraphics[width=0.95\hsize]{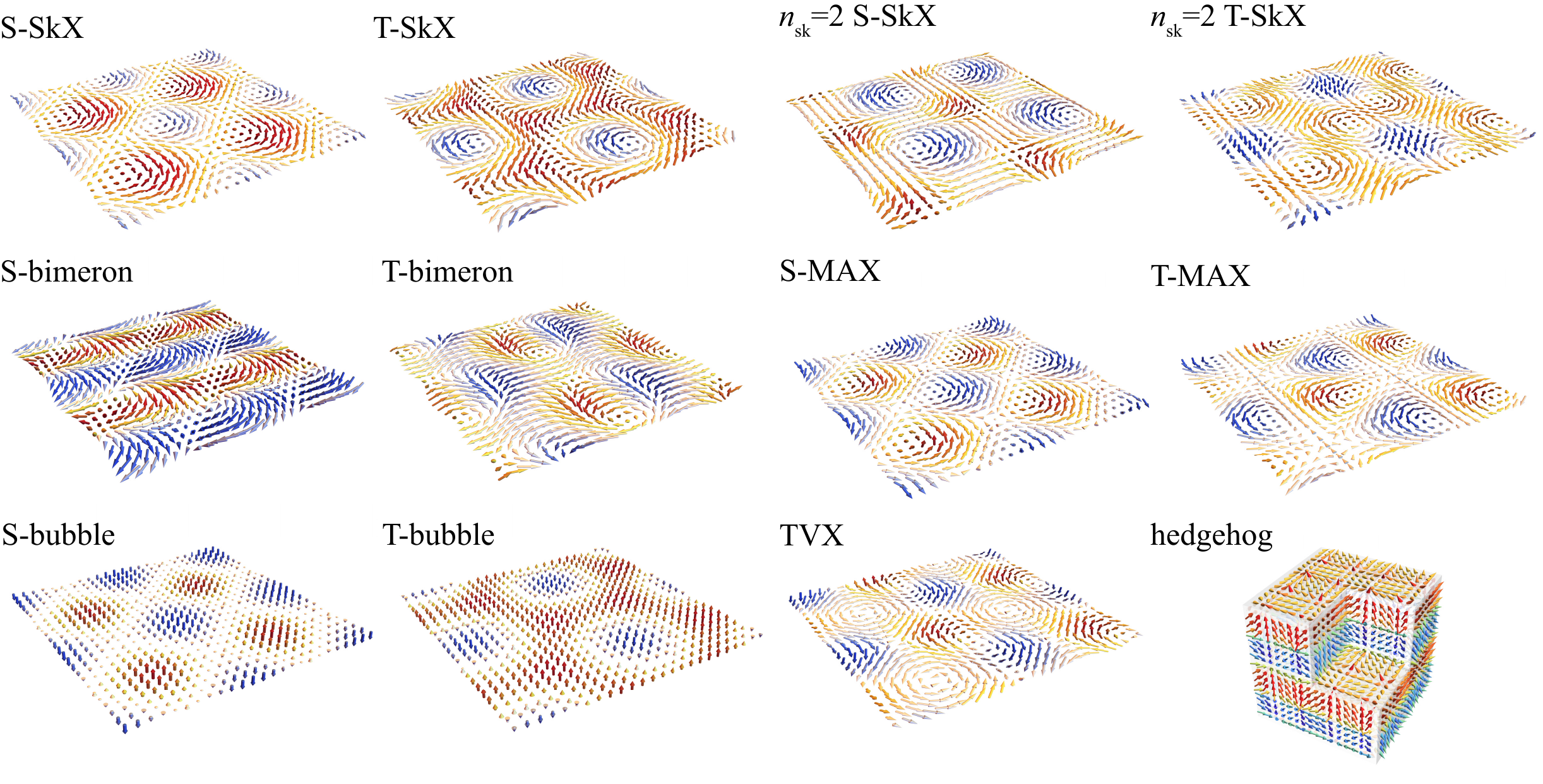} 
\caption{
\label{fig: ponti}
Schematic spin configurations of several topological spin textures. 
The arrows represent the direction of the spin moments and the color of the arrows represents the $z$ spin component. 
The prefixes S and T stand for square and triangular, respectively. 
SkX, MAX, and TVX stand for the skyrmion crystal, meron-antimeron crystal, and tetra-axial vortex crystal, respectively. 
} 
\end{center}
\end{figure}

We show several microscopic origins to induce multiple-$Q$ instability by additionally considering the effect of the multi-spin interactions, anisotropic interactions, and external fields for the spin model with the momentum-resolved isotropic spin interaction in Eq.~(\ref{eq: Ham_iso_q}). 
We briefly discuss why the single-$Q$ spiral state can be replaced by the multiple-$Q$ states under additional effects in this section; we show the detailed phase diagram for a specific model in Secs.~\ref{sec: Stabilization mechanisms in noncentrosymmetric magnets} and \ref{sec: Stabilization mechanisms in centrosymmetric magnets}.
We mainly introduce the effect of an external magnetic field [Sec.~\ref{sec: External magnetic field}] and four microscopic interactions: biquadratic interaction [Sec.~\ref{sec: Biquadratic interaction}], anisotropic interaction [Sec.~\ref{sec: Anisotropic interaction}], sublattice-dependent interaction [Sec.~\ref{sec: Sublattice-dependent interaction}], and high-harmonic wave-vector interaction [Sec.~\ref{sec: High-harmonic wave-vector interaction}]. 
We also list other ingredients in Sec.~\ref{sec: Other mechanisms}. 
We summarize the resultant multiple-$Q$ states including the SkXs under different noncentrosymmetric and centrosymmetric point group symmetries in Tables~\ref{tab:noncentro} and \ref{tab:centro}, respectively. 
The real-space spin configurations of the representative multiple-$Q$ states are schematically shown in Fig.~\ref{fig: ponti}. 
The double-$Q$ (triple-$Q$) states tend to be stabilized in tetragonal (hexagonal and trigonal) systems owing to the fourfold (threefold) rotational symmetry; here and hereafter, we use the prefixes, S- and T-, which represent square and triangular, respectively. 
For example, the S-SkX (T-SkX) represents the square(triangular)-lattice alignment of the skyrmion, which consists of a double-$Q$ (triple-$Q$) superposition of the spiral states in tetragonal (hexagonal or trigonal) systems.

\subsubsection{External magnetic field}
\label{sec: External magnetic field}

First, let us consider the effect of an external magnetic field. 
The Zeeman Hamiltonian is given by 
 \begin{align}
 \label{eq: HamZeeman}
 \mathcal{H}^{\rm Z}=  - \sum_i \bm{H} \cdot \bm{S}_i, 
 \end{align}
where $\bm{H}=(H_x, H_y, H_z)$. 

For the Hamiltonian $\mathcal{H}^{\rm (2)}_{\rm iso}+\mathcal{H}^{\rm Z}$ in centrosymmetric magnets, the ground state is always given by the single-$Q$ conical spiral state rather than the multiple-$Q$ state, where the spin configuration is given by $\bm{S}_i= (\sin \theta \cos \bm{Q}_\nu \cdot \bm{r}_i, \sin \theta \sin \bm{Q}_\nu \cdot \bm{r}_i, \cos \theta)$ with $\cos \theta = H/2J$ in the case of the out-of-plane magnetic field $\bm{H}=(0,0, H)$. 
The absence of the multiple-$Q$ states is because a superposition of multiple spin density waves usually leads to the intensity at high-harmonic wave vectors like $\bm{Q}_\nu + \bm{Q}_{\nu'}$, which results in the energy cost compared to the single-$Q$ state. 

Meanwhile, the situation changes when the effect of the DM interaction is introduced by supposing noncentrosymmetric crystal structures. 
The momentum-resolved DM Hamiltonian for the symmetry-related wave vectors $\bm{Q}_\nu$ is given by 
 \begin{align}
 \label{eq: HamDM}
 \mathcal{H}^{\rm DM}=  - 2 i \sum_{\nu} \bm{D}_\nu \cdot (\bm{S}_{\bm{Q}_\nu} \times \bm{S}_{-\bm{Q}_\nu}).  
 \end{align}
The direction of the DM vector $\bm{D}_\nu$ is determined by the crystal symmetry as well as $\bm{Q}_\nu$, as discussed in Sec.~\ref{sec: Effective spin model in momentum space}. 
For $\bm{Q}_1 = (Q, 0)$ on the triangular (square) lattice, $\bm{D}_1$ is given by $(D, 0, 0)$ under the chiral $D_6$ ($D_4$) symmetry, $\bm{D}_1$ is given by $(0, D, 0)$ under the polar $C_{\rm 6v}$ ($C_{\rm 4v}$) symmetry, and $\bm{D}_1$ is given by $(0, 0, D)$ under the $D_{\rm 3h}$ symmetry. 
It is noted that the in-plane components of $\bm{D}_1$ at $\bm{Q}_1$ are different from those at the other symmetry-related wave vectors so as to satisfy the rotational symmetry around the principal axis, while the out-of-plane component is common. 

The DM interaction fixes the spiral plane of the single-$Q$ state in the plane perpendicular to the DM vector. 
In such a situation, the multiple-$Q$ instability can be expected when the magnetic field is applied in the direction perpendicular to $\bm{D}_\nu$; the out-of-plane magnetic field $\bm{H}=(0, 0, H)$ tends to favor the SkX instead of the single-$Q$ state when the system has either the chiral-type DM interaction $\bm{D}_1=(D, 0, 0)$ or polar-type DM interaction $\bm{D}_1=(0, D, 0)$~\cite{rossler2006spontaneous, Yi_PhysRevB.80.054416}. 
The typical phase diagrams on the triangular and square lattice are shown in Sec.~\ref{sec: Skyrmion crystal by the Dzyaloshinskii-Moriya interaction}.
Meanwhile, the out-of-plane magnetic field does not induce the SkX for $\bm{D}_1=(0, 0, D)$. 

This difference depending on $\bm{D}_1$ is understood from the energetic viewpoint; the magnetic field parallel to the spiral plane (perpendicular to the DM vectors) leads to the elliptical deformation of the spiral plane, which results in a higher-harmonic wave-vector contribution to the spin structure. 
In other words, applying the magnetic field to the single-$Q$ spiral state gives an energy loss in terms of the exchange energy instead of gaining the Zeeman energy. 
In addition, an effective coupling in the form of $(\bm{S}_{\bm{0}}\cdot \bm{S}_{\bm{Q}_1})(\bm{S}_{\bm{Q}_2}\cdot \bm{S}_{\bm{Q}_3})$ with $\bm{Q}_1+\bm{Q}_2+\bm{Q}_3=\bm{0}$ appears in the free energy, which gives the energy gain for the triple-$Q$ state.  
These energy loss for the single-$Q$ state and gain for the multiple-$Q$ state lead to instability toward the multiple-$Q$ states including the SkX; see also Sec.~\ref{sec: Anisotropic interaction} for further discussions. 
On the other hand, the magnetic field perpendicular to the spiral plane (parallel to the DM vectors) leads to the conical spiral spin structure without the higher-harmonic wave-vector contribution; the single-$Q$ state remains stable under the magnetic field. 

With this energetic argument in mind, one notices that the in-plane magnetic field can induce the multiple-$Q$ instability under the $D_{\rm 3h}$ symmetry to possess the DM interaction $D_{1}=(0, 0, D)$. 
In this case, the T-bimeron crystal, whose schematic spin configuration is presented in Fig.~\ref{fig: ponti}, has been found in the intermediate in-plane magnetic field~\cite{Hayami_PhysRevB.105.224423}.

\subsubsection{Biquadratic interaction}
\label{sec: Biquadratic interaction}

Next, we consider the effect of the four-spin interaction, whose general expression is given in Eq.~(\ref{eq:qspace_4th}). 
The real-space counterpart of the isotropic four-spin interaction without the magnetic anisotropy is represented by $(\bm{S}_i \cdot \bm{S}_j)(\bm{S}_k \cdot \bm{S}_l)$, which often leads to a multiple-$Q$ instability~\cite{Momoi_PhysRevLett.79.2081, Kurz_PhysRevLett.86.1106,heinze2011spontaneous, Yoshida_PhysRevLett.108.087205,ueland2012controllable, Mankovsky_PhysRevB.101.174401,paul2020role, Brinker_PhysRevResearch.2.033240,lounis2020multiple, Spethmann_PhysRevLett.124.227203, Simon_PhysRevMaterials.4.084408, Mendive-Tapia_PhysRevB.103.024410}. 
It is noted that the anisotropic four-spin interaction is also possible depending on the symmetry of the system; for example, in noncentrosymmetric magnets, the chiral biquadratic interaction as $(\bm{S}_i \times \bm{S}_j)(\bm{S}_i\cdot \bm{S}_j)$ appears as the higher-order term to the DM interaction, which also leads to the multiple-$Q$ instability~\cite{brinker2019chiral, Laszloffy_PhysRevB.99.184430, Mankovsky_PhysRevB.101.174401, Brinker_PhysRevResearch.2.033240,lounis2020multiple}. 
These real-space four-spin interactions originate from higher-order exchange processes beyond the second-order process for the Heisenberg interaction in the localized spin model with the short-range interaction~\cite{takahashi1977half,yoshimori1978fourth, Momoi_PhysRevLett.79.2081, Bulaevskii_PhysRevB.78.024402,Hoffmann_PhysRevB.101.024418,li2021spin}.

In contrast, we focus on the momentum-resolved four-spin interaction arising from the long-range interaction owing to the itinerant nature of electrons, whose isotropic form is expressed as $(\bm{S}_{\bm{q}_1}\cdot \bm{S}_{\bm{q}_2})(\bm{S}_{\bm{q}_3} \cdot \bm{S}_{\bm{q}_4})$ with $\bm{q}_1+\bm{q}_2+\bm{q}_3 + \bm{q}_4= l\bm{G}$. 
Among them, we consider the biquadratic-type interaction, which is obtained by inserting $\bm{q}_1=\bm{q}_3=\bm{Q}_\nu$ and $\bm{q}_2=\bm{q}_4=-\bm{Q}_\nu$; the interaction is represented by 
 \begin{align}
 \label{eq: Ham4th}
 \mathcal{H}^{\rm (4)}_{\rm BQ}=   \frac{K}{N}\sum_{\nu}  (\bm{S}_{\bm{Q}_\nu} \cdot \bm{S}_{-\bm{Q}_\nu})^2,   
 \end{align}
where $K$ represents the coupling constant and $N$ represents the system size. 
This biquadratic interaction tends to favor noncoplanar spin textures for $K>0$. 
In other words, the positive biquadratic interaction tends to favor the multiple-$Q$ states instead of the single-$Q$ state. 
One of the microscopic origins of the positive biquadratic interaction is the higher-order RKKY effect caused by the partial nesting of the Fermi surfaces~\cite{Akagi_PhysRevLett.108.096401, Hayami_PhysRevB.90.060402, Ozawa_doi:10.7566/JPSJ.85.103703, Hayami_PhysRevB.95.224424}. 
Thanks to effective four-spin interactions, a plethora of multiple-$Q$ states have been found in itinerant magnets under various lattice structures, such as hexagonal systems~\cite{Martin_PhysRevLett.101.156402, Akagi_JPSJ.79.083711, Kato_PhysRevLett.105.266405, Barros_PhysRevB.88.235101, Venderbos_PhysRevLett.108.126405, Jiang_PhysRevLett.114.216402, Venderbos_PhysRevB.93.115108, Barros_PhysRevB.90.245119, Ghosh_PhysRevB.93.024401, Ozawa_PhysRevB.96.094417,  Chern_PhysRevB.97.035120, Kobayashi_PhysRevB.106.L140406} including Y$_3$Co$_8$Sn$_4$~\cite{takagi2018multiple}, tetragonal systems~\cite{Agterberg_PhysRevB.62.13816, Venderbos_PhysRevLett.109.166405, Solenov_PhysRevLett.108.096403, hayami_PhysRevB.91.075104, Ozawa_doi:10.7566/JPSJ.85.103703, Hayami_PhysRevB.94.024424, Shahzad_PhysRevB.96.224402, Okada_PhysRevB.98.224406, Su_PhysRevResearch.2.013160, Hayami_PhysRevB.108.094415} including CeAuSb$_2$~\cite{Marcus_PhysRevLett.120.097201,Seo_PhysRevX.10.011035,seo2021spin}, and cubic systems~\cite{Chern_PhysRevLett.105.226403,hayami2014charge, Hayami_PhysRevB.89.085124}.

Especially, the positive biquadratic interaction becomes the source of various topological spin crystals, such as SkXs~\cite{Ozawa_PhysRevLett.118.147205, Hayami_PhysRevB.95.224424, hayami2021field, Yambe_PhysRevB.107.014417, Hayami_PhysRevB.109.014415}, hedgehog crystals~\cite{Okumura_PhysRevB.101.144416, Okumura_doi:10.7566/JPSJ.91.093702, Shimizu_PhysRevB.103.054427}, and MAXs~\cite{Hayami_PhysRevB.104.094425}, under both noncentrosymmetric and centrosymmetric lattice structures, as detailed for the SkXs in Secs.~\ref{sec: Effect of biquadratic interaction} and \ref{sec: Effect of biquadratic interaction_2}, for the hedgehog crystals in Secs.~\ref{sec: Hedgehog crystal} and \ref{sec: Hedgehog crystal_2}, and for the MAX in Sec.~\ref{sec: Meron-antimeron crystal}.  
It also gives rise to unconventional SkXs, such as the SkX with the skyrmion number of two ($n_{\rm sk}=2$ SkX) in the triangular~\cite{Ozawa_PhysRevLett.118.147205, Hayami_PhysRevB.95.224424} and square~\cite{hayami2022multiple} lattices and multi-sublattice SkXs with different skyrmion numbers for different sublattices~\cite{Yambe_PhysRevB.107.014417, Hayami_PhysRevB.109.014415}. 
Recently, further higher-order multi-spin interactions leading to the multiple-$Q$ instability have been investigated, such as the chiral-chiral interaction in the forms of $[\bm{S}_i \cdot (\bm{S}_j \times \bm{S}_k)]^2$~\cite{grytsiuk2020topological, Bomerich_PhysRevB.102.100408} and $[\bm{S}_{\bm{Q}_1} \cdot (\bm{S}_{\bm{Q}_2} \times \bm{S}_{\bm{Q}_3})]^2+[\bm{S}_{-\bm{Q}_1} \cdot (\bm{S}_{-\bm{Q}_2} \times \bm{S}_{-\bm{Q}_3})]^2$ with $\bm{Q}_1+\bm{Q}_2+\bm{Q}_3=\bm{0}$~\cite{hayami2021phase}, the latter of which induces the tetra-axial vortex crystal (TVX). 
Here, the spin configuration of the TVX is characterized by a superposition of three sinusoidal waves at $\bm{Q}_1$, $\bm{Q}_2$, and $\bm{Q}_3$ on the triangular lattice, which smoothly connects to the $n_{\rm sk}=2$ SkX by changing the relative phase among the constituent sinusoidal waves; the schematic spin configuration is presented in Fig.~\ref{fig: ponti}.

\subsubsection{Anisotropic interaction}
\label{sec: Anisotropic interaction}

\begin{figure}[tp!]
\begin{center}
\includegraphics[width=1.0\hsize]{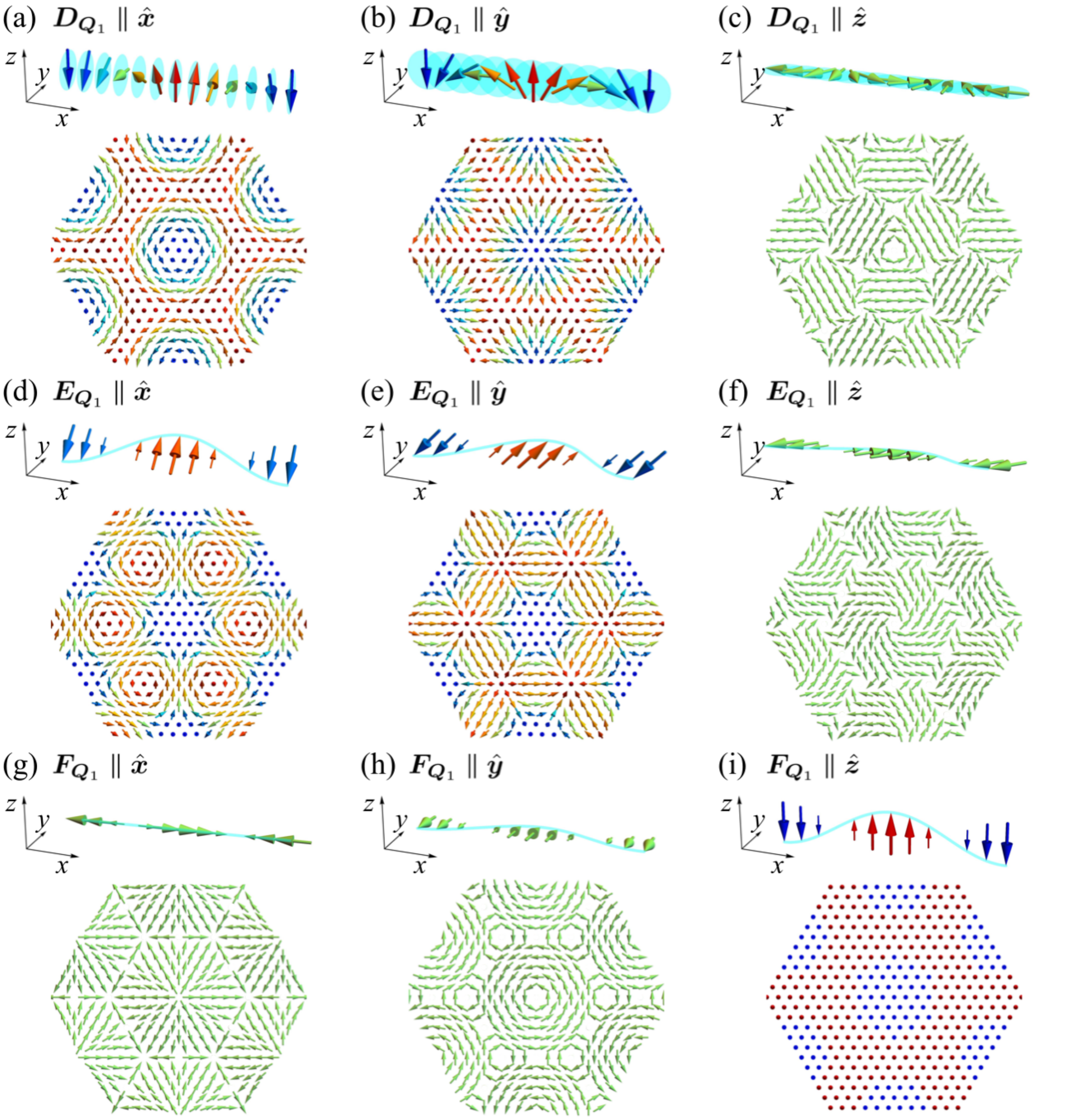} 
\caption{
\label{fig: crystal}
Schematic spin configurations of (upper panel) the single-$Q$ and (lower panel) triple-$Q$ states on the two-dimensional triangular lattice in the presence of the anisotropic interactions:
 (a) $D^x_{\bm{Q}_1}$, 
 (b) $D^y_{\bm{Q}_1}$, 
 (c) $D^z_{\bm{Q}_1}$, 
 (d) $E^x_{\bm{Q}_1}$, 
 (e) $E^y_{\bm{Q}_1}$, 
 (f) $E^z_{\bm{Q}_1}$, 
 (g) $F^x_{\bm{Q}_1}$, 
 (h) $F^y_{\bm{Q}_1}$, 
 and (i) $F^z_{\bm{Q}_1}$ for $\bm{Q}_1=(Q, 0)$. 
For the $\bm{Q}_1$ component, the spiral planes lie on the (a) $yz$, (b) $zx$, and (c) $xy$ planes, while  the oscillating spin directions are (d)  
[011], (e) [101], (f) [110], (g) [100], (h) [010], and (i) [001] directions.
The triple-$Q$ states in the lower panel consist of a superposition of the spin density waves at $\bm{Q}_1$, $\bm{Q}_2=(-Q/2, \sqrt{3}Q/2)$, and $\bm{Q}_3=(-Q/2, -\sqrt{3}Q/2)$.
The red, blue, and green arrows stand for positive, negative, and zero values of the $z$-spin component.
Reprinted figure with permission from~\cite{Yambe_PhysRevB.106.174437}, Copyright (2022) by the American Physical Society.
}
\end{center}
\end{figure}

The anisotropic spin interaction becomes the origin of multiple-$Q$ states even within the bilinear interaction.
Its expression within the single-sublattice system is given by
\begin{align}
\label{eq: Ham_ani_q}
\mathcal{H}^{\rm (2)}_{\rm ani} =- 2 \sum_{\nu, \alpha,\beta} X^{\alpha \beta}_{\bm{Q}_\nu}  S^\alpha_{\bm{Q}_\nu} S^\beta_{-\bm{Q}_\nu}, 
\end{align}
where $X^{\alpha \beta}_{\bm{Q}_\nu}$ consists of the antisymmetric interaction $\bm{D}_{\bm{Q}_\nu}$ and symmetric ones $\bm{E}_{\bm{Q}_\nu}$ and $\bm{F}_{\bm{Q}_\nu}$, as shown in Sec.~\ref{sec: Effective spin model in momentum space}. 
The real-space counterparts of $\bm{D}_{\bm{Q}_\nu}$, $E^x_{\bm{Q}_\nu}$, and $F^x_{\bm{Q}_\nu}$ are given by $\bm{D}_{ij} \cdot (\bm{S}_i \times \bm{S}_j)$, $E^x_{ij}(S^y_i S^z_j+ S^z_i S^y_j)$, and $F^x_{ij}S^x_i S^x_j$, respectively, which originates from the spin-orbit coupling. 
The role of such real-space interactions on the multiple-$Q$ instability has been also studied especially for the  DM interaction~\cite{Binz_PhysRevLett.96.207202, Binz_PhysRevB.74.214408, Park_PhysRevB.83.184406} and bond-dependent interactions in the form of compass and Kitaev type~\cite{Michael_PhysRevB.91.155135,Lee_PhysRevB.91.064407,Lukas_PhysRevLett.117.277202,Rousochatzakis2016,yao2016topological,Chern_PhysRevB.95.144427,Maksimov_PhysRevX.9.021017,amoroso2020spontaneous}. 
In addition, the dipolar interaction, which becomes the origin of the SkX~\cite{lin1973bubble,malozemoff1979magnetic, Garel_PhysRevB.26.325,takao1983study, Ezawa_PhysRevLett.105.197202, Utesov_PhysRevB.103.064414, Utesov_PhysRevB.105.054435}, is the real-space counterpart of $\bm{E}_{\bm{Q}_\nu}$ and $\bm{F}_{\bm{Q}_\nu}$. 

All the anisotropic spin interactions in momentum space, $\bm{D}_{\bm{Q}_\nu}$, $\bm{E}_{\bm{Q}_\nu}$, and $\bm{F}_{\bm{Q}_\nu}$, lead to the multiple-$Q$ instability. 
Meanwhile, their effects are different from each other. 
We show such a difference by taking the two-dimensional triangular-lattice system with the interactions at $\bm{Q}_1=(Q, 0)$ and its symmetry-related wave vectors $\bm{Q}_2$ and $\bm{Q}_3$ connected by the threefold rotation~\cite{Yambe_PhysRevB.106.174437}. 
In the case of $\bm{D}_{\bm{Q}_\eta}$, the model $\mathcal{H}^{\rm (2)}_{\rm iso}+\mathcal{H}^{\rm (2)}_{\rm ani}$ exhibits the single-$Q$ spiral state rather than the multiple-$Q$ state, where the spiral plane is fixed in the plane perpendicular to $\bm{D}_{\bm{Q}_\nu}$. 
In this case, the multiple-$Q$ instability occurs by applying the external magnetic field, as discussed in Sec.~\ref{sec: Microscopic origin of single-$Q$ spiral state}; the Bloch-type (N\'{e}el-type) T-SkX is realized for $D^x_{\bm{Q}_1} \neq 0$ ($D^y_{\bm{Q}_1} \neq 0$) under the out-of-plane magnetic field as a consequence of a superposition of three spiral waves at $\bm{Q}_1$, $\bm{Q}_2$, and $\bm{Q}_3$, whose spin configuration is schematically shown in Fig.~\ref{fig: crystal}(a) [Fig.~\ref{fig: crystal}(b)].  
Meanwhile, the coplanar triple-$Q$ state consisting of three in-plane cycloidal spiral waves at $\bm{Q}_1$, $\bm{Q}_2$, and $\bm{Q}_3$ in Fig.~\ref{fig: crystal}(c) can be realized for $D^z_{\bm{Q}_1} \neq 0$ by combining the biquadratic interaction and other anisotropic interactions. 
 
When considering $\bm{E}_{\bm{Q}_1} \neq \bm{0}$ instead of $\bm{D}_{\bm{Q}_1}$, the spiral plane by the isotropic interaction is elliptically modulated, since $\bm{E}_{\bm{Q}_1}$ tends to favor the collinear sinusoidal modulation perpendicular to $\bm{E}_{\bm{Q}_1}$; $E^x_{\bm{Q}_1}$, $E^y_{\bm{Q}_1}$, and $E^z_{\bm{Q}_1}$ induce the sinusoidal modulation along the [011], [101], and [110] directions in spin space, respectively, as shown in the upper panel of Figs.~\ref{fig: crystal}(d)-\ref{fig: crystal}(f). 
Accordingly, different types of triple-$Q$ states are realized under $\bm{E}_{\bm{Q}_1}$;  $E^x_{\bm{Q}_1}$ and $E^y_{\bm{Q}_1}$ tend to favor the $n_{\rm sk}=2$ SkXs, as shown in the lower panel of Figs.~\ref{fig: crystal}(d) and \ref{fig: crystal}(e), whereas $E^z_{\bm{Q}_z}$ tends to favor the coplanar triple-$Q$ vortex crystal, as shown in the lower panel of Fig.~\ref{fig: crystal}(f). 

Similarly, $\bm{F}_{\bm{Q}_1}$ also favors the sinusoidal modulation, where $F^x_{\bm{Q}_1}$, $F^y_{\bm{Q}_1}$, and $F^z_{\bm{Q}_1}$ induce the sinusoidal modulation along the [100], [010], and [001] directions in spin space, respectively, as shown in the upper panel of Figs.~\ref{fig: crystal}(g)-\ref{fig: crystal}(i). 
The triple-$Q$ superpositions of such sinusoidal waves lead to the coplanar triple-$Q$ vortex crystal in the cases of $F^x_{\bm{Q}_1}$ [Fig.~\ref{fig: crystal}(g)] and $F^y_{\bm{Q}_1}$ [Fig.~\ref{fig: crystal}(h)], while that under $F^z_{\bm{Q}_1}$ leads to the collinear triple-$Q$ state [Fig.~\ref{fig: crystal}(i)], which is referred to as the bubble crystal.

\begin{table}[tb!]
\begin{center}
\caption{
Nonzero components in the anisotropic interactions $\bm{D}_{\bm{Q}}$, $\bm{E}_{\bm{Q}}$, and $\bm{F}_{\bm{Q}}$ under tetragonal, hexagonal, trigonal, and cubic point groups for $\bm{Q}=(Q, 0, 0)$~\cite{Yambe_PhysRevB.106.174437, Yambe_PhysRevB.107.174408}. 
}
\label{tab: PGint}
\begingroup
\renewcommand{\arraystretch}{1.0}
\scalebox{1.0}{
 \begin{tabular}{lcccc}
 \hline \hline
Point group & $\bm{D}_{\bm{Q}}$ & $\bm{E}_{\bm{Q}}$ & $\bm{F}_{\bm{Q}}$ \\ \hline
 \hline\hline
$D_{\rm 4h}$ ($4/mmm$) & -- & --  & $F^x_{\bm{Q}}$, $F^y_{\bm{Q}}$, $F^z_{\bm{Q}}$ \\
$D_{\rm 4}$ ($422$) & $D^x_{\bm{Q}}$ & --  & $F^x_{\bm{Q}}$, $F^y_{\bm{Q}}$, $F^z_{\bm{Q}}$ \\
$D_{\rm 2d}$ ($\bar{4}2m$) & $D^x_{\bm{Q}}$ & --  & $F^x_{\bm{Q}}$, $F^y_{\bm{Q}}$, $F^z_{\bm{Q}}$ \\
$D_{\rm 2d}$ ($\bar{4}m2$) & $D^y_{\bm{Q}}$ & --  & $F^x_{\bm{Q}}$, $F^y_{\bm{Q}}$, $F^z_{\bm{Q}}$ \\
$C_{\rm 4v}$ ($4mm$)  & $D^y_{\bm{Q}}$ & --  & $F^x_{\bm{Q}}$, $F^y_{\bm{Q}}$, $F^z_{\bm{Q}}$ \\
$C_{\rm 4h}$ ($4/m$)  & -- & $E^z_{\bm{Q}}$  & $F^x_{\bm{Q}}$, $F^y_{\bm{Q}}$, $F^z_{\bm{Q}}$ \\
$C_{\rm 4}$ ($4$)  & $D^x_{\bm{Q}}$, $D^y_{\bm{Q}}$ & $E^z_{\bm{Q}}$  & $F^x_{\bm{Q}}$, $F^y_{\bm{Q}}$, $F^z_{\bm{Q}}$ \\
$S_{\rm 4}$ ($\bar{4}$)  & $D^x_{\bm{Q}}$, $D^y_{\bm{Q}}$ & $E^z_{\bm{Q}}$  & $F^x_{\bm{Q}}$, $F^y_{\bm{Q}}$, $F^z_{\bm{Q}}$ \\
\hline 
$D_{\rm 6h}$ ($6/mmm$) & -- & --  & $F^x_{\bm{Q}}$, $F^y_{\bm{Q}}$, $F^z_{\bm{Q}}$ \\
$D_{\rm 6}$ ($622$) & $D^x_{\bm{Q}}$ & --  & $F^x_{\bm{Q}}$, $F^y_{\bm{Q}}$, $F^z_{\bm{Q}}$ \\
$D_{\rm 3h}$ ($\bar{6}m2$) & $D^z_{\bm{Q}}$ & --  & $F^x_{\bm{Q}}$, $F^y_{\bm{Q}}$, $F^z_{\bm{Q}}$ \\
$D_{\rm 3h}$ ($\bar{6}2m$) & -- & --  & $F^x_{\bm{Q}}$, $F^y_{\bm{Q}}$, $F^z_{\bm{Q}}$ \\
$C_{\rm 6v}$ ($6mm$) & $D^y_{\bm{Q}}$ & --  & $F^x_{\bm{Q}}$, $F^y_{\bm{Q}}$, $F^z_{\bm{Q}}$ \\
$C_{\rm 6h}$ ($6/m$) & -- & $E^z_{\bm{Q}}$  & $F^x_{\bm{Q}}$, $F^y_{\bm{Q}}$, $F^z_{\bm{Q}}$ \\
$C_{\rm 3h}$ ($\bar{6}$) & $D^z_{\bm{Q}}$ & $E^z_{\bm{Q}}$  & $F^x_{\bm{Q}}$, $F^y_{\bm{Q}}$, $F^z_{\bm{Q}}$ \\
$C_{\rm 6}$ ($6$) & $D^x_{\bm{Q}}$, $D^y_{\bm{Q}}$ & $E^z_{\bm{Q}}$  & $F^x_{\bm{Q}}$, $F^y_{\bm{Q}}$, $F^z_{\bm{Q}}$ \\
\hline
$D_{\rm 3d}$ ($\bar{3}m1$) & -- & $E^x_{\bm{Q}}$  & $F^x_{\bm{Q}}$, $F^y_{\bm{Q}}$, $F^z_{\bm{Q}}$ \\
$D_{\rm 3d}$ ($\bar{3}1m$) & -- & $E^y_{\bm{Q}}$  & $F^x_{\bm{Q}}$, $F^y_{\bm{Q}}$, $F^z_{\bm{Q}}$ \\
$D_{\rm 3}$ ($321$) & $D^x_{\bm{Q}}$ & $E^x_{\bm{Q}}$  & $F^x_{\bm{Q}}$, $F^y_{\bm{Q}}$, $F^z_{\bm{Q}}$ \\
$D_{\rm 3}$ ($312$) & $D^x_{\bm{Q}}$, $D^z_{\bm{Q}}$ & $E^y_{\bm{Q}}$  & $F^x_{\bm{Q}}$, $F^y_{\bm{Q}}$, $F^z_{\bm{Q}}$ \\
$C_{\rm 3v}$ ($3m1$) & $D^y_{\bm{Q}}$, $D^z_{\bm{Q}}$ & $E^x_{\bm{Q}}$  & $F^x_{\bm{Q}}$, $F^y_{\bm{Q}}$, $F^z_{\bm{Q}}$ \\
$C_{\rm 3v}$ ($31m$) & $D^y_{\bm{Q}}$ & $E^y_{\bm{Q}}$  & $F^x_{\bm{Q}}$, $F^y_{\bm{Q}}$, $F^z_{\bm{Q}}$ \\
$C_{\rm 3i}$ ($\bar{3}$) & -- & $E^x_{\bm{Q}}$, $E^y_{\bm{Q}}$, $E^z_{\bm{Q}}$  & $F^x_{\bm{Q}}$, $F^y_{\bm{Q}}$, $F^z_{\bm{Q}}$ \\
$C_{\rm 3}$ ($3$) & $D^x_{\bm{Q}}$, $D^y_{\bm{Q}}$, $D^z_{\bm{Q}}$ & $E^x_{\bm{Q}}$, $E^y_{\bm{Q}}$, $E^z_{\bm{Q}}$  & $F^x_{\bm{Q}}$, $F^y_{\bm{Q}}$, $F^z_{\bm{Q}}$ \\
\hline
$O_{\rm h}$ ($m\bar{3}m$) & -- & --  & $F^x_{\bm{Q}}$, $F^y_{\bm{Q}}$ \\
$T_{\rm d}$ ($\bar{4}3m$) & -- & --  & $F^x_{\bm{Q}}$, $F^y_{\bm{Q}}$ \\
$O$ ($432$) & $D^x_{\bm{Q}}$ & --  & $F^x_{\bm{Q}}$, $F^y_{\bm{Q}}$ \\
$T_{\rm h}$ ($m\bar{3}$) & -- & --  & $F^x_{\bm{Q}}$, $F^y_{\bm{Q}}$, $F^z_{\bm{Q}}$ \\
$T$ ($23$) & $D^x_{\bm{Q}}$ & --  & $F^x_{\bm{Q}}$, $F^y_{\bm{Q}}$, $F^z_{\bm{Q}}$ \\
 \hline\hline
\end{tabular}
}
\endgroup
\end{center}
\end{table}

The nonzero contributions from these anisotropic interactions depend on the crystal symmetry. 
We summarize the correspondence between the point groups and $(\bm{D}_{\bm{Q}}, \bm{E}_{\bm{Q}}, \bm{F}_{\bm{Q}})$ for $\bm{Q} = (Q,0,0)$ in Table~\ref{tab: PGint}. 
By using the effective spin model incorporating the anisotropic spin interactions in Eq.~(\ref{eq: Ham_ani_q}) and external magnetic field, various multiple-$Q$ instabilities have been revealed, such as the S-SkX under $D_{\rm 4h}$~\cite{Hayami_PhysRevB.103.024439, hayami2023widely}, the S-bubble crystal under $D_{\rm 4h}$~\cite{Hayami_PhysRevB.108.024426}, the hybrid S-SkX under $C_{\rm 4v}$~\cite{Hayami_PhysRevLett.121.137202} and $C_{\rm 4h}$~\cite{Hayami_PhysRevB.105.104428}, the T-SkX under $D_{\rm 6h}$~\cite{Hayami_PhysRevB.103.054422}, the T-MAX under $C_{\rm 6v}$~\cite{Hayami_PhysRevB.104.094425}, the $n_{\rm sk}=2$ SkX under $D_{\rm 3d}$~\cite{yambe2021skyrmion}, the 6$Q$ SkX under $O$~\cite{Hayami_PhysRevB.107.174435}, the hedgehog crystal under $T$~\cite{Kato_PhysRevB.104.224405} and $T_{\rm h}$~\cite{Yambe_PhysRevB.107.174408}, and the distorted T-SkX under $D_{\rm 2h}$~\cite{Hayami_doi:10.7566/JPSJ.91.093701}.
It also accounts for the important ingredients to reproduce the experimental phase diagrams in SkX-hosting materials, such as Gd$_3$Ru$_4$Al$_{12}$~\cite{Hirschberger_10.1088/1367-2630/abdef9}, GdRu$_2$Si$_2$~\cite{khanh2022zoology}, EuPtSi~\cite{hayami2021field}, and EuNiGe$_3$~\cite{singh2023transition}. 
We discuss the effect of the anisotropic interactions on noncentrosymmetric magnets in Sec.~\ref{sec: Effect of symmetric anisotropic interaction} and centrosymmetric magnets in Sec.~\ref{sec: Effect of symmetric anisotropic interaction_2}. 

In addition, it is noted that the anisotropic interactions also depend on the wave-vector symmetry, as shown in the case of the point group $C_{\rm 4v}$ in Figs.~\ref{fig: int_matrix}(a) and \ref{fig: int_matrix}(c). 
Thus, the different types of multiple-$Q$ instabilities can be expected even under the same crystal symmetry when the symmetry of the ordering wave vectors is different~\cite{Hayami_PhysRevB.109.054422}. 
We show such an example in the $C_{\rm 4v}$ system in Sec.~\ref{sec: Effect of low-symmetric ordering wave vectors}, where the DM interaction at low-symmetric ordering wave vectors induces the hybrid S-SkX rather than the N\'eel-type S-SkX.

\subsubsection{Sublattice-dependent interaction}
\label{sec: Sublattice-dependent interaction}

\begin{figure}[tp!]
\begin{center}
\includegraphics[width=0.8\hsize]{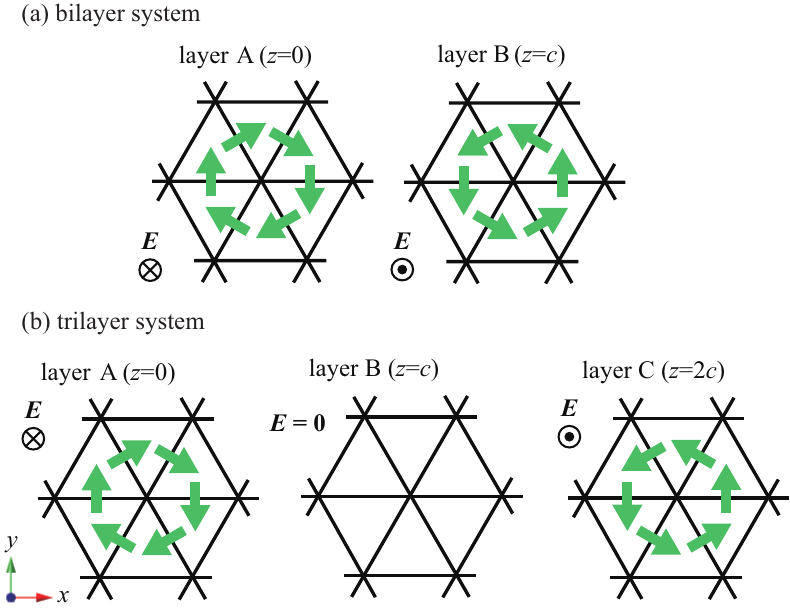} 
\caption{
\label{fig: sublattice}
(a) Bilayer triangular-lattice system consisting of layer A and layer B. 
(b) Trilayer triangular-lattice system consisting of layer A, layer B, and layer C. 
The green arrows denote the DM vectors in each layer, which are induced by the local crystalline electric field $\bm{E}$. 
In (a) and (b), the lattice sites are located at the same $(x, y)$ position for different layers. 
Reprinted figure (a) with permission from~\cite{Hayami_PhysRevB.105.014408}, Copyright (2022) by the American Physical Society. 
Reprinted figure (b) with permission from~\cite{Hayami_PhysRevB.105.184426}, Copyright (2022) by the American Physical Society. 
}
\end{center}
\end{figure}

Sublattice degrees of freedom can also bring about the unconventional multiple-$Q$ states as different stabilization mechanisms. 
One of the situations is the emergence of the SkX under the multi-sublattice structure with the global inversion symmetry but without the local inversion symmetry at each lattice site. 
In this case, the sublattice-dependent DM interaction occurs even in the centrosymmetric lattice structure, as discussed in Sec.~\ref{sec: Effective spin model in momentum space}, which can be the origin of the SkXs~\cite{Hayami_PhysRevB.105.014408, lin2024skyrmion}. 
The bilayer triangular-lattice structure shown in Fig.~\ref{fig: sublattice}(a) is a typical system to have the staggered-type DM interaction owing to the absence of the local inversion center at each layer site; the opposite local crystalline electric field along the $z$ direction in each layer leads to the opposite sign of the DM vector [see also the interaction matrix in Fig.~\ref{fig: int_matrix}(d)]. 
In such a situation, the staggered DM interaction plays a role in inducing the T-SkX as well as the uniform one in noncentrosymmetric systems under the external magnetic field, as discussed in Sec.~\ref{sec: External magnetic field}. 
We discuss the results in detail in Sec.~\ref{sec: Effect of sublattice-dependent interaction}. 
Similarly, the S-SkX is also stabilized by the staggered DM interaction in the centrosymmetric bilayer square-lattice system~\cite{hayami2022square}. 
These sublattice-dependent DM interactions also induce exotic SkXs, such as the SkX with layer-dependent skyrmion numbers in the trilayer system, which consists of two layers with the staggered DM interaction and one layer without the DM interaction, as shown in Fig.~\ref{fig: sublattice}(b)~\cite{Hayami_PhysRevB.105.184426}; see Sec.~\ref{sec: Effect of sublattice-dependent interaction} for details. 
In a similar context, the instability toward the SkX has been investigated in the trilayer system with the threefold screw symmetry but without the threefold rotational one~\cite{Hayami_PhysRevB.105.224411}. 

Another intriguing situation is the emergence of the AF SkX in the multi-sublattice systems. 
The most typical system to induce the AF SkX is bipartite systems with two sublattices A and B, as exemplified by antiferromagnetic bilayer and honeycomb systems, where the skyrmion number exhibits the opposite sign between two sublattices and it is canceled out in the whole system~\cite{Bogdanov_PhysRevB.66.214410, buhl2017topological, Gobel_PhysRevB.96.060406, Akosa_PhysRevLett.121.097204}. 
Such an AF SkX has been clarified in the presence of an external staggered magnetic field~\cite{Gobel_PhysRevB.96.060406} and the absence of the magnetic field~\cite{Yambe_PhysRevB.107.014417} in the two-sublattice honeycomb structure without relying on the DM interaction. 
We discuss the result in the latter situation in Sec.~\ref{sec: Antiferro skyrmion crystal}. 
Furthermore, the AF SkX has been engineered by applying the in-plane uniform magnetic field in the bilayer system with the staggered DM interaction~\cite{Hayami_doi:10.7566/JPSJ.92.084702}. 
It is noted that some of the AF SkX accompany a uniform scalar spin chirality without a perfect cancellation~\cite{Rosales_PhysRevB.92.214439, Diaz_hysRevLett.122.187203, Osorio_PhysRevB.96.024404, liu2020theoretical, Tome_PhysRevB.103.L020403, mukherjee2021antiferromagnetic, Mukherjee_PhysRevB.103.134424, Hayami_PhysRevB.109.014415}; in this case, the AF SkX exhibits the topological Hall effect as in the conventional SkX. 
The latter AF SkX has been observed in MnSc$_2$S$_4$~\cite{Gao2016Spiral, gao2020fractional, Rosales_PhysRevB.105.224402, takeda2024magnon}.

\subsubsection{High-harmonic wave-vector interaction}
\label{sec: High-harmonic wave-vector interaction}

\begin{figure}[tp!]
\begin{center}
\includegraphics[width=1.0\hsize]{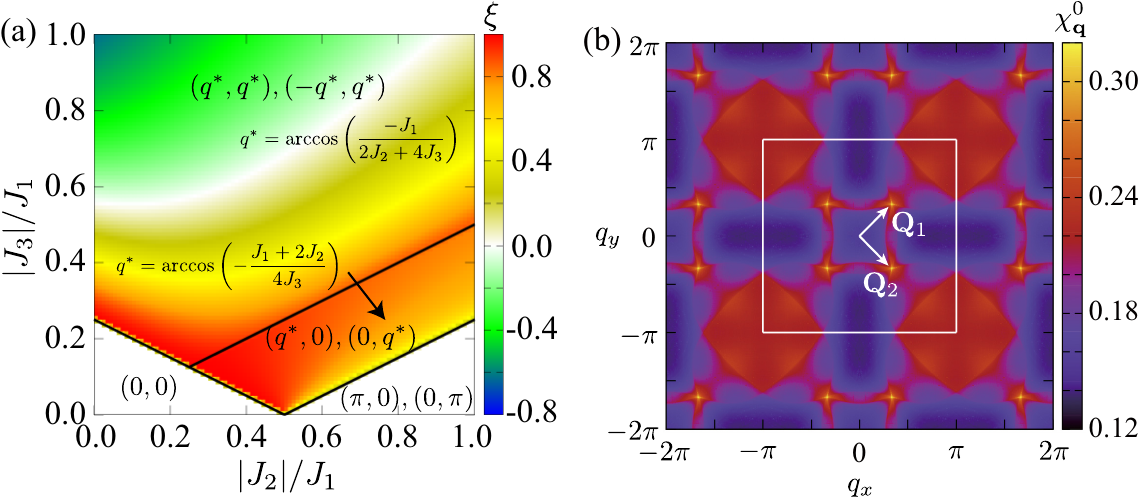} 
\caption{
\label{fig: HHI}
(a) $|J_2|/J_1$ and $|J_3|/J_1$ dependence of $\xi=J_{\bm{Q}_1+\bm{Q}_2}/J_{\bm{Q}_1}$ [$\xi = J_{(\bm{Q}'_1+\bm{Q}'_2)/2}/J_{\bm{Q}'_1} $] in the square-lattice Heisenberg model when the ordering vectors are represented by $\bm{Q}_1=(q^*, q^*)$ and $\bm{Q}_2=(-q^*, q^*)$ [$\bm{Q}'_1=(q^*, 0)$ and $\bm{Q}'_2=(0,q^*)$]. 
(b) $q_x$ and $q_y$ dependence of the bare susceptibility $\chi^0_{\bm{q}}$ in the square-lattice Kondo lattice model with $t_1=1$, $t_3=-0.5$, and $\mu=-3.5$. 
$\bm{Q}_1$ and $\bm{Q}_2$ stand for the ordering wave vectors. 
The white square represents the first Brillouin zone. 
Reprinted figure (a) with permission from~\cite{Hayami_PhysRevB.105.174437}, Copyright (2022) by the American Physical Society.
Reprinted figure (b) with permission from~\cite{Hayami_PhysRevB.95.224424}, Copyright (2017) by the American Physical Society.
}
\end{center}
\end{figure}

In the effective spin model in Eq.~(\ref{eq: Ham_iso_q}) with the biquadratic interaction in Eq.~(\ref{eq: Ham4th}) and/or the bilinear anisotropic interaction in Eq.~(\ref{eq: Ham_ani_q}), we usually consider the contributions from the dominant interactions at a set of symmetry-related wave vectors in order to investigate the multiple-$Q$ instability. 
On the other hand, interactions at high-harmonic wave vectors can give additional energy gains to the multiple-$Q$ states.
For example, the S-SkX consisting of a double-$Q$ superposition of spiral waves at $\bm{Q}_1$ and $\bm{Q}_2$ can naturally lead to the non-negligible energy contribution at high-harmonic wave vectors, such as $\bm{Q}_1+ \bm{Q}_2$, $2\bm{Q}_1$, and so on. 
When the interactions at such high-harmonic wave vectors give negative energy contributions, they tend to favor the multiple-$Q$ states compared to the single-$Q$ state. 

The interactions at high-harmonic wave vectors can be comparable to those at the symmetry-related wave vectors in localized spin systems with frustrated exchange interactions. 
Figure~\ref{fig: HHI}(a) shows the behavior of $\xi=J_{\bm{Q}_1+\bm{Q}_2}/J_{\bm{Q}_1}$ [$\xi = J_{(\bm{Q}'_1+\bm{Q}'_2)/2}/J_{\bm{Q}'_1} $] in the plane of the $|J_2|/J_1$ and $|J_3|/J_1$ in the square-lattice Heisenberg model, where the ordering vectors are given by $\bm{Q}_1=(q^*, q^*)$ and $\bm{Q}_2=(-q^*, q^*)$ [$\bm{Q}'_1=(q^*, 0)$ and $\bm{Q}'_2=(0,q^*)$]; $J_1$, $J_2$, and $J_3$ are the nearest-neighbor, second-nearest-neighbor, and third-nearest-neighbor exchange interactions on the square lattice, and $J_{\bm{Q}_\nu}$ is obtained by their Fourier transformation~\cite{Wang_PhysRevB.103.104408, Hayami_PhysRevB.105.174437}. 
In the region close to the phase boundary with $|J_3|/J_1=0.5|J_2|/J_1$, large $\xi=J_{\bm{Q}_1+\bm{Q}_2}/J_{\bm{Q}_1} \sim 0.8$ means that the high-harmonic wave-vector interaction $J_{\bm{Q}_1+\bm{Q}_2}$ is comparable to the interaction $J_{\bm{Q}_1}$. 
Thus, there is a chance of stabilizing the double-$Q$ state in this region; the emergence of the double-$Q$ S-SkX under this mechanism has been demonstrated in the model calculations~\cite{Wang_PhysRevB.103.104408, Hayami_PhysRevB.105.174437}. 

The contribution of interaction at high-harmonic wave vectors also appears in itinerant electron systems. 
Figure~\ref{fig: HHI}(b) shows the bare susceptibility of the itinerant electrons $\chi^0
_{\bm{q}}$ in the Kondo lattice model with the nearest-neighbor hopping $t_1=1$, the third-nearest-neighbor hopping $t_3=-0.5$, and the chemical potential $\mu=-3.5$~\cite{Hayami_PhysRevB.95.224424}. 
Since the RKKY interaction in the weak-coupling regime in the Kondo lattice model is proportional to $\chi^0_{\bm{q}}$,  the maximum values of $\chi^0_{\bm{q}}$ give the ordering wave vectors; $\bm{Q}_1=(\pi/3,\pi/3)$ and $\bm{Q}_2=(\pi/3,-\pi/3)$ correspond to the ordering wave vectors in the above model parameters. 
Also in this case, the nonzero high-harmonic wave-vector interactions are found at $\bm{Q}_1+\bm{Q}_2$, which indicates that their contributions to the internal energy can play an important role in realizing the double-$Q$ states~\cite{Hayami_doi:10.7566/JPSJ.89.103702, hayami2023widely, Hayami_PhysRevB.108.094416, Hayami_PhysRevB.109.184419}. 
Among them, we detail the magnetic phase diagrams on the square lattice by incorporating the effect of the high-harmonic wave-vector interaction in Sec.~\ref{sec: Effect of high-harmonic wave-vector interaction}.

The importance of the high-harmonic wave-vector interactions has been revealed in the centrosymmetric tetragonal SkX-hosting materials, such as EuAl$_4$~\cite{takagi2022square, hayami2023orthorhombic} and GdRu$_2$Ge$_2$~\cite{yoshimochi2024multi}, where multiple SkXs appear when the external magnetic field is varied.  
Furthermore, the high-harmonic wave-vector interactions can be the origin of the $n_{\rm sk}=2$ S-SkX by combining the biquadratic interaction~\cite{hayami2022multiple}; see Sec.~\ref{sec: Effect of high-harmonic wave-vector interaction} for details.

\subsubsection{Other mechanisms}
\label{sec: Other mechanisms}

In addition to the above five cases, various stabilization mechanisms of the multiple-$Q$ states have been elucidated. 
We briefly discuss them one by one below. 

\paragraph{Thermal fluctuations}
Although we mainly focus on the ground states, the finite-temperature effect brings about the multiple-$Q$ instability~\cite{Okubo_PhysRevB.84.144432, Okubo_PhysRevLett.108.017206, Rosales_PhysRevB.87.104402}. 
A typical example is the SkX by thermal fluctuations in the triangular-lattice spin model with the frustrated isotropic exchange interaction, where the SkX only appears at finite temperatures in the intermediate-field region~\cite{Okubo_PhysRevLett.108.017206}. 
Moreover, thermal fluctuations induce phase transitions between multiple-$Q$ states, such as the transition from the T-SkX to the TVX~\cite{hayami2021phase}, the transition from the T-SkX to the T-bubble crystal~\cite{Hayami_10.1088/1367-2630/ac3683, Utesov_PhysRevB.105.054435}, and the transition from the S-SkX to the S-bubble crystal~\cite{Hayami_PhysRevB.108.024426}; we discuss the result for the last case in Sec.~\ref{sec: Bubble crystal}.  
The numerical method to obtain the thermodynamically stable states in the effective spin model in Eq.~(\ref{eq: Ham}) has been developed~\cite{Kato_PhysRevB.105.174413}, which has clarified the stability of the SkXs and hedgehog crystals and their phase transitions at finite temperatures~\cite{Kato_PhysRevB.107.094437, hayami2023widely}.

\paragraph{Quantum fluctuations}

Fluctuations in quantum magnets also lead to multiple-$Q$ instability, such as the vortex crystal~\cite{Kamiya_PhysRevX.4.011023, Wang_PhysRevLett.115.107201, Marmorini2014, Ueda_PhysRevA.93.021606}. 
In a similar context, a quantum SkX as a stable many-magnon bound state, which is regarded as the quantum analog of the SkXs, has been studied in quantum spin systems with $S=1/2$~\cite{Lohani_PhysRevX.9.041063, Sotnikov_PhysRevB.103.L060404, Siegl_PhysRevResearch.4.023111, Haller_PhysRevResearch.4.043113}.

\paragraph{Nonmagnetic impurity}
Nonmagnetic impurity in magnets can lead to multiple-$Q$ states since they tend to reorient the surrounding spins into a noncollinear way by an effective positive biquadratic interaction $(\bm{S}_i \cdot \bm{S}_j)^2$ around the impurity~\cite{Wollny_PhysRevLett.107.137204, Sen_PhysRevB.86.205134, Maryasin_PhysRevLett.111.247201,maryasin2015collective}. 
Especially, the introduction of a single nonmagnetic impurity into the frustrated Heisenberg model with competing exchange interactions leads to the vortex-type spin texture over a finite range of the magnetic field above the bulk saturation field~\cite{Lin_PhysRevLett.116.187202}.  
Furthermore, a periodic array of nonmagnetic impurities gives rise to a rich multiple-$Q$ phase diagram in the whole magnetic-field range depending on the periodicity of the impurity superlattice~\cite{Hayami_PhysRevB.94.174420}.

\paragraph{Single-ion anisotropy}

Single-ion anisotropy in the form of $-A (S^z_i)^2$ stabilizes the T-SkX as the ground state on the centrosymmetric triangular lattices when the out-of-plane (in-plane) magnetic field is applied for $A>0$ ($A<0$)~~\cite{leonov2015multiply, Lin_PhysRevB.93.064430, Hayami_PhysRevB.93.184413}.
A similar mechanism works on the different lattice structures, such as the square lattice~\cite{Lin_PhysRevB.93.064430}, bilayer triangular lattice~\cite{Okigami_doi:10.7566/JPSJ.91.103701}, and face-centered-cubic lattice~\cite{Lin_PhysRevLett.120.077202}.
It is noted that the S-SkX on the square lattice, which was observed in GdRu$_2$Si$_{2}$~\cite{khanh2022zoology}, EuAl$_4$~\cite{takagi2022square, hayami2023orthorhombic}, and GdRu$_2$Ge$_2$~\cite{yoshimochi2024multi}, is not stabilized by merely the single-ion anisotropy~\cite{Lin_PhysRevB.93.064430}.
When the single-ion anisotropy is characterized by the strong easy-axis one, the bubble crystal is stabilized instead of the SkX~\cite{seo2021spin}. 
The effect of other functional single-ion anisotropies, such as $S_i^z S_i^x [(S_i^x)^2-3(S_i^y)^2]$ and $(S_i^x)^4+(S_i^y)^4-6 (S_i^x)^2(S_i^y)^2$, has also been investigated in hexagonal/trigonal and tetragonal systems, where the former leads to the stabilization of the T-SkX~\cite{hayami2022skyrmion} and the latter leads to that of only the topologically trivial double-$Q$ states~\cite{hayami2023multiple}.

\paragraph{Circularly polarized magnetic field}
The time-dependent magnetic field with the circular polarization, $\bm{H}(t) \propto (\cos \omega t, \sin \omega t, 0)$, also leads to the SkX, since it plays a similar role to the static out-of-plane magnetic field~\cite{mochizuki2018highly}. 
The stability of the SkX has been investigated in the localized spin model with the DM interaction~\cite{Miyake_PhysRevB.101.094419} and the itinerant electron model without the DM interaction~\cite{Eto_PhysRevB.104.104425}

\paragraph{Circularly polarized electric field}
Instead of the external magnetic field, the circularly polarized electric field, $\bm{E}(t) \propto (\cos \omega t, \sin \omega t, 0)$, can be a source of inducing the multiple-$Q$ states by considering an effective coupling between the electric field and spin~\cite{Katsura_PhysRevLett.95.057205, tokura2014multiferroics, matsumoto2017symmetry}. 
Through the analyses based on the symmetry and Floquet formalism, it was clarified that the injection of the circularly polarized electric field can induce an effective three-spin interaction proportional to $|\bm{E}(t)|^2/\omega$ that is directly coupled to the scalar spin chirality $\bm{S}_i \cdot (\bm{S}_j \times \bm{S}_k)$~\cite{Yambe_PhysRevB.108.064420, Yambe_PhysRevB.109.064428}. 
Since the single-$Q$ spiral state does not possess the scalar spin chirality while the SkX does, the injection of the circularly polarized electric field into the single-$Q$ spiral state leads to the instability toward the SkX~\cite{yambe2024dynamical}.

\section{Stabilization mechanisms in noncentrosymmetric magnets}
\label{sec: Stabilization mechanisms in noncentrosymmetric magnets}

In this section, we present the instability toward the SkXs and other multiple-$Q$ states in noncentrosymmetric magnets. 
First, we show the stability of the SkXs by the DM interaction in Sec.~\ref{sec: Skyrmion crystal by the Dzyaloshinskii-Moriya interaction}. 
We also discuss the effects of the biquadratic interaction, symmetric anisotropic interaction, and low-symmetric ordering wave vectors on the SkXs. 
Then, we show the emergence of the hedgehog crystal and MAX in Secs.~\ref{sec: Hedgehog crystal} and \ref{sec: Meron-antimeron crystal}, respectively, in the effective spin model with the DM interaction. 

\subsection{Skyrmion crystal by the Dzyaloshinskii-Moriya interaction}
\label{sec: Skyrmion crystal by the Dzyaloshinskii-Moriya interaction}

\begin{figure}[tp!]
\begin{center}
\includegraphics[width=0.75\hsize]{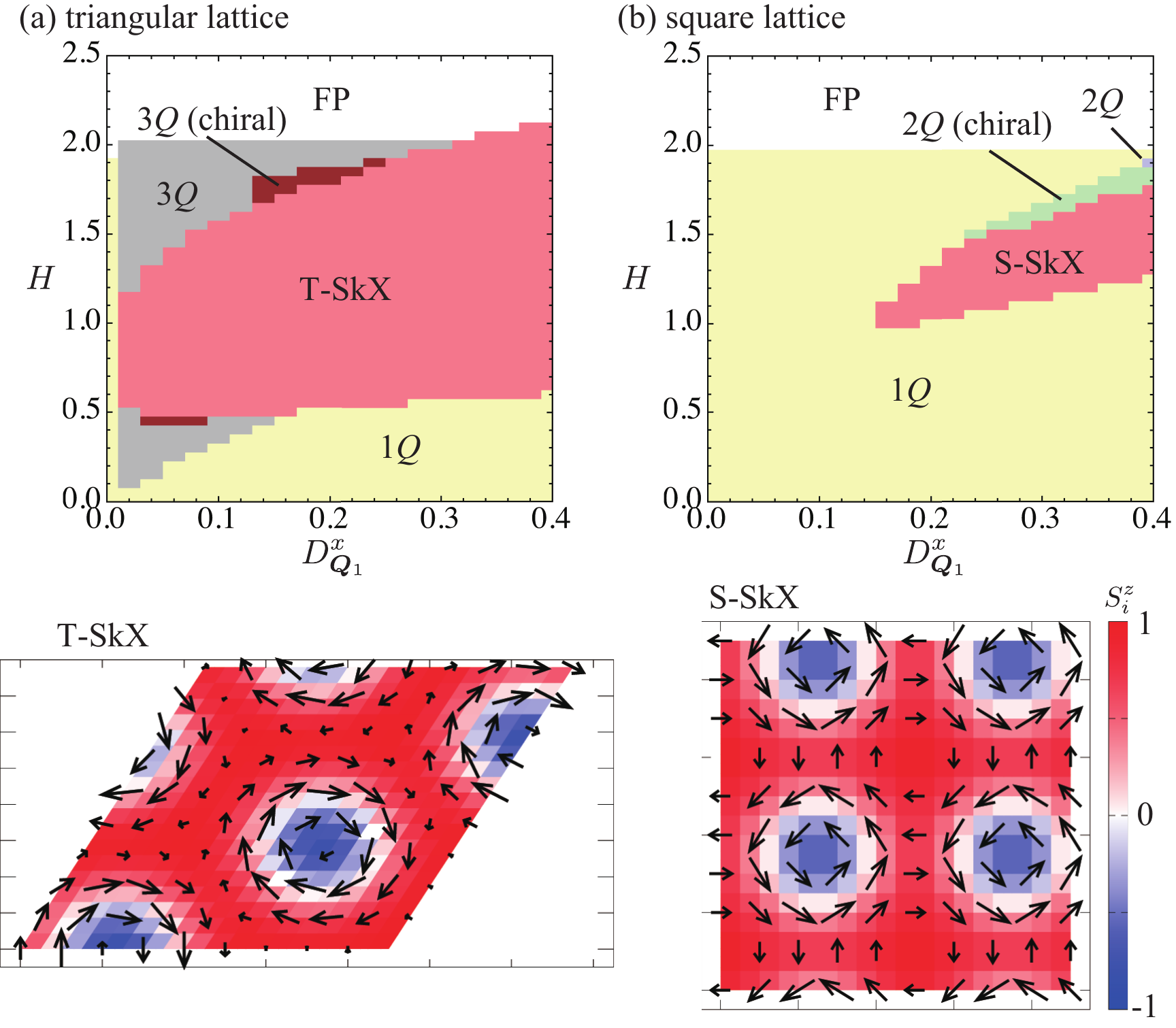} 
\caption{
\label{fig: PD_DM}
Top panels represent the ground-state phase diagrams on (a) the triangular lattice and (b) the square lattice while varying the Dzyaloshinskii-Moriya interaction $D^x_{\bm{Q}_1}$ and the out-of-plane magnetic field $H$. 
1$Q$, 2$Q$, 3$Q$, and FP represent the single-$Q$, double-$Q$, triple-$Q$, and fully polarized state; ``(chiral)" denotes the magnetic states with the scalar chirality $\chi^{\rm sc} \geq 0.01$ but without the integer skyrmion number. 
The bottom panels show the real-space spin configurations of the SkXs. 
The arrow and color represent the in-plane and out-of-plane spin components, respectively. 
}
\end{center}
\end{figure}

The emergence of the SkXs in noncentrosymmetric magnets has been studied in various literature, where Lifshitz invariants that correspond to the free energy contribution by the DM interaction play a significant role~\cite{dzyaloshinskii1964theory, kataoka1981helical, Bogdanov89, Bogdanov94, rossler2006spontaneous}. 
The SkX appears by applying the external magnetic field when the zero-field state corresponds to the single-$Q$ spiral state. 
Such emergence of the SkX and its phase transition from the single-$Q$ spiral state in the magnetic field can be captured by the effective spin model with the momentum-resolved DM interaction. 
A minimum spin model is given by 
\begin{align}
\mathcal{H}= - 2 \sum_{\nu} \left[ J \bm{S}_{\bm{Q}_\nu} \cdot \bm{S}_{-\bm{Q}_\nu} +  \bm{D}_\nu \cdot (\bm{S}_{\bm{Q}_\nu} \times \bm{S}_{-\bm{Q}_\nu})\right]
- H \sum_{i} S_i^z, 
\end{align}
where $\bm{D}_\nu=(D^x_{\bm{Q}_\nu}, D^y_{\bm{Q}_\nu}, D^z_{\bm{Q}_\nu})$. 
In the following calculations, we take $\bm{Q}_1=(Q, 0)$, $\bm{Q}_2=(-Q/2, \sqrt{3}Q/2)$, and $\bm{Q}_3=(-Q/2, -\sqrt{3}Q/2)$ on the two-dimensional triangular lattice and $\bm{Q}_1=(Q, 0)$ and $\bm{Q}_2=(0, Q)$ on the two-dimensional square lattice; $Q=2\pi/5$. 
We set $J=1$ as the energy unit of the model and the lattice constant as unity. 

Figure~\ref{fig: PD_DM}(a) shows the ground-state phase diagram on the triangular lattice in the plane of $D^x_{\bm{Q}_1}$ and $H$, where we set $D^y_{\bm{Q}_1}=D^z_{\bm{Q}_1}=0$; $D^x_{\bm{Q}_1} \neq 0$ but $D^y_{\bm{Q}_1}=D^z_{\bm{Q}_1}=0$ means that the symmetry of the system corresponds to either chiral point group $D_6$ or $D_3$, as shown in Table~\ref{tab: PGint}. 
The phase diagram is obtained by performing the simulated annealing combined with the standard Metropolis local updates and heat bath method following the manner in Ref.~\cite{Hayami_PhysRevB.95.224424}. 
The system size is taken as $N=10^2$ and the periodic boundary condition is adopted. 
We also show the ground-state phase diagram on the square lattice with the chiral point group $D_4$ in Fig.~\ref{fig: PD_DM}(b). 
In the phase diagram, we distinguish the obtained phases by the single-$Q$ (1$Q$) state, double-$Q$ (2$Q$) state, triple-$Q$ (3$Q$) state, SkX, and fully polarized (FP) state, where only the SkX exhibits an integer skyrmion number in the magnetic unit cell. 
We also classify the double-$Q$ and triple-$Q$ states according to the net scalar spin chirality; when the magnetic states accompany nonzero net scalar spin chirality with $\chi^{\rm sc} \geq 0.01$, we denote them as $2Q$(chiral) and $3Q$(chiral).  
Here, the scalar spin chirality for the triangular lattice is given by 
\begin{align}
\label{eq: chiral_TL}
\chi^{\rm sc}=\frac{1}{N}\sum_{\bm{R},\mu}\chi_{\bm{R}}, 
\end{align}
where $\bm{R}$ represents the position vectors at the centers of upward and downward triangles; $\chi_{\bm{R}}= \bm{S}_j \cdot (\bm{S}_k \times \bm{S}_l)$ is the local spin chirality at $\bm{R}$, where $j,k,l$ are the sites on the triangle at $\bm{R}$ in the counterclockwise order. 
Similarly, the scalar spin chirality for the square lattice is given by 
\begin{align}
\chi^{\rm sc}= \frac{1}{N}
\sum_{i,\delta=\pm1}\bm{S}_i \cdot (\bm{S}_{i+\delta \hat{x}}\times \bm{S}_{i+\delta \hat{y}}), 
\end{align}
where $\hat{x}$ ($\hat{y}$) is the unit vector in the $x$ ($y$) direction. 
The skyrmion number $n_{\rm sk}$ can be calculated from $\chi^{\rm sc}$~\cite{BERG1981412}.

As shown in Fig.~\ref{fig: PD_DM}(a), the T-SkX is stabilized for $D_{\bm{Q}_1}^{x}\gtrsim0.01$ in the intermediate-field region, where the skyrmion number is characterized by one, i.e., $|n_{\rm sk}|=1$. 
The region of the T-SkX becomes wider for larger $D_{\bm{Q}_1}^x$, which indicates that the DM interaction tends to favor the T-SkX. 
The real-space snapshot of the spin configuration in the T-SkX is presented in the bottom panel of Fig.~\ref{fig: PD_DM}(a), where the Bloch-type winding is found owing to the chiral-type DM interaction. 

When the effect of $D^y_{\bm{Q}_1}$ is considered instead of $D^x_{\bm{Q}_1}$ by supposing that the symmetry of the system is either polar point group $C_{\rm 6v}$ or $C_{\rm 3v}$, the N\'eel-type SkX is realized rather than the Bloch-type one, although the overall phase boundaries are unchanged. 
Meanwhile, for $D^z_{\bm{Q}_1}\neq 0$ and $D^x_{\bm{Q}_1}=D^y_{\bm{Q}_1}=0$ in the case of $D_{\rm 3h}$ or $C_{\rm 3h}$, the single-$Q$ state always appear in the phase diagram and the SkX does not appear. 
In this case, the T-bimeron crystal can be stabilized by changing the magnetic field direction from the out-of-plane direction to in-plane one, as discussed in Sec.~\ref{sec: External magnetic field}~\cite{Hayami_PhysRevB.105.224423}.

In the square-lattice case in Fig.~\ref{fig: PD_DM}(b) with $D^x_{\bm{Q}_1} \neq 0$ and $D^y_{\bm{Q}_1} =D^z_{\bm{Q}_1} = 0$ under the chiral point group $D_4$ symmetry, the S-SkX also appears in the presence of $D^x_{\bm{Q}_1}$, although the larger DM interaction with $D_{\bm{Q}_1}^{x}\gtrsim0.15$ is required to stabilize the S-SkX compared to the T-SkX in Fig.~\ref{fig: PD_DM}(a). 
This indicates that the T-SkX on the triangular lattice is more easily stabilized than the S-SkX on the square lattice. 
Owing to the chiral DM interaction, the S-SkX corresponds to the Bloch-type one, whose real-space spin configuration is shown in the bottom panel of Fig.~\ref{fig: PD_DM}(b). 
The skyrmion core at $S_i^z=-1$ is located at the interstitial site~\cite{Hayami_PhysRevResearch.3.043158}. 

Similarly to the triangular-lattice case, the N\'eel-type SkX is stabilized by considering the polar point group $C_{\rm 4v}$ with $D_{\bm{Q}_1}^{y} \neq 0$. 
Meanwhile, there is no situation where $D_{\bm{Q}_1}^{z}$ is symmetry-allowed in the tetragonal system, as shown in Table~\ref{tab: PGint}. 
In addition, another difference appears when the $D_{\rm 2d}$ symmetry is considered; anti-type SkX is stabilized owing to the improper fourfold rotational symmetry, which is not stabilized in the presence of the hexagonal/trigonal DM interaction.  
It is noted that the overall phase boundaries are not altered when considering the DM interaction arising from the $C_{\rm 4v}$ and $D_{\rm 2d}$ symmetry.

The above stability tendency and the nature of the SkXs also depend on other interactions, such as the biquadratic interactions and symmetric anisotropic interactions. 
In the following subsections, we discuss the effect of the biquadratic interaction in Sec.~\ref{sec: Effect of biquadratic interaction}, symmetric anisotropic exchange interaction in Sec.~\ref{sec: Effect of symmetric anisotropic interaction}, and low-symmetric ordering wave vectors in Sec.~\ref{sec: Effect of low-symmetric ordering wave vectors}.

\subsubsection{Effect of biquadratic interaction}
\label{sec: Effect of biquadratic interaction}

\begin{figure}[tp!]
\begin{center}
\includegraphics[width=0.8\hsize]{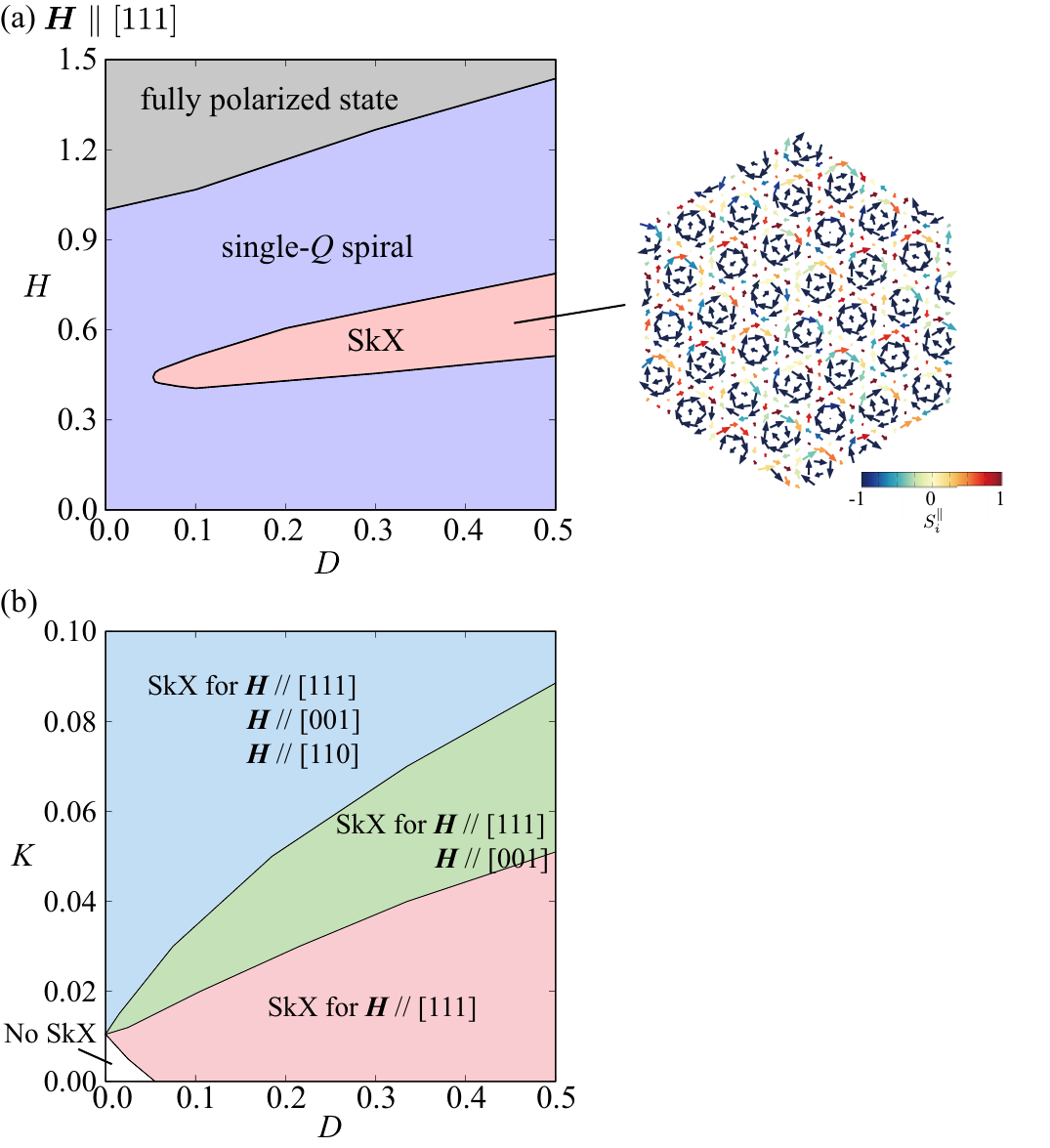} 
\caption{
\label{fig: EuPtSi}
(a) Magnetic phase diagram in the plane of the Dzyaloshinskii-Moriya interaction $D$ and the external magnetic field $H$ along the [111] direction for $K=0$. 
A snapshot of the spin configuration in the SkX viewed from the [111] direction is shown in the right panel. 
The arrows represent the averaged spin moments projected onto the (111) plane and the contour represents the parallel component ($S_i^{\parallel}$). 
(b) Phase diagram representing the instability toward the SkX in an applied magnetic field along the [111], [001], and [110] directions while varying $D$ and the biquadratic interaction $K$. 
``No SkX" means the absence of the SkX phase irrespective of the magnetic field directions.
Reproduced with permission from~\cite{hayami2021field}. Copyright 2021 by the Physical Society of Japan.
}
\end{center}
\end{figure}

We here focus on the interplay between the DM interaction and the biquadratic interaction by considering the following spin Hamiltonian on a simple cubic lattice under the chiral point group $O$ or $T$ as~\cite{hayami2021field} 
\begin{align}
\label{eq:Ham_EuPtSi}
\mathcal{H} = \sum_{\nu=1}^{12}
&\left[-J\bm{S}_{\bm{Q}_\nu}\cdot\bm{S}_{-\bm{Q}_\nu}+\frac{K}{N}\left({\bm S}_{\bm{Q}_\nu}\cdot{\bm S}_{-\bm{Q}_\nu}\right)^2 \right.
\nonumber \\
&  \left.-i{\bm D}_\nu\cdot\left({\bm S}_{\bm{Q}_\nu}\times{\bm S}_{-\bm{Q}_{\nu}}\right)
\right]-\sum_{i}\bm{H}\cdot \bm{S}_i,
\end{align}
where the ordering wave vectors are set as $\bm{Q}_1=(-Q_a,-Q_b,-Q_c)$, $\bm{Q}_2=(-Q_c,-Q_a,-Q_b)$, $\bm{Q}_3=(-Q_b,-Q_c,-Q_a)$, $(\bm{Q}_{4},\bm{Q}_8, \bm{Q}_{12})=R^z(\bm{Q}_1, \bm{Q}_2, \bm{Q}_3)$, $(\bm{Q}_7, \bm{Q}_{11}, \bm{Q}_6)=R^y(\bm{Q}_1, \bm{Q}_2, \bm{Q}_3)$, and $(\bm{Q}_{10}, \bm{Q}_5, \bm{Q}_9)=R^x(\bm{Q}_1, \bm{Q}_2, \bm{Q}_3)$ with $Q_a=2\pi/15$, $Q_b=2\pi/5$, and $Q_c=8\pi/15$, where $R^{\alpha}$ represents $\pi$ rotation around the $\alpha=x,y,z$ axis. 
The DM interaction is set as $\bm{D}_\nu \parallel \bm{Q}_\nu$ and $|\bm{D}_\nu|=D$. 
The effect of the symmetric anisotropic interaction is neglected in the model.  
The isotropic interaction $J=1$ is set as the energy scale. 

Figure~\ref{fig: EuPtSi}(a) shows the ground-state phase diagram while varying the DM interaction $D$ and the magnetic field $H$ along the [111] direction for $K=0$, which is obtained by the simulated annealing~\cite{hayami2021field}.  
Similar to the phase diagrams in Figs.~\ref{fig: PD_DM}(a) and \ref{fig: PD_DM}(b), the SkX consisting of the triple-$Q$ ordering wave vectors appears in the intermediate-field region even without the biquadratic interaction; the real-space spin configuration of the SkX onto the (111) plane is shown in Fig.~\ref{fig: EuPtSi}, where the SkX forms the hexagonal structure. 
Meanwhile, there is no SkX phase when the magnetic field is applied along the [001] and [110] directions. 
Thus, the stability of the SkX is sensitive to the magnetic field directions, which is attributed to a particular set of ordering wave vectors; the SkX tends to be stabilized when triple-$Q$ ordering wave vectors lie on the same plane perpendicular to the magnetic field. 

When the biquadratic interaction $K$ is turned on, the SkXs are induced for $\bm{H}\parallel [001]$ and $\bm{H} \parallel [110]$. 
Figure~\ref{fig: EuPtSi}(b) shows the phase diagram to denote the appearance of the SkXs for $\bm{H} \parallel [111]$, $[001]$, and $[110]$. 
Although the SkX for $\bm{H}\parallel [111]$ appears in almost all the regions including $K=0$ or $D=0$, while that for $\bm{H}\parallel [001]$ and $\bm{H}\parallel [110]$ emerges only for $K \neq 0$. 
The critical values of $K$ for the [001] field are smaller than those for the [110] field, which means that the [110] axis is ``the hard axis" to realize the SkX. 
In the end, the DM interaction tends to favor the SkXs in particular field directions, while the biquadratic interaction stabilizes the SkXs irrespective of field directions. 

This result indicates that the emergence of the SkX induced by the DM interaction is sensitive to the magnetic field direction, which is related to the SkX-hosting compound EuPtSi~\cite{kakihana2018giant,kaneko2019unique,tabata2019magnetic,kakihana2019unique}. 
In this compound, the SkX was observed in the [111] and suggested in the [001] fields, while it was not observed in the [110] field. 
These anisotropic features of the phase diagrams against the field direction are explained by the model parameters in the green region in Fig.~\ref{fig: EuPtSi}(b), which suggests the importance of the interplay between Eu-$4f$ localized spins and Pt-5$d$ itinerant electrons with the strong spin-orbit coupling, since the microscopic origins of $D$ and $K$ are spin-orbit and spin-charge couplings in noncentrosymmetric itinerant magnets, respectively.

\subsubsection{Effect of symmetric anisotropic interaction}
\label{sec: Effect of symmetric anisotropic interaction}

\begin{figure}[tp!]
\begin{center}
\includegraphics[width=0.8\hsize]{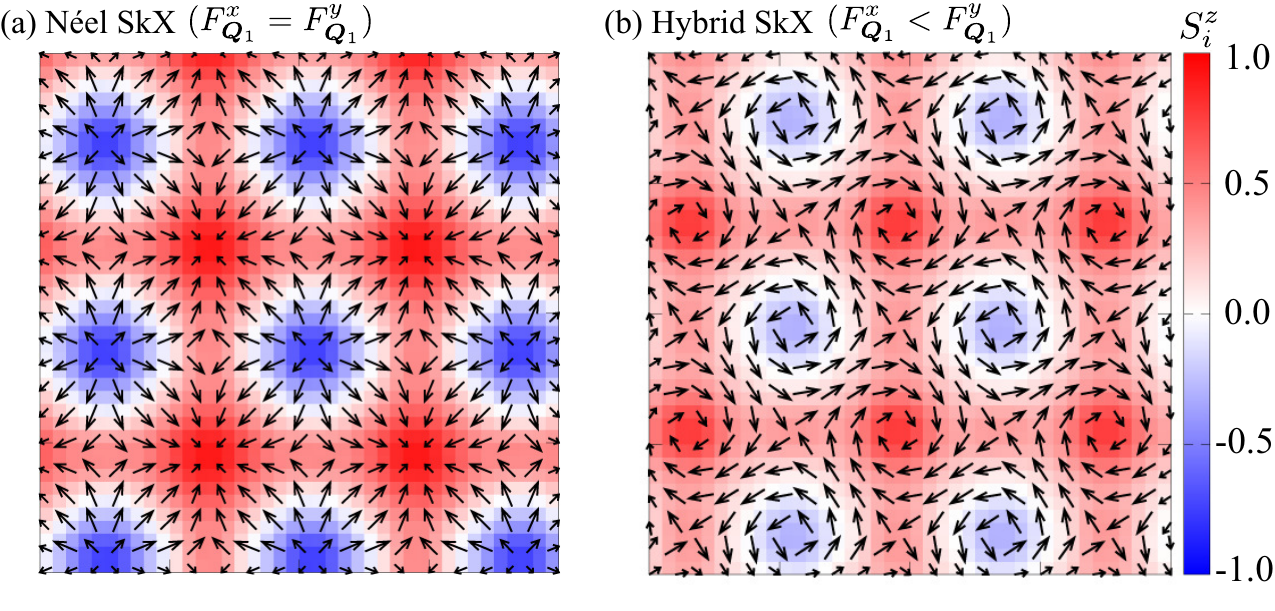} 
\caption{
\label{fig: DM_aniso}
Real-space spin configurations of the square SkXs: (a) N\'eel-type SkX for $F^x_{\bm{Q}_1} = F^y_{\bm{Q}_1}$ and (b) hybrid SkX for $F^x_{\bm{Q}_1} < F^y_{\bm{Q}_1}$. 
The arrows represent the direction of the in-plane spin moments and the color shows its $z$ component. 
Reprinted figure with permission from~\cite{Hayami_PhysRevLett.121.137202}, Copyright (2018) by the American Physical Society.
}
\end{center}
\end{figure}

As shown in Table~\ref{tab: PGint}, the symmetric anisotropic interactions, $\bm{E}_{\bm{Q}_\nu}$ and $\bm{F}_{\bm{Q}_\nu}$, become nonzero in noncentrosymmetric systems. 
We consider the effect of such symmetric anisotropic interactions by considering the square-lattice model under the $C_{\rm 4v}$ symmetry~\cite{Hayami_PhysRevLett.121.137202}. 
The spin Hamiltonian is given by 
\begin{eqnarray}
\label{eq:HamC4v}
\mathcal{H} = -2\sum_{\nu} 
\left[\sum_{\alpha=x,y,z}
F_{\bm{Q}_\nu}^{\alpha} S^{\alpha}_{\bm{Q}_{\nu}} S^{\alpha}_{-\bm{Q}_{\nu}}+i \bm{D}_{\nu} \cdot \left(\bm{S}_{\bm{Q}_{\nu}} \times \bm{S}_{-\bm{Q}_\nu}\right)
\right] - H \sum_i S_i^z, 
\end{eqnarray}
where the anisotropic interactions for $\bm{Q}_1=(\pi/4, 0)$ and $\bm{Q}_2=(0, \pi/4)$ are given by $F_{\bm{Q}_1}^{x}=F_{\bm{Q}_2}^{y}$, $F_{\bm{Q}_1}^{y}=F_{\bm{Q}_2}^{x}$, and $F_{\bm{Q}_1}^{z}=F_{\bm{Q}_2}^{z}$ so as to satisfy the fourfold rotational symmetry. 
For the DM interaction, $D_1^y=-D_{2}^x$ is present owing to the polar symmetry (all other components are zero). 
The biquadratic interaction $K$ is ignored in this section. 

When the $xy$ component of the symmetric anisotropic exchange interaction is isotropic, i.e., $F_{\bm{Q}_1}^{x} = F_{\bm{Q}_1}^{y}$, the N\'eel-type S-SkX is stabilized in the intermediate magnetic field owing to the polar-type DM interaction, as discussed in Sec.~\ref{sec: Skyrmion crystal by the Dzyaloshinskii-Moriya interaction}.  
The spin configuration of the N\'eel-type S-SkX is shown in Fig.~\ref{fig: DM_aniso}(a). 
Meanwhile, the hybrid S-SkX, which is represented by a linear combination of the N\'eel-type S-SkX and Bloch-type S-SkX, can be realized for $F^x_{\bm{Q}_1} < F^y_{\bm{Q}_1}$, as shown in Fig.~\ref{fig: DM_aniso}(b); the Bloch-type winding is dominant around the skyrmion core even in the presence of the polar-type DM interaction. 
This is because $F^y_{\bm{Q}_1}$ tends to favor the sinusoidal modulation along the $y$ direction, whereas $D_{1}^y$ tends to favor the spiral modulation in the $zx$ plane.  
In other words, there is a frustration between the symmetric anisotropic interaction $F^y_{\bm{Q}_1}$ and antisymmetric anisotropic interaction $D_{1}^y$, which gives rise to different types of SkXs from the DM-only system.

\subsubsection{Effect of low-symmetric ordering wave vectors}
\label{sec: Effect of low-symmetric ordering wave vectors}

\begin{figure}[tp!]
\begin{center}
\includegraphics[width=0.8\hsize]{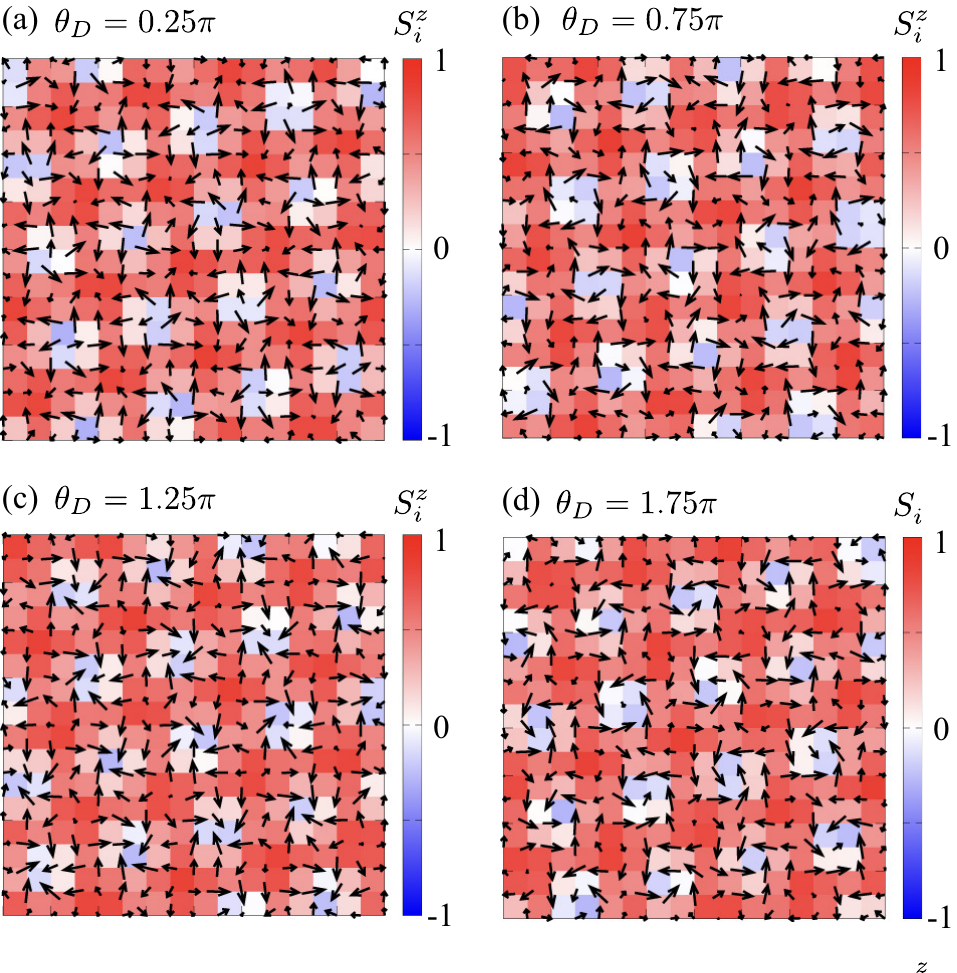} 
\caption{
\label{fig: hybrid_lowsymmetric}
Real-space spin configurations in the S-SkX for $H=1.1$ at (a) $\theta_{D}=0.25 \pi$, (b) $\theta_{D}=0.75 \pi$, (c) $\theta_{D}=1.25 \pi$, and (d) $\theta_{D}=1.75 \pi$. 
The arrows represent the direction of the in-plane spin moments and the color shows its $z$ component. 
Reprinted figure with permission from~\cite{Hayami_PhysRevB.109.054422}, Copyright (2024) by the American Physical Society.
}
\end{center}
\end{figure}

Another way to obtain the unconventional SkX has been clarified by focusing on the situation, where the ordering wave vectors lie at low-symmetric ones in the Brillouin zone, since additional components in the anisotropic interaction are induced compared to high-symmetric ordering wave vectors, as discussed in Sec.~\ref{sec: Effective spin model in momentum space}. 
In such a situation, the hybrid SkX can be obtained even without the symmetric anisotropic interaction, as shown in the previous section. 
Specifically, the spin model on the square lattice under the $C_{\rm 4v}$ symmetry is considered again, which is given by 
\begin{align}
\label{eq: Ham_lowsymmetry}
\mathcal{H}^{\rm eff}=&-2\sum_{\nu}  \left[J \bm{S}_{\bm{Q}_\nu} \cdot  \bm{S}_{-\bm{Q}_\nu} 
+ i \bm{D}_{\nu} \cdot (\bm{S}_{\bm{Q}_\nu} \times  \bm{S}_{-\bm{Q}_\nu})\right]  
-H\sum_i  S^z_i,  
\end{align}
where the ordering wave vectors are located at low-symmetric ones: $\pm \bm{Q}_1=\pm (Q_a, Q_b)$, $\pm \bm{Q}_2=\pm (-Q_b, Q_a)$, $\pm \bm{Q}_3=\pm (Q_a, -Q_b)$, and $\pm \bm{Q}_4=\pm (Q_b, Q_a)$ with $Q_a=13\pi/25$ and $Q_b=3\pi/25$;  $\bm{Q}_1$--$\bm{Q}_4$ are connected by the fourfold rotational and mirror symmetries of the square lattice under the $C_{\rm 4v}$ point group. 
In contrast to the DM interaction for the high-symmetric lines along the [100] and [110] directions, the direction of the DM vector at $\bm{Q}_1$ is not restricted by the symmetry, which is given by $\bm{D}_{\bm{Q}_1}=D(-\cos \theta_D, \sin \theta_D)$. 
In other words, the emergence of the N\'eel-type SkX is not necessarily favored even in polar crystals. 
The symmetric anisotropic interaction is neglected in this model. 

Figure~\ref{fig: hybrid_lowsymmetric} shows the real-space spin configurations in the S-SkX, which consists of double-$Q$ ordering wave vectors at $\bm{Q}_1$ and $\bm{Q}_2$ [or $\bm{Q}_3$ and $\bm{Q}_4$], for $\theta_{D}=0.25 \pi$ [Fig.~\ref{fig: hybrid_lowsymmetric}(a)], $\theta_{D}=0.75 \pi$ [Fig.~\ref{fig: hybrid_lowsymmetric}(b)], $\theta_{D}=1.25 \pi$ [Fig.~\ref{fig: hybrid_lowsymmetric}(c)], and $\theta_{D}=1.75 \pi$ [Fig.~\ref{fig: hybrid_lowsymmetric}(d)]~\cite{Hayami_PhysRevB.109.054422}. 
The other parameters are set as $J=1$, $D=0.2$, and $H=1.1$. 
As seen in the spin configuration around the skyrmion core, the helicity is characterized by neither Bloch-type nor N\'eel-type winding. 
Thus, the hybrid SkX is possible by considering the DM interactions at the low-symmetric wave vectors. 
Such an effect of the low-symmetric ordering wave vectors has been experimentally found in the SkX-hosting noncentrosymmetric material EuNiGe$_3$~\cite{singh2023transition, matsumura2023distorted}. 

Furthermore, the specific values of $\theta_{D}$ lead to another type of the SkX: the anti-type SkX appears for $-0.01 \pi \lesssim \theta \lesssim 0.01 \pi$ and  $0.99 \pi \lesssim \theta \lesssim 1.01 \pi$ and the rectangular SkX for $0.49 \pi \lesssim \theta \lesssim 0.51 \pi$ and $1.49 \pi \lesssim \theta \lesssim 1.51 \pi$, both of which are characterized by the double-$Q$ superposition at $\bm{Q}_2$ and $\bm{Q}_3$ [or $\bm{Q}_1$ and $\bm{Q}_4$] in contrast to the above S-SkX~\cite{Hayami_PhysRevB.109.054422}. 
In particular, the anti-type SkX exhibits a positive skyrmion number, which is different from the hybrid SkX with a negative skyrmion number. 
This indicates that the engineering of the anti-type SkX is possible even in polar magnets when the ordering wave vectors are located at low-symmetric wave vectors in the Brillouin zone.

\subsection{Hedgehog crystal}
\label{sec: Hedgehog crystal}

\begin{figure}[tp!]
\begin{center}
\includegraphics[width=1.0\hsize]{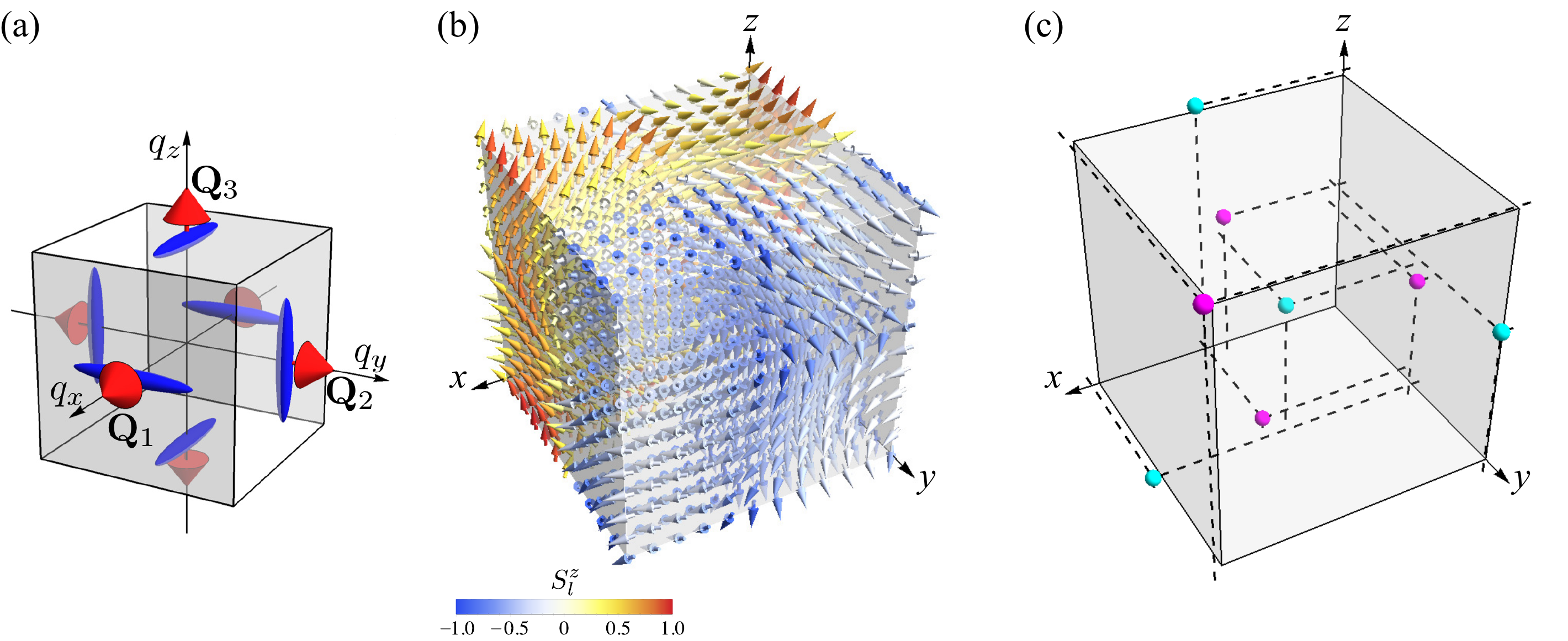} 
\caption{
\label{fig: hedgehog_DM}
(a) Schematic picture of the ordering wave vectors and anisotropic interactions for the cubic-lattice model in Eq.~(\ref{eq:HamT}). 
The red arrows represent $\bm{D}_{\bm{Q}_\nu}$. 
The blue ellipsoids represent $F^\alpha_{\bm{Q}_\nu}$, where the lengths along the principal [100], [010], and [001] directions stand for the amplitudes of $F^x_{\bm{Q}_\nu}$, $F^y_{\bm{Q}_\nu}$, and $F^z_{\bm{Q}_\nu}$, respectively. 
(b) Spin configuration with $Q=\pi/6$ at $D/J=0.3$ and $\Delta=0.3$. 
(c) Positions of the hedgehogs denoted by magenta spheres and the antihedgehogs denoted by the cyan spheres in the hedgehog spin texture in (b), where the dashed lines are the guides for eyes. 
Reprinted figure with permission from~\cite{Kato_PhysRevB.104.224405}, Copyright (2021) by the American Physical Society. 
}
\end{center}
\end{figure}

The effective spin model in Eq.~(\ref{eq: Ham}) also captures the essence of the stabilization of the hedgehog crystal with the three-dimensional topological spin textures~\cite{Binz_PhysRevB.74.214408, Park_PhysRevB.83.184406, Yang2016,tanigaki2015real,kanazawa2017noncentrosymmetric,fujishiro2019topological, Kanazawa_PhysRevLett.125.137202, Aoyama_PhysRevB.103.014406, Aoyama_PhysRevB.106.064412, Eto_PhysRevLett.132.226705}. 
One of the spin configurations in the hedgehog crystal is characterized by a triple-$Q$ superposition at $\bm{Q}_1=(Q, 0, 0)$, $\bm{Q}_2=(0, Q, 0)$, and $\bm{Q}_3=(0, 0, Q)$ on the cubic lattice, which is schematically shown in Fig.~\ref{fig: ponti}. 
Although only the DM interaction does not lead to instability toward the zero-field hedgehog crystal, incorporating additional interactions can make its realization possible. 
We show two types of interactions to induce the hedgehog crystal without the magnetic field. 

One is the symmetric anisotropic interaction that arises from the point group $T$, i.e., $F^x_{\bm{Q}_1} \neq F^y_{\bm{Q}_1} \neq F^z_{\bm{Q}_1}$ for $\bm{Q}_1 =(Q, 0, 0)$~\cite{Kato_PhysRevB.104.224405}. 
The spin model on the cubic lattice is described by 
\begin{align}
\label{eq:HamT}
\mathcal{H} = -2\sum_{\nu} 
\left[\sum_{\alpha=x,y,z}
F_{\bm{Q}_\nu}^{\alpha} S^{\alpha}_{\bm{Q}_{\nu}} S^{\alpha}_{-\bm{Q}_{\nu}}+i \bm{D}_{\nu} \cdot \left(\bm{S}_{\bm{Q}_{\nu}} \times \bm{S}_{-\bm{Q}_\nu}\right)
\right], 
\end{align}
where $\bm{Q}_1=(Q, 0, 0)$, $\bm{Q}_{2}=(0, Q, 0)$, and $\bm{Q}_3=(0, 0, Q)$ with $Q=\pi/6$. 
Reflecting the $T$ symmetry, the symmetric anisotropic interaction is taken by $(F_{\bm{Q}_1}^{x}, F_{\bm{Q}_1}^{y}, F_{\bm{Q}_1}^{z})=[J(1-\Delta), J(1+2\Delta), J(1-\Delta)]$ and the DM interaction is taken by $\bm{D}_{\bm{Q}_1}= (D, 0, 0)$, as schematically shown in Fig.~\ref{fig: hedgehog_DM}(a).  
The interactions at the other symmetry-related ordering wave vectors are obtained by taking the threefold rotation around the [111] axis. 

When the effects of both $\Delta$ and $D$ are taken into account, the hedgehog crystal, which consists of the triple-$Q$ spin density waves at $\bm{Q}_1$, $\bm{Q}_2$, and $\bm{Q}_3$, is stabilized in the zero-field ground state instead of the single-$Q$ spiral state~\cite{Kato_PhysRevB.104.224405}. 
Figure~\ref{fig: hedgehog_DM}(b) shows the spin configuration in the hedgehog crystals at $D/J=0.3$ and $\Delta=0.3$. 
Although this spin configuration seems complicated, it is characterized by a periodic alignment of the topological defects called hedgehogs and antihedgehogs, which become sources of the emergent magnetic field, as shown in  Fig.~\ref{fig: hedgehog_DM}(c); the magenta and cyan spheres represent the positions of hedgehogs and antihedgehogs, respectively.

The instability toward the hedgehog crystal is understood by the competition between the symmetric anisotropic interaction $\Delta$ and the DM interaction $D$. 
For the $\bm{Q}_1$ component, the former $\Delta$ tends to favor the sinusoidal modulation along the $y$ direction, while the latter $D$ tends to favor the spiral modulation in the $yz$ plane. 
Thus, the spiral plane is elliptically modulated by introducing $\Delta$, which leads to the intensities at high-harmonic wave vectors and results in the energy cost in the single-$Q$ spiral state. 
This is why the multiple-$Q$ hedgehog crystal is stabilized even at zero field. 

Another mechanism of the hedgehog crystal is the positive biquadratic interaction instead of the symmetric anisotropic interaction~\cite{Okumura_PhysRevB.101.144416, Shimizu_PhysRevB.103.054427}. 
Since the positive biquadratic interaction tends to favor the multiple-$Q$ states, as discussed in Sec.~\ref{sec: Biquadratic interaction}, the triple-$Q$ hedgehog crystal is stabilized in the ground state for a larger biquadratic interaction. 

\subsection{Meron-antimeron crystal}
\label{sec: Meron-antimeron crystal}

\begin{figure}[tp!]
\begin{center}
\includegraphics[width=1.0\hsize]{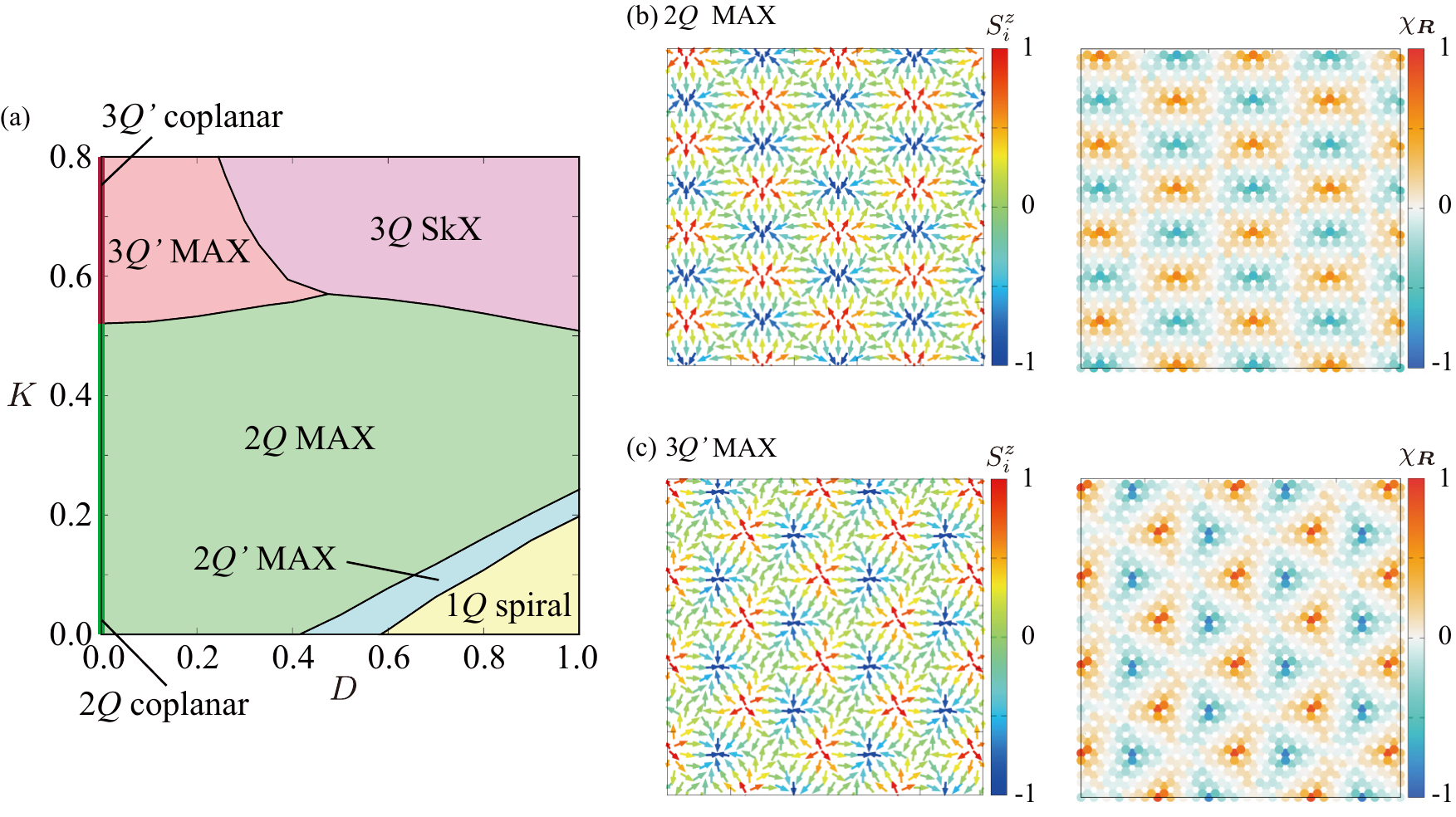} 
\caption{
\label{fig: 3Qmeron}
(a) Magnetic phase diagram of the model in Eq.~(\ref{eq:HamMAX}) while varying $D$ and $K$. 
(b,c) The left panels show the real-space spin configurations in (b) the 2$Q$ MAX for $D=0.2$ and $K=0.2$ and (c) the 3$Q'$ MAX for $D=0.2$ and $K=0.6$. 
The right panels show the real-space scalar spin chirality configurations. 
Reprinted figure with permission from~\cite{Hayami_PhysRevB.104.094425}, Copyright (2021) by the American Physical Society. 
}
\end{center}
\end{figure}

The MAX, which consists of the meron and antimeron with a half-integer skyrmion number, is also stabilized in noncentrosymmetric magnets at zero magnetic field~\cite{brey1996skyrme, Ezawa_PhysRevB.83.100408, Lin_PhysRevB.91.224407, Tan_PhysRevB.94.014433}. 
The square MAX has been observed in chiral magnets Co$_8$Zn$_9$Mn$_3$~\cite{yu2018transformation} and the triangular one has been suggested in centrosymmetric magnets Gd$_2$PdSi$_3$~\cite{kurumaji2019skyrmion}; the schematic spin configurations of the square and triangular MAXs are shown in Fig.~\ref{fig: ponti}. 

We present the stability of two-type MAXs on the noncentrosymmetric two-dimensional triangular lattice under the polar point group $C_\mathrm{6v}$~\cite{Hayami_PhysRevB.104.094425}. 
The effective spin model is given by 
\begin{align}
\label{eq:HamMAX}
\mathcal{H} = \sum_{\nu}
&\left[-J\sum_{\alpha,\beta}I^{\alpha \beta}_{\bm{Q}_{\nu}} S^\alpha_{\bm{Q}_{\nu}} S^\beta_{-\bm{Q}_{\nu}}+\frac{K}{N}\left(\sum_{\alpha,\beta}I^{\alpha \beta}_{\bm{Q}_{\nu}} S^\alpha_{\bm{Q}_{\nu}} S^\beta_{-\bm{Q}_{\nu}}\right)^2 \right.
\nonumber \\
&  \left.-i{\bm D}_\nu\cdot\left({\bm S}_{\bm{Q}_\nu}\times{\bm S}_{-\bm{Q}_{\nu}}\right)
\right]-A\sum_{i}(S^z_i)^2,  
\end{align}
where $\bm{Q}_1=(Q,0)$, $\bm{Q}_2=(-Q/2,\sqrt{3}Q/2)$, and $\bm{Q}_3=(-Q/2,-\sqrt{3}Q/2)$ with $Q=\pi/3$. 
The first, second, and third terms in the square bracket stand for the symmetric, biquadratic, and DM interactions, respectively. 
The form factor $I^{\alpha\beta}_{\bm{Q}_{\nu}}$ consists of the isotropic contribution described by the Kronecker delta $\delta^{\alpha\beta}$ ($I^{xx}_{\bm{Q}_{\nu}}=I^{yy}_{\bm{Q}_{\nu}}=I^{zz}_{\bm{Q}_{\nu}}$) and the anisotropic contribution $\tilde{I}^{\alpha\beta}_{\bm{Q}_{\nu}}$, $I^{\alpha\beta}_{\bm{Q}_{\nu}}=\delta^{\alpha\beta}+\tilde{I}^{\alpha\beta}_{\bm{Q}_{\nu}}$ to satisfy $
-\tilde{I}^{xx}_{\bm{Q}_{1}}=\tilde{I}^{yy}_{\bm{Q}_{1}}=2\tilde{I}^{xx}_{\bm{Q}_{2}}=-2\tilde{I}^{yy}_{\bm{Q}_{2}}=2\tilde{I}^{xy}_{\bm{Q}_{2}}/\sqrt{3}=2\tilde{I}^{yx}_{\bm{Q}_{2}}/\sqrt{3}=2\tilde{I}^{xx}_{\bm{Q}_{3}}=-2\tilde{I}^{yy}_{\bm{Q}_{3}}=-2\tilde{I}^{xy}_{\bm{Q}_{3}}/\sqrt{3}=-2\tilde{I}^{yx}_{\bm{Q}_{3}}/\sqrt{3} \equiv I^{\rm A}$. 
The DM vector with the magnitude of $D=|\bm{D}_{\nu}|$ is polar type along the $\bm{Q}_\nu \times \hat{\bm{z}}$ direction. 
The model also includes the easy-plane single-ion anisotropy $A<0$. 

Figure~\ref{fig: 3Qmeron}(a) shows the ground-state phase diagram in the plane of $D$ and $K$ at $J=1$, $I^{\rm A}=-0.2$ and $A=-0.8$, which is obtained by the simulated annealing~\cite{Hayami_PhysRevB.104.094425}. 
In the phase diagram, $Q'$ means the different intensities of the spin structure factor at multiple-$Q$ ordering wave vectors. 
For $D=0$, two phases with coplanar spin configurations appear depending on $K$: One is the 2$Q$ coplanar state and the other is the 3$Q'$ coplanar state, whose stabilization is attributed to the synergy between the bond-dependent anisotropy $I^{\rm A}$ and easy-plane anisotropy $A$~\cite{Hayami_PhysRevB.103.054422}. 
This is because $I^{\rm A}$ tends to favor the inplane sinusoidal modulation, which leads to the elliptical deformation of the spiral plane resulting in the energy cost of the single-$Q$ spiral state. 
As $K$ further tends to favor the multiple-$Q$ states, the 2$Q$ coplanar state is replaced by the 3$Q'$ coplanar state for large $K$. 

When the effect of $D$ is turned on, two magnetic phases with coplanar spin textures are modulated so as to have noncoplanar ones, since the DM interaction under the $C_{\rm 6v}$ symmetry favors the out-of-plane cycloidal spiral structure. 
For small $K$, the 2$Q$ coplanar state turns into the 2$Q$ MAX, which consists of the two vortices with the same vorticity but different chirality, as shown by the real-space spin and scalar spin chirality configurations in Fig.~\ref{fig: 3Qmeron}(b). 
As the (anti)vortices with the negative (positive) scalar spin chirality have (anti)meron-like spin textures, this spin configuration is regarded as the periodic alignment of the meron-antimeron pairs. 
The rectangle-lattice alignment of such meron-antimeron pairs indicates the double-$Q$ modulation on the triangular lattice. 
The net scalar chirality is canceled out owing to the equivalence of the positive and negative contributions. 

For large $K$, the 3$Q'$ MAX is stabilized instead of the 2$Q$ MAX. 
This state is characterized by the different amplitudes of the spin structure factor in the triple-$Q$ ordering wave vectors in both $xy$ and $z$ spin components. 
Accordingly, the vortices with positive scalar spin chirality (antimeron) and those with negative scalar spin chirality (meron) form a distorted triangular lattice shown in Fig.~\ref{fig: 3Qmeron}(c); the uniform component of the scalar spin chirality is zero. 
In the real-space picture, the spin configuration is characterized by the superposition of the three cycloidal waves at $\bm{Q}_1$, $\bm{Q}_2$, and $\bm{Q}_3$, which is similar to the case of the triple-$Q$ SkX with nonzero scalar spin chirality. 
The main difference between them is represented by the relative phase degree of freedom among the constituent waves~\cite{kurumaji2019skyrmion,hayami2021phase}; in the triple-$Q$ SkX, the sum of the phase in the three cycloidal waves becomes zero, $\sum_\nu \theta_{\nu} =0$, where the spiral spin texture is given by $\bm{S}_i = [\cos (\bm{Q}_1\cdot \bm{r}_i+\theta_1), 0,  \sin (\bm{Q}_1\cdot \bm{r}_i+\theta_1)]$ along the $\bm{Q}_{1}$ direction for instance, while that leads to $\sum_\nu \theta_{\nu} =\pi/2$ in the 3$Q'$ MAX without the net scalar spin chirality~\cite{hayami2021phase}.

\section{Stabilization mechanisms in centrosymmetric magnets}
\label{sec: Stabilization mechanisms in centrosymmetric magnets}

In this section, we present the stabilization mechanisms of the SkXs and other multiple-$Q$ states in centrosymmetric magnets. 
First, we discuss the case of the SkX in Sec.~\ref{sec: Skyrmion crystal}. 
Even without the conventional DM interaction under noncentrosymmetric crystal structures, several mechanisms have been recognized in inducing SkXs, such as the biquadratic interaction, symmetric anisotropic interaction, high-harmonic wave-vector interaction, and sublattice-dependent interaction. 
We introduce their role one by one. 
Then, we show the instabilities toward other multiple-$Q$ states including the hedgehog crystal in Sec.~\ref{sec: Hedgehog crystal_2}, AF skyrmion crystal in Sec.~\ref{sec: Antiferro skyrmion crystal}, and bubble crystal in Sec.~\ref{sec: Bubble crystal}. 

\subsection{Skyrmion crystal}
\label{sec: Skyrmion crystal}

To discuss the stability of the SkX in centrosymmetric magnets, we show the effects of the biquadratic interaction in Sec.~\ref{sec: Effect of biquadratic interaction_2}, symmetric anisotropic interaction in Sec.~\ref{sec: Effect of symmetric anisotropic interaction_2}, high-harmonic wave-vector interaction in Sec.~\ref{sec: Effect of high-harmonic wave-vector interaction}, and sublattice-dependent interaction in Sec.~\ref{sec: Effect of sublattice-dependent interaction}. 
We also discuss the similarities and differences in the stability between the S-SkX on the square lattice and the T-SkX on the triangular lattice. 

\subsubsection{Effect of biquadratic interaction}
\label{sec: Effect of biquadratic interaction_2}

\begin{figure}[tp!]
\begin{center}
\includegraphics[width=0.97\hsize]{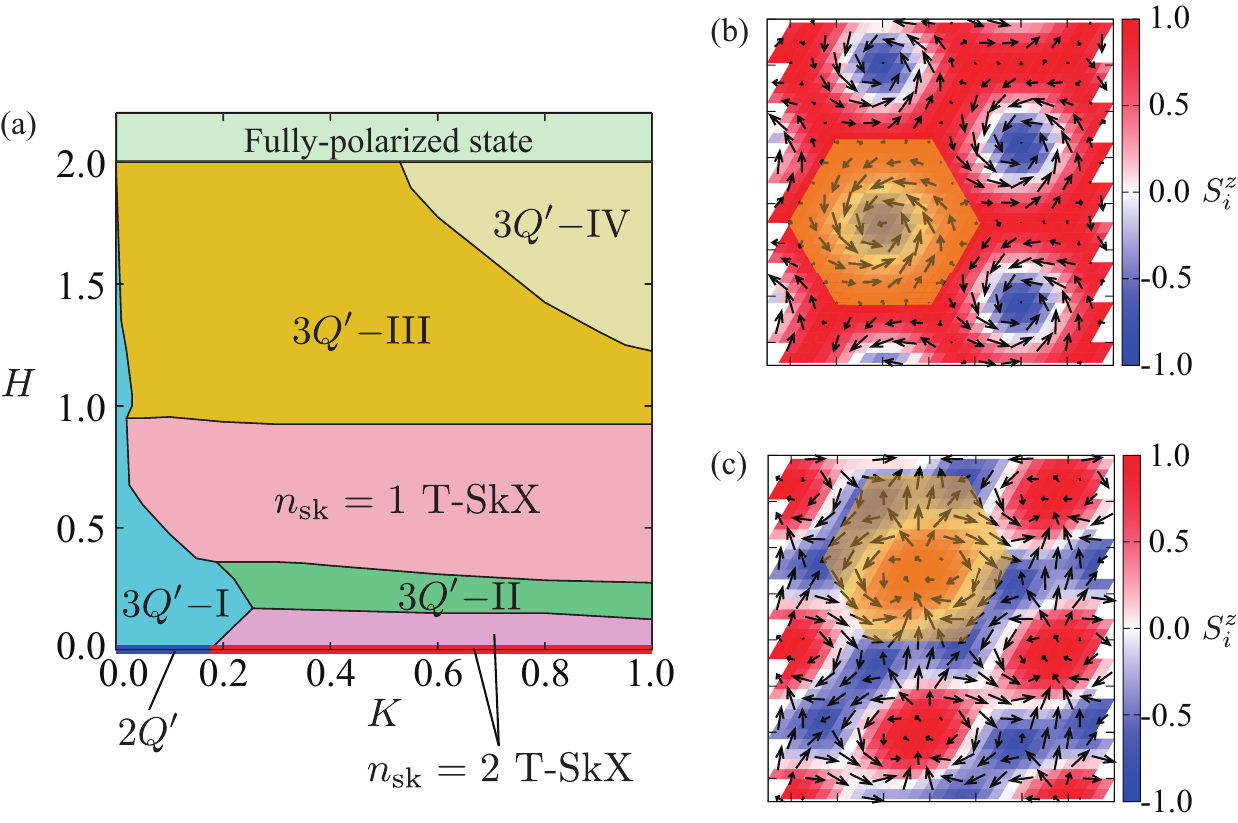} 
\caption{
\label{fig: BQ}
(a) Magnetic phase diagram for the model in Eq.~(\ref{eq: Ham_BQ_cen}) on the triangular lattice with $\bm{Q}_1 = (\pi/3,0)$ under the $D_{\rm 6h}$ symmetry. 
(b, c) Snapshots of the spin configurations in (b) the $n_{\rm sk}=1$ T-SkX for $K=0.6$ and $H=0.8$ and (c) the $n_{\rm sk}=2$ T-SkX for $K=0.4$ and $H=0.2$. 
The yellow hexagons show the magnetic unit cells. 
Reprinted figure (a) with permission from~\cite{Hayami_PhysRevB.98.019903}, Copyright (2018) by the American Physical Society. 
Reprinted figures (b) and (c) with permission from~\cite{Hayami_PhysRevB.95.224424}, Copyright (2017) by the American Physical Society. 
}
\end{center}
\end{figure}

The effect of the biquadratic interaction is taken into account in the following spin model on the two-dimensional triangular lattice under the $D_{\rm 6h}$ symmetry, which is given by 
\begin{align}
\label{eq: Ham_BQ_cen}
\mathcal{H} = \sum_{\nu}
&\left[-J\bm{S}_{\bm{Q}_\nu}\cdot\bm{S}_{-\bm{Q}_\nu}+\frac{K}{N}\left({\bm S}_{\bm{Q}_\nu}\cdot{\bm S}_{-\bm{Q}_\nu}\right)^2\right] - H \sum_{i} S^z_i,
\end{align}
where $\bm{Q}_1=(Q,0)$, $\bm{Q}_2=(-Q/2,\sqrt{3}Q/2)$, and $\bm{Q}_3=(-Q/2,-\sqrt{3}Q/2)$ with $Q=\pi/3$. 
The symmetric anisotropic interaction is neglected. 
For $K=0$, the single-$Q$ spiral state is realized irrespective of $H$ at zero temperature. 

By performing the simulated annealing by changing $K$ and $H$ at $J=1$, the ground-state phase diagram is constructed in Fig.~\ref{fig: BQ}(a)~\cite{Hayami_PhysRevB.95.224424}.  
The two types of T-SkXs have been identified: One is the $n_{\rm sk}=2$ T-SkX in the low-field region and the other is the $n_{\rm sk}=1$ T-SkX in the intermediate-field region. 
The spin configuration of the $n_{\rm sk}=1$ T-SkX shown in Fig.~\ref{fig: BQ}(b) is similar to that in noncentrosymmetric magnets [For example, please see Fig.~\ref{fig: PD_DM}(a)]. 
Meanwhile, the $n_{\rm sk}=1$ T-SkX induced by the biquadratic interaction shows the degeneracy with respect to the spin rotational symmetry along the $z$ axis; all the Bloch-type SkX, N\'eel-type SkX, and anti-type SkX have the same energy. 

The characteristic point in this mechanism is the emergence of the $n_{\rm sk}=2$ T-SkX~\cite{Ozawa_PhysRevLett.118.147205}. 
Similarly to the $n_{\rm sk}=1$ T-SkX, the $n_{\rm sk}=2$ T-SkX corresponds to the triple-$Q$ state, whose spin configuration is shown in Fig.~\ref{fig: BQ}(c). 
Meanwhile, the constituent spin density waves are different from each other; the $n_{\rm sk}=1$ T-SkX consists of the superposition of the three spiral waves, while the $n_{\rm sk}=2$ T-SkX consists of the superposition of the three sinusoidal waves.  
Reflecting such a difference, the topological number $n_{\rm sk}$ is two per magnetic unit cell in the $n_{\rm sk}=2$ T-SkX.

In contrast to the triangular-lattice spin model, the spin model on the square lattice does not exhibit instabilities toward both $n_{\rm sk}=1$ S-SkX and $n_{\rm sk}=2$ S-SkX~\cite{Hayami_PhysRevB.95.224424}. 
This difference is attributed to the constituent multiple-$Q$ ordering wave vectors. 
Since the T-SkX is constructed by the superposition of the three spin density waves with $\bm{Q}_1$, $\bm{Q}_2$, and $\bm{Q}_3$ to satisfy $\bm{Q}_1+\bm{Q}_2+\bm{Q}_3=\bm{0}$, an effective interaction in the form of $(\bm{S}_{\bm{0}}\cdot \bm{S}_{\bm{Q}_1})(\bm{S}_{\bm{Q}_2}\cdot \bm{S}_{\bm{Q}_3})$ appears in the Ginzburg-Landau free energy. 
Meanwhile, in the case of the S-SkX consisting of the two spin density waves with $\bm{Q}_1$ and $\bm{Q}_2$ to satisfy $\bm{Q}_1\perp\bm{Q}_2$, there are no such effective interactions owing to $\bm{Q}_1+\bm{Q}_2 \neq \bm{0}$.
In addition, the energy cost by the exchange interaction at higher-harmonic wave vectors owing to multiple-$Q$ superpositions becomes larger unless $\bm{Q}_1+\bm{Q}_2+\bm{Q}_3=\bm{0}$, indicating that the S-SkX with $\bm{Q}_1+\bm{Q}_2 \neq \bm{0}$ is difficult to stabilize.
To realize the S-SkX, additional interactions, such as the symmetric anisotropic interaction and high-harmonic wave-vector interaction, are required, as discussed in the following subsections.

\subsubsection{Effect of symmetric anisotropic interaction}
\label{sec: Effect of symmetric anisotropic interaction_2}

\begin{figure}[tp!]
\begin{center}
\includegraphics[width=1.0\hsize]{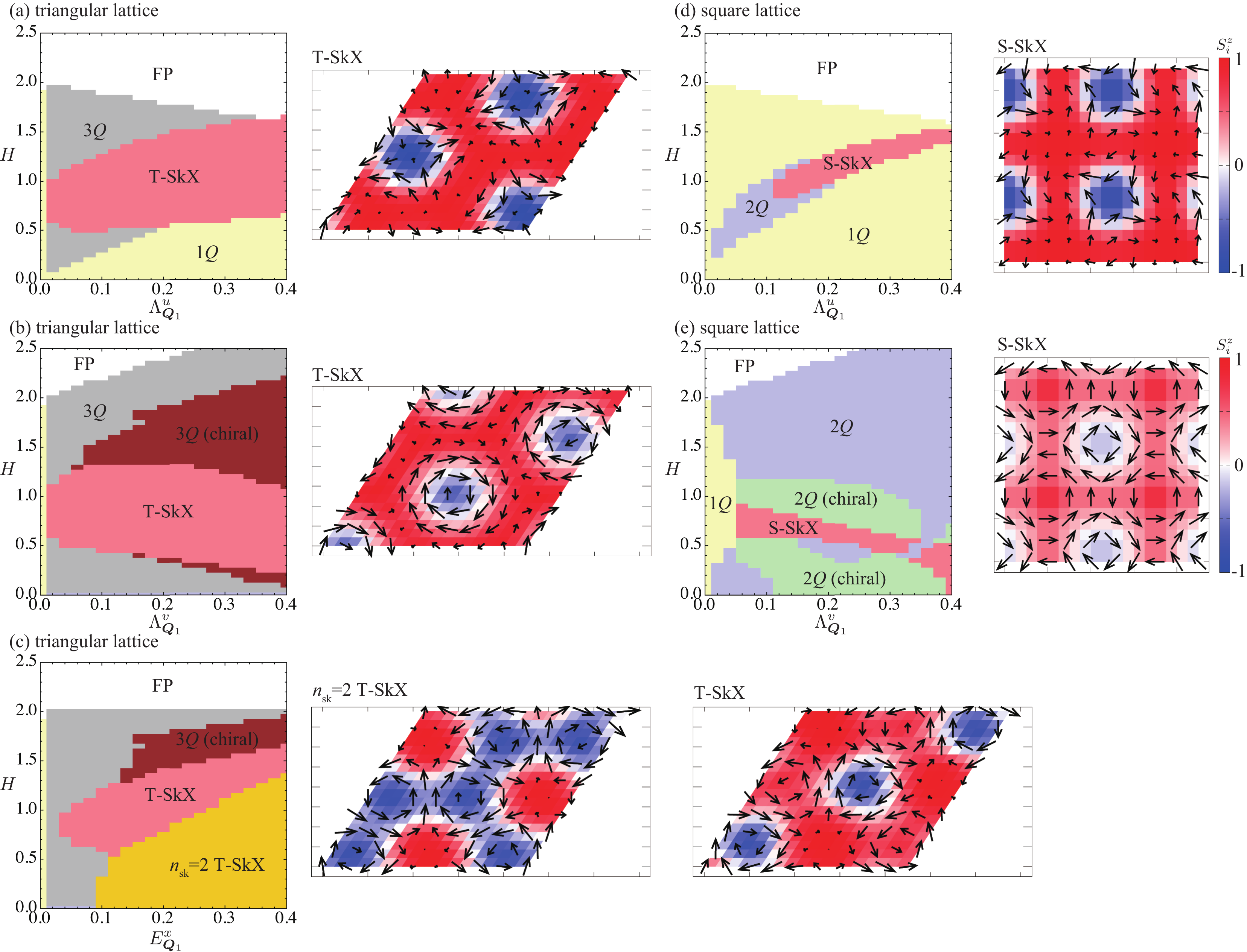} 
\caption{
\label{fig: PD}
The left panels represent the ground-state phase diagrams on (a,b,c) the triangular lattice and (d,e) the square lattice while varying the symmetric anisotropic interactions (a,d) $\Lambda^u_{\bm{Q}_1}$, (b,e) $\Lambda^v_{\bm{Q}_1}$, and (c) $E^x_{\bm{Q}_1}$ as well as the out-of-plane magnetic field $H$. 
1$Q$, 2$Q$, 3$Q$, and FP represent the single-$Q$, double-$Q$, triple-$Q$, and fully polarized state, respectively; ``(chiral)" denotes the magnetic states with the scalar spin chirality $\chi^{\rm sc} \geq 0.01$ but without the integer skyrmion number. 
In (d) and (e), the isotropic interactions at $\bm{Q}_1+\bm{Q}_2$ with the coupling constant $J'=0.6$ are also introduced. 
The right panels show the real-space spin configurations of the SkXs. 
The arrow and color represent the in-plane and out-of-plane spin components, respectively. 
}
\end{center}
\end{figure}

We consider the effect of the symmetric anisotropic interaction on the stabilization of the SkX in centrosymmetric magnets. 
The spin model with the symmetric anisotropic interaction is generally given by 
\begin{align}
\label{eq:qspace_model_ani}
\mathcal{H}
= - 2 \sum_{\nu, \alpha,\beta} X^{\alpha \beta}_{\bm{Q}_\nu}  S^\alpha_{\bm{Q}_\nu} S^\beta_{-\bm{Q}_\nu}-H\sum_i S_i^z,
\end{align} 
with
\begin{align}
X_{\bm{Q}_1}&=
\begin{pmatrix}
F_{\bm{Q}_1}^{x} & E_{\bm{Q}_1}^{z} & E_{\bm{Q}_1}^{y} \\
E_{\bm{Q}_1}^{z} & F_{\bm{Q}_1}^{y}  & E_{\bm{Q}_1}^{x} \\
E_{\bm{Q}_1}^{y} & E_{\bm{Q}_1}^{x}  & F_{\bm{Q}_1}^{z}  
\end{pmatrix}\nonumber\\
&=
\label{eq:qspace_int}
\begin{pmatrix}
J_{\bm{Q}_1}^{\rm iso} -\frac{1}{2}\Lambda_{\bm{Q}_1}^{u} -\Lambda_{\bm{Q}_1}^{v}  & E_{\bm{Q}_1}^{z} & E_{\bm{Q}_1}^{y} \\
E_{\bm{Q}_1}^{z} & J_{\bm{Q}_1}^{\rm iso} -\frac{1}{2}\Lambda_{\bm{Q}_1}^{u} +\Lambda_{\bm{Q}_1}^{v}   & E_{\bm{Q}_1}^{x} \\
E_{\bm{Q}_1}^{y} & E_{\bm{Q}_1}^{x}  &J_{\bm{Q}_1}^{\rm iso} + \Lambda_{\bm{Q}_1}^{u} 
\end{pmatrix}, 
\end{align}
where we rewrite $\bm{F}_{\bm{Q}_1}$ so as to decouple the isotropic contribution $J^{\rm iso}_{\bm{Q}_1}$ and anisotropic contributions $\Lambda_{\bm{Q}_1}^{u}$ and $\Lambda_{\bm{Q}_1}^{v}$; $\Lambda_{\bm{Q}_1}^{u}$ 
represents the bond-independent-type interaction, while $\Lambda_{\bm{Q}_1}^{v}$ represents the bond-dependent-type interaction. 
We consider two lattice structures: One is the square lattice with the ordering wave vectors $\bm{Q}_1=(Q,0)$ and $\bm{Q}_2=(0, Q)$ at $Q=2\pi/5$ and the other is the triangular lattice with $\bm{Q}_1=(Q,0)$, $\bm{Q}_2=(-Q/2,\sqrt{3}Q/2)$, and $\bm{Q}_3=(-Q/2,-\sqrt{3}Q/2)$ at $Q=2\pi/5$; the lattice constant is set as unity in each lattice structure. 
Hereafter in this section, we take $J^{\rm iso}_{\bm{Q}_1}=1$ as the energy unit of the spin model. 

First, let us discuss the triangular-lattice case. 
The ground-state phase diagrams are calculated by performing the simulated annealing combined with the standard Metropolis local updates and heat bath method.
The system size is set as $N=10^2$ under the periodic boundary condition. 
In each case, only one anisotropic component included in $X_{\bm{Q}_1}$ is considered in addition to $J^{\rm iso}$ in order to focus on the role of each magnetic anisotropy. 
Figures~\ref{fig: PD}(a), \ref{fig: PD}(b), and \ref{fig: PD}(c) show the ground-state phase diagram of the triangular-lattice spin model by changing $\Lambda_{\bm{Q}_1}^{u}$, $\Lambda_{\bm{Q}_1}^{v}$, and $E_{\bm{Q}_1}^{x}$, respectively; the vertical axis represents the magnetic field $H$. 
It is noted that the phase diagrams for $E^y_{Q_1}$ and $E^z_{Q_1}$ are the same as those for $E^x_{Q_1}$ and $\Lambda^v_{Q_1}$, respectively, except that the ground-state spin structures are globally rotated.

In Fig.~\ref{fig: PD}(a), the T-SkX appears in the intermediate-field region, where $\Lambda_{\bm{Q}_1}^{u}>0$ ($\Lambda_{\bm{Q}_1}^{u}<0$) represents the easy-axis-type (easy-plane-type) interaction. 
This result indicates that the small easy-axis-type interaction ($\Lambda_{\bm{Q}_1}^{u}\gtrsim0.01$) is enough to stabilize the T-SkX. 
Meanwhile, the easy-plane-type interaction ($\Lambda_{\bm{Q}_1}^{u}<0$) does not favor the T-SkX under the out-of-plane magnetic field (not shown). 
The real-space spin configuration of the T-SkX is shown in Fig.~\ref{fig: PD}(a). 
Owing to the spin rotational symmetry around the $z$ axis, the Bloch-type, N\'eel-type, and anti-type T-SkXs are energetically degenerate under the easy-axis-type interaction.

In Fig.~\ref{fig: PD}(b), $\Lambda_{\bm{Q}_1}^{v}\gtrsim0.01$ stabilizes the Bloch-type T-SkX, as shown by the real-space spin configuration in the right panel.
In this case, the opposite sign of $\Lambda_{\bm{Q}_1}^{v} (<0)$ leads to different types of T-SkX, i.e, N\'eel-type T-SkX. 
It is noted that the anti-type T-SkX is not stabilized in the presence of $\Lambda_{\bm{Q}_1}^{v}$. 
The qualitatively same phase diagram is obtained for $E_{\bm{Q}_1}^{z}$. 

In Fig.~\ref{fig: PD}(c), $E_{\bm{Q}_1}^{x}$ stabilizes two types of T-SkXs: T-SkX with the skyrmion number of one and the $n_{\rm sk}=2$ T-SkX with the skyrmion number of two depending on the magnetic field~\cite{yambe2021skyrmion}.
The real-space spin configurations of two T-SkXs are shown in the middle and right panels of Fig.~\ref{fig: PD}(c). 
Thus, $E_{\bm{Q}_1}^{x}$ can become another origin of the $n_{\rm sk}=2$ T-SkX, which is also stabilized in the bilinear-biquadratic spin model in Eq.~(\ref{eq: Ham_BQ_cen}) in Sec.~\ref{sec: Effect of biquadratic interaction_2}.
The qualitatively same phase diagram is obtained for $E_{\bm{Q}_1}^{y}$ and negative $E_{\bm{Q}_1}^{x}$.

Next, we show the result of the square-lattice model by performing the simulated annealing in the same condition as the triangular-lattice model. 
In the case of the square-lattice model, $E_{\bm{Q}_1}^{x}$ and $E_{\bm{Q}_1}^{y}$ are identically zero owing to the presence of the twofold rotation around the $z$ axis, as shown in Table~\ref{tab: PGint} and Fig.~\ref{fig: interaction}(c). 
In contrast to the triangular-lattice model, the S-SkX is not stabilized by introducing solely $\Lambda_{\bm{Q}_1}^{u}$ and $\Lambda_{\bm{Q}_1}^{v}$. 
Thus, we additionally consider the effect of the high-harmonic wave-vector interaction $J'=0.6$ at $\bm{Q}_1+\bm{Q}_2$ and $\bm{Q}_1- \bm{Q}_2$ in the form of $J' (\bm{S}_{\bm{Q}_1+\bm{Q}_2} \cdot\bm{S}_{-\bm{Q}_1-\bm{Q}_2}+\bm{S}_{\bm{Q}_1-\bm{Q}_2} \cdot\bm{S}_{-\bm{Q}_1+\bm{Q}_2})$ in order to enhance the instability toward the S-SkX. 
The resulting phase diagrams are shown for $\Lambda_{\bm{Q}_1}^{u} \neq 0$ in Fig.~\ref{fig: PD}(d) and for $\Lambda_{\bm{Q}_1}^{v} \neq 0$ in Fig.~\ref{fig: PD}(e). 
The result for $E^z_{\bm{Q}_1}$ is the same as that for $\Lambda^v_{\bm{Q}_1}$ except that the spin coordinate is globally rotated.
The role of the high-harmonic wave-vector interaction is discussed in the next section.

In Fig.~\ref{fig: PD}(d), the S-SkX is stabilized by introducing the easy-axis-type interaction with $\Lambda_{\bm{Q}_1}^{u}\gtrsim0.11$ in the intermediate-field region. 
The real-space spin configuration is shown in the right panel of Fig.~\ref{fig: PD}(d). 
Similarly to the triangular-lattice model, the Bloch-type, N\'eel-type, and anti-type S-SkXs are energetically degenerate. 
Meanwhile, the easy-plane-type interaction ($\Lambda_{\bm{Q}_1}^{u}<0$) does not stabilize the S-SkX phase even when considering the effect of high-harmonic wave-vector interaction. 

In Fig.~\ref{fig: PD}(e), the S-SkX is also stabilized by the bond-dependent interaction $\Lambda_{\bm{Q}_1}^{v}$; $\Lambda_{\bm{Q}_1}^{v}>0$ leads to the Bloch-type S-SkX, while $\Lambda_{\bm{Q}_1}^{v}<0$ leads to the N\'eel-type S-SkX. 
We show the real-space spin configuration of the S-SkX for $\Lambda_{\bm{Q}_1}^{v}>0$ in the right panel of Fig.~\ref{fig: PD}(e). 
It is noted that anti-type S-SkX has the same energy as the Bloch-type or N\'eel-type S-SkX in contrast to the triangular-lattice model. 
Their degeneracy is lifted by introducing the anisotropic interaction at $\bm{Q}_1\pm\bm{Q}_2$~\cite{Hayami_doi:10.7566/JPSJ.89.103702}. 

Compared to the phase diagrams in the triangular-lattice model with those in the square-lattice model, one finds the difference in terms of the stability tendency between the T-SkX and S-SkX; the stability region of the T-SkX is wider than that of the S-SkX. 
In addition, the critical value of the anisotropic interactions to induce the T-SkX is smaller for the triangular-lattice model. 
The difference between their stability is attributed to the constituent ordering wave vectors, as discussed in the last paragraph in the previous section.

The above results indicate that it is difficult to identify the dominant interactions for the SkXs in real materials, since they emerge in the presence of various types of symmetric anisotropic exchange interactions. 
Meanwhile, such information can be extracted by investigating the phases around the SkXs. 
For example, the different magnetic states, $1Q$, $3Q$, and $n_{\rm sk}=2$ T-SkX, are stabilized in the low-field region under $\Lambda^u_{\bm{Q}_1}$ [Fig.~\ref{fig: PD}(a)], $\Lambda^v_{\bm{Q}_1}$ [Fig.~\ref{fig: PD}(b)], and $E^x_{\bm{Q}_1}$ [Fig.~\ref{fig: PD}(c)] in the triangular-lattice case. 
Thus, when the low-field phase corresponds to $1Q$, $3Q$, and $n_{\rm sk}=2$ T-SkX in real materials, the dominant symmetric anisotropic interactions are given by $\Lambda^u_{\bm{Q}_1}$, $\Lambda^v_{\bm{Q}_1}$, and $E^x_{\bm{Q}_1}$, respectively. 
A similar argument also holds for the square-lattice case in Figs.~\ref{fig: PD}(d) and \ref{fig: PD}(e).

\subsubsection{Effect of high-harmonic wave-vector interaction}
\label{sec: Effect of high-harmonic wave-vector interaction}

\begin{figure}[tp!]
\begin{center}
\includegraphics[width=1.0\hsize]{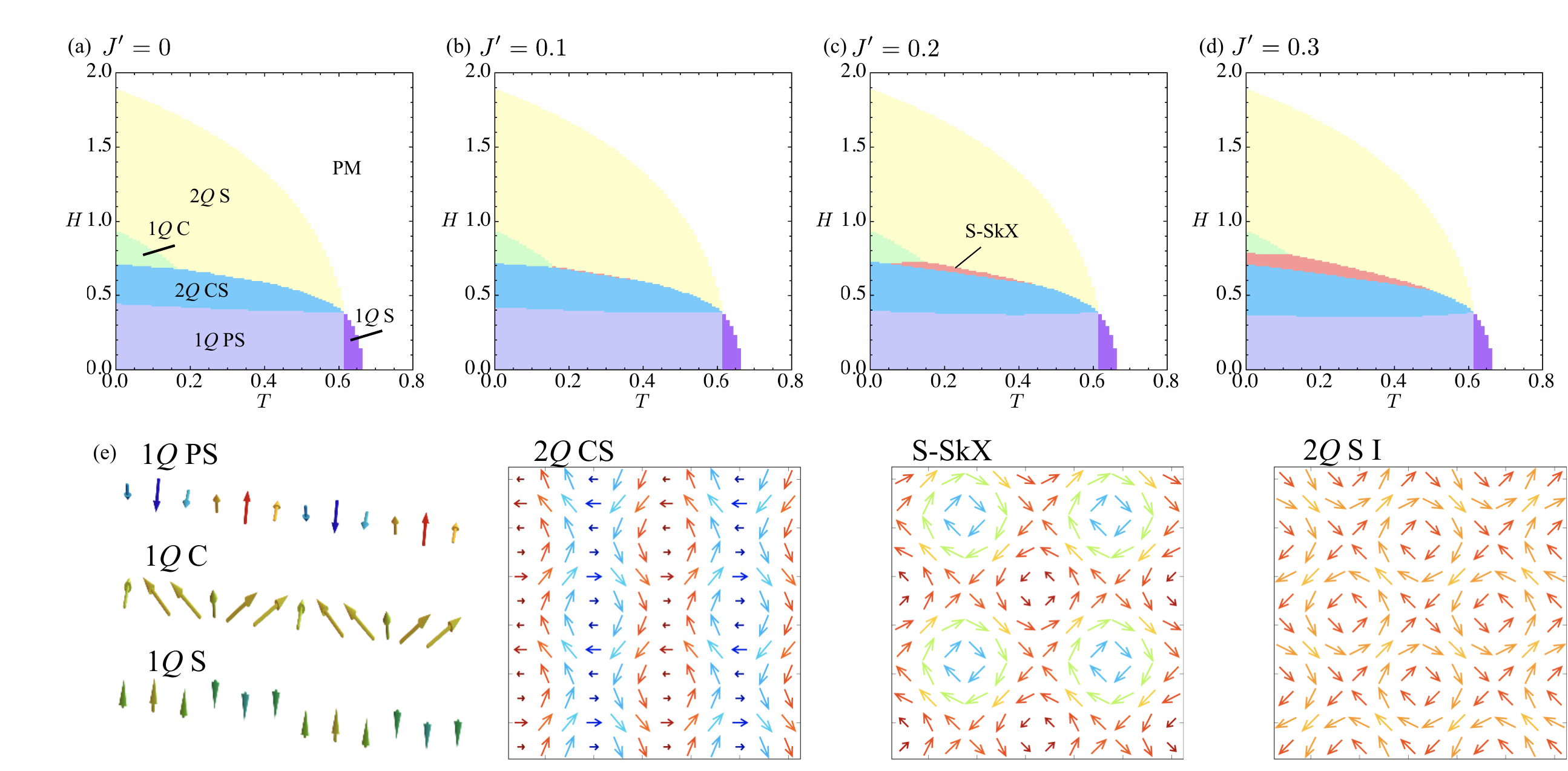} 
\caption{
\label{fig: PD_HHI}
(a)--(d) Magnetic field($H$)--temperature($T$) phase diagram of the model in Eq.~(\ref{eq: Ham_cen_hhi}) on the square lattice under the $D_{\rm 4h}$ symmetry for different values of (a) $J'=0$, (b) $0.1$, (c) $0.2$, and (d) $0.3$. 
(e) The schematic spin configurations appearing in the phase diagrams. PS, C, S, CS, and S-SkX stand for proper-screw, conical, sinusoidal, chiral stripe, and square skyrmion crystal, respectively. 
Reprinted figure with permission from~\cite{hayami2023widely}, Copyright (2023) by Elsevier. 
}
\end{center}
\end{figure}

As discussed in Secs.~\ref{sec: High-harmonic wave-vector interaction} and \ref{sec: Effect of symmetric anisotropic interaction_2}, the high-harmonic wave-vector interaction plays an important role in inducing the multiple-$Q$ states including the SkXs. 
Especially, such a high-harmonic wave-vector interaction becomes important when the instability toward the double-$Q$ S-SkX is investigated. 
This is because the high-harmonic wave vectors, one of which corresponds to $\bm{Q}_h=\bm{Q}_1+\bm{Q}_2$, leads to the relation of $\bm{Q}_1 + \bm{Q}_2 -\bm{Q}_h = \bm{0}$. 
In other words, an effective coupling of $(\bm{S}_{\bm{0}}\cdot \bm{S}_{\bm{Q}_1})(\bm{S}_{\bm{Q}_2}\cdot \bm{S}_{\bm{Q}_h})$ becomes non-negligible when the effect of the high-harmonic wave-vector interaction is taken into account, which results in a similar situation to the triangular-lattice model with symmetry-related wave vectors $\bm{Q}_1$, $\bm{Q}_2$, and $\bm{Q}_3$ satisfying $\bm{Q}_1 + \bm{Q}_2 + \bm{Q}_3 = \bm{0}$. 
In the following, we present two situations where the high-harmonic wave-vector interaction becomes the origin of the S-SkXs on the square lattice under the $D_{\rm 4h}$ symmetry. 

The first spin model is given by~\cite{hayami2023widely} 
\begin{align}
\label{eq: Ham_cen_hhi}
\mathcal{H}=  &-J \sum_{\nu,\alpha,\beta}\Gamma^{\alpha\beta}_{\bm{Q}_\nu}S^{\alpha}_{\bm{Q}_\nu}S^{\beta}_{-\bm{Q}_\nu} - H \sum_{i}   S^z_{i}, 
\end{align}
where $\{\pm \bm{Q}_1=\pm(Q,0), \pm \bm{Q}_2=\pm(0,Q), \pm \bm{Q}_3=\pm (\bm{Q}_1+\bm{Q}_2), \pm\bm{Q}_4=\pm(-\bm{Q}_1+\bm{Q}_2)\}$ with $Q=\pi/3$; $\bm{Q}_3$ and $\bm{Q}_4$ correspond to the high-harmonic wave vectors of $\bm{Q}_1$ and $\bm{Q}_2$. 
$J=1$ is the energy unit of the model. 
We suppose that the interactions at $\bm{Q}_1$ and $\bm{Q}_2$ are dominant and those at $\bm{Q}_3$ and $\bm{Q}_4$ are subdominant.  
For $\bm{Q}_1$ and $\bm{Q}_2$, $\Gamma^{\alpha\beta}_{\bm{Q}_\nu}$ is given as follows: $\bm{\Gamma}_{\bm{Q}_1}\equiv \Gamma^{\alpha\alpha}_{\bm{Q}_1}=(\Gamma_x, \Gamma_y, \Gamma_z)$ and $\bm{\Gamma}_{\bm{Q}_2}=(\Gamma_y, \Gamma_x, \Gamma_z)$, where $\Gamma^{\alpha\beta}_{\bm{Q}_\nu}=0$ for $\alpha \neq \beta$; we set $\Gamma_x=0.855$, $\Gamma_y=0.95$, and $\Gamma_z=1$. 
Meanwhile, $\Gamma^{\alpha\beta}_{\bm{Q}_\nu}$ for $\bm{Q}_3$ and $\bm{Q}_4$ is set as $\bm{\Gamma}_{\bm{Q}_3}=\bm{\Gamma}_{\bm{Q}_4}=(J', J', J')$ and $\Gamma^{\alpha\beta}_{\bm{Q}_\nu}=0$ for $\alpha \neq \beta$. 

Figures~\ref{fig: PD_HHI}(a)--\ref{fig: PD_HHI}(d) show the magnetic-field($H$)--temperature($T$) phase diagrams for $J'=0$--$0.3$. 
The phase diagrams are calculated by using the steepest descent method~\cite{Kato_PhysRevB.105.174413}, which provides a numerically exact solution in the thermodynamic limit at any temperature. 
The spin configurations in each phase are shown in Fig.~\ref{fig: PD_HHI}(e). 
When $\Gamma'=0$, the S-SkX does not appear in the phase diagram, as shown in Fig.~\ref{fig: PD_HHI}(a). 

On the other hand, the S-SkX appears in the vicinity region among the $2Q$ CS, $1Q$ C, and $2Q$ S when $J'$ is introduced, and its region becomes wider with increasing $J'$, as shown in Figs.~\ref{fig: PD_HHI}(b)--\ref{fig: PD_HHI}(d). 
Thus, the high-harmonic wave-vector interaction plays an important role in stabilizing the S-SkX. 
Furthermore, one finds that the instability of the S-SkX is found at finite temperatures rather than zero temperature even in centrosymmetric itinerant magnets similar to noncentrosymmetric DM-based magnets; for example, the S-SkX only appears at finite temperatures for small $\Gamma'=0.1$ and $\Gamma'=0.2$. 
This indicates that thermal fluctuations tend to favor the S-SkX in the effective spin model with the momentum-resolved interactions~\cite{Kato_PhysRevB.105.174413}, similar to that in the frustrated spin model with the real-space competing interactions~\cite{Okubo_PhysRevLett.108.017206, Mitsumoto_PhysRevB.104.184432, Mitsumoto_PhysRevB.105.094427}.

\begin{figure}[tp!]
\begin{center}
\includegraphics[width=1.0\hsize]{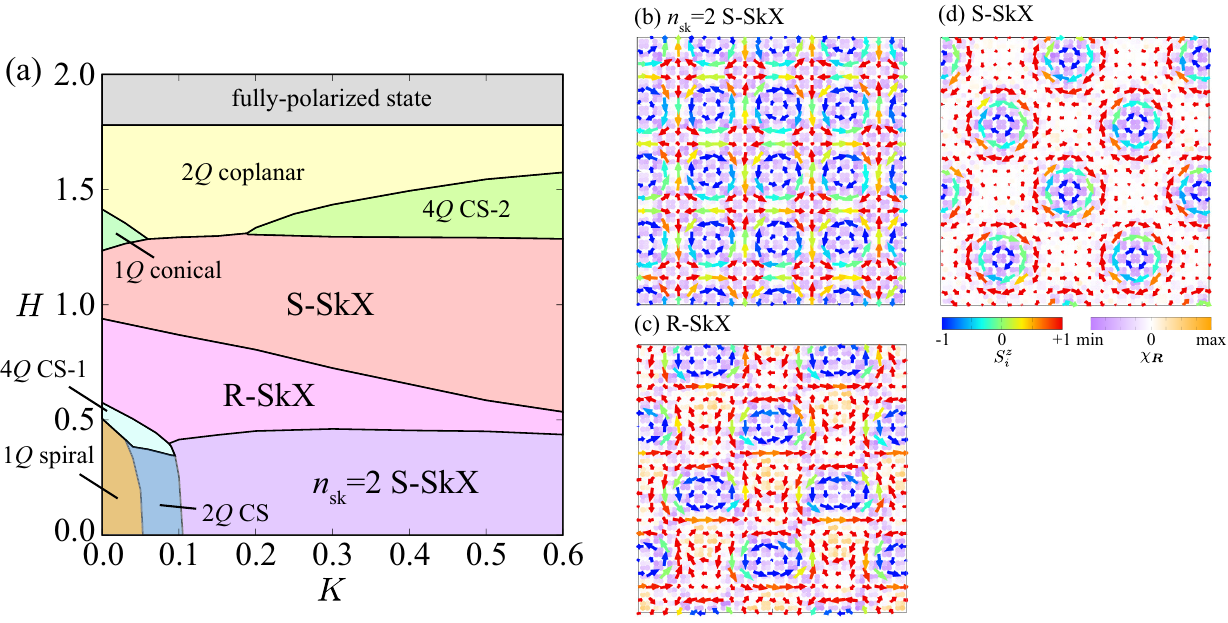} 
\caption{
\label{fig: nsk=2_SL_SkX}
(a) Magnetic phase diagram of the model in Eq.~(\ref{eq: Ham_cen_hhi2}) on the square lattice under the $D_{\rm 4h}$ symmetry while varying $K$ and $H$. 
SkX and CS represent the skyrmion crystal and chiral stripe, respectively. 
(b)--(d) Snapshots of the spin configurations in (b) the $n_{\rm sk}=2$ S-SkX for $K=0.2$ and $H=0.2$, (c) the R-SkX for $K=0.2$ and $H=0.6$, and (d) the S-SkX for $K=0.2$ and $H=1$. 
The arrows and their color show the $xy$ and $z$ spin components, respectively. 
The background color stands for the scalar spin chirality denoted by $\chi_{\bm{R}}$. 
Reproduced with permission from~\cite{hayami2022multiple}. Copyright 2022 by the Physical Society of Japan.
}
\end{center}
\end{figure}

The second spin model is given by~\cite{hayami2022multiple} 
\begin{align}
\label{eq: Ham_cen_hhi2}
\mathcal{H}=  2\sum_\nu \left( -J  \lambda_\nu +\frac{K}{N} \lambda_\nu^2 \right)-H \sum_i S_i^z,
\end{align}
where $\lambda_\nu=\sum_{\alpha,\beta}\Gamma^{\alpha\beta}_{\bm{Q}_{\nu}} S^\alpha_{\bm{Q}_{\nu}} S^\beta_{-\bm{Q_{\nu}}} $ for $\alpha,\beta=x,y,z$; $J=1$ is the energy unit of the model. 
The ordering wave vectors that give the dominant interaction are supposed to be $\bm{Q}_1=(2\pi/5,0)$ and $\bm{Q}_2=(0, 2\pi/5)$ and those that give the subdominant interaction are $\bm{Q}_3=(\pi/5,\pi/5)$ and $\bm{Q}_4=(-\pi/5, \pi/5)$, where $\bm{Q}_1=\bm{Q}_3-\bm{Q}_4$ and $\bm{Q}_2=\bm{Q}_3+\bm{Q}_4$. 
In contrast to the first spin model in Eq.~(\ref{eq: Ham_cen_hhi}), the high-harmonic wave vectors correspond to the wave vectors $\bm{Q}_1$ and $\bm{Q}_2$ giving the dominant interaction. 
The anisotropic form factor $\Gamma^{\alpha\beta}_{\bm{Q}_{\nu}} $ is given by $\Gamma^{yy}_{\bm{Q}_1}=\Gamma^{xx}_{\bm{Q}_2}=\gamma_1$, 
$\Gamma^{xx}_{\bm{Q}_1}=\Gamma^{yy}_{\bm{Q}_2}=\gamma_2$, 
$\Gamma^{zz}_{\bm{Q}_1}=\Gamma^{zz}_{\bm{Q}_2}=\gamma_3$, 
$\Gamma^{xx}_{\bm{Q}_3}=\Gamma^{yy}_{\bm{Q}_3}=\Gamma^{xx}_{\bm{Q}_4}=\Gamma^{yy}_{\bm{Q}_4}=\gamma_4$, 
$-\Gamma^{xy}_{\bm{Q}_3}=-\Gamma^{yx}_{\bm{Q}_3}=\Gamma^{xy}_{\bm{Q}_4}=\Gamma^{yx}_{\bm{Q}_4}=\gamma_5$, 
$\Gamma^{zz}_{\bm{Q}_3}=\Gamma^{zz}_{\bm{Q}_4}=\gamma_6$ (the others are zero); $\gamma_1=0.9$, $\gamma_2=0.855$, $\gamma_3=1$, $\gamma_4=0.81$, $\gamma_5=0.06525$, and $\gamma_6=0.9$, where $\gamma_1$ and $\gamma_2$ ($\gamma_4$ and $\gamma_5$) stand for the in-plane bond-dependent anisotropy, while $\gamma_3$ ($\gamma_6$) denotes the easy-axis anisotropy at $\bm{Q}_1$ and $\bm{Q}_2$ ($\bm{Q}_3$ and $\bm{Q}_4$). 
We also consider the effect of the biquadratic interaction. 

By performing the simulated annealing for the spin model, the magnetic phase diagram against $K$ and $H$ is constructed, as shown in Fig.~\ref{fig: nsk=2_SL_SkX}(a)~\cite{hayami2022multiple}. 
In contrast to the result in Fig.~\ref{fig: PD_HHI}, three SkX phases emerge in the phase diagram. 
The first SkX appears in the region for $K \gtrsim 0.1$ and $0 \leq H \lesssim 0.45$, which is denoted as the $n_{\rm sk}=2$ S-SkX. 
The spin configuration in this state is mainly characterized by the double-$Q$ peaks with equal intensity in the $xy$ component of the spin structure factor at $\bm{Q}_3$ and $\bm{Q}_4$ and the $z$ component of the spin structure factor at $\bm{Q}_1$ and $\bm{Q}_2$~\cite{Wang_PhysRevB.103.104408, hayami2022multiple}. 
Hence, this spin configuration is regarded as the superposition of four sinusoidal waves as $\bm{S}_i=(1/N_i) [a_{xy} (-\sin \mathcal{Q}_3 +\sin \mathcal{Q}_4), a_{xy}( \sin \mathcal{Q}_3 +\sin \mathcal{Q}_4),a_z (\cos \mathcal{Q}_1 +\cos \mathcal{Q}_2)]$, where $\mathcal{Q}_\nu=\bm{Q}_\nu \cdot \bm{r}_i+\theta_\nu$ ($\theta_\nu$ is the phase of waves), $a_{xy}$ and $a_z$ are the numerical coefficients, and $N_i$ is the normalization constant satisfying $|\bm{S}_i|=1$.  
In the real-space picture in Fig.~\ref{fig: nsk=2_SL_SkX}(b), there are two pairs of merons with the opposite vorticity but the same scalar spin chirality in a magnetic unit cell, which results in $n_{\rm sk}=\pm 2$ in the magnetic unit cell. 
The contour of the scalar spin chirality $\chi_{\bm{R}}=\bm{S}_i \cdot (\bm{S}_j \times \bm{S}_k)$ is also shown in Fig.~\ref{fig: nsk=2_SL_SkX}(b), whose summation in the magnetic unit cell is related to $n_{\rm sk}$. 
It is noted that the $n_{\rm sk}=2$ S-SkX is not stabilized by only $K$ as discussed in Sec.~\ref{sec: Effect of biquadratic interaction_2}.

The second SkX phase is stabilized in the intermediate-$H$ region, next to the $n_{\rm sk}=2$ S-SkX phase upon increasing $H$. 
This state exhibits the dominant peaks with equal intensity at $\bm{Q}_3$ and $\bm{Q}_4$ in both $xy$ and $z$ spin components. 
Meanwhile, the intensities at $\bm{Q}_1$ and $\bm{Q}_2$ are different, which indicates the breaking of fourfold rotational symmetry. 
We call this state the rectangular SkX (R-SkX). 
The breaking of the fourfold rotational symmetry is also found in the real-space spin configuration in Fig.~\ref{fig: nsk=2_SL_SkX}(c). 
This state exhibits $n_{\rm sk}= -1$, where the sign of $n_{\rm sk}$ is determined by $\gamma_1$, $\gamma_2$, and $\gamma_5$~\cite{Hayami_doi:10.7566/JPSJ.89.103702}. 

The third SkX phase appears with a further increase of $H$ in the R-SkX phase. 
In contrast to the R-SkX, the spin configuration is characterized by the fourfold-symmetric double-$Q$ structures with $n_{\rm sk}=-1$, which means the emergence of the S-SkX. 
The spin configuration in Fig.~\ref{fig: nsk=2_SL_SkX}(d) is characterized by a superposition of four proper-screw spirals at $\bm{Q}_1$-$\bm{Q}_4$. 

In this way, the high-harmonic wave-vector interaction gives rise to rich SkXs and multiple-$Q$ states. 
Indeed, the multiple SkXs found in EuAl$_4$~\cite{takagi2022square} and GdRu$_2$Ge$_2$~\cite{yoshimochi2024multi} have been accounted for by similar effective spin models incorporating the effect of the high-harmonic wave-vector interactions. 

\subsubsection{Effect of sublattice-dependent interaction}
\label{sec: Effect of sublattice-dependent interaction}

\begin{figure}[tp!]
\begin{center}
\includegraphics[width=1.0\hsize]{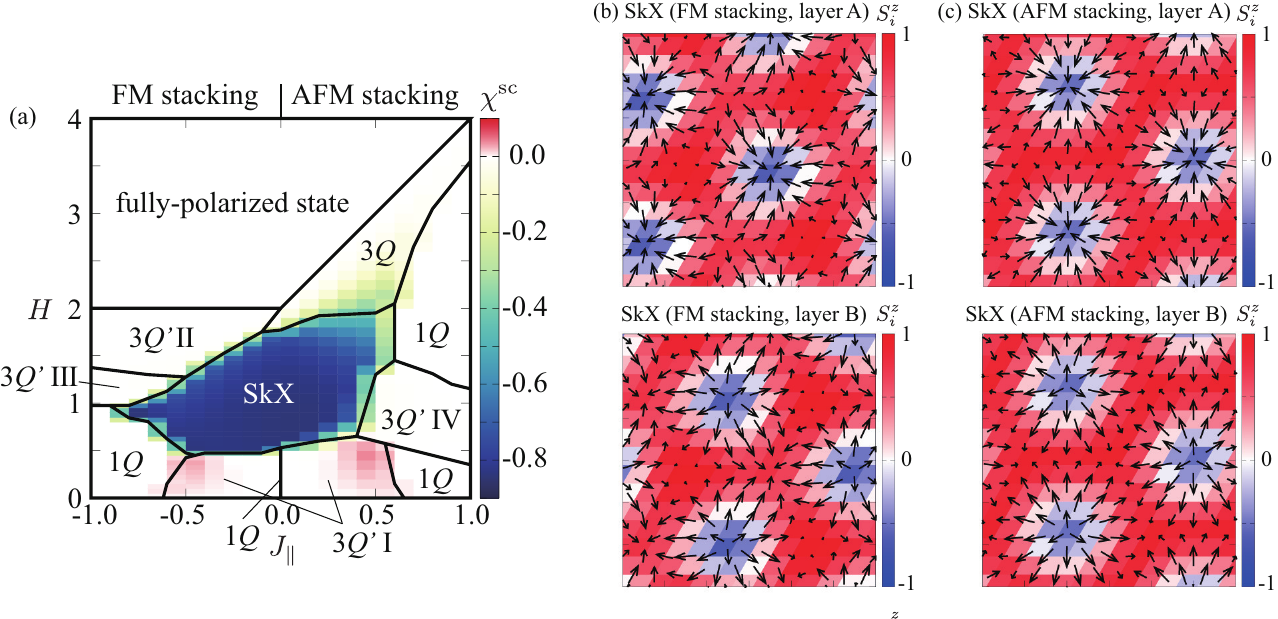} 
\caption{
\label{fig: PD_SDI}
(a) Magnetic phase diagram of the model in Eq.~(\ref{eq: Ham_staDM_bilayer}) in the bilayer triangular-lattice system while the interlayer exchange interaction $J_{\parallel}$ and the external magnetic field $H$ are varied. 
The contour shows the scalar spin chirality $\chi^{\rm sc}$. 
$J_{\parallel}>0$ ($J_{\parallel}<0$) represents the AFM (FM) exchange interaction. 
(b,c) Real-space spin configurations of the SkXs on (top panel) layer A and (bottom panel) layer B in (b) the FM stacking and (c) the AFM stacking. 
Reprinted figure with permission from~\cite{Hayami_PhysRevB.105.014408}, Copyright (2022) by the American Physical Society. 
}
\end{center}
\end{figure}

In this section, we discuss the effect of the sublattice-dependent interactions introduced in Sec.~\ref{sec: Sublattice-dependent interaction}. 
Among them, we focus on the role of the staggered DM interaction in stabilizing the SkX by considering the centrosymmetric bilayer and trilayer triangular-lattice systems. 
First, we show the result in the bilayer triangular-lattice system, which consists of layers A and B under the site symmetry $C_{\rm 6v}$, while the global symmetry in the whole lattice structure is $D_{\rm 6h}$ [see Fig.~\ref{fig: sublattice}(a)]. 
The spin model is given by 
\begin{align}
\label{eq: Ham_staDM_bilayer}
\mathcal{H}=&\mathcal{H}^{\perp}+\mathcal{H}^{\parallel}+\mathcal{H}^{{\rm Z}}, \\
\label{eq:Ham_perp}
\mathcal{H}^{\perp}=&  \sum_{\nu} \sum_{\eta} \Big[ -J \bm{S}^{(\eta)}_{\bm{Q}_{\nu}} \cdot \bm{S}^{(\eta)}_{-\bm{Q}_{\nu}}- i   \bm{D}^{(\eta)}_\nu \cdot ( \bm{S}^{(\eta)}_{\bm{Q}_{\nu}} \times \bm{S}^{(\eta)}_{-\bm{Q}_{\nu}}) \Big],  \\
\label{eq:Ham_parallel}
\mathcal{H}^{\parallel}=& J_{\parallel} \sum_{i} \bm{S}_i \cdot \bm{S}_{i+\hat{z}},\\
\label{eq:Ham_Zeeman}
\mathcal{H}^{{\rm Z}}=&-H \sum_i S_i^z, 
\end{align}
where $\eta={\rm A}, {\rm B}$ is the layer index. 
The total Hamiltonian $\mathcal{H}$ consists of the intralayer Hamiltonian $\mathcal{H}^{\perp}$, the interlayer Hamiltonian $\mathcal{H}^{\parallel}$, and the Zeeman Hamiltonian $\mathcal{H}^{{\rm Z}}$. 
In $\mathcal{H}^{\perp}$, the isotropic interaction $J$ and the layer-dependent staggered DM interaction $\bm{D}^{(\eta)}_\nu$ at the ordering wave vectors $\bm{Q}_1=(\pi/3,0)$, $\bm{Q}_2=(-\pi/6,\sqrt{3}\pi/6)$, $\bm{Q}_3=(-\pi/6,-\sqrt{3}\pi/6)$, $\bm{Q}_4=-\bm{Q}_1$, $\bm{Q}_5=-\bm{Q}_2$, and $\bm{Q}_6=-\bm{Q}_3$ are considered. 
The directions of the DM vector are perpendicular to the bond and $z$ directions owing to the polar symmetry, as shown in Fig.~\ref{fig: sublattice}(a); the relation of $\bm{D}_{\nu}^{({\rm A})}=-\bm{D}_{\nu}^{({\rm B})}$ is satisfied owing to the presence of the inversion center at the bonds between layers A and B. 
For $\mathcal{H}^{\parallel}$, the positive (negative) $J_{\parallel}$ represents the AFM(FM)-stacked case.

Figure~\ref{fig: PD_SDI}(a) shows the magnetic phase diagram obtained by simulated annealing down to $T=0.001$ while changing the interlayer exchange coupling $J_{\parallel}$ and the magnetic field $H$ at $J=1$ and $|\bm{D}^{(\eta)}_{\bm{Q}_\nu}|=D=0.2$~\cite{Hayami_PhysRevB.105.014408}.
When $J_{\parallel}=0$, the system reduces to the single-layer system, as demonstrated in Fig.~\ref{fig: PD_DM}(a); the single-$Q$ spiral (1$Q$) state, the SkX, the triple-$Q$ (3$Q$) state, and the fully polarized state are realized while increasing $H$, which have been found in various spin models with the FM and the DM interaction in the single-layer system~\cite{Yi_PhysRevB.80.054416, Mochizuki_PhysRevLett.108.017601}. 

When $J_{\parallel}$ is introduced, a variety of multiple-$Q$ states appear in the phase diagram, as shown in Fig.~\ref{fig: PD_SDI}(a); the SkX remains stable for both $J_{\parallel}>0$ and $J_{\parallel}<0$. 
As shown by the contour plot in terms of the scalar spin chirality $\chi^{\rm sc}$ in the total system, the SkX exhibits a large scalar spin chirality owing to its topological spin texture. 
Indeed, only the SkX phase has the quantized skyrmion number $-1$ in each layer in the magnetic unit cell. 
It is noted that the scalar spin chirality as well as the skyrmion number are the same for layers A and B. 

Owing to the staggered DM interaction, the skyrmion spin textures, which consist of three cycloidal spirals with $\bm{Q}_1$, $\bm{Q}_2$, and $\bm{Q}_3$, are characterized by the opposite helicities in each layer for both $J_{\parallel}$, as shown in Figs.~\ref{fig: PD_SDI}(b) and \ref{fig: PD_SDI}(c); the direction of the inplane spins around the skyrmion core is inward for layer A [top panel of Figs.~\ref{fig: PD_SDI}(b) and \ref{fig: PD_SDI}(c)], while that is outward for layer B [bottom panel of Figs.~\ref{fig: PD_SDI}(b) and \ref{fig: PD_SDI}(c)]

Meanwhile, a clear difference between the FM and AFM interlayer interactions appears in local spin configurations in a real-space picture. 
The skyrmion cores lie at the different positions on layers A and B under the FM stacking as shown in Fig.~\ref{fig: PD_SDI}(b), while those lie at the same positions under the AFM stacking as shown in Fig.~\ref{fig: PD_SDI}(c).
More specifically, the SkXs are stacked so that the inplane spins on the two layers are aligned (anti)parallel to each other in the FM (AFM) interaction, which is attributed to the fact that the $xy$ component of the spin structure factor is larger than the $z$ component in the case of the single-layer SkX; the SkX can gain more exchange energy by aligning the $xy$ spin component in a parallel (antiparallel) way for the FM (AFM) interaction.

\begin{figure}[tp!]
\begin{center}
\includegraphics[width=1.0\hsize]{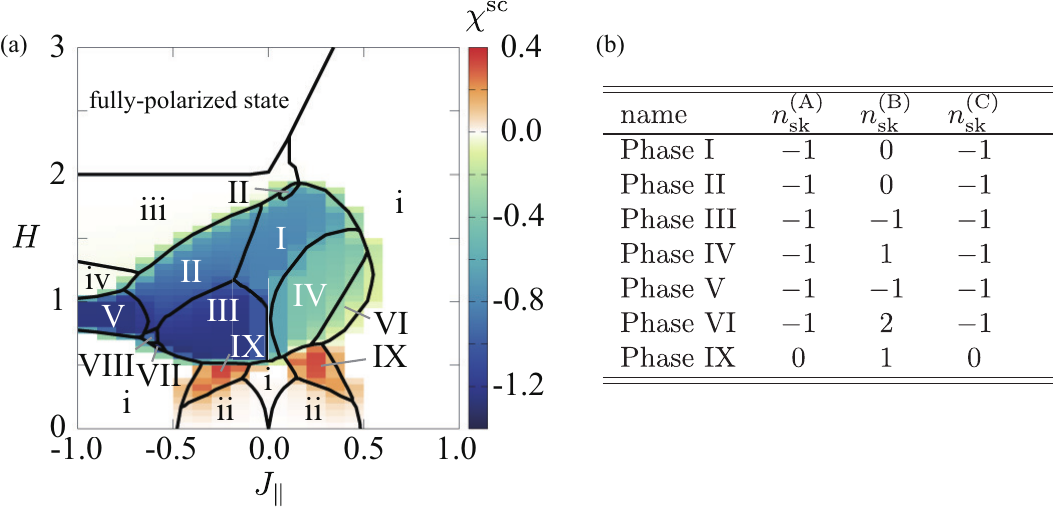} 
\caption{
\label{fig: PD_3layer}
(a) Magnetic phase diagram of the trilayer triangular-lattice model in Eq.~(\ref{eq: Ham_staDM_bilayer}) with $\bm{D}_{\nu}^{({\rm A})}=-\bm{D}_{\nu}^{({\rm C})}$ and $\bm{D}_{\nu}^{({\rm B})}=\bm{0}$ in the plane of $J_{\parallel}$ and $H$. 
The color map shows the total scalar spin chirality $\chi^{\rm sc}$. 
(b) Table of the skyrmion number for layers $\eta=$A, B, and C, $n_{\rm sk}^{\rm (\eta)}$, in the main phases appearing in (a).  
Reprinted figure with permission from~\cite{Hayami_PhysRevB.105.184426}, Copyright (2022) by the American Physical Society. 
}
\end{center}
\end{figure}

Another intriguing situation is a trilayer triangular-lattice model under the point group $D_{\rm 6h}$, where the DM interaction with the opposite sign is present for the top and bottom layers (layers A and C), while no DM interaction for the middle layer (layer B) owing to the presence of the inversion center; the lattice structure and the DM vector are shown in Fig.~\ref{fig: sublattice}(b)~\cite{Hayami_PhysRevB.105.184426}. 
In other words, the site symmetry for layers A, B, and C is $C_{\rm 6v}$, $D_{\rm 6h}$, and $C_{\rm 6v}$, respectively. 
The spin model is the same as that with the same ordering wave vectors in Eq.~(\ref{eq: Ham_staDM_bilayer}), where only the layer-dependent DM interaction is changed as $\bm{D}_{\nu}^{({\rm A})}=-\bm{D}_{\nu}^{({\rm C})}$ and $\bm{D}_{\nu}^{({\rm B})}=\bm{0}$. 

Figure~\ref{fig: PD_3layer}(a) shows the magnetic phase diagram in the plane of $J_{\parallel}$ and $H$ at $J=1$ and $D=0.2$, where the contour represents the total scalar spin chirality $\chi^{\rm sc}$ over the trilayer; the phase diagram is obtained by performing the simulated annealing~\cite{Hayami_PhysRevB.105.184426}.   
There are thirteen phases with distinct spin and scalar spin chirality configurations in addition to the fully polarized state. 
Among them, nine out of thirteen phases possess a quantized skyrmion number for any of the layers; the uppercase Roman numerals as ``Phase I", ``Phase II", $\cdots$, ``Phase IX" denote the SkXs, while the lowercase roman numerals as ``Phase i", ``Phase ii", ``Phase iii", and ``Phase iv" denote the topologically trivial states.  
Thus, the trilayer system exhibits rich SkX phases compared to the bilayer system, where only the single SkX phase is realized, as shown in Fig.~\ref{fig: PD_SDI}(a); one finds that such a variety of SkX phases are distinguished from the skyrmion number in layer B, as shown in Fig.~\ref{fig: PD_3layer}(b), where $n^{(\eta)}_{\rm sk}$ represents the skyrmion number for layer $\eta=$ A, B, C. 
In other words, the inversion-symmetric middle layer (layer B) is a source of multiple SkX phases.

\subsection{Hedgehog crystal}
\label{sec: Hedgehog crystal_2}

\begin{figure}[tp!]
\begin{center}
\includegraphics[width=1.0\hsize]{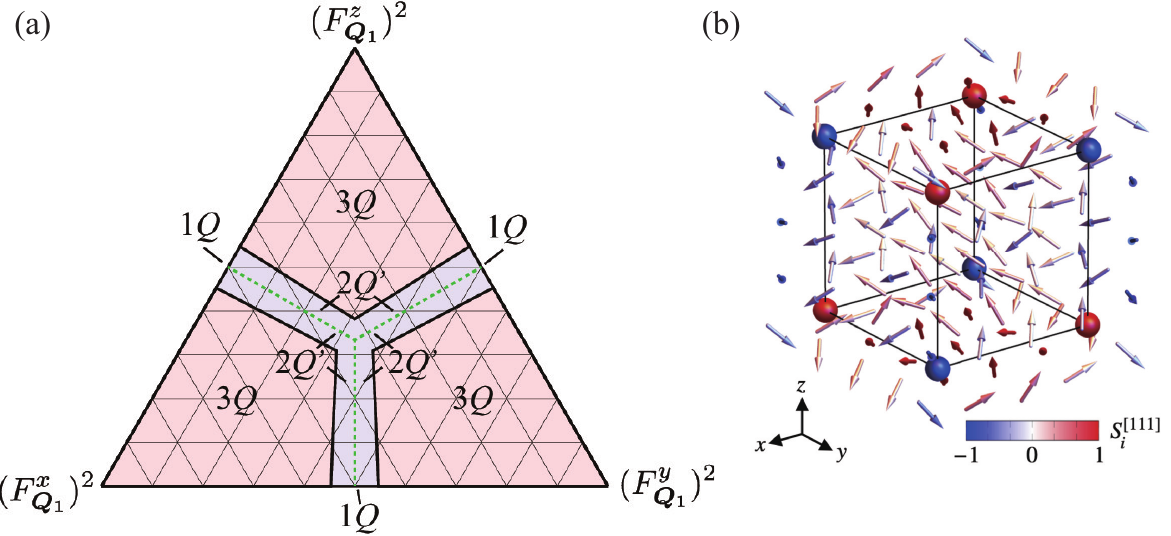} 
\caption{
\label{fig: PD_hedgehog}
(a) Magnetic phase diagram of the spin model in Eq.~(\ref{eq: model_Th}) on the cubic lattice under the $T_{\rm h}$ symmetry in the plane of $(F^x_{\bm{Q}_1})^2+(F^y_{\bm{Q}_1})^2+(F^z_{\bm{Q}_1})^2=1$. 
The dashed green lines represent the region where the 1$Q$ state is stabilized. 
3$Q$ state corresponds to the hedgehog state. 
(b) Real-space spin configuration for the 3$Q$ (hedgehog) state. 
The arrows represent the spin and their color stands for $S^{{\rm [111]}}_i=(S_i^x+S_i^y+S_i^z)/\sqrt{3}$. 
The red and blue spheres represent the hedgehogs and antihedgehogs, respectively. 
Reprinted figure with permission from~\cite{Yambe_PhysRevB.107.174408}, Copyright (2023) by the American Physical Society. 
}
\end{center}
\end{figure}

The effective spin model in centrosymmetric magnets describes the instability toward the hedgehog crystal similar to that in noncentrosymmetric magnets. 
For example, one of the models to stabilize the hedgehog crystal is constructed on the simple cubic lattice under the cubic point group $T_{\rm h}$, which is given by~\cite{Yambe_PhysRevB.107.174408} 
\begin{align}
\label{eq: model_Th}
\mathcal{H}= -2\sum_{\nu} \sum_{\alpha} F^\alpha_{\bm{Q}_\nu}S^\alpha_{\bm{Q}_\nu}   S^\alpha_{-\bm{Q}_\nu},
\end{align}
where $\bm{Q}_1=(Q, 0, 0)$, $\bm{Q}_{2}=(0, Q, 0)$, and $\bm{Q}_{3}=(0, 0, Q)$ with $Q=\pi/3$. 
From the cubic symmetry, $F_{\bm{Q}_1}^{x}=F_{\bm{Q}_2}^{y}=F_{\bm{Q}_3}^{z}$, $F_{\bm{Q}_1}^{y}=F_{\bm{Q}_2}^{z}=F_{\bm{Q}_3}^{x}$, and $F_{\bm{Q}_1}^{z}=F_{\bm{Q}_2}^{x}=F_{\bm{Q}_3}^{y}$. 

The ground-state phase diagram obtained by the simulated annealing is shown in Fig.~\ref{fig: PD_hedgehog}(a), where the parameters are set as $(F^x_{\bm{Q}_1})^2+(F^y_{\bm{Q}_1})^2+(F^z_{\bm{Q}_1})^2=1$ and  $F^\alpha_{\bm{Q}_1}\ge0$~\cite{Yambe_PhysRevB.107.174408}. 
The phase diagram is threefold symmetric at $F^x_{\bm{Q}_1}=F^y_{\bm{Q}_1}=F^z_{\bm{Q}_1}$ and twofold symmetric for $F^x_{\bm{Q}_1} \neq F^y_{\bm{Q}_1}=F^z_{\bm{Q}_1}$, $F^y_{\bm{Q}_1} \neq F^z_{\bm{Q}_1}=F^x_{\bm{Q}_1}$, and $F^z_{\bm{Q}_1} \neq F^x_{\bm{Q}_1}=F^y_{\bm{Q}_1}$.
There are three phases in the phase diagram, where the triple-$Q$ phase denoted as $3Q$ corresponds to the hedgehog crystal; the real-space spin configuration for $\bm{F}_{\bm{Q}_1}=(0,0,1)$ is presented in Fig.~\ref{fig: PD_hedgehog}(b). 
The hedgehog crystal is regarded as a superposition of the three sinusoidal waves along the $x$ direction at $\bm{Q}_2$, the $y$ direction at $\bm{Q}_3$, and the $z$ direction at $\bm{Q}_1$. 
The hedgehog crystal accompanies the periodic alignment of the hedgehogs (red spheres) and antihedgehogs (blue spheres), forming the simple cubic lattice, as shown in Fig.~\ref{fig: PD_hedgehog}(b)~\cite{Kato_PhysRevB.105.174413, Yambe_PhysRevB.107.174408}.  
Thus, the anisotropic interaction originating from $\bm{F}_{\bm{Q}_\nu}$ becomes the source of the hedgehog lattice even in centrosymmetric magnets without the DM interaction. 

In addition, the biquadratic interaction also induces the hedgehog crystal in centrosymmetric magnets even without the symmetric anisotropic interaction~\cite{Okumura_doi:10.7566/JPSJ.91.093702}. 
These mechanisms are relevant to the emergence of the hedgehog crystal in the centrosymmetric cubic material SrFeO$_3$~\cite{Ishiwata_PhysRevB.84.054427, Ishiwata_PhysRevB.101.134406,Rogge_PhysRevMaterials.3.084404, Onose_PhysRevMaterials.4.114420}.

\subsection{Antiferro skyrmion crystal}
\label{sec: Antiferro skyrmion crystal}

\begin{figure}[tp!]
\begin{center}
\includegraphics[width=0.7\hsize]{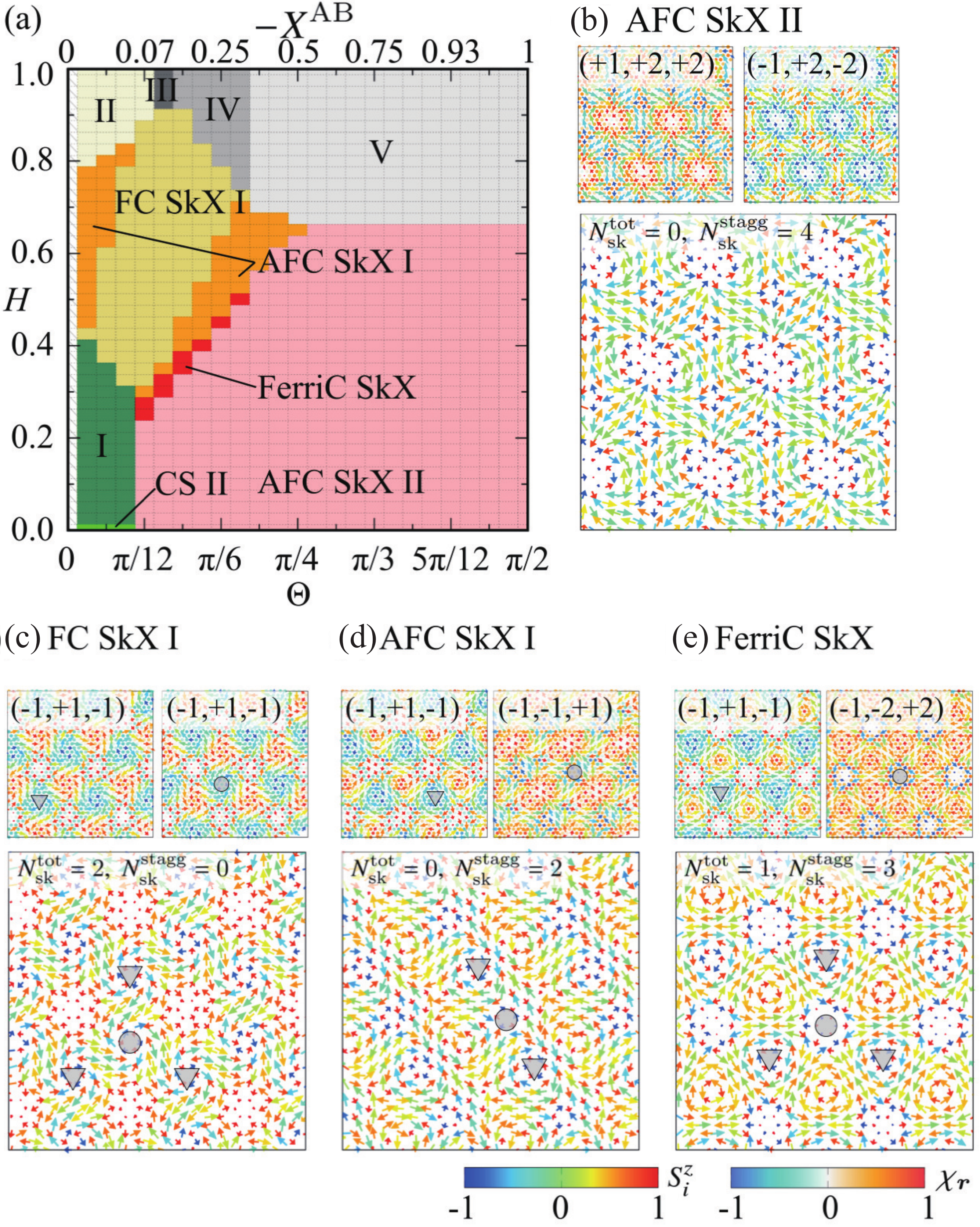} 
\caption{
\label{fig: PD_honeycomb}
(a) Magnetic phase diagram of the spin model in Eq.~(\ref{eq: BBQM}) in the plane of $\Theta$ and $H$ at $X^{\rm AB}<0$, $K=0.4$, and $T=0.01$. 
FC SkX, AFC SkX, and FerriC SkX stand for the ferrochiral SkX, antiferrochiral SkX, and ferrichiral SkX, respectively. 
I--V corresponds to nontopological phases. 
(b)--(e) Snapshots of the spin and scalar spin chirality configurations in (b) the AFC SkX II at $\Theta=\pi/12$ and $H=0$, (c) the FC SkX I at $\Theta = \pi/8$ and $H=0.5$, (d) the AFC SkX I at $\Theta=\pi/8$ and $H=0.4$, and (e) the FerriC SkX at $\Theta= \pi/8$ and $H=0.375$. 
In (b)--(e), the upper left (right) panels show the data for sublattice A (B), while the lower panels show the data for the two-sublattice honeycomb structure. 
The directions and the color of the arrows show the $xy$ and $z$ spin components, respectively, and the background contours of the circles in the upper panels show the scalar spin chirality $(\chi_{\bm{r}})$. 
$(p^\alpha, v^\alpha, N^\alpha_{\rm sk})$ in the upper panel stand for the polarity, vorticity, and topological number of the skyrmion at sublattice $\alpha$, respectively.  
$(N^{\rm tot}_{\rm sk}, N^{\rm stagg}_{\rm sk})$ in the lower panel stand for the total and staggered skyrmion numbers, respectively. 
Reprinted figure with permission from~\cite{Yambe_PhysRevB.107.014417}, Copyright (2023) by the American Physical Society. 
}
\end{center}
\end{figure}

The emergence of the AF SkXs with the different skyrmion numbers in each sublattice is also described by the effective spin model. 
The instability toward the AF SkXs is investigated in the momentum-resolved spin model under the honeycomb structure consisting of two sublattices A and B, which is given by~\cite{Yambe_PhysRevB.107.014417} 
\begin{align}
\label{eq: BBQM}
\mathcal{H}^\mathrm{eff}&=-2J\sum_{\eta}\Gamma_\eta (X)
+2\frac{K}{N}\sum_{\eta}\Gamma_\eta(X)^2-H\sum_{\alpha,i}S_{\alpha i}^z,
\end{align} 
where $\Gamma_\eta(X)=\sum_{\alpha,\beta}X^{\alpha\beta}\bm{S}_{\alpha \bm{Q}_\eta}\cdot\bm{S}_{\beta -\bm{Q}_\eta}$ with $X^{\alpha\beta}=(X^{\alpha\beta})^*$; $\bm{S}_{\alpha \bm{q}}$ with the wave vector $\bm{q}$ and the sublattice $\alpha={\rm A}, {\rm B}$ is the Fourier transform of the localized spin $\bm{S}_{\alpha i}$, $X^{\alpha\beta}$ represents the form factor of the interaction in terms of the sublattice, $X^{{\rm AA}}=X^{{\rm BB}}$ and $X^{{\rm AB}}=(X^{{\rm BA}})^*$, and $N$ is the number of unit cells. 
The ordering wave vectors are chosen as $\bm{Q}_1=(0,\pi/3)$, $\bm{Q}_2=(-\sqrt{3}\pi/6,-\pi/6)$, and $\bm{Q}_3=(\sqrt{3}\pi/6,-\pi/6)$. 
This model is regarded as an extension of the bilinear-biquadratic model in Eq.~(\ref{eq: Ham_BQ_cen}) incorporating the sublattice degree of freedom.

Figure~\ref{fig: PD_honeycomb}(a) shows the magnetic phase diagram at $T=0.01$, $J=1$, $K=0.4$, $X^{{\rm AA}} \equiv \cos^2 \Theta$, and $X^{\rm AB} \equiv - \sin^2 \Theta$, which is obtained by the simulated annealing for the system size with $N=36^2$~\cite{Yambe_PhysRevB.107.014417}. 
There are four types of the SkX phases: AFC SkX II, FC SkX I, AFC SkX I, and FerriC SkX, where AFC, FC, and FerriC mean antiferrochiral, ferrochiral, and ferrichiral, respectively; AFC SkX corresponds to the AF SkX. 
We show the snapshot of the real-space spin configurations in each SkX phase in Figs.~\ref{fig: PD_honeycomb}(b)--\ref{fig: PD_honeycomb}(e).  
In each figure, the upper two panels show the spin configurations for sublattices A and B, while the lower panel shows the spin configurations for both sublattices; the upper panels also show the distributions of the scalar spin chirality by the color map.
 
In the low-field region, the AFC SkX II appears. 
The spin configurations in both sublattices are characterized by the $n_{\rm sk}=2$ T-SkX with different skyrmion numbers; $(p^{\mathrm{A}},v^{\mathrm{A}}, N^{\mathrm{A}}_\mathrm{sk})=(+1,+2,+2)$ and $(p^{\mathrm{B}},v^{\mathrm{B}}, N^{\mathrm{B}}_\mathrm{sk})=(-1,+2,-2)$, where $p^\alpha$, $v^\alpha$, and $N_{\rm sk}^\alpha$ stand for the polarity, vorticity, and the topological (skyrmion) number of the skyrmion for sublattice $\alpha=$ A and B, respectively, as shown in Fig.~\ref{fig: PD_honeycomb}(b). 
Thus, this state has the staggered skyrmion number as  $N^\mathrm{tot}_\mathrm{sk}\equiv |N^{\mathrm{A}}+N^{\mathrm{B}}|=0$ and $N^\mathrm{stagg}_\mathrm{sk} \equiv |N^{\mathrm{A}}-N^{\mathrm{B}}|=4$. 
The key ingredients for the AFC SkX II are the antiferromagnetically coupled bipartite structure and the biquadratic interaction $K$ arising from the itinerant nature of electrons. 
The present AF SkX does not require the multi-layer structure with the staggered magnetic field and the staggered DM interaction~\cite{Gobel_PhysRevB.96.060406, Mukherjee_PhysRevB.105.075102, Hayami_PhysRevB.105.184426, Hayami_doi:10.7566/JPSJ.92.084702, Hayami_PhysRevB.109.014415}. 

When the magnetic field $H$ increases, the other three SkX phases are stabilized. 
The FC SkX I consists of the skyrmion with  $(p^{\alpha},v^{\alpha},N^{\alpha}_\mathrm{sk})=(-1,+1,-1)$ for both sublattices ($\alpha=$A, B), as shown in Fig.~\ref{fig: PD_honeycomb}(c). 
It is noted that the FC SkX I is stabilized irrespective of the sign of $X^\mathrm{AB}$; the antiferromagnetic coupling $X^\mathrm{AB}<0$ also induces the FC SkX I. 
This is because the skyrmion cores for sublattices A and B are located at different positions; the skyrmion core at sublattice B represented by the circle is separated from that at sublattice A by the triangle so that the honeycomb network is formed, as shown in Fig.~\ref{fig: PD_honeycomb}(c). 
In such a situation, the energy gain in terms of $J X^{\mathrm{AB}}$ occurs.

The AFC SkX I is characterized by $N^\mathrm{tot}_\mathrm{sk}=0$ and $N^\mathrm{stagg}_\mathrm{sk}=2$, which consists of the skyrmions with $(p^{\mathrm{A}},v^{\mathrm{A}},N^{\mathrm{A}}_\mathrm{sk})=(-1,+1,-1)$ and $(p^{\mathrm{B}},v^{\mathrm{B}},N^{\mathrm{B}}_\mathrm{sk})=(-1,-1,+1)$, as shown in Fig.~\ref{fig: PD_honeycomb}(d).
Similarly to the AFC SkX II, the opposite sign of $N^{\alpha}_\mathrm{sk}$ happens for each sublattice in the AFC SkX I. 
A different point from the AFC SkX II is that the spin configuration consists of the skyrmion with the opposite vorticity rather than the polarity. 
Accordingly, the AFC SkX I is regarded as the coexisting state of the skyrmion and anti-skyrmion with different vorticities. 
Owing to the one-dimensional alignment of the (anti-)skyrmion cores on the honeycomb network, this state breaks the threefold rotational symmetry.

The FerriC SkX has both uniform and staggered skyrmion numbers, $N^\mathrm{tot}_\mathrm{sk}=1$ and $N^\mathrm{stagg}_\mathrm{sk}=3$, which consists of the skyrmions with $(p^{\mathrm{A}},v^{\mathrm{A}},N^{\mathrm{A}}_\mathrm{sk})=(-1,+1,-1)$ and $(p^{\mathrm{B}},v^{\mathrm{B}},N^{\mathrm{B}}_\mathrm{sk})=(-1,-2,+2)$, as shown in Fig.~\ref{fig: PD_honeycomb}(e). 
Thus, there is no perfect cancellation of the topological number, which results in the topological physical phenomena as seen in the conventional SkX. 
The appearance of the FerriC SkX might be owing to the competition between the biquadratic interaction and magnetic field; the effect of the biquadratic interaction is more important for the skyrmions with $|N^{\alpha}_{\rm sk}|=2$, while the effect of the magnetic field is more important for 
the skyrmions with $|N^{\alpha}_{\rm sk}|=1$.

\subsection{Bubble crystal}
\label{sec: Bubble crystal}

\begin{figure}[tp!]
\begin{center}
\includegraphics[width=1.0\hsize]{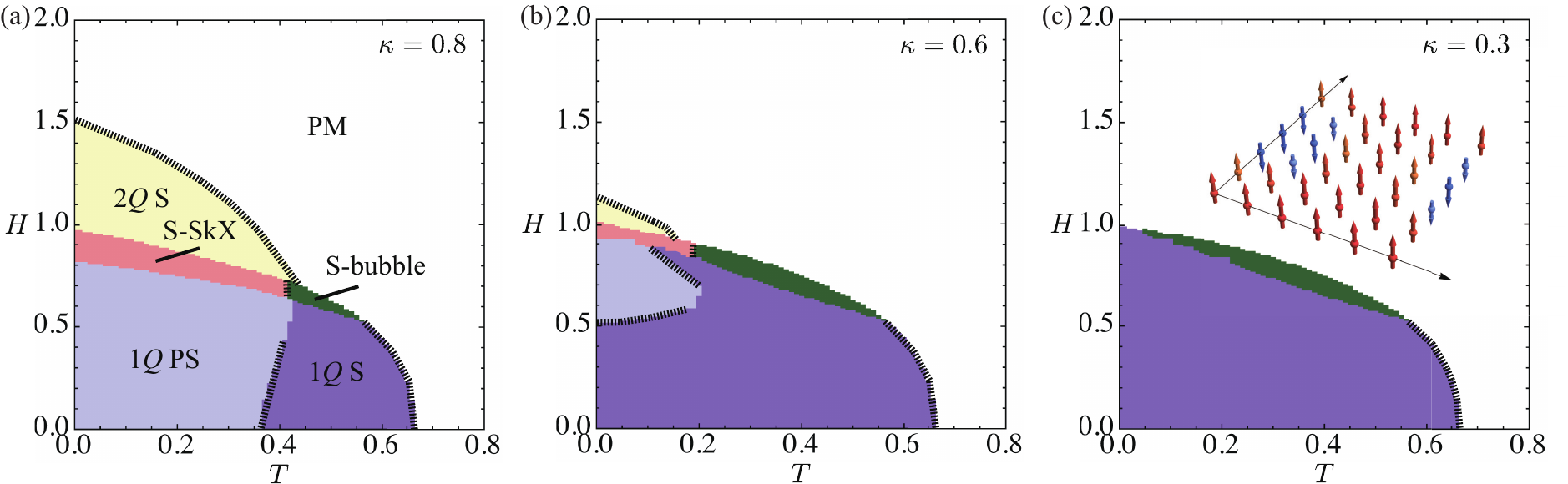} 
\caption{
\label{fig: PD_bubble}
Magnetic-field ($H$)-temperature ($T$) phase diagram of the model in Eq.~(\ref{eq: Ham_bubble}) at $J'=0.3$ for (a) $\kappa=0.8$, (b) $\kappa=0.6$, and (c) $\kappa=0.3$. 
PS, S, S-SkX, and PM stand for proper-screw, sinusoidal, square skyrmion crystal, and paramagnetic states, respectively. 
The real-space spin configuration of the S-bubble state is shown in the inset of (c). 
Reprinted figure with permission from~\cite{Hayami_PhysRevB.108.024426}, Copyright (2023) by the American Physical Society. 
}
\end{center}
\end{figure}

Finally, let us introduce a typical phase diagram hosting the magnetic bubble crystal. 
The effective spin model on the square lattice under the $D_{\rm 4h}$ symmetry is investigated, which is given by 
\begin{align}
\label{eq: Ham_bubble}
\mathcal{H}=  &-J \sum_{\nu,\alpha}\Gamma^{\alpha}_{\bm{Q}_\nu}S^{\alpha}_{\bm{Q}_\nu}S^{\alpha}_{-\bm{Q}_\nu} -  H \sum_{i}   S^z_{i},
\end{align}
where $\pm \bm{Q}_1=\pm(Q,0)$ and $ \pm \bm{Q}_2=\pm(0,Q)$ with $Q=\pi/3$. 
The anisotropic form factor $\Gamma^{\alpha}_{\bm{Q}_\nu}$ under the $D_{\rm 4h}$ symmetry is given by $\bm{\Gamma}_{\bm{Q}_1}\equiv (\Gamma^{x}_{\bm{Q}_1}, \Gamma^{y}_{\bm{Q}_1}, \Gamma^{z}_{\bm{Q}_1})=(\Gamma_x, \Gamma_y, \Gamma_z)$ and $\bm{\Gamma}_{\bm{Q}_2}\equiv (\Gamma^{x}_{\bm{Q}_2}, \Gamma^{y}_{\bm{Q}_2}, \Gamma^{z}_{\bm{Q}_2})=(\Gamma_y, \Gamma_x, \Gamma_z)$.  
We set $J=1$ as the energy unit of the model and take $\Gamma_y=  0.95 \kappa$, $\Gamma_x = 0.95 \Gamma_y$, and $\Gamma_z=1$, where $ \kappa \geq 0$ represents the parameter for the easy-axis anisotropic two-spin interaction; $\kappa<1$ enhances the degree of the easy-axis interaction (the results for $\kappa=1$ is discussed in Sec.~\ref{sec: Effect of high-harmonic wave-vector interaction}). 
In addition, the interactions at high-harmonic wave vectors $\pm \bm{Q}_3=\pm (\bm{Q}_1+\bm{Q}_2)$ and $\pm\bm{Q}_4=\pm(-\bm{Q}_1+\bm{Q}_2)$ are considered as $\bm{\Gamma}_{\bm{Q}_3}=\bm{\Gamma}_{\bm{Q}_4}=(\kappa J', \kappa J', J')$.

Figures~\ref{fig: PD_bubble}(a)--\ref{fig: PD_bubble}(c) show the magnetic field($H$)--temperature($T$) phase diagrams at $J'=0.3$ for $\kappa=0.8$ [Fig.~\ref{fig: PD_bubble}(a)], $\kappa=0.6$ [Fig.~\ref{fig: PD_bubble}(b)], and $\kappa=0.3$ [Fig.~\ref{fig: PD_bubble}(c)], which is obtained by the steepest descent method~\cite{Kato_PhysRevB.105.174413}.  
The S-bubble crystal appears in the finite-temperature region next to the S-SkX at $\kappa=0.8$, as shown in Fig.~\ref{fig: PD_bubble}(a); the obtained spin configuration in the S-bubble crystal is shown in the inset of Fig.~\ref{fig: PD_bubble}(c), which is mainly represented by a collinear superposition of two sinusoidal waves with the $z$-spin component at $\bm{Q}_1$ and $\bm{Q}_2$. 
When $\kappa$ decreases, the region where the S-bubble crystal is stabilized extends to a lower-$T$ region close to a zero temperature, as shown in Figs.~\ref{fig: PD_bubble}(b) and \ref{fig: PD_bubble}(c). 

The emergence of the S-bubble crystal is attributed to the interplay between the easy-axis anisotropic two-spin interaction with $\kappa <1$ and the high-harmonic wave-vector interaction $J'$ under the magnetic field. 
This is understood from the effective coupling in the form of $S^z_{\bm{Q}_1}S^z_{\bm{Q}_2}S^z_{-\bm{Q}_3} S^z_{\bm{0}}$ and $S^z_{-\bm{Q}_1} S^z_{\bm{Q}_2}S^z_{-\bm{Q}_4} S^z_{\bm{0}}$ appearing in the free energy, where the $z$ component of the spins is favored for $\kappa<1$ and the magnitude of $S^z_{\bm{Q}_3}$ and $S^z_{\bm{Q}_4}$ depends on $J'$.
Since larger $J'$ tends to make $S^z_{\bm{Q}_3}$ and $S^z_{\bm{Q}_4}$ larger, it results in the stabilization of the S-bubble crystal. 
The above effective coupling indicates that uniform magnetization also plays an important role in stabilizing the S-bubble crystal. 
This is why the S-bubble crystal appears only for nonzero $H$.

\section{Summary and perspective}
\label{sec: Summary and perspective}

To summarize, we have reviewed the stabilization mechanisms of the SkXs and other multiple-$Q$ states from the viewpoint of the theoretical modeling. 
On the basis of the momentum-resolved spin interactions, we raised a plethora of microscopic origins for the SkXs in both noncentrosymmetric and centrosymmetric magnets, such as the DM interaction, biquadratic interaction, symmetric anisotropic interaction, high-harmonic wave-vector interaction, sublattice-dependent interaction, and so on. 
Since the effective spin model with momentum-resolved spin interactions is simpler than the conventional spin model with real-space spin interactions, it enables us to explore rich multiple-$Q$ structures efficiently with a small computational cost. 
Indeed, analyzing the effective spin model has clarified the stabilization conditions of complicated multiple-$Q$ states other than the SkXs, such as the AF SkX, hedgehog crystal, meron-antimeron crystal, bubble crystal, and other vortex crystals. 
In addition, the effective spin model can be used to reproduce the experimental phase diagram as demonstrated in SkX-hosting materials like GdRu$_2$Si$_2$~\cite{khanh2022zoology}, Gd$_3$Ru$_4$Al$_{12}$~\cite{Hirschberger_10.1088/1367-2630/abdef9}, EuAl$_4$~\cite{takagi2022square}, EuPtSi~\cite{hayami2021field}, EuNiGe$_3$~\cite{singh2023transition}, and GdRu$_2$Ge$_2$~\cite{yoshimochi2024multi}, and other materials like Y$_3$Co$_8$Sn$_4$~\cite{takagi2018multiple} and CeAuSb$_2$~\cite{seo2021spin}. 
In this way, the effective spin model serves as a useful model not only to understand the origin of the exotic multiple-$Q$ states found in experiments but also to discover new types of topological spin textures that have never been clarified in theoretical models. 

Finally, we would like to raise two research directions to achieve a further fundamental understanding of multiple-$Q$ physics based on the effective spin model with the momentum-resolved spin interaction. 
The first one is to explore the multiple-$Q$ states accompanying charge and quadrupole density waves. 
Since the effective spin model can be applied to the $S=1$ spin system with the quadrupole degree of freedom, one can investigate the instability toward the multiple-$Q$ states consisting of the coexistence of dipole and quadrupole moments~\cite{hayami2023multipleJPSJ}. 
Indeed, recent studies have shown the possibility of exotic multiple-$Q$ states, which have never been realized in the dipole-only system, in $f$-electron compounds, such as PrV$_2$Al$_{20}$~\cite{Ishitobi_PhysRevB.104.L241110} and UNi$_4$B~\cite{Ishitobi_PhysRevB.107.104413}. 
In addition, the CP$^2$ SkX is another example to exhibit the multiple-$Q$ superstructure consisting of both dipole and quadrupole density waves~\cite{Garaud_PhysRevB.87.014507, Akagi_PhysRevD.103.065008, Amari_PhysRevB.106.L100406, zhang2023cp2}. 
A multipolar SkX in $f^2$ non-Kramers doublet systems also belongs to this category~\cite{zhang2024multipolar}. 

The second one is to derive an effective spin model with the real-space spin interactions from that with the momentum-resolved spin interactions. 
Since the present effective spin model with the momentum-resolved spin interactions is characterized by a few dominant-channel interactions in momentum space, it is difficult to derive the real-space spin interactions by directly performing the Fourier transformation owing to ambiguity. 
To tackle this issue, an approach based on machine learning might be appropriate~\cite{sharma2023machine, okigami2024exploring}. 
Once the real-space spin model is obtained, one can investigate the possibility of the isolated skyrmion and its dynamics, which will be useful for future spintronics devices.

The authors thank T.-h. Arima, R. Arita, C. D. Batista, T. Hanaguri, K. Hattori, M. Hirschberger, N. Kanazawa, Y. Kato, N. D. Khanh, K. Kobayashi, S.-Z. Lin, Y. Motome, T. Nakajima, T. Nomoto, K. Okigami, T. Okubo, S. Okumura, R. Ozawa, P. F. S. Rosa, S. Seki, S. Seo, K. Shimizu, D. Singh, R. Takagi, J. S. White, Y. Yasui, and H. Yoshimochi for fruitful collaborations and constructive discussions. 
This research was supported by JSPS KAKENHI Grants Numbers JP21H01037, JP22H04468, JP22H00101, JP22H01183, JP23KJ0557, JP23H04869, JP23K03288, JP23K20827, and by JST PRESTO (JPMJPR20L8) and JST CREST (JPMJCR23O4). 
Parts of the numerical calculations were performed in the supercomputing systems in ISSP, the University of Tokyo.

\bibliographystyle{elsarticle-num-names} 
\bibliography{ref.bib}

\begin{thebibliography}{338}
\expandafter\ifx\csname natexlab\endcsname\relax\def\natexlab#1{#1}\fi
\providecommand{\url}[1]{\texttt{#1}}
\providecommand{\href}[2]{#2}
\providecommand{\path}[1]{#1}
\providecommand{\DOIprefix}{doi:}
\providecommand{\ArXivprefix}{arXiv:}
\providecommand{\URLprefix}{URL: }
\providecommand{\Pubmedprefix}{pmid:}
\providecommand{\doi}[1]{\href{http://dx.doi.org/#1}{\path{#1}}}
\providecommand{\Pubmed}[1]{\href{pmid:#1}{\path{#1}}}
\providecommand{\bibinfo}[2]{#2}
\ifx\xfnm\relax \def\xfnm[#1]{\unskip,\space#1}\fi
\bibitem[{Karplus and Luttinger(1954)}]{Karplus_PhysRev.95.1154}
\bibinfo{author}{R.~Karplus}, \bibinfo{author}{J.~M. Luttinger},
\newblock \bibinfo{title}{{Hall} effect in ferromagnetics},
\newblock \bibinfo{journal}{Phys. Rev.} \bibinfo{volume}{95}
  (\bibinfo{year}{1954}) \bibinfo{pages}{1154--1160}.
  \DOIprefix\doi{10.1103/PhysRev.95.1154}.
\bibitem[{Smit(1958)}]{smit1958spontaneous}
\bibinfo{author}{J.~Smit},
\newblock \bibinfo{title}{The spontaneous {Hall} effect in ferromagnetics
  {II}},
\newblock \bibinfo{journal}{Physica (Amsterdam)} \bibinfo{volume}{24}
  (\bibinfo{year}{1958}) \bibinfo{pages}{39--51}.
\bibitem[{Nagaosa et~al.(2010)Nagaosa, Sinova, Onoda, MacDonald, and
  Ong}]{Nagaosa_RevModPhys.82.1539}
\bibinfo{author}{N.~Nagaosa}, \bibinfo{author}{J.~Sinova},
  \bibinfo{author}{S.~Onoda}, \bibinfo{author}{A.~H. MacDonald},
  \bibinfo{author}{N.~P. Ong},
\newblock \bibinfo{title}{Anomalous {Hall} effect},
\newblock \bibinfo{journal}{Rev. Mod. Phys.} \bibinfo{volume}{82}
  (\bibinfo{year}{2010}) \bibinfo{pages}{1539--1592}.
  \DOIprefix\doi{10.1103/RevModPhys.82.1539}.
\bibitem[{Kimura et~al.(2003)Kimura, Goto, Shintani, Ishizaka, Arima, and
  Tokura}]{kimura2003magnetic}
\bibinfo{author}{T.~Kimura}, \bibinfo{author}{T.~Goto},
  \bibinfo{author}{H.~Shintani}, \bibinfo{author}{K.~Ishizaka},
  \bibinfo{author}{T.~Arima}, \bibinfo{author}{Y.~Tokura},
\newblock \bibinfo{title}{Magnetic control of ferroelectric polarization},
\newblock \bibinfo{journal}{Nature} \bibinfo{volume}{426}
  (\bibinfo{year}{2003}) \bibinfo{pages}{55--58}.
  \DOIprefix\doi{10.1038/nature02018}.
\bibitem[{Katsura et~al.(2005)Katsura, Nagaosa, and
  Balatsky}]{Katsura_PhysRevLett.95.057205}
\bibinfo{author}{H.~Katsura}, \bibinfo{author}{N.~Nagaosa},
  \bibinfo{author}{A.~V. Balatsky},
\newblock \bibinfo{title}{Spin current and magnetoelectric effect in
  noncollinear magnets},
\newblock \bibinfo{journal}{Phys. Rev. Lett.} \bibinfo{volume}{95}
  (\bibinfo{year}{2005}) \bibinfo{pages}{057205}.
  \DOIprefix\doi{10.1103/PhysRevLett.95.057205}.
\bibitem[{Mostovoy(2006)}]{Mostovoy_PhysRevLett.96.067601}
\bibinfo{author}{M.~Mostovoy},
\newblock \bibinfo{title}{Ferroelectricity in spiral magnets},
\newblock \bibinfo{journal}{Phys. Rev. Lett.} \bibinfo{volume}{96}
  (\bibinfo{year}{2006}) \bibinfo{pages}{067601}.
  \DOIprefix\doi{10.1103/PhysRevLett.96.067601}.
\bibitem[{Sergienko and Dagotto(2006)}]{SergienkoPhysRevB.73.094434}
\bibinfo{author}{I.~A. Sergienko}, \bibinfo{author}{E.~Dagotto},
\newblock \bibinfo{title}{Role of the {Dzyaloshinskii}-{Moriya} interaction in
  multiferroic perovskites},
\newblock \bibinfo{journal}{Phys. Rev. B} \bibinfo{volume}{73}
  (\bibinfo{year}{2006}) \bibinfo{pages}{094434}.
  \DOIprefix\doi{10.1103/PhysRevB.73.094434}.
\bibitem[{Harris et~al.(2006)Harris, Yildirim, Aharony, and
  Entin-Wohlman}]{Harris_PhysRevB.73.184433}
\bibinfo{author}{A.~B. Harris}, \bibinfo{author}{T.~Yildirim},
  \bibinfo{author}{A.~Aharony}, \bibinfo{author}{O.~Entin-Wohlman},
\newblock \bibinfo{title}{{Towards a microscopic model of magnetoelectric
  interactions in ${\mathrm{Ni}}_{3}{\mathrm{V}}_{2}{\mathrm{O}}_{8}$}},
\newblock \bibinfo{journal}{Phys. Rev. B} \bibinfo{volume}{73}
  (\bibinfo{year}{2006}) \bibinfo{pages}{184433}.
  \DOIprefix\doi{10.1103/PhysRevB.73.184433}.
\bibitem[{Tokura et~al.(2014)Tokura, Seki, and
  Nagaosa}]{tokura2014multiferroics}
\bibinfo{author}{Y.~Tokura}, \bibinfo{author}{S.~Seki},
  \bibinfo{author}{N.~Nagaosa},
\newblock \bibinfo{title}{Multiferroics of spin origin},
\newblock \bibinfo{journal}{Rep. Prog. Phys.} \bibinfo{volume}{77}
  (\bibinfo{year}{2014}) \bibinfo{pages}{076501}.
  \DOIprefix\doi{10.1088/0034-4885/77/7/076501}.
\bibitem[{Cardias et~al.(2020)Cardias, Szilva, Bezerra-Neto, Ribeiro, Bergman,
  Kvashnin, Fransson, Klautau, Eriksson, and Nordstr{\"o}m}]{cardias2020first}
\bibinfo{author}{R.~Cardias}, \bibinfo{author}{A.~Szilva},
  \bibinfo{author}{M.~Bezerra-Neto}, \bibinfo{author}{M.~Ribeiro},
  \bibinfo{author}{A.~Bergman}, \bibinfo{author}{Y.~O. Kvashnin},
  \bibinfo{author}{J.~Fransson}, \bibinfo{author}{A.~Klautau},
  \bibinfo{author}{O.~Eriksson}, \bibinfo{author}{L.~Nordstr{\"o}m},
\newblock \bibinfo{title}{First-principles {Dzyaloshinskii}--{Moriya}
  interaction in a non-collinear framework},
\newblock \bibinfo{journal}{Sci. Rep.} \bibinfo{volume}{10}
  (\bibinfo{year}{2020}) \bibinfo{pages}{20339}.
  \DOIprefix\doi{10.1038/s41598-020-77219-3}.
\bibitem[{Hayami and Yatsushiro(2022)}]{Hayami_PhysRevB.106.014420}
\bibinfo{author}{S.~Hayami}, \bibinfo{author}{M.~Yatsushiro},
\newblock \bibinfo{title}{Nonlinear nonreciprocal transport in antiferromagnets
  free from spin-orbit coupling},
\newblock \bibinfo{journal}{Phys. Rev. B} \bibinfo{volume}{106}
  (\bibinfo{year}{2022}) \bibinfo{pages}{014420}.
  \DOIprefix\doi{10.1103/PhysRevB.106.014420}.
\bibitem[{Batista et~al.(2016)Batista, Lin, Hayami, and
  Kamiya}]{batista2016frustration}
\bibinfo{author}{C.~D. Batista}, \bibinfo{author}{S.-Z. Lin},
  \bibinfo{author}{S.~Hayami}, \bibinfo{author}{Y.~Kamiya},
\newblock \bibinfo{title}{Frustration and chiral orderings in correlated
  electron systems},
\newblock \bibinfo{journal}{Rep. Prog. Phys.} \bibinfo{volume}{79}
  (\bibinfo{year}{2016}) \bibinfo{pages}{084504}.
  \DOIprefix\doi{10.1088/0034-4885/79/8/084504}.
\bibitem[{Skyrme(1962)}]{skyrme1962unified}
\bibinfo{author}{T.~H.~R. Skyrme},
\newblock \bibinfo{title}{A unified field theory of mesons and baryons},
\newblock \bibinfo{journal}{Nucl. Phys.} \bibinfo{volume}{31}
  (\bibinfo{year}{1962}) \bibinfo{pages}{556--569}.
\bibitem[{Bogdanov and Yablonskii(1989)}]{Bogdanov89}
\bibinfo{author}{A.~N. Bogdanov}, \bibinfo{author}{D.~A. Yablonskii},
\newblock \bibinfo{title}{Thermodynamically stable ``vortices" in magnetically
  ordered crystals: The mixed state of magnets},
\newblock \bibinfo{journal}{Sov. Phys. JETP} \bibinfo{volume}{68}
  (\bibinfo{year}{1989}) \bibinfo{pages}{101}.
\bibitem[{Bogdanov and Hubert(1994)}]{Bogdanov94}
\bibinfo{author}{A.~Bogdanov}, \bibinfo{author}{A.~Hubert},
\newblock \bibinfo{title}{Thermodynamically stable magnetic vortex states in
  magnetic crystals},
\newblock \bibinfo{journal}{J. Magn. Magn. Mater.} \bibinfo{volume}{138}
  (\bibinfo{year}{1994}) \bibinfo{pages}{255 -- 269}.
  \DOIprefix\doi{http://dx.doi.org/10.1016/0304-8853(94)90046-9}.
\bibitem[{R{\"o}{\ss}ler et~al.(2006)R{\"o}{\ss}ler, Bogdanov, and
  Pfleiderer}]{rossler2006spontaneous}
\bibinfo{author}{U.~K. R{\"o}{\ss}ler}, \bibinfo{author}{A.~N. Bogdanov},
  \bibinfo{author}{C.~Pfleiderer},
\newblock \bibinfo{title}{Spontaneous skyrmion ground states in magnetic
  metals},
\newblock \bibinfo{journal}{Nature} \bibinfo{volume}{442}
  (\bibinfo{year}{2006}) \bibinfo{pages}{797--801}.
  \DOIprefix\doi{10.1038/nature05056}.
\bibitem[{Nagaosa and Tokura(2013)}]{nagaosa2013topological}
\bibinfo{author}{N.~Nagaosa}, \bibinfo{author}{Y.~Tokura},
\newblock \bibinfo{title}{Topological properties and dynamics of magnetic
  skyrmions},
\newblock \bibinfo{journal}{Nat. Nanotechnol.} \bibinfo{volume}{8}
  (\bibinfo{year}{2013}) \bibinfo{pages}{899--911}.
  \DOIprefix\doi{10.1038/nnano.2013.243}.
\bibitem[{Ohgushi et~al.(2000)Ohgushi, Murakami, and
  Nagaosa}]{Ohgushi_PhysRevB.62.R6065}
\bibinfo{author}{K.~Ohgushi}, \bibinfo{author}{S.~Murakami},
  \bibinfo{author}{N.~Nagaosa},
\newblock \bibinfo{title}{Spin anisotropy and quantum {Hall} effect in the
  \textit{kagom\'e} lattice: Chiral spin state based on a ferromagnet},
\newblock \bibinfo{journal}{Phys. Rev. B} \bibinfo{volume}{62}
  (\bibinfo{year}{2000}) \bibinfo{pages}{R6065--R6068}.
  \DOIprefix\doi{10.1103/PhysRevB.62.R6065}.
\bibitem[{Kurz et~al.(2001)Kurz, Bihlmayer, Hirai, and
  Bl\"ugel}]{Kurz_PhysRevLett.86.1106}
\bibinfo{author}{P.~Kurz}, \bibinfo{author}{G.~Bihlmayer},
  \bibinfo{author}{K.~Hirai}, \bibinfo{author}{S.~Bl\"ugel},
\newblock \bibinfo{title}{{Three-Dimensional Spin Structure on a
  Two-Dimensional Lattice: Mn$/$Cu(111)}},
\newblock \bibinfo{journal}{Phys. Rev. Lett.} \bibinfo{volume}{86}
  (\bibinfo{year}{2001}) \bibinfo{pages}{1106--1109}.
  \DOIprefix\doi{10.1103/PhysRevLett.86.1106}.
\bibitem[{Martin and Batista(2008)}]{Martin_PhysRevLett.101.156402}
\bibinfo{author}{I.~Martin}, \bibinfo{author}{C.~D. Batista},
\newblock \bibinfo{title}{Itinerant electron-driven chiral magnetic ordering
  and spontaneous quantum {Hall} effect in triangular lattice models},
\newblock \bibinfo{journal}{Phys. Rev. Lett.} \bibinfo{volume}{101}
  (\bibinfo{year}{2008}) \bibinfo{pages}{156402}.
  \DOIprefix\doi{10.1103/PhysRevLett.101.156402}.
\bibitem[{Akagi and Motome(2010)}]{Akagi_JPSJ.79.083711}
\bibinfo{author}{Y.~Akagi}, \bibinfo{author}{Y.~Motome},
\newblock \bibinfo{title}{Spin chirality ordering and anomalous {Hall} effect
  in the ferromagnetic {Kondo} lattice model on a triangular lattice},
\newblock \bibinfo{journal}{J. Phys. Soc. Jpn.} \bibinfo{volume}{79}
  (\bibinfo{year}{2010}) \bibinfo{pages}{083711}.
  \DOIprefix\doi{10.1143/JPSJ.79.083711}.
\bibitem[{Kumar and van~den Brink(2010)}]{Kumar_PhysRevLett.105.216405}
\bibinfo{author}{S.~Kumar}, \bibinfo{author}{J.~van~den Brink},
\newblock \bibinfo{title}{Frustration-induced insulating chiral spin state in
  itinerant triangular-lattice magnets},
\newblock \bibinfo{journal}{Phys. Rev. Lett.} \bibinfo{volume}{105}
  (\bibinfo{year}{2010}) \bibinfo{pages}{216405}.
  \DOIprefix\doi{10.1103/PhysRevLett.105.216405}.
\bibitem[{Berry(1984)}]{berry1984quantal}
\bibinfo{author}{M.~V. Berry},
\newblock \bibinfo{title}{Quantal phase factors accompanying adiabatic
  changes},
\newblock \bibinfo{journal}{Proceedings of the Royal Society of London A:
  Mathematical, Physical and Engineering Sciences} \bibinfo{volume}{392}
  (\bibinfo{year}{1984}) \bibinfo{pages}{45--57}.
  \DOIprefix\doi{10.1098/rspa.1984.0023}.
\bibitem[{Loss and Goldbart(1992)}]{Loss_PhysRevB.45.13544}
\bibinfo{author}{D.~Loss}, \bibinfo{author}{P.~M. Goldbart},
\newblock \bibinfo{title}{Persistent currents from {Berry's} phase in
  mesoscopic systems},
\newblock \bibinfo{journal}{Phys. Rev. B} \bibinfo{volume}{45}
  (\bibinfo{year}{1992}) \bibinfo{pages}{13544--13561}.
  \DOIprefix\doi{10.1103/PhysRevB.45.13544}.
\bibitem[{Ye et~al.(1999)Ye, Kim, Millis, Shraiman, Majumdar, and Te\ifmmode
  \check{s}\else \v{s}\fi{}anovi\ifmmode~\acute{c}\else
  \'{c}\fi{}}]{Ye_PhysRevLett.83.3737}
\bibinfo{author}{J.~Ye}, \bibinfo{author}{Y.~B. Kim}, \bibinfo{author}{A.~J.
  Millis}, \bibinfo{author}{B.~I. Shraiman}, \bibinfo{author}{P.~Majumdar},
  \bibinfo{author}{Z.~Te\ifmmode \check{s}\else
  \v{s}\fi{}anovi\ifmmode~\acute{c}\else \'{c}\fi{}},
\newblock \bibinfo{title}{Berry phase theory of the anomalous {Hall} effect:
  Application to colossal magnetoresistance manganites},
\newblock \bibinfo{journal}{Phys. Rev. Lett.} \bibinfo{volume}{83}
  (\bibinfo{year}{1999}) \bibinfo{pages}{3737--3740}.
  \DOIprefix\doi{10.1103/PhysRevLett.83.3737}.
\bibitem[{Shindou and Nagaosa(2001)}]{Shindou_PhysRevLett.87.116801}
\bibinfo{author}{R.~Shindou}, \bibinfo{author}{N.~Nagaosa},
\newblock \bibinfo{title}{Orbital ferromagnetism and anomalous {Hall} effect in
  antiferromagnets on the distorted fcc lattice},
\newblock \bibinfo{journal}{Phys. Rev. Lett.} \bibinfo{volume}{87}
  (\bibinfo{year}{2001}) \bibinfo{pages}{116801}.
  \DOIprefix\doi{10.1103/PhysRevLett.87.116801}.
\bibitem[{Xiao et~al.(2010)Xiao, Chang, and Niu}]{Xiao_RevModPhys.82.1959}
\bibinfo{author}{D.~Xiao}, \bibinfo{author}{M.-C. Chang},
  \bibinfo{author}{Q.~Niu},
\newblock \bibinfo{title}{Berry phase effects on electronic properties},
\newblock \bibinfo{journal}{Rev. Mod. Phys.} \bibinfo{volume}{82}
  (\bibinfo{year}{2010}) \bibinfo{pages}{1959--2007}.
  \DOIprefix\doi{10.1103/RevModPhys.82.1959}.
\bibitem[{Taguchi et~al.(2001)Taguchi, Oohara, Yoshizawa, Nagaosa, and
  Tokura}]{taguchi2001spin}
\bibinfo{author}{Y.~Taguchi}, \bibinfo{author}{Y.~Oohara},
  \bibinfo{author}{H.~Yoshizawa}, \bibinfo{author}{N.~Nagaosa},
  \bibinfo{author}{Y.~Tokura},
\newblock \bibinfo{title}{Spin chirality, {Berry} phase, and anomalous {Hall}
  effect in a frustrated ferromagnet},
\newblock \bibinfo{journal}{Science} \bibinfo{volume}{291}
  (\bibinfo{year}{2001}) \bibinfo{pages}{2573--2576}.
  \DOIprefix\doi{10.1126/science.1058161}.
\bibitem[{Tatara and Kawamura(2002)}]{tatara2002chirality}
\bibinfo{author}{G.~Tatara}, \bibinfo{author}{H.~Kawamura},
\newblock \bibinfo{title}{Chirality-driven anomalous {Hall} effect in weak
  coupling regime},
\newblock \bibinfo{journal}{J. Phys. Soc. Jpn.} \bibinfo{volume}{71}
  (\bibinfo{year}{2002}) \bibinfo{pages}{2613--2616}.
  \DOIprefix\doi{10.1143/JPSJ.71.2613}.
\bibitem[{Neubauer et~al.(2009)Neubauer, Pfleiderer, Binz, Rosch, Ritz,
  Niklowitz, and B\"oni}]{Neubauer_PhysRevLett.102.186602}
\bibinfo{author}{A.~Neubauer}, \bibinfo{author}{C.~Pfleiderer},
  \bibinfo{author}{B.~Binz}, \bibinfo{author}{A.~Rosch},
  \bibinfo{author}{R.~Ritz}, \bibinfo{author}{P.~G. Niklowitz},
  \bibinfo{author}{P.~B\"oni},
\newblock \bibinfo{title}{Topological {Hall} effect in the ${A}$ phase of
  {MnSi}},
\newblock \bibinfo{journal}{Phys. Rev. Lett.} \bibinfo{volume}{102}
  (\bibinfo{year}{2009}) \bibinfo{pages}{186602}.
  \DOIprefix\doi{10.1103/PhysRevLett.102.186602}.
\bibitem[{Shiomi et~al.(2013)Shiomi, Kanazawa, Shibata, Onose, and
  Tokura}]{Shiomi_PhysRevB.88.064409}
\bibinfo{author}{Y.~Shiomi}, \bibinfo{author}{N.~Kanazawa},
  \bibinfo{author}{K.~Shibata}, \bibinfo{author}{Y.~Onose},
  \bibinfo{author}{Y.~Tokura},
\newblock \bibinfo{title}{Topological nernst effect in a three-dimensional
  skyrmion-lattice phase},
\newblock \bibinfo{journal}{Phys. Rev. B} \bibinfo{volume}{88}
  (\bibinfo{year}{2013}) \bibinfo{pages}{064409}.
  \DOIprefix\doi{10.1103/PhysRevB.88.064409}.
\bibitem[{Hamamoto et~al.(2015)Hamamoto, Ezawa, and
  Nagaosa}]{Hamamoto_PhysRevB.92.115417}
\bibinfo{author}{K.~Hamamoto}, \bibinfo{author}{M.~Ezawa},
  \bibinfo{author}{N.~Nagaosa},
\newblock \bibinfo{title}{Quantized topological {Hall} effect in skyrmion
  crystal},
\newblock \bibinfo{journal}{Phys. Rev. B} \bibinfo{volume}{92}
  (\bibinfo{year}{2015}) \bibinfo{pages}{115417}.
  \DOIprefix\doi{10.1103/PhysRevB.92.115417}.
\bibitem[{Nakazawa et~al.(2018)Nakazawa, Bibes, and
  Kohno}]{nakazawa2018topological}
\bibinfo{author}{K.~Nakazawa}, \bibinfo{author}{M.~Bibes},
  \bibinfo{author}{H.~Kohno},
\newblock \bibinfo{title}{Topological {Hall} effect from strong to weak
  coupling},
\newblock \bibinfo{journal}{J. Phys. Soc. Jpn.} \bibinfo{volume}{87}
  (\bibinfo{year}{2018}) \bibinfo{pages}{033705}.
  \DOIprefix\doi{10.7566/JPSJ.87.033705}.
\bibitem[{Jonietz et~al.(2010)Jonietz, M$\ddot{\rm u}$hlbauer, Pfleiderer,
  Neubauer, M$\ddot{\rm u}$nzer, Bauer, Adams, Georgii, B$\ddot{\rm o}$ni,
  Duine, Everschor, Garst, and Rosch}]{Jonietz_skyrmion}
\bibinfo{author}{F.~Jonietz}, \bibinfo{author}{S.~M$\ddot{\rm u}$hlbauer},
  \bibinfo{author}{C.~Pfleiderer}, \bibinfo{author}{A.~Neubauer},
  \bibinfo{author}{W.~M$\ddot{\rm u}$nzer}, \bibinfo{author}{A.~Bauer},
  \bibinfo{author}{T.~Adams}, \bibinfo{author}{R.~Georgii},
  \bibinfo{author}{P.~B$\ddot{\rm o}$ni}, \bibinfo{author}{R.~A. Duine},
  \bibinfo{author}{K.~Everschor}, \bibinfo{author}{M.~Garst},
  \bibinfo{author}{A.~Rosch},
\newblock \bibinfo{title}{{Spin Transfer Torques in MnSi at Ultralow Current
  Densities}},
\newblock \bibinfo{journal}{Science} \bibinfo{volume}{330}
  (\bibinfo{year}{2010}) \bibinfo{pages}{1648}.
  \DOIprefix\doi{10.1126/science.1195709}.
\bibitem[{Yu et~al.(2012)Yu, Kanazawa, Zhang, Nagai, Hara, Kimoto, Matsui,
  Onose, and Tokura}]{yu2012skyrmion}
\bibinfo{author}{X.~Z. Yu}, \bibinfo{author}{N.~Kanazawa},
  \bibinfo{author}{W.~Zhang}, \bibinfo{author}{T.~Nagai},
  \bibinfo{author}{T.~Hara}, \bibinfo{author}{K.~Kimoto},
  \bibinfo{author}{Y.~Matsui}, \bibinfo{author}{Y.~Onose},
  \bibinfo{author}{Y.~Tokura},
\newblock \bibinfo{title}{Skyrmion flow near room temperature in an ultralow
  current density},
\newblock \bibinfo{journal}{Nat. Commun.} \bibinfo{volume}{3}
  (\bibinfo{year}{2012}) \bibinfo{pages}{988}.
  \DOIprefix\doi{10.1038/ncomms1990}.
\bibitem[{Jiang et~al.(2017)Jiang, Zhang, Yu, Zhang, Wang,
  Benjamin~Jungfleisch, Pearson, Cheng, Heinonen, Wang, Zhou, Hoffmann, and
  te~Velthuis}]{jiang2017direct}
\bibinfo{author}{W.~Jiang}, \bibinfo{author}{X.~Zhang},
  \bibinfo{author}{G.~Yu}, \bibinfo{author}{W.~Zhang},
  \bibinfo{author}{X.~Wang}, \bibinfo{author}{M.~Benjamin~Jungfleisch},
  \bibinfo{author}{J.~E. Pearson}, \bibinfo{author}{X.~Cheng},
  \bibinfo{author}{O.~Heinonen}, \bibinfo{author}{K.~L. Wang},
  \bibinfo{author}{Y.~Zhou}, \bibinfo{author}{A.~Hoffmann},
  \bibinfo{author}{G.~E. te~Velthuis},
\newblock \bibinfo{title}{Direct observation of the skyrmion {Hall} effect},
\newblock \bibinfo{journal}{Nat. Phys.} \bibinfo{volume}{13}
  (\bibinfo{year}{2017}) \bibinfo{pages}{162--169}.
  \DOIprefix\doi{https://doi.org/10.1038/nphys3883}.
\bibitem[{Yu et~al.(2020)Yu, Morikawa, Nakajima, Shibata, Kanazawa, Arima,
  Nagaosa, and Tokura}]{yu2020motion}
\bibinfo{author}{X.~Yu}, \bibinfo{author}{D.~Morikawa},
  \bibinfo{author}{K.~Nakajima}, \bibinfo{author}{K.~Shibata},
  \bibinfo{author}{N.~Kanazawa}, \bibinfo{author}{T.-h. Arima},
  \bibinfo{author}{N.~Nagaosa}, \bibinfo{author}{Y.~Tokura},
\newblock \bibinfo{title}{Motion tracking of 80-nm-size skyrmions upon
  directional current injections},
\newblock \bibinfo{journal}{Sci. Adv.} \bibinfo{volume}{6}
  (\bibinfo{year}{2020}) \bibinfo{pages}{eaaz9744}.
  \DOIprefix\doi{10.1126/sciadv.aaz9744}.
\bibitem[{Okamura et~al.(2013)Okamura, Kagawa, Mochizuki, Kubota, Seki,
  Ishiwata, Kawasaki, Onose, and Tokura}]{okamura2013microwave}
\bibinfo{author}{Y.~Okamura}, \bibinfo{author}{F.~Kagawa},
  \bibinfo{author}{M.~Mochizuki}, \bibinfo{author}{M.~Kubota},
  \bibinfo{author}{S.~Seki}, \bibinfo{author}{S.~Ishiwata},
  \bibinfo{author}{M.~Kawasaki}, \bibinfo{author}{Y.~Onose},
  \bibinfo{author}{Y.~Tokura},
\newblock \bibinfo{title}{Microwave magnetoelectric effect via skyrmion
  resonance modes in a helimagnetic multiferroic},
\newblock \bibinfo{journal}{Nat. Commun.} \bibinfo{volume}{4}
  (\bibinfo{year}{2013}) \bibinfo{pages}{2391}.
  \DOIprefix\doi{https://doi.org/10.1038/ncomms3391}.
\bibitem[{Mochizuki(2012)}]{Mochizuki_PhysRevLett.108.017601}
\bibinfo{author}{M.~Mochizuki},
\newblock \bibinfo{title}{Spin-wave modes and their intense excitation effects
  in skyrmion crystals},
\newblock \bibinfo{journal}{Phys. Rev. Lett.} \bibinfo{volume}{108}
  (\bibinfo{year}{2012}) \bibinfo{pages}{017601}.
  \DOIprefix\doi{10.1103/PhysRevLett.108.017601}.
\bibitem[{Onose et~al.(2012)Onose, Okamura, Seki, Ishiwata, and
  Tokura}]{Onose_PhysRevLett.109.037603}
\bibinfo{author}{Y.~Onose}, \bibinfo{author}{Y.~Okamura},
  \bibinfo{author}{S.~Seki}, \bibinfo{author}{S.~Ishiwata},
  \bibinfo{author}{Y.~Tokura},
\newblock \bibinfo{title}{{Observation of Magnetic Excitations of Skyrmion
  Crystal in a Helimagnetic Insulator ${\mathrm{Cu}}_{2}{\mathrm{OSeO}}_{3}$}},
\newblock \bibinfo{journal}{Phys. Rev. Lett.} \bibinfo{volume}{109}
  (\bibinfo{year}{2012}) \bibinfo{pages}{037603}.
  \DOIprefix\doi{10.1103/PhysRevLett.109.037603}.
\bibitem[{Jiang et~al.(2015)Jiang, Upadhyaya, Zhang, Yu, Jungfleisch, Fradin,
  Pearson, Tserkovnyak, Wang, Heinonen et~al.}]{jiang2015blowing}
\bibinfo{author}{W.~Jiang}, \bibinfo{author}{P.~Upadhyaya},
  \bibinfo{author}{W.~Zhang}, \bibinfo{author}{G.~Yu}, \bibinfo{author}{M.~B.
  Jungfleisch}, \bibinfo{author}{F.~Y. Fradin}, \bibinfo{author}{J.~E.
  Pearson}, \bibinfo{author}{Y.~Tserkovnyak}, \bibinfo{author}{K.~L. Wang},
  \bibinfo{author}{O.~Heinonen}, et~al.,
\newblock \bibinfo{title}{Blowing magnetic skyrmion bubbles},
\newblock \bibinfo{journal}{Science} \bibinfo{volume}{349}
  (\bibinfo{year}{2015}) \bibinfo{pages}{283--286}.
  \DOIprefix\doi{10.1126/science.aaa1442}.
\bibitem[{B{\"u}ttner et~al.(2017)B{\"u}ttner, Lemesh, Schneider, Pfau,
  G{\"u}nther, Hessing, Geilhufe, Caretta, Engel, Kr{\"u}ger
  et~al.}]{buttner2017field}
\bibinfo{author}{F.~B{\"u}ttner}, \bibinfo{author}{I.~Lemesh},
  \bibinfo{author}{M.~Schneider}, \bibinfo{author}{B.~Pfau},
  \bibinfo{author}{C.~M. G{\"u}nther}, \bibinfo{author}{P.~Hessing},
  \bibinfo{author}{J.~Geilhufe}, \bibinfo{author}{L.~Caretta},
  \bibinfo{author}{D.~Engel}, \bibinfo{author}{B.~Kr{\"u}ger}, et~al.,
\newblock \bibinfo{title}{Field-free deterministic ultrafast creation of
  magnetic skyrmions by spin--orbit torques},
\newblock \bibinfo{journal}{Nat. Nanotech.} \bibinfo{volume}{12}
  (\bibinfo{year}{2017}) \bibinfo{pages}{1040--1044}.
  \DOIprefix\doi{https://doi.org/10.1038/nnano.2017.178}.
\bibitem[{Hrabec et~al.(2017)Hrabec, Sampaio, Belmeguenai, Gross, Weil,
  Ch{\'e}rif, Stashkevich, Jacques, Thiaville, and Rohart}]{hrabec2017current}
\bibinfo{author}{A.~Hrabec}, \bibinfo{author}{J.~Sampaio},
  \bibinfo{author}{M.~Belmeguenai}, \bibinfo{author}{I.~Gross},
  \bibinfo{author}{R.~Weil}, \bibinfo{author}{S.~M. Ch{\'e}rif},
  \bibinfo{author}{A.~Stashkevich}, \bibinfo{author}{V.~Jacques},
  \bibinfo{author}{A.~Thiaville}, \bibinfo{author}{S.~Rohart},
\newblock \bibinfo{title}{Current-induced skyrmion generation and dynamics in
  symmetric bilayers},
\newblock \bibinfo{journal}{Nat. Commun.} \bibinfo{volume}{8}
  (\bibinfo{year}{2017}). \DOIprefix\doi{10.1038/ncomms15765}.
\bibitem[{Fert et~al.(2013)Fert, Cros, and Sampaio}]{fert2013skyrmions}
\bibinfo{author}{A.~Fert}, \bibinfo{author}{V.~Cros},
  \bibinfo{author}{J.~Sampaio},
\newblock \bibinfo{title}{Skyrmions on the track},
\newblock \bibinfo{journal}{Nat. Nanotechnol.} \bibinfo{volume}{8}
  (\bibinfo{year}{2013}) \bibinfo{pages}{152}.
  \DOIprefix\doi{10.1038/nnano.2013.29}.
\bibitem[{Fert et~al.(2017)Fert, Reyren, and Cros}]{fert2017magnetic}
\bibinfo{author}{A.~Fert}, \bibinfo{author}{N.~Reyren},
  \bibinfo{author}{V.~Cros},
\newblock \bibinfo{title}{Magnetic skyrmions: advances in physics and potential
  applications},
\newblock \bibinfo{journal}{Nat. Rev. Mater.} \bibinfo{volume}{2}
  (\bibinfo{year}{2017}) \bibinfo{pages}{17031}.
  \DOIprefix\doi{10.1038/natrevmats.2017.31}.
\bibitem[{Zhang et~al.(2020)Zhang, Zhou, Song, Park, Xia, Ezawa, Liu, Zhao,
  Zhao, and Woo}]{zhang2020skyrmion}
\bibinfo{author}{X.~Zhang}, \bibinfo{author}{Y.~Zhou}, \bibinfo{author}{K.~M.
  Song}, \bibinfo{author}{T.-E. Park}, \bibinfo{author}{J.~Xia},
  \bibinfo{author}{M.~Ezawa}, \bibinfo{author}{X.~Liu},
  \bibinfo{author}{W.~Zhao}, \bibinfo{author}{G.~Zhao},
  \bibinfo{author}{S.~Woo},
\newblock \bibinfo{title}{Skyrmion-electronics: writing, deleting, reading and
  processing magnetic skyrmions toward spintronic applications},
\newblock \bibinfo{journal}{J. Phys.: Condens. Matter} \bibinfo{volume}{32}
  (\bibinfo{year}{2020}) \bibinfo{pages}{143001}.
  \DOIprefix\doi{10.1088/1361-648X/ab5488}.
\bibitem[{Zhang et~al.(2015)Zhang, Ezawa, and Zhou}]{zhang2015magnetic}
\bibinfo{author}{X.~Zhang}, \bibinfo{author}{M.~Ezawa},
  \bibinfo{author}{Y.~Zhou},
\newblock \bibinfo{title}{Magnetic skyrmion logic gates: conversion,
  duplication and merging of skyrmions},
\newblock \bibinfo{journal}{Sci. Rep.} \bibinfo{volume}{5}
  (\bibinfo{year}{2015}) \bibinfo{pages}{9400}.
  \DOIprefix\doi{https://doi.org/10.1038/srep09400}.
\bibitem[{Luo et~al.(2018)Luo, Song, Li, Zhang, Hong, Yang, Zou, Xu, and
  You}]{luo2018reconfigurable}
\bibinfo{author}{S.~Luo}, \bibinfo{author}{M.~Song}, \bibinfo{author}{X.~Li},
  \bibinfo{author}{Y.~Zhang}, \bibinfo{author}{J.~Hong},
  \bibinfo{author}{X.~Yang}, \bibinfo{author}{X.~Zou}, \bibinfo{author}{N.~Xu},
  \bibinfo{author}{L.~You},
\newblock \bibinfo{title}{Reconfigurable skyrmion logic gates},
\newblock \bibinfo{journal}{Nano Lett.} \bibinfo{volume}{18}
  (\bibinfo{year}{2018}) \bibinfo{pages}{1180--1184}.
  \DOIprefix\doi{10.1021/acs.nanolett.7b04722}.
\bibitem[{Chauwin et~al.(2019)Chauwin, Hu, Garcia-Sanchez, Betrabet, Paler,
  Moutafis, and Friedman}]{Chauwin_PhysRevApplied.12.064053}
\bibinfo{author}{M.~Chauwin}, \bibinfo{author}{X.~Hu},
  \bibinfo{author}{F.~Garcia-Sanchez}, \bibinfo{author}{N.~Betrabet},
  \bibinfo{author}{A.~Paler}, \bibinfo{author}{C.~Moutafis},
  \bibinfo{author}{J.~S. Friedman},
\newblock \bibinfo{title}{Skyrmion logic system for large-scale reversible
  computation},
\newblock \bibinfo{journal}{Phys. Rev. Appl.} \bibinfo{volume}{12}
  (\bibinfo{year}{2019}) \bibinfo{pages}{064053}.
  \DOIprefix\doi{10.1103/PhysRevApplied.12.064053}.
\bibitem[{Tokura and Kanazawa(2021)}]{Tokura_doi:10.1021/acs.chemrev.0c00297}
\bibinfo{author}{Y.~Tokura}, \bibinfo{author}{N.~Kanazawa},
\newblock \bibinfo{title}{Magnetic skyrmion materials},
\newblock \bibinfo{journal}{Chem. Rev.} \bibinfo{volume}{121}
  (\bibinfo{year}{2021}) \bibinfo{pages}{2857}.
  \DOIprefix\doi{10.1021/acs.chemrev.0c00297}.
\bibitem[{M{\"u}hlbauer et~al.(2009)M{\"u}hlbauer, Binz, Jonietz, Pfleiderer,
  Rosch, Neubauer, Georgii, and B{\"o}ni}]{Muhlbauer_2009skyrmion}
\bibinfo{author}{S.~M{\"u}hlbauer}, \bibinfo{author}{B.~Binz},
  \bibinfo{author}{F.~Jonietz}, \bibinfo{author}{C.~Pfleiderer},
  \bibinfo{author}{A.~Rosch}, \bibinfo{author}{A.~Neubauer},
  \bibinfo{author}{R.~Georgii}, \bibinfo{author}{P.~B{\"o}ni},
\newblock \bibinfo{title}{Skyrmion lattice in a chiral magnet},
\newblock \bibinfo{journal}{Science} \bibinfo{volume}{323}
  (\bibinfo{year}{2009}) \bibinfo{pages}{915--919}.
  \DOIprefix\doi{10.1126/science.1166767}.
\bibitem[{Adams et~al.(2011)Adams, M\"uhlbauer, Pfleiderer, Jonietz, Bauer,
  Neubauer, Georgii, B\"oni, Keiderling, Everschor, Garst, and
  Rosch}]{Adams_PhysRevLett.107.217206}
\bibinfo{author}{T.~Adams}, \bibinfo{author}{S.~M\"uhlbauer},
  \bibinfo{author}{C.~Pfleiderer}, \bibinfo{author}{F.~Jonietz},
  \bibinfo{author}{A.~Bauer}, \bibinfo{author}{A.~Neubauer},
  \bibinfo{author}{R.~Georgii}, \bibinfo{author}{P.~B\"oni},
  \bibinfo{author}{U.~Keiderling}, \bibinfo{author}{K.~Everschor},
  \bibinfo{author}{M.~Garst}, \bibinfo{author}{A.~Rosch},
\newblock \bibinfo{title}{Long-range crystalline nature of the skyrmion lattice
  in {MnSi}},
\newblock \bibinfo{journal}{Phys. Rev. Lett.} \bibinfo{volume}{107}
  (\bibinfo{year}{2011}) \bibinfo{pages}{217206}.
  \DOIprefix\doi{10.1103/PhysRevLett.107.217206}.
\bibitem[{Bauer and Pfleiderer(2012)}]{Bauer_PhysRevB.85.214418}
\bibinfo{author}{A.~Bauer}, \bibinfo{author}{C.~Pfleiderer},
\newblock \bibinfo{title}{Magnetic phase diagram of {MnSi} inferred from
  magnetization and ac susceptibility},
\newblock \bibinfo{journal}{Phys. Rev. B} \bibinfo{volume}{85}
  (\bibinfo{year}{2012}) \bibinfo{pages}{214418}.
  \DOIprefix\doi{10.1103/PhysRevB.85.214418}.
\bibitem[{Bauer et~al.(2013)Bauer, Garst, and
  Pfleiderer}]{Bauer_PhysRevLett.110.177207}
\bibinfo{author}{A.~Bauer}, \bibinfo{author}{M.~Garst},
  \bibinfo{author}{C.~Pfleiderer},
\newblock \bibinfo{title}{Specific heat of the skyrmion lattice phase and
  field-induced tricritical point in {MnSi}},
\newblock \bibinfo{journal}{Phys. Rev. Lett.} \bibinfo{volume}{110}
  (\bibinfo{year}{2013}) \bibinfo{pages}{177207}.
  \DOIprefix\doi{10.1103/PhysRevLett.110.177207}.
\bibitem[{Chacon et~al.(2015)Chacon, Bauer, Adams, Rucker, Brandl, Georgii,
  Garst, and Pfleiderer}]{Chacon_PhysRevLett.115.267202}
\bibinfo{author}{A.~Chacon}, \bibinfo{author}{A.~Bauer},
  \bibinfo{author}{T.~Adams}, \bibinfo{author}{F.~Rucker},
  \bibinfo{author}{G.~Brandl}, \bibinfo{author}{R.~Georgii},
  \bibinfo{author}{M.~Garst}, \bibinfo{author}{C.~Pfleiderer},
\newblock \bibinfo{title}{Uniaxial pressure dependence of magnetic order in
  {MnSi}},
\newblock \bibinfo{journal}{Phys. Rev. Lett.} \bibinfo{volume}{115}
  (\bibinfo{year}{2015}) \bibinfo{pages}{267202}.
  \DOIprefix\doi{10.1103/PhysRevLett.115.267202}.
\bibitem[{M{\"u}hlbauer et~al.(2016)M{\"u}hlbauer, Kindervater, Adams, Bauer,
  Keiderling, and Pfleiderer}]{muhlbauer2016kinetic}
\bibinfo{author}{S.~M{\"u}hlbauer}, \bibinfo{author}{J.~Kindervater},
  \bibinfo{author}{T.~Adams}, \bibinfo{author}{A.~Bauer},
  \bibinfo{author}{U.~Keiderling}, \bibinfo{author}{C.~Pfleiderer},
\newblock \bibinfo{title}{Kinetic small angle neutron scattering of the
  skyrmion lattice in {MnSi}},
\newblock \bibinfo{journal}{New J. Phys.} \bibinfo{volume}{18}
  (\bibinfo{year}{2016}) \bibinfo{pages}{075017}.
  \DOIprefix\doi{10.1088/1367-2630/18/7/075017}.
\bibitem[{Reiner et~al.(2016)Reiner, Bauer, Leitner, Gigl, Anwand, Butterling,
  Wagner, Kudejova, Pfleiderer, and Hugenschmidt}]{reiner2016positron}
\bibinfo{author}{M.~Reiner}, \bibinfo{author}{A.~Bauer},
  \bibinfo{author}{M.~Leitner}, \bibinfo{author}{T.~Gigl},
  \bibinfo{author}{W.~Anwand}, \bibinfo{author}{M.~Butterling},
  \bibinfo{author}{A.~Wagner}, \bibinfo{author}{P.~Kudejova},
  \bibinfo{author}{C.~Pfleiderer}, \bibinfo{author}{C.~Hugenschmidt},
\newblock \bibinfo{title}{Positron spectroscopy of point defects in the
  skyrmion-lattice compound {MnSi}},
\newblock \bibinfo{journal}{Sci. Rep.} \bibinfo{volume}{6}
  (\bibinfo{year}{2016}) \bibinfo{pages}{29109}.
  \DOIprefix\doi{https://doi.org/10.1038/srep29109}.
\bibitem[{Yu et~al.(2010)Yu, Onose, Kanazawa, Park, Han, Matsui, Nagaosa, and
  Tokura}]{yu2010real}
\bibinfo{author}{X.~Z. Yu}, \bibinfo{author}{Y.~Onose},
  \bibinfo{author}{N.~Kanazawa}, \bibinfo{author}{J.~H. Park},
  \bibinfo{author}{J.~H. Han}, \bibinfo{author}{Y.~Matsui},
  \bibinfo{author}{N.~Nagaosa}, \bibinfo{author}{Y.~Tokura},
\newblock \bibinfo{title}{Real-space observation of a two-dimensional skyrmion
  crystal},
\newblock \bibinfo{journal}{Nature} \bibinfo{volume}{465}
  (\bibinfo{year}{2010}) \bibinfo{pages}{901--904}.
  \DOIprefix\doi{10.1038/nature09124}.
\bibitem[{M\"unzer et~al.(2010)M\"unzer, Neubauer, Adams, M\"uhlbauer, Franz,
  Jonietz, Georgii, B\"oni, Pedersen, Schmidt, Rosch, and
  Pfleiderer}]{Munzer_PhysRevB.81.041203}
\bibinfo{author}{W.~M\"unzer}, \bibinfo{author}{A.~Neubauer},
  \bibinfo{author}{T.~Adams}, \bibinfo{author}{S.~M\"uhlbauer},
  \bibinfo{author}{C.~Franz}, \bibinfo{author}{F.~Jonietz},
  \bibinfo{author}{R.~Georgii}, \bibinfo{author}{P.~B\"oni},
  \bibinfo{author}{B.~Pedersen}, \bibinfo{author}{M.~Schmidt},
  \bibinfo{author}{A.~Rosch}, \bibinfo{author}{C.~Pfleiderer},
\newblock \bibinfo{title}{Skyrmion lattice in the doped semiconductor
  {Fe$_{1-x}$Co$_{x}$Si}},
\newblock \bibinfo{journal}{Phys. Rev. B} \bibinfo{volume}{81}
  (\bibinfo{year}{2010}) \bibinfo{pages}{041203}.
  \DOIprefix\doi{10.1103/PhysRevB.81.041203}.
\bibitem[{Adams et~al.(2010)Adams, M{\"u}hlbauer, Neubauer, M{\"u}nzer,
  Jonietz, Georgii, Pedersen, B{\"o}ni, Rosch, and
  Pfleiderer}]{adams2010skyrmion}
\bibinfo{author}{T.~Adams}, \bibinfo{author}{S.~M{\"u}hlbauer},
  \bibinfo{author}{A.~Neubauer}, \bibinfo{author}{W.~M{\"u}nzer},
  \bibinfo{author}{F.~Jonietz}, \bibinfo{author}{R.~Georgii},
  \bibinfo{author}{B.~Pedersen}, \bibinfo{author}{P.~B{\"o}ni},
  \bibinfo{author}{A.~Rosch}, \bibinfo{author}{C.~Pfleiderer},
\newblock \bibinfo{title}{Skyrmion lattice domains in {Fe$_{1- x}$Co$_x$Si}},
\newblock in: \bibinfo{booktitle}{J. Phys.: Conf. Ser.}, volume
  \bibinfo{volume}{200}, \bibinfo{organization}{IOP Publishing},
  \bibinfo{year}{2010}, p. \bibinfo{pages}{032001}.
  \DOIprefix\doi{10.1088/1742-6596/200/3/032001}.
\bibitem[{Yu et~al.(2011)Yu, Kanazawa, Onose, Kimoto, Zhang, Ishiwata, Matsui,
  and Tokura}]{yu2011near}
\bibinfo{author}{X.~Z. Yu}, \bibinfo{author}{N.~Kanazawa},
  \bibinfo{author}{Y.~Onose}, \bibinfo{author}{K.~Kimoto},
  \bibinfo{author}{W.~Zhang}, \bibinfo{author}{S.~Ishiwata},
  \bibinfo{author}{Y.~Matsui}, \bibinfo{author}{Y.~Tokura},
\newblock \bibinfo{title}{Near room-temperature formation of a skyrmion crystal
  in thin-films of the helimagnet {FeGe}},
\newblock \bibinfo{journal}{Nat. Mater.} \bibinfo{volume}{10}
  (\bibinfo{year}{2011}) \bibinfo{pages}{106--109}.
  \DOIprefix\doi{10.1038/nmat2916}.
\bibitem[{Gallagher et~al.(2017)Gallagher, Meng, Brangham, Wang, Esser, McComb,
  and Yang}]{Gallagher_PhysRevLett.118.027201}
\bibinfo{author}{J.~C. Gallagher}, \bibinfo{author}{K.~Y. Meng},
  \bibinfo{author}{J.~T. Brangham}, \bibinfo{author}{H.~L. Wang},
  \bibinfo{author}{B.~D. Esser}, \bibinfo{author}{D.~W. McComb},
  \bibinfo{author}{F.~Y. Yang},
\newblock \bibinfo{title}{{Robust Zero-Field Skyrmion Formation in FeGe
  Epitaxial Thin Films}},
\newblock \bibinfo{journal}{Phys. Rev. Lett.} \bibinfo{volume}{118}
  (\bibinfo{year}{2017}) \bibinfo{pages}{027201}.
  \DOIprefix\doi{10.1103/PhysRevLett.118.027201}.
\bibitem[{Turgut et~al.(2018)Turgut, Paik, Nguyen, Muller, Schlom, and
  Fuchs}]{Turgut_PhysRevMaterials.2.074404}
\bibinfo{author}{E.~Turgut}, \bibinfo{author}{H.~Paik},
  \bibinfo{author}{K.~Nguyen}, \bibinfo{author}{D.~A. Muller},
  \bibinfo{author}{D.~G. Schlom}, \bibinfo{author}{G.~D. Fuchs},
\newblock \bibinfo{title}{Engineering {Dzyaloshinskii-Moriya} interaction in
  {B20} thin-film chiral magnets},
\newblock \bibinfo{journal}{Phys. Rev. Mater.} \bibinfo{volume}{2}
  (\bibinfo{year}{2018}) \bibinfo{pages}{074404}.
  \DOIprefix\doi{10.1103/PhysRevMaterials.2.074404}.
\bibitem[{Spencer et~al.(2018)Spencer, Gayles, Porter, Sugimoto, Aslam, Kinane,
  Charlton, Freimuth, Chadov, Langridge, Sinova, Felser, Bl\"ugel, Mokrousov,
  and Marrows}]{Spencer_PhysRevB.97.214406}
\bibinfo{author}{C.~S. Spencer}, \bibinfo{author}{J.~Gayles},
  \bibinfo{author}{N.~A. Porter}, \bibinfo{author}{S.~Sugimoto},
  \bibinfo{author}{Z.~Aslam}, \bibinfo{author}{C.~J. Kinane},
  \bibinfo{author}{T.~R. Charlton}, \bibinfo{author}{F.~Freimuth},
  \bibinfo{author}{S.~Chadov}, \bibinfo{author}{S.~Langridge},
  \bibinfo{author}{J.~Sinova}, \bibinfo{author}{C.~Felser},
  \bibinfo{author}{S.~Bl\"ugel}, \bibinfo{author}{Y.~Mokrousov},
  \bibinfo{author}{C.~H. Marrows},
\newblock \bibinfo{title}{{Helical magnetic structure and the anomalous and
  topological Hall effects in epitaxial B20
  ${\mathrm{Fe}}_{1\ensuremath{-}y}{\mathrm{Co}}_{y}\mathrm{Ge}$ films}},
\newblock \bibinfo{journal}{Phys. Rev. B} \bibinfo{volume}{97}
  (\bibinfo{year}{2018}) \bibinfo{pages}{214406}.
  \DOIprefix\doi{10.1103/PhysRevB.97.214406}.
\bibitem[{Balasubramanian et~al.(2020)Balasubramanian, Manchanda, Pahari, Chen,
  Zhang, Valloppilly, Li, Sarella, Yue, Ullah, Dev, Muller, Skomski,
  Hadjipanayis, and Sellmyer}]{Balasubramanian_PhysRevLett.124.057201}
\bibinfo{author}{B.~Balasubramanian}, \bibinfo{author}{P.~Manchanda},
  \bibinfo{author}{R.~Pahari}, \bibinfo{author}{Z.~Chen},
  \bibinfo{author}{W.~Zhang}, \bibinfo{author}{S.~R. Valloppilly},
  \bibinfo{author}{X.~Li}, \bibinfo{author}{A.~Sarella},
  \bibinfo{author}{L.~Yue}, \bibinfo{author}{A.~Ullah},
  \bibinfo{author}{P.~Dev}, \bibinfo{author}{D.~A. Muller},
  \bibinfo{author}{R.~Skomski}, \bibinfo{author}{G.~C. Hadjipanayis},
  \bibinfo{author}{D.~J. Sellmyer},
\newblock \bibinfo{title}{{Chiral Magnetism and High-Temperature Skyrmions in
  B20-Ordered Co-Si}},
\newblock \bibinfo{journal}{Phys. Rev. Lett.} \bibinfo{volume}{124}
  (\bibinfo{year}{2020}) \bibinfo{pages}{057201}.
  \DOIprefix\doi{10.1103/PhysRevLett.124.057201}.
\bibitem[{Borisov et~al.(2022)Borisov, Xu, Ntallis, Clulow, Shtender,
  Cedervall, Sahlberg, Wikfeldt, Thonig, Pereiro, Bergman, Delin, and
  Eriksson}]{Borisov_PhysRevMaterials.6.084401}
\bibinfo{author}{V.~Borisov}, \bibinfo{author}{Q.~Xu},
  \bibinfo{author}{N.~Ntallis}, \bibinfo{author}{R.~Clulow},
  \bibinfo{author}{V.~Shtender}, \bibinfo{author}{J.~Cedervall},
  \bibinfo{author}{M.~Sahlberg}, \bibinfo{author}{K.~T. Wikfeldt},
  \bibinfo{author}{D.~Thonig}, \bibinfo{author}{M.~Pereiro},
  \bibinfo{author}{A.~Bergman}, \bibinfo{author}{A.~Delin},
  \bibinfo{author}{O.~Eriksson},
\newblock \bibinfo{title}{{Tuning skyrmions in B20 compounds by $4d$ and $5d$
  doping}},
\newblock \bibinfo{journal}{Phys. Rev. Mater.} \bibinfo{volume}{6}
  (\bibinfo{year}{2022}) \bibinfo{pages}{084401}.
  \DOIprefix\doi{10.1103/PhysRevMaterials.6.084401}.
\bibitem[{Dzyaloshinsky(1958)}]{dzyaloshinsky1958thermodynamic}
\bibinfo{author}{I.~Dzyaloshinsky},
\newblock \bibinfo{title}{A thermodynamic theory of “weak” ferromagnetism
  of antiferromagnetics},
\newblock \bibinfo{journal}{J. Phys. Chem. Solids} \bibinfo{volume}{4}
  (\bibinfo{year}{1958}) \bibinfo{pages}{241--255}.
\bibitem[{Moriya(1960)}]{moriya1960anisotropic}
\bibinfo{author}{T.~Moriya},
\newblock \bibinfo{title}{Anisotropic superexchange interaction and weak
  ferromagnetism},
\newblock \bibinfo{journal}{Phys. Rev.} \bibinfo{volume}{120}
  (\bibinfo{year}{1960}) \bibinfo{pages}{91}.
  \DOIprefix\doi{https://doi.org/10.1103/PhysRev.120.91}.
\bibitem[{Yi et~al.(2009)Yi, Onoda, Nagaosa, and Han}]{Yi_PhysRevB.80.054416}
\bibinfo{author}{S.~D. Yi}, \bibinfo{author}{S.~Onoda},
  \bibinfo{author}{N.~Nagaosa}, \bibinfo{author}{J.~H. Han},
\newblock \bibinfo{title}{Skyrmions and anomalous {Hall} effect in a
  {Dzyaloshinskii}-{Moriya} spiral magnet},
\newblock \bibinfo{journal}{Phys. Rev. B} \bibinfo{volume}{80}
  (\bibinfo{year}{2009}) \bibinfo{pages}{054416}.
  \DOIprefix\doi{10.1103/PhysRevB.80.054416}.
\bibitem[{Butenko et~al.(2010)Butenko, Leonov, R\"o\ss{}ler, and
  Bogdanov}]{Butenko_PhysRevB.82.052403}
\bibinfo{author}{A.~B. Butenko}, \bibinfo{author}{A.~A. Leonov},
  \bibinfo{author}{U.~K. R\"o\ss{}ler}, \bibinfo{author}{A.~N. Bogdanov},
\newblock \bibinfo{title}{Stabilization of skyrmion textures by uniaxial
  distortions in noncentrosymmetric cubic helimagnets},
\newblock \bibinfo{journal}{Phys. Rev. B} \bibinfo{volume}{82}
  (\bibinfo{year}{2010}) \bibinfo{pages}{052403}.
  \DOIprefix\doi{10.1103/PhysRevB.82.052403}.
\bibitem[{Wilson et~al.(2014)Wilson, Butenko, Bogdanov, and
  Monchesky}]{Wilson_PhysRevB.89.094411}
\bibinfo{author}{M.~N. Wilson}, \bibinfo{author}{A.~B. Butenko},
  \bibinfo{author}{A.~N. Bogdanov}, \bibinfo{author}{T.~L. Monchesky},
\newblock \bibinfo{title}{Chiral skyrmions in cubic helimagnet films: The role
  of uniaxial anisotropy},
\newblock \bibinfo{journal}{Phys. Rev. B} \bibinfo{volume}{89}
  (\bibinfo{year}{2014}) \bibinfo{pages}{094411}.
  \DOIprefix\doi{10.1103/PhysRevB.89.094411}.
\bibitem[{Banerjee et~al.(2014)Banerjee, Rowland, Erten, and
  Randeria}]{Banerjee_PhysRevX.4.031045}
\bibinfo{author}{S.~Banerjee}, \bibinfo{author}{J.~Rowland},
  \bibinfo{author}{O.~Erten}, \bibinfo{author}{M.~Randeria},
\newblock \bibinfo{title}{Enhanced stability of skyrmions in two-dimensional
  chiral magnets with {R}ashba spin-orbit coupling},
\newblock \bibinfo{journal}{Phys. Rev. X} \bibinfo{volume}{4}
  (\bibinfo{year}{2014}) \bibinfo{pages}{031045}.
  \DOIprefix\doi{10.1103/PhysRevX.4.031045}.
\bibitem[{G\"ung\"ord\"u et~al.(2016)G\"ung\"ord\"u, Nepal, Tretiakov,
  Belashchenko, and Kovalev}]{Gungordu_PhysRevB.93.064428}
\bibinfo{author}{U.~G\"ung\"ord\"u}, \bibinfo{author}{R.~Nepal},
  \bibinfo{author}{O.~A. Tretiakov}, \bibinfo{author}{K.~Belashchenko},
  \bibinfo{author}{A.~A. Kovalev},
\newblock \bibinfo{title}{Stability of skyrmion lattices and symmetries of
  quasi-two-dimensional chiral magnets},
\newblock \bibinfo{journal}{Phys. Rev. B} \bibinfo{volume}{93}
  (\bibinfo{year}{2016}) \bibinfo{pages}{064428}.
  \DOIprefix\doi{10.1103/PhysRevB.93.064428}.
\bibitem[{Rowland et~al.(2016)Rowland, Banerjee, and
  Randeria}]{Rowland_PhysRevB.93.020404}
\bibinfo{author}{J.~Rowland}, \bibinfo{author}{S.~Banerjee},
  \bibinfo{author}{M.~Randeria},
\newblock \bibinfo{title}{Skyrmions in chiral magnets with {R}ashba and
  {D}resselhaus spin-orbit coupling},
\newblock \bibinfo{journal}{Phys. Rev. B} \bibinfo{volume}{93}
  (\bibinfo{year}{2016}) \bibinfo{pages}{020404}.
  \DOIprefix\doi{10.1103/PhysRevB.93.020404}.
\bibitem[{Leonov and K\'ezsm\'arki(2017)}]{Leonov_PhysRevB.96.014423}
\bibinfo{author}{A.~O. Leonov}, \bibinfo{author}{I.~K\'ezsm\'arki},
\newblock \bibinfo{title}{Asymmetric isolated skyrmions in polar magnets with
  easy-plane anisotropy},
\newblock \bibinfo{journal}{Phys. Rev. B} \bibinfo{volume}{96}
  (\bibinfo{year}{2017}) \bibinfo{pages}{014423}.
  \DOIprefix\doi{10.1103/PhysRevB.96.014423}.
\bibitem[{Tokunaga et~al.(2015)Tokunaga, Yu, White, R{\o}nnow, Morikawa,
  Taguchi, and Tokura}]{tokunaga2015new}
\bibinfo{author}{Y.~Tokunaga}, \bibinfo{author}{X.~Yu},
  \bibinfo{author}{J.~White}, \bibinfo{author}{H.~M. R{\o}nnow},
  \bibinfo{author}{D.~Morikawa}, \bibinfo{author}{Y.~Taguchi},
  \bibinfo{author}{Y.~Tokura},
\newblock \bibinfo{title}{A new class of chiral materials hosting magnetic
  skyrmions beyond room temperature},
\newblock \bibinfo{journal}{Nat. Commun.} \bibinfo{volume}{6}
  (\bibinfo{year}{2015}) \bibinfo{pages}{7638}.
  \DOIprefix\doi{10.1038/ncomms8638}.
\bibitem[{Karube et~al.(2016)Karube, White, Reynolds, Gavilano, Oike, Kikkawa,
  Kagawa, Tokunaga, R{\o}nnow, Tokura, and Taguchi}]{karube2016robust}
\bibinfo{author}{K.~Karube}, \bibinfo{author}{J.~White},
  \bibinfo{author}{N.~Reynolds}, \bibinfo{author}{J.~Gavilano},
  \bibinfo{author}{H.~Oike}, \bibinfo{author}{A.~Kikkawa},
  \bibinfo{author}{F.~Kagawa}, \bibinfo{author}{Y.~Tokunaga},
  \bibinfo{author}{H.~M. R{\o}nnow}, \bibinfo{author}{Y.~Tokura},
  \bibinfo{author}{Y.~Taguchi},
\newblock \bibinfo{title}{Robust metastable skyrmions and their
  triangular--square lattice structural transition in a high-temperature chiral
  magnet},
\newblock \bibinfo{journal}{Nat. Mater.} \bibinfo{volume}{15}
  (\bibinfo{year}{2016}) \bibinfo{pages}{1237}.
  \DOIprefix\doi{10.1038/nmat4752}.
\bibitem[{Li et~al.(2016)Li, Jin, Che, Wei, Lin, Zhang, Du, Tian, and
  Zang}]{Li_PhysRevB.93.060409}
\bibinfo{author}{W.~Li}, \bibinfo{author}{C.~Jin}, \bibinfo{author}{R.~Che},
  \bibinfo{author}{W.~Wei}, \bibinfo{author}{L.~Lin},
  \bibinfo{author}{L.~Zhang}, \bibinfo{author}{H.~Du},
  \bibinfo{author}{M.~Tian}, \bibinfo{author}{J.~Zang},
\newblock \bibinfo{title}{Emergence of skyrmions from rich parent phases in the
  molybdenum nitrides},
\newblock \bibinfo{journal}{Phys. Rev. B} \bibinfo{volume}{93}
  (\bibinfo{year}{2016}) \bibinfo{pages}{060409}.
  \DOIprefix\doi{10.1103/PhysRevB.93.060409}.
\bibitem[{Karube et~al.(2018{\natexlab{a}})Karube, White, Morikawa, Dewhurst,
  Cubitt, Kikkawa, Yu, Tokunaga, Arima, R{\o}nnow, Tokura, and
  Taguchi}]{karube2018disordered}
\bibinfo{author}{K.~Karube}, \bibinfo{author}{J.~S. White},
  \bibinfo{author}{D.~Morikawa}, \bibinfo{author}{C.~D. Dewhurst},
  \bibinfo{author}{R.~Cubitt}, \bibinfo{author}{A.~Kikkawa},
  \bibinfo{author}{X.~Yu}, \bibinfo{author}{Y.~Tokunaga},
  \bibinfo{author}{T.-h. Arima}, \bibinfo{author}{H.~M. R{\o}nnow},
  \bibinfo{author}{Y.~Tokura}, \bibinfo{author}{Y.~Taguchi},
\newblock \bibinfo{title}{Disordered skyrmion phase stabilized by magnetic
  frustration in a chiral magnet},
\newblock \bibinfo{journal}{Sci. Adv.} \bibinfo{volume}{4}
  (\bibinfo{year}{2018}{\natexlab{a}}) \bibinfo{pages}{eaar7043}.
  \DOIprefix\doi{10.1126/sciadv.aar7043}.
\bibitem[{Karube et~al.(2018{\natexlab{b}})Karube, Shibata, White, Koretsune,
  Yu, Tokunaga, R\o{}nnow, Arita, Arima, Tokura, and
  Taguchi}]{Karube_PhysRevB.98.155120}
\bibinfo{author}{K.~Karube}, \bibinfo{author}{K.~Shibata},
  \bibinfo{author}{J.~S. White}, \bibinfo{author}{T.~Koretsune},
  \bibinfo{author}{X.~Z. Yu}, \bibinfo{author}{Y.~Tokunaga},
  \bibinfo{author}{H.~M. R\o{}nnow}, \bibinfo{author}{R.~Arita},
  \bibinfo{author}{T.~Arima}, \bibinfo{author}{Y.~Tokura},
  \bibinfo{author}{Y.~Taguchi},
\newblock \bibinfo{title}{Controlling the helicity of magnetic skyrmions in a
  $\ensuremath{\beta}$-mn-type high-temperature chiral magnet},
\newblock \bibinfo{journal}{Phys. Rev. B} \bibinfo{volume}{98}
  (\bibinfo{year}{2018}{\natexlab{b}}) \bibinfo{pages}{155120}.
  \DOIprefix\doi{10.1103/PhysRevB.98.155120}.
\bibitem[{Seki et~al.(2012)Seki, Yu, Ishiwata, and
  Tokura}]{seki2012observation}
\bibinfo{author}{S.~Seki}, \bibinfo{author}{X.~Z. Yu},
  \bibinfo{author}{S.~Ishiwata}, \bibinfo{author}{Y.~Tokura},
\newblock \bibinfo{title}{Observation of skyrmions in a multiferroic material},
\newblock \bibinfo{journal}{Science} \bibinfo{volume}{336}
  (\bibinfo{year}{2012}) \bibinfo{pages}{198--201}.
  \DOIprefix\doi{10.1126/science.1214143}.
\bibitem[{Adams et~al.(2012)Adams, Chacon, Wagner, Bauer, Brandl, Pedersen,
  Berger, Lemmens, and Pfleiderer}]{Adams2012}
\bibinfo{author}{T.~Adams}, \bibinfo{author}{A.~Chacon},
  \bibinfo{author}{M.~Wagner}, \bibinfo{author}{A.~Bauer},
  \bibinfo{author}{G.~Brandl}, \bibinfo{author}{B.~Pedersen},
  \bibinfo{author}{H.~Berger}, \bibinfo{author}{P.~Lemmens},
  \bibinfo{author}{C.~Pfleiderer},
\newblock \bibinfo{title}{Long-wavelength helimagnetic order and skyrmion
  lattice phase in {$\mathrm{Cu_2OSeO_3}$}},
\newblock \bibinfo{journal}{Phys. Rev. Lett.} \bibinfo{volume}{108}
  (\bibinfo{year}{2012}) \bibinfo{pages}{237204}.
  \DOIprefix\doi{10.1103/PhysRevLett.108.237204}.
\bibitem[{Seki et~al.(2012)Seki, Kim, Inosov, Georgii, Keimer, Ishiwata, and
  Tokura}]{Seki_PhysRevB.85.220406}
\bibinfo{author}{S.~Seki}, \bibinfo{author}{J.-H. Kim}, \bibinfo{author}{D.~S.
  Inosov}, \bibinfo{author}{R.~Georgii}, \bibinfo{author}{B.~Keimer},
  \bibinfo{author}{S.~Ishiwata}, \bibinfo{author}{Y.~Tokura},
\newblock \bibinfo{title}{{Formation and rotation of skyrmion crystal in the
  chiral-lattice insulator Cu${}_{2}$OSeO${}_{3}$}},
\newblock \bibinfo{journal}{Phys. Rev. B} \bibinfo{volume}{85}
  (\bibinfo{year}{2012}) \bibinfo{pages}{220406}.
  \DOIprefix\doi{10.1103/PhysRevB.85.220406}.
\bibitem[{Kurumaji et~al.(2017)Kurumaji, Nakajima, Ukleev, Feoktystov, Arima,
  Kakurai, and Tokura}]{Kurumaji_PhysRevLett.119.237201}
\bibinfo{author}{T.~Kurumaji}, \bibinfo{author}{T.~Nakajima},
  \bibinfo{author}{V.~Ukleev}, \bibinfo{author}{A.~Feoktystov},
  \bibinfo{author}{T.-h. Arima}, \bibinfo{author}{K.~Kakurai},
  \bibinfo{author}{Y.~Tokura},
\newblock \bibinfo{title}{{{N}\'eel-Type Skyrmion Lattice in the Tetragonal
  Polar Magnet ${\mathrm{VOSe}}_{2}{\mathrm{O}}_{5}$}},
\newblock \bibinfo{journal}{Phys. Rev. Lett.} \bibinfo{volume}{119}
  (\bibinfo{year}{2017}) \bibinfo{pages}{237201}.
  \DOIprefix\doi{10.1103/PhysRevLett.119.237201}.
\bibitem[{K{\'e}zsm{\'a}rki et~al.(2015)K{\'e}zsm{\'a}rki, Bord{\'a}cs, Milde,
  Neuber, Eng, White, R{\o}nnow, Dewhurst, Mochizuki, Yanai, Nakamura, Ehlers,
  Tsurkan, and Loidl}]{kezsmarki_neel-type_2015}
\bibinfo{author}{I.~K{\'e}zsm{\'a}rki}, \bibinfo{author}{S.~Bord{\'a}cs},
  \bibinfo{author}{P.~Milde}, \bibinfo{author}{E.~Neuber},
  \bibinfo{author}{L.~M. Eng}, \bibinfo{author}{J.~S. White},
  \bibinfo{author}{H.~M. R{\o}nnow}, \bibinfo{author}{C.~D. Dewhurst},
  \bibinfo{author}{M.~Mochizuki}, \bibinfo{author}{K.~Yanai},
  \bibinfo{author}{H.~Nakamura}, \bibinfo{author}{D.~Ehlers},
  \bibinfo{author}{V.~Tsurkan}, \bibinfo{author}{A.~Loidl},
\newblock \bibinfo{title}{Neel-type skyrmion lattice with confined orientation
  in the polar magnetic semiconductor {GaV$_4$S$_8$}},
\newblock \bibinfo{journal}{Nat. Mater.} \bibinfo{volume}{14}
  (\bibinfo{year}{2015}) \bibinfo{pages}{1116--1122}.
  \DOIprefix\doi{10.1038/nmat4402}.
\bibitem[{Heinze et~al.(2011)Heinze, von Bergmann, Menzel, Brede, Kubetzka,
  Wiesendanger, Bihlmayer, and Bl{\"u}gel}]{heinze2011spontaneous}
\bibinfo{author}{S.~Heinze}, \bibinfo{author}{K.~von Bergmann},
  \bibinfo{author}{M.~Menzel}, \bibinfo{author}{J.~Brede},
  \bibinfo{author}{A.~Kubetzka}, \bibinfo{author}{R.~Wiesendanger},
  \bibinfo{author}{G.~Bihlmayer}, \bibinfo{author}{S.~Bl{\"u}gel},
\newblock \bibinfo{title}{Spontaneous atomic-scale magnetic skyrmion lattice in
  two dimensions},
\newblock \bibinfo{journal}{Nat. Phys.} \bibinfo{volume}{7}
  (\bibinfo{year}{2011}) \bibinfo{pages}{713--718}.
  \DOIprefix\doi{10.1038/nphys2045}.
\bibitem[{Romming et~al.(2013)Romming, Hanneken, Menzel, Bickel, Wolter, von
  Bergmann, Kubetzka, and Wiesendanger}]{romming2013writing}
\bibinfo{author}{N.~Romming}, \bibinfo{author}{C.~Hanneken},
  \bibinfo{author}{M.~Menzel}, \bibinfo{author}{J.~E. Bickel},
  \bibinfo{author}{B.~Wolter}, \bibinfo{author}{K.~von Bergmann},
  \bibinfo{author}{A.~Kubetzka}, \bibinfo{author}{R.~Wiesendanger},
\newblock \bibinfo{title}{Writing and deleting single magnetic skyrmions},
\newblock \bibinfo{journal}{Science} \bibinfo{volume}{341}
  (\bibinfo{year}{2013}) \bibinfo{pages}{636--639}.
  \DOIprefix\doi{10.1126/science.1240573}.
\bibitem[{Nayak et~al.(2017)Nayak, Kumar, Ma, Werner, Pippel, Sahoo, Damay,
  R{\"o}{\ss}ler, Felser, and Parkin}]{nayak2017discovery}
\bibinfo{author}{A.~K. Nayak}, \bibinfo{author}{V.~Kumar},
  \bibinfo{author}{T.~Ma}, \bibinfo{author}{P.~Werner},
  \bibinfo{author}{E.~Pippel}, \bibinfo{author}{R.~Sahoo},
  \bibinfo{author}{F.~Damay}, \bibinfo{author}{U.~K. R{\"o}{\ss}ler},
  \bibinfo{author}{C.~Felser}, \bibinfo{author}{S.~S. Parkin},
\newblock \bibinfo{title}{Magnetic antiskyrmions above room temperature in
  tetragonal heusler materials},
\newblock \bibinfo{journal}{Nature} \bibinfo{volume}{548}
  (\bibinfo{year}{2017}) \bibinfo{pages}{561--566}.
  \DOIprefix\doi{10.1038/nature23466}.
\bibitem[{Peng et~al.(2020)Peng, Takagi, Koshibae, Shibata, Nakajima, Arima,
  Nagaosa, Seki, Yu, and Tokura}]{peng2020controlled}
\bibinfo{author}{L.~Peng}, \bibinfo{author}{R.~Takagi},
  \bibinfo{author}{W.~Koshibae}, \bibinfo{author}{K.~Shibata},
  \bibinfo{author}{K.~Nakajima}, \bibinfo{author}{T.-h. Arima},
  \bibinfo{author}{N.~Nagaosa}, \bibinfo{author}{S.~Seki},
  \bibinfo{author}{X.~Yu}, \bibinfo{author}{Y.~Tokura},
\newblock \bibinfo{title}{Controlled transformation of skyrmions and
  antiskyrmions in a non-centrosymmetric magnet},
\newblock \bibinfo{journal}{Nat. Nanotechnol.} \bibinfo{volume}{15}
  (\bibinfo{year}{2020}) \bibinfo{pages}{181--186}.
  \DOIprefix\doi{10.1038/s41565-019-0616-6}.
\bibitem[{Kakihana et~al.(2018)Kakihana, Aoki, Nakamura, Honda, Nakashima,
  Amako, Nakamura, Sakakibara, Hedo, Nakama, and Onuki}]{kakihana2018giant}
\bibinfo{author}{M.~Kakihana}, \bibinfo{author}{D.~Aoki},
  \bibinfo{author}{A.~Nakamura}, \bibinfo{author}{F.~Honda},
  \bibinfo{author}{M.~Nakashima}, \bibinfo{author}{Y.~Amako},
  \bibinfo{author}{S.~Nakamura}, \bibinfo{author}{T.~Sakakibara},
  \bibinfo{author}{M.~Hedo}, \bibinfo{author}{T.~Nakama},
  \bibinfo{author}{Y.~Onuki},
\newblock \bibinfo{title}{Giant {Hall} resistivity and magnetoresistance in
  cubic chiral antiferromagnet {EuPtSi}},
\newblock \bibinfo{journal}{J. Phys. Soc. Jpn.} \bibinfo{volume}{87}
  (\bibinfo{year}{2018}) \bibinfo{pages}{023701}.
  \DOIprefix\doi{10.7566/JPSJ.87.023701}.
\bibitem[{Kaneko et~al.(2019)Kaneko, Frontzek, Matsuda, Nakao, Munakata,
  Ohhara, Kakihana, Haga, Hedo, Nakama, and Onuki}]{kaneko2019unique}
\bibinfo{author}{K.~Kaneko}, \bibinfo{author}{M.~D. Frontzek},
  \bibinfo{author}{M.~Matsuda}, \bibinfo{author}{A.~Nakao},
  \bibinfo{author}{K.~Munakata}, \bibinfo{author}{T.~Ohhara},
  \bibinfo{author}{M.~Kakihana}, \bibinfo{author}{Y.~Haga},
  \bibinfo{author}{M.~Hedo}, \bibinfo{author}{T.~Nakama},
  \bibinfo{author}{Y.~Onuki},
\newblock \bibinfo{title}{Unique helical magnetic order and field-induced phase
  in trillium lattice antiferromagnet {EuPtSi}},
\newblock \bibinfo{journal}{J. Phys. Soc. Jpn.} \bibinfo{volume}{88}
  (\bibinfo{year}{2019}) \bibinfo{pages}{013702}.
  \DOIprefix\doi{10.7566/JPSJ.88.013702}.
\bibitem[{Tabata et~al.(2019)Tabata, Matsumura, Nakao, Michimura, Kakihana,
  Inami, Kaneko, Hedo, Nakama, and {\=O}nuki}]{tabata2019magnetic}
\bibinfo{author}{C.~Tabata}, \bibinfo{author}{T.~Matsumura},
  \bibinfo{author}{H.~Nakao}, \bibinfo{author}{S.~Michimura},
  \bibinfo{author}{M.~Kakihana}, \bibinfo{author}{T.~Inami},
  \bibinfo{author}{K.~Kaneko}, \bibinfo{author}{M.~Hedo},
  \bibinfo{author}{T.~Nakama}, \bibinfo{author}{Y.~{\=O}nuki},
\newblock \bibinfo{title}{Magnetic field induced triple-$q$ magnetic order in
  trillium lattice antiferromagnet {EuPtSi} studied by resonant {X}-ray
  scattering},
\newblock \bibinfo{journal}{J. Phys. Soc. Jpn.} \bibinfo{volume}{88}
  (\bibinfo{year}{2019}) \bibinfo{pages}{093704}.
  \DOIprefix\doi{10.7566/JPSJ.88.093704}.
\bibitem[{Kakihana et~al.(2019)Kakihana, Aoki, Nakamura, Honda, Nakashima,
  Amako, Takeuchi, Harima, Hedo, Nakama, and Onuki}]{kakihana2019unique}
\bibinfo{author}{M.~Kakihana}, \bibinfo{author}{D.~Aoki},
  \bibinfo{author}{A.~Nakamura}, \bibinfo{author}{F.~Honda},
  \bibinfo{author}{M.~Nakashima}, \bibinfo{author}{Y.~Amako},
  \bibinfo{author}{T.~Takeuchi}, \bibinfo{author}{H.~Harima},
  \bibinfo{author}{M.~Hedo}, \bibinfo{author}{T.~Nakama},
  \bibinfo{author}{Y.~Onuki},
\newblock \bibinfo{title}{Unique magnetic phases in the skyrmion lattice and
  {Fermi} surface properties in cubic chiral antiferromagnet {EuPtSi}},
\newblock \bibinfo{journal}{J. Phys. Soc. Jpn.} \bibinfo{volume}{88}
  (\bibinfo{year}{2019}) \bibinfo{pages}{094705}.
  \DOIprefix\doi{10.7566/JPSJ.88.094705}.
\bibitem[{Mishra and Ganesan(2019)}]{Mishra_PhysRevB.100.125113}
\bibinfo{author}{A.~K. Mishra}, \bibinfo{author}{V.~Ganesan},
\newblock \bibinfo{title}{{$A$}-phase, field-induced tricritical point, and
  universal magnetocaloric scaling in {EuPtSi}},
\newblock \bibinfo{journal}{Phys. Rev. B} \bibinfo{volume}{100}
  (\bibinfo{year}{2019}) \bibinfo{pages}{125113}.
  \DOIprefix\doi{10.1103/PhysRevB.100.125113}.
\bibitem[{Takeuchi et~al.(2019)Takeuchi, Kakihana, Hedo, Nakama, and
  {\=O}nuki}]{takeuchi2019magnetic}
\bibinfo{author}{T.~Takeuchi}, \bibinfo{author}{M.~Kakihana},
  \bibinfo{author}{M.~Hedo}, \bibinfo{author}{T.~Nakama},
  \bibinfo{author}{Y.~{\=O}nuki},
\newblock \bibinfo{title}{Magnetic field versus temperature phase diagram for
  {H}$\parallel$[001] in the trillium lattice antiferromagnet {EuPtSi}},
\newblock \bibinfo{journal}{J. Phys. Soc. Jpn.} \bibinfo{volume}{88}
  (\bibinfo{year}{2019}) \bibinfo{pages}{053703}.
  \DOIprefix\doi{10.7566/JPSJ.88.053703}.
\bibitem[{Matsumura et~al.(2024)Matsumura, Tabata, Kaneko, Nakao, Kakihana,
  Hedo, Nakama, and \ifmmode~\bar{O}\else
  \={O}\fi{}nuki}]{Matsumura_PhysRevB.109.174437}
\bibinfo{author}{T.~Matsumura}, \bibinfo{author}{C.~Tabata},
  \bibinfo{author}{K.~Kaneko}, \bibinfo{author}{H.~Nakao},
  \bibinfo{author}{M.~Kakihana}, \bibinfo{author}{M.~Hedo},
  \bibinfo{author}{T.~Nakama}, \bibinfo{author}{Y.~\ifmmode~\bar{O}\else
  \={O}\fi{}nuki},
\newblock \bibinfo{title}{Single helicity of the triple-$q$ triangular skyrmion
  lattice state in the cubic chiral helimagnet {EuPtSi}},
\newblock \bibinfo{journal}{Phys. Rev. B} \bibinfo{volume}{109}
  (\bibinfo{year}{2024}) \bibinfo{pages}{174437}.
  \DOIprefix\doi{10.1103/PhysRevB.109.174437}.
\bibitem[{Goetsch et~al.(2013)Goetsch, Anand, and
  Johnston}]{Goetsch_PhysRevB.87.064406}
\bibinfo{author}{R.~J. Goetsch}, \bibinfo{author}{V.~K. Anand},
  \bibinfo{author}{D.~C. Johnston},
\newblock \bibinfo{title}{Antiferromagnetism in {EuNiGe${}_{3}$}},
\newblock \bibinfo{journal}{Phys. Rev. B} \bibinfo{volume}{87}
  (\bibinfo{year}{2013}) \bibinfo{pages}{064406}.
  \DOIprefix\doi{10.1103/PhysRevB.87.064406}.
\bibitem[{Fabr\`eges et~al.(2016)Fabr\`eges, Gukasov, Bonville, Maurya,
  Thamizhavel, and Dhar}]{Fabreges_PhysRevB.93.214414}
\bibinfo{author}{X.~Fabr\`eges}, \bibinfo{author}{A.~Gukasov},
  \bibinfo{author}{P.~Bonville}, \bibinfo{author}{A.~Maurya},
  \bibinfo{author}{A.~Thamizhavel}, \bibinfo{author}{S.~K. Dhar},
\newblock \bibinfo{title}{Exploring metamagnetism of single crystalline
  ${{\mathrm{EuNiGe}}_{3}}$ by neutron scattering},
\newblock \bibinfo{journal}{Phys. Rev. B} \bibinfo{volume}{93}
  (\bibinfo{year}{2016}) \bibinfo{pages}{214414}.
  \DOIprefix\doi{10.1103/PhysRevB.93.214414}.
\bibitem[{Singh et~al.(2023)Singh, Fujishiro, Hayami, Moody, Nomoto, Baral,
  Ukleev, Cubitt, Steinke, Gawryluk, Pomjakushina, {\=O}nuki, Arita, Tokura,
  Kanazawa, and White}]{singh2023transition}
\bibinfo{author}{D.~Singh}, \bibinfo{author}{Y.~Fujishiro},
  \bibinfo{author}{S.~Hayami}, \bibinfo{author}{S.~H. Moody},
  \bibinfo{author}{T.~Nomoto}, \bibinfo{author}{P.~R. Baral},
  \bibinfo{author}{V.~Ukleev}, \bibinfo{author}{R.~Cubitt},
  \bibinfo{author}{N.-J. Steinke}, \bibinfo{author}{D.~J. Gawryluk},
  \bibinfo{author}{E.~Pomjakushina}, \bibinfo{author}{Y.~{\=O}nuki},
  \bibinfo{author}{R.~Arita}, \bibinfo{author}{Y.~Tokura},
  \bibinfo{author}{N.~Kanazawa}, \bibinfo{author}{J.~S. White},
\newblock \bibinfo{title}{Transition between distinct hybrid skyrmion textures
  through their hexagonal-to-square crystal transformation in a polar magnet},
\newblock \bibinfo{journal}{Nat. Commun.} \bibinfo{volume}{14}
  (\bibinfo{year}{2023}) \bibinfo{pages}{8050}.
  \DOIprefix\doi{https://doi.org/10.1038/s41467-023-43814-x}.
\bibitem[{Matsumura et~al.(2024)Matsumura, Kurauchi, Tsukagoshi, Higa, Nakao,
  Kakihana, Hedo, Nakama, and {\=O}nuki}]{matsumura2023distorted}
\bibinfo{author}{T.~Matsumura}, \bibinfo{author}{K.~Kurauchi},
  \bibinfo{author}{M.~Tsukagoshi}, \bibinfo{author}{N.~Higa},
  \bibinfo{author}{H.~Nakao}, \bibinfo{author}{M.~Kakihana},
  \bibinfo{author}{M.~Hedo}, \bibinfo{author}{T.~Nakama},
  \bibinfo{author}{Y.~{\=O}nuki},
\newblock \bibinfo{title}{{Helicity Unification by Triangular Skyrmion Lattice
  Formation in the Noncentrosymmetric Tetragonal Magnet EuNiGe$_3$}},
\newblock \bibinfo{journal}{J. Phys. Soc. Jpn.} \bibinfo{volume}{93}
  (\bibinfo{year}{2024}) \bibinfo{pages}{074705}.
  \DOIprefix\doi{https://doi.org/10.7566/JPSJ.93.074705}.
\bibitem[{Saha et~al.(1999)Saha, Sugawara, Matsuda, Sato, Mallik, and
  Sampathkumaran}]{Saha_PhysRevB.60.12162}
\bibinfo{author}{S.~R. Saha}, \bibinfo{author}{H.~Sugawara},
  \bibinfo{author}{T.~D. Matsuda}, \bibinfo{author}{H.~Sato},
  \bibinfo{author}{R.~Mallik}, \bibinfo{author}{E.~V. Sampathkumaran},
\newblock \bibinfo{title}{{Magnetic anisotropy, first-order-like metamagnetic
  transitions, and large negative magnetoresistance in single-crystal
  ${\mathrm{Gd}}_{2}{\mathrm{PdSi}}_{3}$}},
\newblock \bibinfo{journal}{Phys. Rev. B} \bibinfo{volume}{60}
  (\bibinfo{year}{1999}) \bibinfo{pages}{12162--12165}.
  \DOIprefix\doi{10.1103/PhysRevB.60.12162}.
\bibitem[{Kurumaji et~al.(2019)Kurumaji, Nakajima, Hirschberger, Kikkawa,
  Yamasaki, Sagayama, Nakao, Taguchi, Arima, and Tokura}]{kurumaji2019skyrmion}
\bibinfo{author}{T.~Kurumaji}, \bibinfo{author}{T.~Nakajima},
  \bibinfo{author}{M.~Hirschberger}, \bibinfo{author}{A.~Kikkawa},
  \bibinfo{author}{Y.~Yamasaki}, \bibinfo{author}{H.~Sagayama},
  \bibinfo{author}{H.~Nakao}, \bibinfo{author}{Y.~Taguchi},
  \bibinfo{author}{T.-h. Arima}, \bibinfo{author}{Y.~Tokura},
\newblock \bibinfo{title}{Skyrmion lattice with a giant topological {Hall}
  effect in a frustrated triangular-lattice magnet},
\newblock \bibinfo{journal}{Science} \bibinfo{volume}{365}
  (\bibinfo{year}{2019}) \bibinfo{pages}{914--918}.
  \DOIprefix\doi{10.1126/science.aau0968}.
\bibitem[{Hirschberger et~al.(2020)Hirschberger, Spitz, Nomoto, Kurumaji, Gao,
  Masell, Nakajima, Kikkawa, Yamasaki, Sagayama, Nakao, Taguchi, Arita, Arima,
  and Tokura}]{Hirschberger_PhysRevLett.125.076602}
\bibinfo{author}{M.~Hirschberger}, \bibinfo{author}{L.~Spitz},
  \bibinfo{author}{T.~Nomoto}, \bibinfo{author}{T.~Kurumaji},
  \bibinfo{author}{S.~Gao}, \bibinfo{author}{J.~Masell},
  \bibinfo{author}{T.~Nakajima}, \bibinfo{author}{A.~Kikkawa},
  \bibinfo{author}{Y.~Yamasaki}, \bibinfo{author}{H.~Sagayama},
  \bibinfo{author}{H.~Nakao}, \bibinfo{author}{Y.~Taguchi},
  \bibinfo{author}{R.~Arita}, \bibinfo{author}{T.-h. Arima},
  \bibinfo{author}{Y.~Tokura},
\newblock \bibinfo{title}{{Topological Nernst Effect of the Two-Dimensional
  Skyrmion Lattice}},
\newblock \bibinfo{journal}{Phys. Rev. Lett.} \bibinfo{volume}{125}
  (\bibinfo{year}{2020}) \bibinfo{pages}{076602}.
  \DOIprefix\doi{10.1103/PhysRevLett.125.076602}.
\bibitem[{Nomoto et~al.(2020)Nomoto, Koretsune, and
  Arita}]{Nomoto_PhysRevLett.125.117204}
\bibinfo{author}{T.~Nomoto}, \bibinfo{author}{T.~Koretsune},
  \bibinfo{author}{R.~Arita},
\newblock \bibinfo{title}{{Formation Mechanism of the Helical $\bm{Q}$
  Structure in Gd-Based Skyrmion Materials}},
\newblock \bibinfo{journal}{Phys. Rev. Lett.} \bibinfo{volume}{125}
  (\bibinfo{year}{2020}) \bibinfo{pages}{117204}.
  \DOIprefix\doi{10.1103/PhysRevLett.125.117204}.
\bibitem[{Kumar et~al.(2020)Kumar, Iyer, Paulose, and
  Sampathkumaran}]{Kumar_PhysRevB.101.144440}
\bibinfo{author}{R.~Kumar}, \bibinfo{author}{K.~K. Iyer},
  \bibinfo{author}{P.~L. Paulose}, \bibinfo{author}{E.~V. Sampathkumaran},
\newblock \bibinfo{title}{{Magnetic and transport anomalies in
  ${R}_{2}\mathrm{RhS}{\mathrm{i}}_{3}\phantom{\rule{4pt}{0ex}}(R=\mathrm{Gd}$,
  Tb, and Dy) resembling those of the exotic magnetic material
  $\mathrm{G}{\mathrm{d}}_{2}\mathrm{PdS}{\mathrm{i}}_{3}$}},
\newblock \bibinfo{journal}{Phys. Rev. B} \bibinfo{volume}{101}
  (\bibinfo{year}{2020}) \bibinfo{pages}{144440}.
  \DOIprefix\doi{10.1103/PhysRevB.101.144440}.
\bibitem[{Spachmann et~al.(2021)Spachmann, Elghandour, Frontzek, L\"oser, and
  Klingeler}]{Spachmann_PhysRevB.103.184424}
\bibinfo{author}{S.~Spachmann}, \bibinfo{author}{A.~Elghandour},
  \bibinfo{author}{M.~Frontzek}, \bibinfo{author}{W.~L\"oser},
  \bibinfo{author}{R.~Klingeler},
\newblock \bibinfo{title}{{Magnetoelastic coupling and phases in the skyrmion
  lattice magnet ${\mathrm{Gd}}_{2}{\mathrm{PdSi}}_{3}$ discovered by
  high-resolution dilatometry}},
\newblock \bibinfo{journal}{Phys. Rev. B} \bibinfo{volume}{103}
  (\bibinfo{year}{2021}) \bibinfo{pages}{184424}.
  \DOIprefix\doi{10.1103/PhysRevB.103.184424}.
\bibitem[{Nakamura et~al.(2018)Nakamura, Kabeya, Kobayashi, Araki, Katoh, and
  Ochiai}]{Nakamura_PhysRevB.98.054410}
\bibinfo{author}{S.~Nakamura}, \bibinfo{author}{N.~Kabeya},
  \bibinfo{author}{M.~Kobayashi}, \bibinfo{author}{K.~Araki},
  \bibinfo{author}{K.~Katoh}, \bibinfo{author}{A.~Ochiai},
\newblock \bibinfo{title}{Spin trimer formation in the metallic compound
  {${\mathrm{Gd}}_{3}{\mathrm{Ru}}_{4}{\mathrm{Al}}_{12}$} with a distorted
  kagome lattice structure},
\newblock \bibinfo{journal}{Phys. Rev. B} \bibinfo{volume}{98}
  (\bibinfo{year}{2018}) \bibinfo{pages}{054410}.
  \DOIprefix\doi{10.1103/PhysRevB.98.054410}.
\bibitem[{Hirschberger et~al.(2019)Hirschberger, Nakajima, Gao, Peng, Kikkawa,
  Kurumaji, Kriener, Yamasaki, Sagayama, Nakao, Ohishi, Kakurai, Taguchi, Yu,
  Arima, and Tokura}]{hirschberger2019skyrmion}
\bibinfo{author}{M.~Hirschberger}, \bibinfo{author}{T.~Nakajima},
  \bibinfo{author}{S.~Gao}, \bibinfo{author}{L.~Peng},
  \bibinfo{author}{A.~Kikkawa}, \bibinfo{author}{T.~Kurumaji},
  \bibinfo{author}{M.~Kriener}, \bibinfo{author}{Y.~Yamasaki},
  \bibinfo{author}{H.~Sagayama}, \bibinfo{author}{H.~Nakao},
  \bibinfo{author}{K.~Ohishi}, \bibinfo{author}{K.~Kakurai},
  \bibinfo{author}{Y.~Taguchi}, \bibinfo{author}{X.~Yu}, \bibinfo{author}{T.-h.
  Arima}, \bibinfo{author}{Y.~Tokura},
\newblock \bibinfo{title}{Skyrmion phase and competing magnetic orders on a
  breathing kagome lattice},
\newblock \bibinfo{journal}{Nat. Commun.} \bibinfo{volume}{10}
  (\bibinfo{year}{2019}) \bibinfo{pages}{5831}.
  \DOIprefix\doi{10.1038/s41467-019-13675-4}.
\bibitem[{Hirschberger et~al.(2021)Hirschberger, Hayami, and
  Tokura}]{Hirschberger_10.1088/1367-2630/abdef9}
\bibinfo{author}{M.~Hirschberger}, \bibinfo{author}{S.~Hayami},
  \bibinfo{author}{Y.~Tokura},
\newblock \bibinfo{title}{Nanometric skyrmion lattice from anisotropic exchange
  interactions in a centrosymmetric host},
\newblock \bibinfo{journal}{New J. Phys.} \bibinfo{volume}{23}
  (\bibinfo{year}{2021}) \bibinfo{pages}{023039}.
  \DOIprefix\doi{10.1088/1367-2630/abdef9}.
\bibitem[{Khanh et~al.(2020)Khanh, Nakajima, Yu, Gao, Shibata, Hirschberger,
  Yamasaki, Sagayama, Nakao, Peng, Nakajima, Takagi, Arima, Tokura, and
  Seki}]{khanh2020nanometric}
\bibinfo{author}{N.~D. Khanh}, \bibinfo{author}{T.~Nakajima},
  \bibinfo{author}{X.~Yu}, \bibinfo{author}{S.~Gao},
  \bibinfo{author}{K.~Shibata}, \bibinfo{author}{M.~Hirschberger},
  \bibinfo{author}{Y.~Yamasaki}, \bibinfo{author}{H.~Sagayama},
  \bibinfo{author}{H.~Nakao}, \bibinfo{author}{L.~Peng},
  \bibinfo{author}{K.~Nakajima}, \bibinfo{author}{R.~Takagi},
  \bibinfo{author}{T.-h. Arima}, \bibinfo{author}{Y.~Tokura},
  \bibinfo{author}{S.~Seki},
\newblock \bibinfo{title}{Nanometric square skyrmion lattice in a
  centrosymmetric tetragonal magnet},
\newblock \bibinfo{journal}{Nat. Nanotechnol.} \bibinfo{volume}{15}
  (\bibinfo{year}{2020}) \bibinfo{pages}{444}.
  \DOIprefix\doi{10.1038/s41565-020-0684-7}.
\bibitem[{Yasui et~al.(2020)Yasui, Butler, Khanh, Hayami, Nomoto, Hanaguri,
  Motome, Arita, h.~Arima, Tokura, and Seki}]{Yasui2020imaging}
\bibinfo{author}{Y.~Yasui}, \bibinfo{author}{C.~J. Butler},
  \bibinfo{author}{N.~D. Khanh}, \bibinfo{author}{S.~Hayami},
  \bibinfo{author}{T.~Nomoto}, \bibinfo{author}{T.~Hanaguri},
  \bibinfo{author}{Y.~Motome}, \bibinfo{author}{R.~Arita},
  \bibinfo{author}{T.~h.~Arima}, \bibinfo{author}{Y.~Tokura},
  \bibinfo{author}{S.~Seki},
\newblock \bibinfo{title}{Imaging the coupling between itinerant electrons and
  localised moments in the centrosymmetric skyrmion magnet {GdRu$_2$Si$_2$}},
\newblock \bibinfo{journal}{Nat. Commun.} \bibinfo{volume}{11}
  (\bibinfo{year}{2020}) \bibinfo{pages}{5925}.
  \DOIprefix\doi{10.1038/s41467-020-19751-4}.
\bibitem[{Khanh et~al.(2022)Khanh, Nakajima, Hayami, Gao, Yamasaki, Sagayama,
  Nakao, Takagi, Motome, Tokura, Arima, and Seki}]{khanh2022zoology}
\bibinfo{author}{N.~D. Khanh}, \bibinfo{author}{T.~Nakajima},
  \bibinfo{author}{S.~Hayami}, \bibinfo{author}{S.~Gao},
  \bibinfo{author}{Y.~Yamasaki}, \bibinfo{author}{H.~Sagayama},
  \bibinfo{author}{H.~Nakao}, \bibinfo{author}{R.~Takagi},
  \bibinfo{author}{Y.~Motome}, \bibinfo{author}{Y.~Tokura},
  \bibinfo{author}{T.-h. Arima}, \bibinfo{author}{S.~Seki},
\newblock \bibinfo{title}{Zoology of {M}ultiple-${Q}$ {S}pin {T}extures in a
  {C}entrosymmetric {T}etragonal {M}agnet with {I}tinerant {E}lectrons},
\newblock \bibinfo{journal}{Adv. Sci.} \bibinfo{volume}{9}
  (\bibinfo{year}{2022}) \bibinfo{pages}{2105452}.
  \DOIprefix\doi{10.1002/advs.202105452}.
\bibitem[{Matsuyama et~al.(2023)Matsuyama, Nomura, Imajo, Nomoto, Arita, Sudo,
  Kimata, Khanh, Takagi, Tokura, Seki, Kindo, and
  Kohama}]{Matsuyama_PhysRevB.107.104421}
\bibinfo{author}{N.~Matsuyama}, \bibinfo{author}{T.~Nomura},
  \bibinfo{author}{S.~Imajo}, \bibinfo{author}{T.~Nomoto},
  \bibinfo{author}{R.~Arita}, \bibinfo{author}{K.~Sudo},
  \bibinfo{author}{M.~Kimata}, \bibinfo{author}{N.~D. Khanh},
  \bibinfo{author}{R.~Takagi}, \bibinfo{author}{Y.~Tokura},
  \bibinfo{author}{S.~Seki}, \bibinfo{author}{K.~Kindo},
  \bibinfo{author}{Y.~Kohama},
\newblock \bibinfo{title}{Quantum oscillations in the centrosymmetric
  skyrmion-hosting magnet ${{\mathrm{GdRu}}_{2}{\mathrm{Si}}_{2}}$},
\newblock \bibinfo{journal}{Phys. Rev. B} \bibinfo{volume}{107}
  (\bibinfo{year}{2023}) \bibinfo{pages}{104421}.
  \DOIprefix\doi{10.1103/PhysRevB.107.104421}.
\bibitem[{Wood et~al.(2023)Wood, Khalyavin, Mayoh, Bouaziz, Hall, Holt,
  Orlandi, Manuel, Bl\"ugel, Staunton, Petrenko, Lees, and
  Balakrishnan}]{Wood_PhysRevB.107.L180402}
\bibinfo{author}{G.~D.~A. Wood}, \bibinfo{author}{D.~D. Khalyavin},
  \bibinfo{author}{D.~A. Mayoh}, \bibinfo{author}{J.~Bouaziz},
  \bibinfo{author}{A.~E. Hall}, \bibinfo{author}{S.~J.~R. Holt},
  \bibinfo{author}{F.~Orlandi}, \bibinfo{author}{P.~Manuel},
  \bibinfo{author}{S.~Bl\"ugel}, \bibinfo{author}{J.~B. Staunton},
  \bibinfo{author}{O.~A. Petrenko}, \bibinfo{author}{M.~R. Lees},
  \bibinfo{author}{G.~Balakrishnan},
\newblock \bibinfo{title}{Double-$q$ ground state with topological charge
  stripes in the centrosymmetric skyrmion candidate
  ${{\mathrm{GdRu}}_{2}{\mathrm{Si}}_{2}}$},
\newblock \bibinfo{journal}{Phys. Rev. B} \bibinfo{volume}{107}
  (\bibinfo{year}{2023}) \bibinfo{pages}{L180402}.
  \DOIprefix\doi{10.1103/PhysRevB.107.L180402}.
\bibitem[{Hayami and Kato(2023)}]{hayami2023widely}
\bibinfo{author}{S.~Hayami}, \bibinfo{author}{Y.~Kato},
\newblock \bibinfo{title}{Widely-sweeping magnetic field--temperature phase
  diagrams for skyrmion-hosting centrosymmetric tetragonal magnets},
\newblock \bibinfo{journal}{J. Magn. Magn. Mater.} \bibinfo{volume}{571}
  (\bibinfo{year}{2023}) \bibinfo{pages}{170547}.
  \DOIprefix\doi{https://doi.org/10.1016/j.jmmm.2023.170547}.
\bibitem[{Eremeev et~al.(2023)Eremeev, Glazkova, Poelchen, Kraiker, Ali,
  Tarasov, Schulz, Kliemt, Chulkov, Stolyarov et~al.}]{eremeev2023insight}
\bibinfo{author}{S.~Eremeev}, \bibinfo{author}{D.~Glazkova},
  \bibinfo{author}{G.~Poelchen}, \bibinfo{author}{A.~Kraiker},
  \bibinfo{author}{K.~Ali}, \bibinfo{author}{A.~V. Tarasov},
  \bibinfo{author}{S.~Schulz}, \bibinfo{author}{K.~Kliemt},
  \bibinfo{author}{E.~V. Chulkov}, \bibinfo{author}{V.~Stolyarov}, et~al.,
\newblock \bibinfo{title}{{Insight into the electronic structure of the
  centrosymmetric skyrmion magnet GdRu$_2$Si$_2$}},
\newblock \bibinfo{journal}{Nanoscale Adv.} \bibinfo{volume}{5}
  (\bibinfo{year}{2023}) \bibinfo{pages}{6678--6687}.
  \DOIprefix\doi{https://doi.org/10.1039/D3NA00435J}.
\bibitem[{Nomoto and Arita(2023)}]{nomoto2023ab}
\bibinfo{author}{T.~Nomoto}, \bibinfo{author}{R.~Arita},
\newblock \bibinfo{title}{Ab initio exploration of short-pitch skyrmion
  materials: {R}ole of orbital frustration},
\newblock \bibinfo{journal}{J. Appl. Phys.} \bibinfo{volume}{133}
  (\bibinfo{year}{2023}). \DOIprefix\doi{https://doi.org/10.1063/5.0141628}.
\bibitem[{Spethmann et~al.(2024)Spethmann, Khanh, Yoshimochi, Takagi, Hayami,
  Motome, Wiesendanger, Seki, and von
  Bergmann}]{Spethmann_PhysRevMaterials.8.064404}
\bibinfo{author}{J.~Spethmann}, \bibinfo{author}{N.~D. Khanh},
  \bibinfo{author}{H.~Yoshimochi}, \bibinfo{author}{R.~Takagi},
  \bibinfo{author}{S.~Hayami}, \bibinfo{author}{Y.~Motome},
  \bibinfo{author}{R.~Wiesendanger}, \bibinfo{author}{S.~Seki},
  \bibinfo{author}{K.~von Bergmann},
\newblock \bibinfo{title}{{SP-STM study of the multi-Q phases in
  ${\mathrm{GdRu}}_{2}{\mathrm{Si}}_{2}$}},
\newblock \bibinfo{journal}{Phys. Rev. Mater.} \bibinfo{volume}{8}
  (\bibinfo{year}{2024}) \bibinfo{pages}{064404}.
  \DOIprefix\doi{10.1103/PhysRevMaterials.8.064404}.
\bibitem[{Shang et~al.(2021)Shang, Xu, Gawryluk, Ma, Shiroka, Shi, and
  Pomjakushina}]{Shang_PhysRevB.103.L020405}
\bibinfo{author}{T.~Shang}, \bibinfo{author}{Y.~Xu}, \bibinfo{author}{D.~J.
  Gawryluk}, \bibinfo{author}{J.~Z. Ma}, \bibinfo{author}{T.~Shiroka},
  \bibinfo{author}{M.~Shi}, \bibinfo{author}{E.~Pomjakushina},
\newblock \bibinfo{title}{Anomalous {Hall} resistivity and possible topological
  {Hall} effect in the {${\mathrm{EuAl}}_{4}$} antiferromagnet},
\newblock \bibinfo{journal}{Phys. Rev. B} \bibinfo{volume}{103}
  (\bibinfo{year}{2021}) \bibinfo{pages}{L020405}.
  \DOIprefix\doi{10.1103/PhysRevB.103.L020405}.
\bibitem[{Kaneko et~al.(2021)Kaneko, Kawasaki, Nakamura, Munakata, Nakao,
  Hanashima, Kiyanagi, Ohhara, Hedo, Nakama, and Onuki}]{kaneko2021charge}
\bibinfo{author}{K.~Kaneko}, \bibinfo{author}{T.~Kawasaki},
  \bibinfo{author}{A.~Nakamura}, \bibinfo{author}{K.~Munakata},
  \bibinfo{author}{A.~Nakao}, \bibinfo{author}{T.~Hanashima},
  \bibinfo{author}{R.~Kiyanagi}, \bibinfo{author}{T.~Ohhara},
  \bibinfo{author}{M.~Hedo}, \bibinfo{author}{T.~Nakama},
  \bibinfo{author}{Y.~Onuki},
\newblock \bibinfo{title}{Charge-density-wave order and multiple magnetic
  transitions in divalent europium compound {EuAl$_4$}},
\newblock \bibinfo{journal}{J. Phys. Soc. Jpn.} \bibinfo{volume}{90}
  (\bibinfo{year}{2021}) \bibinfo{pages}{064704}.
  \DOIprefix\doi{10.7566/JPSJ.90.064704}.
\bibitem[{Takagi et~al.(2022)Takagi, Matsuyama, Ukleev, Yu, White, Francoual,
  Mardegan, Hayami, Saito, Kaneko, Ohishi, {\=O}nuki, Arima, Tokura, Nakajima,
  and Seki}]{takagi2022square}
\bibinfo{author}{R.~Takagi}, \bibinfo{author}{N.~Matsuyama},
  \bibinfo{author}{V.~Ukleev}, \bibinfo{author}{L.~Yu}, \bibinfo{author}{J.~S.
  White}, \bibinfo{author}{S.~Francoual}, \bibinfo{author}{J.~R.~L. Mardegan},
  \bibinfo{author}{S.~Hayami}, \bibinfo{author}{H.~Saito},
  \bibinfo{author}{K.~Kaneko}, \bibinfo{author}{K.~Ohishi},
  \bibinfo{author}{Y.~{\=O}nuki}, \bibinfo{author}{T.-h. Arima},
  \bibinfo{author}{Y.~Tokura}, \bibinfo{author}{T.~Nakajima},
  \bibinfo{author}{S.~Seki},
\newblock \bibinfo{title}{Square and rhombic lattices of magnetic skyrmions in
  a centrosymmetric binary compound},
\newblock \bibinfo{journal}{Nat. Commun.} \bibinfo{volume}{13}
  (\bibinfo{year}{2022}) \bibinfo{pages}{1472}.
  \DOIprefix\doi{10.1038/s41467-022-29131-9}.
\bibitem[{Zhu et~al.(2022)Zhu, Zhang, Gawryluk, Zhen, Yu, Ju, Xie, Jiang,
  Cheng, Xu, Shi, Pomjakushina, Zhan, Shiroka, and
  Shang}]{Zhu_PhysRevB.105.014423}
\bibinfo{author}{X.~Y. Zhu}, \bibinfo{author}{H.~Zhang}, \bibinfo{author}{D.~J.
  Gawryluk}, \bibinfo{author}{Z.~X. Zhen}, \bibinfo{author}{B.~C. Yu},
  \bibinfo{author}{S.~L. Ju}, \bibinfo{author}{W.~Xie}, \bibinfo{author}{D.~M.
  Jiang}, \bibinfo{author}{W.~J. Cheng}, \bibinfo{author}{Y.~Xu},
  \bibinfo{author}{M.~Shi}, \bibinfo{author}{E.~Pomjakushina},
  \bibinfo{author}{Q.~F. Zhan}, \bibinfo{author}{T.~Shiroka},
  \bibinfo{author}{T.~Shang},
\newblock \bibinfo{title}{Spin order and fluctuations in the
  ${{\mathrm{EuAl}}_{4}}$ and ${{\mathrm{EuGa}}_{4}}$ topological
  antiferromagnets: A $\ensuremath{\mu}\mathrm{SR}$ study},
\newblock \bibinfo{journal}{Phys. Rev. B} \bibinfo{volume}{105}
  (\bibinfo{year}{2022}) \bibinfo{pages}{014423}.
  \DOIprefix\doi{10.1103/PhysRevB.105.014423}.
\bibitem[{Gen et~al.(2023)Gen, Takagi, Watanabe, Kitou, Sagayama, Matsuyama,
  Kohama, Ikeda, \ifmmode~\bar{O}\else \={O}\fi{}nuki, Kurumaji, Arima, and
  Seki}]{Gen_PhysRevB.107.L020410}
\bibinfo{author}{M.~Gen}, \bibinfo{author}{R.~Takagi},
  \bibinfo{author}{Y.~Watanabe}, \bibinfo{author}{S.~Kitou},
  \bibinfo{author}{H.~Sagayama}, \bibinfo{author}{N.~Matsuyama},
  \bibinfo{author}{Y.~Kohama}, \bibinfo{author}{A.~Ikeda},
  \bibinfo{author}{Y.~\ifmmode~\bar{O}\else \={O}\fi{}nuki},
  \bibinfo{author}{T.~Kurumaji}, \bibinfo{author}{T.-h. Arima},
  \bibinfo{author}{S.~Seki},
\newblock \bibinfo{title}{Rhombic skyrmion lattice coupled with orthorhombic
  structural distortion in ${{\mathrm{EuAl}}_{4}}$},
\newblock \bibinfo{journal}{Phys. Rev. B} \bibinfo{volume}{107}
  (\bibinfo{year}{2023}) \bibinfo{pages}{L020410}.
  \DOIprefix\doi{10.1103/PhysRevB.107.L020410}.
\bibitem[{Moya et~al.(2023)Moya, Huang, Lei, Allen, Gao, Sun, Yi, and
  Morosan}]{Moya_PhysRevB.108.064436}
\bibinfo{author}{J.~M. Moya}, \bibinfo{author}{J.~Huang},
  \bibinfo{author}{S.~Lei}, \bibinfo{author}{K.~Allen},
  \bibinfo{author}{Y.~Gao}, \bibinfo{author}{Y.~Sun}, \bibinfo{author}{M.~Yi},
  \bibinfo{author}{E.~Morosan},
\newblock \bibinfo{title}{Real-space and reciprocal-space topology in the
  {${{\mathrm{Eu}(\mathrm{Ga}}_{1\ensuremath{-}x}{\mathrm{Al}}_{x})}_{4}$}
  square net system},
\newblock \bibinfo{journal}{Phys. Rev. B} \bibinfo{volume}{108}
  (\bibinfo{year}{2023}) \bibinfo{pages}{064436}.
  \DOIprefix\doi{10.1103/PhysRevB.108.064436}.
\bibitem[{Yang et~al.(2024)Yang, Le, Zhu, Wang, Shang, Dai, Hu, and
  Dressel}]{Yang_PhysRevB.109.L041113}
\bibinfo{author}{R.~Yang}, \bibinfo{author}{C.-C. Le},
  \bibinfo{author}{P.~Zhu}, \bibinfo{author}{Z.-W. Wang},
  \bibinfo{author}{T.~Shang}, \bibinfo{author}{Y.-M. Dai},
  \bibinfo{author}{J.-P. Hu}, \bibinfo{author}{M.~Dressel},
\newblock \bibinfo{title}{Charge density wave transition in the magnetic
  topological semimetal {${{\mathrm{EuAl}}_{4}}$}},
\newblock \bibinfo{journal}{Phys. Rev. B} \bibinfo{volume}{109}
  (\bibinfo{year}{2024}) \bibinfo{pages}{L041113}.
  \DOIprefix\doi{10.1103/PhysRevB.109.L041113}.
\bibitem[{Miao et~al.(2024)Miao, Bouaziz, Fabbris, Meier, Yang, Li, Nelson,
  Vescovo, Zhang, Christianson, Lee, Zhang, Batista, and
  Bl\"ugel}]{Miao_PhysRevX.14.011053}
\bibinfo{author}{H.~Miao}, \bibinfo{author}{J.~Bouaziz},
  \bibinfo{author}{G.~Fabbris}, \bibinfo{author}{W.~R. Meier},
  \bibinfo{author}{F.~Z. Yang}, \bibinfo{author}{H.~X. Li},
  \bibinfo{author}{C.~Nelson}, \bibinfo{author}{E.~Vescovo},
  \bibinfo{author}{S.~Zhang}, \bibinfo{author}{A.~D. Christianson},
  \bibinfo{author}{H.~N. Lee}, \bibinfo{author}{Y.~Zhang},
  \bibinfo{author}{C.~D. Batista}, \bibinfo{author}{S.~Bl\"ugel},
\newblock \bibinfo{title}{Spontaneous chirality flipping in an orthogonal
  spin-charge ordered topological magnet},
\newblock \bibinfo{journal}{Phys. Rev. X} \bibinfo{volume}{14}
  (\bibinfo{year}{2024}) \bibinfo{pages}{011053}.
  \DOIprefix\doi{10.1103/PhysRevX.14.011053}.
\bibitem[{Yoshimochi et~al.(2024)Yoshimochi, Takagi, Ju, Khanh, Saito,
  Sagayama, Nakao, Itoh, Tokura, Arima, Hayami, Nakajima, and
  Seki}]{yoshimochi2024multi}
\bibinfo{author}{H.~Yoshimochi}, \bibinfo{author}{R.~Takagi},
  \bibinfo{author}{J.~Ju}, \bibinfo{author}{N.~Khanh},
  \bibinfo{author}{H.~Saito}, \bibinfo{author}{H.~Sagayama},
  \bibinfo{author}{H.~Nakao}, \bibinfo{author}{S.~Itoh},
  \bibinfo{author}{Y.~Tokura}, \bibinfo{author}{T.~Arima},
  \bibinfo{author}{S.~Hayami}, \bibinfo{author}{T.~Nakajima},
  \bibinfo{author}{S.~Seki},
\newblock \bibinfo{title}{Multistep topological transitions among meron and
  skyrmion crystals in a centrosymmetric magnet},
\newblock \bibinfo{journal}{Nat. Phys.}  (\bibinfo{year}{2024}).
  \DOIprefix\doi{https://doi.org/10.1038/s41567-024-02445-9}.
\bibitem[{G{\"o}bel et~al.(2021)G{\"o}bel, Mertig, and
  Tretiakov}]{gobel2021beyond}
\bibinfo{author}{B.~G{\"o}bel}, \bibinfo{author}{I.~Mertig},
  \bibinfo{author}{O.~A. Tretiakov},
\newblock \bibinfo{title}{Beyond skyrmions: Review and perspectives of
  alternative magnetic quasiparticles},
\newblock \bibinfo{journal}{Phys. Rep.} \bibinfo{volume}{895}
  (\bibinfo{year}{2021}) \bibinfo{pages}{1}.
  \DOIprefix\doi{10.1016/j.physrep.2020.10.001}.
\bibitem[{{Gao} et~al.(2017){Gao}, {Zaharko}, {Tsurkan}, {Su}, {White},
  {Tucker}, {Roessli}, {Bourdarot}, {Sibille}, {Chernyshov}, {Fennell},
  {Loidl}, and {R{\"u}egg}}]{Gao2016Spiral}
\bibinfo{author}{S.~{Gao}}, \bibinfo{author}{O.~{Zaharko}},
  \bibinfo{author}{V.~{Tsurkan}}, \bibinfo{author}{Y.~{Su}},
  \bibinfo{author}{J.~S. {White}}, \bibinfo{author}{G.~S. {Tucker}},
  \bibinfo{author}{B.~{Roessli}}, \bibinfo{author}{F.~{Bourdarot}},
  \bibinfo{author}{R.~{Sibille}}, \bibinfo{author}{D.~{Chernyshov}},
  \bibinfo{author}{T.~{Fennell}}, \bibinfo{author}{A.~{Loidl}},
  \bibinfo{author}{C.~{R{\"u}egg}},
\newblock \bibinfo{title}{{Spiral spin-liquid and the emergence of a
  vortex-like state in MnSc$_2$S$_4$}},
\newblock \bibinfo{journal}{Nat. Phys.} \bibinfo{volume}{13}
  (\bibinfo{year}{2017}) \bibinfo{pages}{157--161}.
  \DOIprefix\doi{10.1038/nphys3914}.
\bibitem[{Gao et~al.(2020)Gao, Rosales, Albarrac{\'\i}n, Tsurkan, Kaur,
  Fennell, Steffens, Boehm, {\v{C}}erm{\'a}k, Schneidewind, Ressouche,
  C.~Cabra, R{\"u}egg, and Oksana}]{gao2020fractional}
\bibinfo{author}{S.~Gao}, \bibinfo{author}{H.~D. Rosales},
  \bibinfo{author}{F.~A.~G. Albarrac{\'\i}n}, \bibinfo{author}{V.~Tsurkan},
  \bibinfo{author}{G.~Kaur}, \bibinfo{author}{T.~Fennell},
  \bibinfo{author}{P.~Steffens}, \bibinfo{author}{M.~Boehm},
  \bibinfo{author}{P.~{\v{C}}erm{\'a}k}, \bibinfo{author}{A.~Schneidewind},
  \bibinfo{author}{E.~Ressouche}, \bibinfo{author}{D.~C.~Cabra},
  \bibinfo{author}{C.~R{\"u}egg}, \bibinfo{author}{Z.~Oksana},
\newblock \bibinfo{title}{Fractional antiferromagnetic skyrmion lattice induced
  by anisotropic couplings},
\newblock \bibinfo{journal}{Nature} \bibinfo{volume}{586}
  (\bibinfo{year}{2020}) \bibinfo{pages}{37--41}.
  \DOIprefix\doi{10.1038/s41586-020-2716-8}.
\bibitem[{Rosales et~al.(2022)Rosales, Albarrac\'{\i}n, Guratinder, Tsurkan,
  Prodan, Ressouche, and Zaharko}]{Rosales_PhysRevB.105.224402}
\bibinfo{author}{H.~D. Rosales}, \bibinfo{author}{F.~A.~G. Albarrac\'{\i}n},
  \bibinfo{author}{K.~Guratinder}, \bibinfo{author}{V.~Tsurkan},
  \bibinfo{author}{L.~Prodan}, \bibinfo{author}{E.~Ressouche},
  \bibinfo{author}{O.~Zaharko},
\newblock \bibinfo{title}{{Anisotropy-driven response of the fractional
  antiferromagnetic skyrmion lattice in ${\mathrm{MnSc}}_{2}{\mathrm{S}}_{4}$
  to applied magnetic fields}},
\newblock \bibinfo{journal}{Phys. Rev. B} \bibinfo{volume}{105}
  (\bibinfo{year}{2022}) \bibinfo{pages}{224402}.
  \DOIprefix\doi{10.1103/PhysRevB.105.224402}.
\bibitem[{Takeda et~al.(2024)Takeda, Kawano, Tamura, Akazawa, Yan, Waki,
  Nakamura, Sato, Narumi, Hagiwara, Yamashita, and Hotta}]{takeda2024magnon}
\bibinfo{author}{H.~Takeda}, \bibinfo{author}{M.~Kawano},
  \bibinfo{author}{K.~Tamura}, \bibinfo{author}{M.~Akazawa},
  \bibinfo{author}{J.~Yan}, \bibinfo{author}{T.~Waki},
  \bibinfo{author}{H.~Nakamura}, \bibinfo{author}{K.~Sato},
  \bibinfo{author}{Y.~Narumi}, \bibinfo{author}{M.~Hagiwara},
  \bibinfo{author}{M.~Yamashita}, \bibinfo{author}{C.~Hotta},
\newblock \bibinfo{title}{Magnon thermal {Hall} effect via emergent {SU(3)}
  flux on the antiferromagnetic skyrmion lattice},
\newblock \bibinfo{journal}{Nat. Commun.} \bibinfo{volume}{15}
  (\bibinfo{year}{2024}) \bibinfo{pages}{566}.
  \DOIprefix\doi{https://doi.org/10.1038/s41467-024-44793-3}.
\bibitem[{Tanigaki et~al.(2015)Tanigaki, Shibata, Kanazawa, Yu, Onose, Park,
  Shindo, and Tokura}]{tanigaki2015real}
\bibinfo{author}{T.~Tanigaki}, \bibinfo{author}{K.~Shibata},
  \bibinfo{author}{N.~Kanazawa}, \bibinfo{author}{X.~Yu},
  \bibinfo{author}{Y.~Onose}, \bibinfo{author}{H.~S. Park},
  \bibinfo{author}{D.~Shindo}, \bibinfo{author}{Y.~Tokura},
\newblock \bibinfo{title}{{Real-space observation of short-period cubic lattice
  of skyrmions in MnGe}},
\newblock \bibinfo{journal}{Nano Lett.} \bibinfo{volume}{15}
  (\bibinfo{year}{2015}) \bibinfo{pages}{5438--5442}.
  \DOIprefix\doi{10.1021/acs.nanolett.5b02653}.
\bibitem[{Kanazawa et~al.(2017)Kanazawa, Seki, and
  Tokura}]{kanazawa2017noncentrosymmetric}
\bibinfo{author}{N.~Kanazawa}, \bibinfo{author}{S.~Seki},
  \bibinfo{author}{Y.~Tokura},
\newblock \bibinfo{title}{Noncentrosymmetric magnets hosting magnetic
  skyrmions},
\newblock \bibinfo{journal}{Adv. Mater.} \bibinfo{volume}{29}
  (\bibinfo{year}{2017}) \bibinfo{pages}{1603227}.
  \DOIprefix\doi{10.1002/adma.201603227}.
\bibitem[{Fujishiro et~al.(2019)Fujishiro, Kanazawa, Nakajima, Yu, Ohishi,
  Kawamura, Kakurai, Arima, Mitamura, Miyake, Akiba, Tokunaga, Matsuo, Kindo,
  Koretsune, Arita, and Tokura}]{fujishiro2019topological}
\bibinfo{author}{Y.~Fujishiro}, \bibinfo{author}{N.~Kanazawa},
  \bibinfo{author}{T.~Nakajima}, \bibinfo{author}{X.~Z. Yu},
  \bibinfo{author}{K.~Ohishi}, \bibinfo{author}{Y.~Kawamura},
  \bibinfo{author}{K.~Kakurai}, \bibinfo{author}{T.~Arima},
  \bibinfo{author}{H.~Mitamura}, \bibinfo{author}{A.~Miyake},
  \bibinfo{author}{K.~Akiba}, \bibinfo{author}{M.~Tokunaga},
  \bibinfo{author}{A.~Matsuo}, \bibinfo{author}{K.~Kindo},
  \bibinfo{author}{T.~Koretsune}, \bibinfo{author}{R.~Arita},
  \bibinfo{author}{Y.~Tokura},
\newblock \bibinfo{title}{Topological transitions among skyrmion-and
  hedgehog-lattice states in cubic chiral magnets},
\newblock \bibinfo{journal}{Nat. Commun.} \bibinfo{volume}{10}
  (\bibinfo{year}{2019}) \bibinfo{pages}{1059}.
  \DOIprefix\doi{10.1038/s41467-019-08985-6}.
\bibitem[{Kanazawa et~al.(2020)Kanazawa, Kitaori, White, Ukleev, R\o{}nnow,
  Tsukazaki, Ichikawa, Kawasaki, and Tokura}]{Kanazawa_PhysRevLett.125.137202}
\bibinfo{author}{N.~Kanazawa}, \bibinfo{author}{A.~Kitaori},
  \bibinfo{author}{J.~S. White}, \bibinfo{author}{V.~Ukleev},
  \bibinfo{author}{H.~M. R\o{}nnow}, \bibinfo{author}{A.~Tsukazaki},
  \bibinfo{author}{M.~Ichikawa}, \bibinfo{author}{M.~Kawasaki},
  \bibinfo{author}{Y.~Tokura},
\newblock \bibinfo{title}{Direct observation of the statics and dynamics of
  emergent magnetic monopoles in a chiral magnet},
\newblock \bibinfo{journal}{Phys. Rev. Lett.} \bibinfo{volume}{125}
  (\bibinfo{year}{2020}) \bibinfo{pages}{137202}.
  \DOIprefix\doi{10.1103/PhysRevLett.125.137202}.
\bibitem[{Ishiwata et~al.(2011)Ishiwata, Tokunaga, Kaneko, Okuyama, Tokunaga,
  Wakimoto, Kakurai, Arima, Taguchi, and Tokura}]{Ishiwata_PhysRevB.84.054427}
\bibinfo{author}{S.~Ishiwata}, \bibinfo{author}{M.~Tokunaga},
  \bibinfo{author}{Y.~Kaneko}, \bibinfo{author}{D.~Okuyama},
  \bibinfo{author}{Y.~Tokunaga}, \bibinfo{author}{S.~Wakimoto},
  \bibinfo{author}{K.~Kakurai}, \bibinfo{author}{T.~Arima},
  \bibinfo{author}{Y.~Taguchi}, \bibinfo{author}{Y.~Tokura},
\newblock \bibinfo{title}{Versatile helimagnetic phases under magnetic fields
  in cubic perovskite ${SrFeO}_{3}$},
\newblock \bibinfo{journal}{Phys. Rev. B} \bibinfo{volume}{84}
  (\bibinfo{year}{2011}) \bibinfo{pages}{054427}.
  \DOIprefix\doi{10.1103/PhysRevB.84.054427}.
\bibitem[{Ishiwata et~al.(2020)Ishiwata, Nakajima, Kim, Inosov, Kanazawa,
  White, Gavilano, Georgii, Seemann, Brandl, Manuel, Khalyavin, Seki, Tokunaga,
  Kinoshita, Long, Kaneko, Taguchi, Arima, Keimer, and
  Tokura}]{Ishiwata_PhysRevB.101.134406}
\bibinfo{author}{S.~Ishiwata}, \bibinfo{author}{T.~Nakajima},
  \bibinfo{author}{J.-H. Kim}, \bibinfo{author}{D.~S. Inosov},
  \bibinfo{author}{N.~Kanazawa}, \bibinfo{author}{J.~S. White},
  \bibinfo{author}{J.~L. Gavilano}, \bibinfo{author}{R.~Georgii},
  \bibinfo{author}{K.~M. Seemann}, \bibinfo{author}{G.~Brandl},
  \bibinfo{author}{P.~Manuel}, \bibinfo{author}{D.~D. Khalyavin},
  \bibinfo{author}{S.~Seki}, \bibinfo{author}{Y.~Tokunaga},
  \bibinfo{author}{M.~Kinoshita}, \bibinfo{author}{Y.~W. Long},
  \bibinfo{author}{Y.~Kaneko}, \bibinfo{author}{Y.~Taguchi},
  \bibinfo{author}{T.~Arima}, \bibinfo{author}{B.~Keimer},
  \bibinfo{author}{Y.~Tokura},
\newblock \bibinfo{title}{{Emergent topological spin structures in the
  centrosymmetric cubic perovskite ${\mathrm{SrFeO}}_{3}$}},
\newblock \bibinfo{journal}{Phys. Rev. B} \bibinfo{volume}{101}
  (\bibinfo{year}{2020}) \bibinfo{pages}{134406}.
  \DOIprefix\doi{10.1103/PhysRevB.101.134406}.
\bibitem[{Rogge et~al.(2019)Rogge, Green, Sutarto, and
  May}]{Rogge_PhysRevMaterials.3.084404}
\bibinfo{author}{P.~C. Rogge}, \bibinfo{author}{R.~J. Green},
  \bibinfo{author}{R.~Sutarto}, \bibinfo{author}{S.~J. May},
\newblock \bibinfo{title}{{Itinerancy-dependent noncollinear spin textures in
  ${\mathrm{SrFeO}}_{3}, {\mathrm{CaFeO}}_{3}$, and
  ${\mathrm{CaFeO}}_{3}/{\mathrm{SrFeO}}_{3}$ heterostructures probed via
  resonant x-ray scattering}},
\newblock \bibinfo{journal}{Phys. Rev. Materials} \bibinfo{volume}{3}
  (\bibinfo{year}{2019}) \bibinfo{pages}{084404}.
  \DOIprefix\doi{10.1103/PhysRevMaterials.3.084404}.
\bibitem[{Onose et~al.(2020)Onose, Takahashi, Sagayama, Yamasaki, and
  Ishiwata}]{Onose_PhysRevMaterials.4.114420}
\bibinfo{author}{M.~Onose}, \bibinfo{author}{H.~Takahashi},
  \bibinfo{author}{H.~Sagayama}, \bibinfo{author}{Y.~Yamasaki},
  \bibinfo{author}{S.~Ishiwata},
\newblock \bibinfo{title}{{Complete phase diagram of
  ${\mathrm{Sr}}_{1-x}{\mathrm{La}}_{x}\mathrm{Fe}{\mathrm{O}}_{3}$ with
  versatile magnetic and charge ordering}},
\newblock \bibinfo{journal}{Phys. Rev. Materials} \bibinfo{volume}{4}
  (\bibinfo{year}{2020}) \bibinfo{pages}{114420}.
  \DOIprefix\doi{10.1103/PhysRevMaterials.4.114420}.
\bibitem[{Yu et~al.(2018)Yu, Koshibae, Tokunaga, Shibata, Taguchi, Nagaosa, and
  Tokura}]{yu2018transformation}
\bibinfo{author}{X.~Z. Yu}, \bibinfo{author}{W.~Koshibae},
  \bibinfo{author}{Y.~Tokunaga}, \bibinfo{author}{K.~Shibata},
  \bibinfo{author}{Y.~Taguchi}, \bibinfo{author}{N.~Nagaosa},
  \bibinfo{author}{Y.~Tokura},
\newblock \bibinfo{title}{Transformation between meron and skyrmion topological
  spin textures in a chiral magnet},
\newblock \bibinfo{journal}{Nature} \bibinfo{volume}{564}
  (\bibinfo{year}{2018}) \bibinfo{pages}{95--98}.
  \DOIprefix\doi{10.1038/s41586-018-0745-3}.
\bibitem[{Zhang et~al.(2016)Zhang, Xia, Zhou, Wang, Liu, Zhao, and
  Ezawa}]{Zhang_PhysRevB.94.094420}
\bibinfo{author}{X.~Zhang}, \bibinfo{author}{J.~Xia},
  \bibinfo{author}{Y.~Zhou}, \bibinfo{author}{D.~Wang},
  \bibinfo{author}{X.~Liu}, \bibinfo{author}{W.~Zhao},
  \bibinfo{author}{M.~Ezawa},
\newblock \bibinfo{title}{Control and manipulation of a magnetic skyrmionium in
  nanostructures},
\newblock \bibinfo{journal}{Phys. Rev. B} \bibinfo{volume}{94}
  (\bibinfo{year}{2016}) \bibinfo{pages}{094420}.
  \DOIprefix\doi{10.1103/PhysRevB.94.094420}.
\bibitem[{Pylypovskyi et~al.(2018)Pylypovskyi, Makarov, Kravchuk, Gaididei,
  Saxena, and Sheka}]{Pylypovskyi_PhysRevApplied.10.064057}
\bibinfo{author}{O.~V. Pylypovskyi}, \bibinfo{author}{D.~Makarov},
  \bibinfo{author}{V.~P. Kravchuk}, \bibinfo{author}{Y.~Gaididei},
  \bibinfo{author}{A.~Saxena}, \bibinfo{author}{D.~D. Sheka},
\newblock \bibinfo{title}{{Chiral Skyrmion and Skyrmionium States Engineered by
  the Gradient of Curvature}},
\newblock \bibinfo{journal}{Phys. Rev. Appl.} \bibinfo{volume}{10}
  (\bibinfo{year}{2018}) \bibinfo{pages}{064057}.
  \DOIprefix\doi{10.1103/PhysRevApplied.10.064057}.
\bibitem[{Zhang et~al.(2018)Zhang, Kronast, van~der Laan, and
  Hesjedal}]{zhang2018real}
\bibinfo{author}{S.~Zhang}, \bibinfo{author}{F.~Kronast},
  \bibinfo{author}{G.~van~der Laan}, \bibinfo{author}{T.~Hesjedal},
\newblock \bibinfo{title}{Real-space observation of skyrmionium in a
  ferromagnet-magnetic topological insulator heterostructure},
\newblock \bibinfo{journal}{Nano Lett.} \bibinfo{volume}{18}
  (\bibinfo{year}{2018}) \bibinfo{pages}{1057--1063}.
\bibitem[{Barts and Mostovoy(2021)}]{barts2021magnetic}
\bibinfo{author}{E.~Barts}, \bibinfo{author}{M.~Mostovoy},
\newblock \bibinfo{title}{Magnetic particles and strings in iron langasite},
\newblock \bibinfo{journal}{npj Quantum Mater.} \bibinfo{volume}{6}
  (\bibinfo{year}{2021}) \bibinfo{pages}{104}.
  \DOIprefix\doi{https://doi.org/10.1038/s41535-021-00408-4}.
\bibitem[{Liu et~al.(2018)Liu, Lake, and Zang}]{Liu_PhysRevB.98.174437}
\bibinfo{author}{Y.~Liu}, \bibinfo{author}{R.~K. Lake},
  \bibinfo{author}{J.~Zang},
\newblock \bibinfo{title}{Binding a hopfion in a chiral magnet nanodisk},
\newblock \bibinfo{journal}{Phys. Rev. B} \bibinfo{volume}{98}
  (\bibinfo{year}{2018}) \bibinfo{pages}{174437}.
  \DOIprefix\doi{10.1103/PhysRevB.98.174437}.
\bibitem[{Sutcliffe(2017)}]{Sutcliffe_PhysRevLett.118.247203}
\bibinfo{author}{P.~Sutcliffe},
\newblock \bibinfo{title}{Skyrmion knots in frustrated magnets},
\newblock \bibinfo{journal}{Phys. Rev. Lett.} \bibinfo{volume}{118}
  (\bibinfo{year}{2017}) \bibinfo{pages}{247203}.
  \DOIprefix\doi{10.1103/PhysRevLett.118.247203}.
\bibitem[{Sutcliffe(2018)}]{sutcliffe2018hopfions}
\bibinfo{author}{P.~Sutcliffe},
\newblock \bibinfo{title}{Hopfions in chiral magnets},
\newblock \bibinfo{journal}{J. Phys. A} \bibinfo{volume}{51}
  (\bibinfo{year}{2018}) \bibinfo{pages}{375401}.
  \DOIprefix\doi{10.1088/1751-8121/aad521}.
\bibitem[{Tai and Smalyukh(2018)}]{Tai_PhysRevLett.121.187201}
\bibinfo{author}{J.-S.~B. Tai}, \bibinfo{author}{I.~I. Smalyukh},
\newblock \bibinfo{title}{{Static Hopf Solitons and Knotted Emergent Fields in
  Solid-State Noncentrosymmetric Magnetic Nanostructures}},
\newblock \bibinfo{journal}{Phys. Rev. Lett.} \bibinfo{volume}{121}
  (\bibinfo{year}{2018}) \bibinfo{pages}{187201}.
  \DOIprefix\doi{10.1103/PhysRevLett.121.187201}.
\bibitem[{G\"obel et~al.(2020)G\"obel, Akosa, Tatara, and
  Mertig}]{Gobel_PhysRevResearch.2.013315}
\bibinfo{author}{B.~G\"obel}, \bibinfo{author}{C.~A. Akosa},
  \bibinfo{author}{G.~Tatara}, \bibinfo{author}{I.~Mertig},
\newblock \bibinfo{title}{Topological {Hall} signatures of magnetic hopfions},
\newblock \bibinfo{journal}{Phys. Rev. Research} \bibinfo{volume}{2}
  (\bibinfo{year}{2020}) \bibinfo{pages}{013315}.
  \DOIprefix\doi{10.1103/PhysRevResearch.2.013315}.
\bibitem[{Liu et~al.(2020)Liu, Hou, Han, and Zang}]{Liu_PhysRevLett.124.127204}
\bibinfo{author}{Y.~Liu}, \bibinfo{author}{W.~Hou}, \bibinfo{author}{X.~Han},
  \bibinfo{author}{J.~Zang},
\newblock \bibinfo{title}{Three-dimensional dynamics of a magnetic hopfion
  driven by spin transfer torque},
\newblock \bibinfo{journal}{Phys. Rev. Lett.} \bibinfo{volume}{124}
  (\bibinfo{year}{2020}) \bibinfo{pages}{127204}.
  \DOIprefix\doi{10.1103/PhysRevLett.124.127204}.
\bibitem[{Raftrey and Fischer(2021)}]{Raftrey_PhysRevLett.127.257201}
\bibinfo{author}{D.~Raftrey}, \bibinfo{author}{P.~Fischer},
\newblock \bibinfo{title}{Field-driven dynamics of magnetic hopfions},
\newblock \bibinfo{journal}{Phys. Rev. Lett.} \bibinfo{volume}{127}
  (\bibinfo{year}{2021}) \bibinfo{pages}{257201}.
  \DOIprefix\doi{10.1103/PhysRevLett.127.257201}.
\bibitem[{Kent et~al.(2021)Kent, Reynolds, Raftrey, Campbell, Virasawmy, Dhuey,
  Chopdekar, Hierro-Rodriguez, Sorrentino, Pereiro, Ferrer, Hellman, Sutcliffe,
  and Fischer}]{kent2021creation}
\bibinfo{author}{N.~Kent}, \bibinfo{author}{N.~Reynolds},
  \bibinfo{author}{D.~Raftrey}, \bibinfo{author}{I.~T. Campbell},
  \bibinfo{author}{S.~Virasawmy}, \bibinfo{author}{S.~Dhuey},
  \bibinfo{author}{R.~V. Chopdekar}, \bibinfo{author}{A.~Hierro-Rodriguez},
  \bibinfo{author}{A.~Sorrentino}, \bibinfo{author}{E.~Pereiro},
  \bibinfo{author}{S.~Ferrer}, \bibinfo{author}{F.~Hellman},
  \bibinfo{author}{P.~Sutcliffe}, \bibinfo{author}{P.~Fischer},
\newblock \bibinfo{title}{Creation and observation of {Hopfions} in magnetic
  multilayer systems},
\newblock \bibinfo{journal}{Nat. Commun.} \bibinfo{volume}{12}
  (\bibinfo{year}{2021}) \bibinfo{pages}{1562}.
  \DOIprefix\doi{https://doi.org/10.1038/s41467-021-21846-5}.
\bibitem[{Garaud et~al.(2013)Garaud, Carlstr\"om, Babaev, and
  Speight}]{Garaud_PhysRevB.87.014507}
\bibinfo{author}{J.~Garaud}, \bibinfo{author}{J.~Carlstr\"om},
  \bibinfo{author}{E.~Babaev}, \bibinfo{author}{M.~Speight},
\newblock \bibinfo{title}{Chiral $\mathbb{C}{P}^{2}$ skyrmions in three-band
  superconductors},
\newblock \bibinfo{journal}{Phys. Rev. B} \bibinfo{volume}{87}
  (\bibinfo{year}{2013}) \bibinfo{pages}{014507}.
  \DOIprefix\doi{10.1103/PhysRevB.87.014507}.
\bibitem[{Akagi et~al.(2021)Akagi, Amari, Sawado, and
  Shnir}]{Akagi_PhysRevD.103.065008}
\bibinfo{author}{Y.~Akagi}, \bibinfo{author}{Y.~Amari},
  \bibinfo{author}{N.~Sawado}, \bibinfo{author}{Y.~Shnir},
\newblock \bibinfo{title}{Isolated skyrmions in the ${CP}^{2}$ nonlinear sigma
  model with a {Dzyaloshinskii}-{Moriya} type interaction},
\newblock \bibinfo{journal}{Phys. Rev. D} \bibinfo{volume}{103}
  (\bibinfo{year}{2021}) \bibinfo{pages}{065008}.
  \DOIprefix\doi{10.1103/PhysRevD.103.065008}.
\bibitem[{Amari et~al.(2022)Amari, Akagi, Gudnason, Nitta, and
  Shnir}]{Amari_PhysRevB.106.L100406}
\bibinfo{author}{Y.~Amari}, \bibinfo{author}{Y.~Akagi}, \bibinfo{author}{S.~B.
  Gudnason}, \bibinfo{author}{M.~Nitta}, \bibinfo{author}{Y.~Shnir},
\newblock \bibinfo{title}{$\mathbb{C}{P}^{2}$ skyrmion crystals in an {SU(3)}
  magnet with a generalized {Dzyaloshinskii}-{Moriya} interaction},
\newblock \bibinfo{journal}{Phys. Rev. B} \bibinfo{volume}{106}
  (\bibinfo{year}{2022}) \bibinfo{pages}{L100406}.
  \DOIprefix\doi{10.1103/PhysRevB.106.L100406}.
\bibitem[{Zhang et~al.(2023)Zhang, Wang, Dahlbom, Barros, and
  Batista}]{zhang2023cp2}
\bibinfo{author}{H.~Zhang}, \bibinfo{author}{Z.~Wang},
  \bibinfo{author}{D.~Dahlbom}, \bibinfo{author}{K.~Barros},
  \bibinfo{author}{C.~D. Batista},
\newblock \bibinfo{title}{{$CP^{2}$} skyrmions and skyrmion crystals in
  realistic quantum magnets},
\newblock \bibinfo{journal}{Nat. Commun.} \bibinfo{volume}{14}
  (\bibinfo{year}{2023}) \bibinfo{pages}{3626}.
\bibitem[{Hayami and Motome(2021)}]{hayami2021topological}
\bibinfo{author}{S.~Hayami}, \bibinfo{author}{Y.~Motome},
\newblock \bibinfo{title}{Topological spin crystals by itinerant frustration},
\newblock \bibinfo{journal}{J. Phys.: Condens. Matter} \bibinfo{volume}{33}
  (\bibinfo{year}{2021}) \bibinfo{pages}{443001}.
  \DOIprefix\doi{10.1088/1361-648x/ac1a30}.
\bibitem[{Hayami et~al.(2017)Hayami, Ozawa, and
  Motome}]{Hayami_PhysRevB.95.224424}
\bibinfo{author}{S.~Hayami}, \bibinfo{author}{R.~Ozawa},
  \bibinfo{author}{Y.~Motome},
\newblock \bibinfo{title}{Effective bilinear-biquadratic model for noncoplanar
  ordering in itinerant magnets},
\newblock \bibinfo{journal}{Phys. Rev. B} \bibinfo{volume}{95}
  (\bibinfo{year}{2017}) \bibinfo{pages}{224424}.
  \DOIprefix\doi{10.1103/PhysRevB.95.224424}.
\bibitem[{Hayami and Motome(2021{\natexlab{a}})}]{Hayami_PhysRevB.103.024439}
\bibinfo{author}{S.~Hayami}, \bibinfo{author}{Y.~Motome},
\newblock \bibinfo{title}{Square skyrmion crystal in centrosymmetric itinerant
  magnets},
\newblock \bibinfo{journal}{Phys. Rev. B} \bibinfo{volume}{103}
  (\bibinfo{year}{2021}{\natexlab{a}}) \bibinfo{pages}{024439}.
  \DOIprefix\doi{10.1103/PhysRevB.103.024439}.
\bibitem[{Hayami and Motome(2021{\natexlab{b}})}]{Hayami_PhysRevB.103.054422}
\bibinfo{author}{S.~Hayami}, \bibinfo{author}{Y.~Motome},
\newblock \bibinfo{title}{Noncoplanar multiple-{$Q$} spin textures by itinerant
  frustration: Effects of single-ion anisotropy and bond-dependent anisotropy},
\newblock \bibinfo{journal}{Phys. Rev. B} \bibinfo{volume}{103}
  (\bibinfo{year}{2021}{\natexlab{b}}) \bibinfo{pages}{054422}.
  \DOIprefix\doi{10.1103/PhysRevB.103.054422}.
\bibitem[{Yambe and Hayami(2021)}]{yambe2021skyrmion}
\bibinfo{author}{R.~Yambe}, \bibinfo{author}{S.~Hayami},
\newblock \bibinfo{title}{Skyrmion crystals in centrosymmetric itinerant
  magnets without horizontal mirror plane},
\newblock \bibinfo{journal}{Sci. Rep.} \bibinfo{volume}{11}
  (\bibinfo{year}{2021}) \bibinfo{pages}{11184}.
  \DOIprefix\doi{10.1038/s41598-021-90308-1}.
\bibitem[{Hayami(2022{\natexlab{a}})}]{Hayami_PhysRevB.105.014408}
\bibinfo{author}{S.~Hayami},
\newblock \bibinfo{title}{Skyrmion crystal and spiral phases in centrosymmetric
  bilayer magnets with staggered {Dzyaloshinskii}-{Moriya} interaction},
\newblock \bibinfo{journal}{Phys. Rev. B} \bibinfo{volume}{105}
  (\bibinfo{year}{2022}{\natexlab{a}}) \bibinfo{pages}{014408}.
  \DOIprefix\doi{10.1103/PhysRevB.105.014408}.
\bibitem[{Hayami(2022{\natexlab{b}})}]{Hayami_PhysRevB.105.174437}
\bibinfo{author}{S.~Hayami},
\newblock \bibinfo{title}{Rectangular and square skyrmion crystals on a
  centrosymmetric square lattice with easy-axis anisotropy},
\newblock \bibinfo{journal}{Phys. Rev. B} \bibinfo{volume}{105}
  (\bibinfo{year}{2022}{\natexlab{b}}) \bibinfo{pages}{174437}.
  \DOIprefix\doi{10.1103/PhysRevB.105.174437}.
\bibitem[{Hayami(2022{\natexlab{c}})}]{hayami2022multiple}
\bibinfo{author}{S.~Hayami},
\newblock \bibinfo{title}{Multiple skyrmion crystal phases by itinerant
  frustration in centrosymmetric tetragonal magnets},
\newblock \bibinfo{journal}{J. Phys. Soc. Jpn.} \bibinfo{volume}{91}
  (\bibinfo{year}{2022}{\natexlab{c}}) \bibinfo{pages}{023705}.
  \DOIprefix\doi{10.7566/JPSJ.91.023705}.
\bibitem[{Hayami and Yambe(2021)}]{hayami2021field}
\bibinfo{author}{S.~Hayami}, \bibinfo{author}{R.~Yambe},
\newblock \bibinfo{title}{Field-direction sensitive skyrmion crystals in cubic
  chiral systems: Implication to $4f$-electron compound {EuPtSi}},
\newblock \bibinfo{journal}{J. Phys. Soc. Jpn.} \bibinfo{volume}{90}
  (\bibinfo{year}{2021}) \bibinfo{pages}{073705}.
  \DOIprefix\doi{10.7566/JPSJ.90.073705}.
\bibitem[{Okumura et~al.(2020)Okumura, Hayami, Kato, and
  Motome}]{Okumura_PhysRevB.101.144416}
\bibinfo{author}{S.~Okumura}, \bibinfo{author}{S.~Hayami},
  \bibinfo{author}{Y.~Kato}, \bibinfo{author}{Y.~Motome},
\newblock \bibinfo{title}{Magnetic hedgehog lattices in noncentrosymmetric
  metals},
\newblock \bibinfo{journal}{Phys. Rev. B} \bibinfo{volume}{101}
  (\bibinfo{year}{2020}) \bibinfo{pages}{144416}.
  \DOIprefix\doi{10.1103/PhysRevB.101.144416}.
\bibitem[{Hayami and Yambe(2021)}]{Hayami_PhysRevB.104.094425}
\bibinfo{author}{S.~Hayami}, \bibinfo{author}{R.~Yambe},
\newblock \bibinfo{title}{Meron-antimeron crystals in noncentrosymmetric
  itinerant magnets on a triangular lattice},
\newblock \bibinfo{journal}{Phys. Rev. B} \bibinfo{volume}{104}
  (\bibinfo{year}{2021}) \bibinfo{pages}{094425}.
  \DOIprefix\doi{10.1103/PhysRevB.104.094425}.
\bibitem[{Hayami(2021)}]{Hayami_PhysRevB.103.224418}
\bibinfo{author}{S.~Hayami},
\newblock \bibinfo{title}{In-plane magnetic field-induced skyrmion crystal in
  frustrated magnets with easy-plane anisotropy},
\newblock \bibinfo{journal}{Phys. Rev. B} \bibinfo{volume}{103}
  (\bibinfo{year}{2021}) \bibinfo{pages}{224418}.
  \DOIprefix\doi{10.1103/PhysRevB.103.224418}.
\bibitem[{Hayami and Kato(2023)}]{Hayami_PhysRevB.108.024426}
\bibinfo{author}{S.~Hayami}, \bibinfo{author}{Y.~Kato},
\newblock \bibinfo{title}{Magnetic bubble crystal in tetragonal magnets},
\newblock \bibinfo{journal}{Phys. Rev. B} \bibinfo{volume}{108}
  (\bibinfo{year}{2023}) \bibinfo{pages}{024426}.
  \DOIprefix\doi{10.1103/PhysRevB.108.024426}.
\bibitem[{Yambe and Hayami(2023)}]{Yambe_PhysRevB.107.014417}
\bibinfo{author}{R.~Yambe}, \bibinfo{author}{S.~Hayami},
\newblock \bibinfo{title}{Ferrochiral, antiferrochiral, and ferrichiral
  skyrmion crystals in an itinerant honeycomb magnet},
\newblock \bibinfo{journal}{Phys. Rev. B} \bibinfo{volume}{107}
  (\bibinfo{year}{2023}) \bibinfo{pages}{014417}.
  \DOIprefix\doi{10.1103/PhysRevB.107.014417}.
\bibitem[{Hayami(2023)}]{Hayami_doi:10.7566/JPSJ.92.084702}
\bibinfo{author}{S.~Hayami},
\newblock \bibinfo{title}{Antiferro skyrmion crystal phases in a synthetic
  bilayer antiferromagnet under an in-plane magnetic field},
\newblock \bibinfo{journal}{J. Phys. Soc. Jpn.} \bibinfo{volume}{92}
  (\bibinfo{year}{2023}) \bibinfo{pages}{084702}.
  \DOIprefix\doi{10.7566/JPSJ.92.084702}.
\bibitem[{Hayami et~al.(2021)Hayami, Okubo, and Motome}]{hayami2021phase}
\bibinfo{author}{S.~Hayami}, \bibinfo{author}{T.~Okubo},
  \bibinfo{author}{Y.~Motome},
\newblock \bibinfo{title}{Phase shift in skyrmion crystals},
\newblock \bibinfo{journal}{Nat. Commun.} \bibinfo{volume}{12}
  (\bibinfo{year}{2021}) \bibinfo{pages}{6927}.
  \DOIprefix\doi{10.1038/s41467-021-27083-0}.
\bibitem[{Takagi et~al.(2018)Takagi, White, Hayami, Arita, Honecker, R{\o}nnow,
  Tokura, and Seki}]{takagi2018multiple}
\bibinfo{author}{R.~Takagi}, \bibinfo{author}{J.~White},
  \bibinfo{author}{S.~Hayami}, \bibinfo{author}{R.~Arita},
  \bibinfo{author}{D.~Honecker}, \bibinfo{author}{H.~R{\o}nnow},
  \bibinfo{author}{Y.~Tokura}, \bibinfo{author}{S.~Seki},
\newblock \bibinfo{title}{Multiple-$q$ noncollinear magnetism in an itinerant
  hexagonal magnet},
\newblock \bibinfo{journal}{Sci. Adv.} \bibinfo{volume}{4}
  (\bibinfo{year}{2018}) \bibinfo{pages}{eaau3402}.
  \DOIprefix\doi{10.1126/sciadv.aau3402}.
\bibitem[{Seo et~al.(2021)Seo, Hayami, Su, Thomas, Ronning, Bauer, Thompson,
  Lin, and Rosa}]{seo2021spin}
\bibinfo{author}{S.~Seo}, \bibinfo{author}{S.~Hayami}, \bibinfo{author}{Y.~Su},
  \bibinfo{author}{S.~M. Thomas}, \bibinfo{author}{F.~Ronning},
  \bibinfo{author}{E.~D. Bauer}, \bibinfo{author}{J.~D. Thompson},
  \bibinfo{author}{S.-Z. Lin}, \bibinfo{author}{P.~F. Rosa},
\newblock \bibinfo{title}{Spin-texture-driven electrical transport in multi-{Q}
  antiferromagnets},
\newblock \bibinfo{journal}{Commun. Phys.} \bibinfo{volume}{4}
  (\bibinfo{year}{2021}) \bibinfo{pages}{58}.
  \DOIprefix\doi{10.1038/s42005-021-00558-8}.
\bibitem[{Yambe and Hayami(2022)}]{Yambe_PhysRevB.106.174437}
\bibinfo{author}{R.~Yambe}, \bibinfo{author}{S.~Hayami},
\newblock \bibinfo{title}{Effective spin model in momentum space: Toward a
  systematic understanding of multiple-{$Q$} instability by momentum-resolved
  anisotropic exchange interactions},
\newblock \bibinfo{journal}{Phys. Rev. B} \bibinfo{volume}{106}
  (\bibinfo{year}{2022}) \bibinfo{pages}{174437}.
  \DOIprefix\doi{10.1103/PhysRevB.106.174437}.
\bibitem[{Yambe and Hayami(2023)}]{Yambe_PhysRevB.107.174408}
\bibinfo{author}{R.~Yambe}, \bibinfo{author}{S.~Hayami},
\newblock \bibinfo{title}{Anisotropic spin model and multiple-{$Q$} states in
  cubic systems},
\newblock \bibinfo{journal}{Phys. Rev. B} \bibinfo{volume}{107}
  (\bibinfo{year}{2023}) \bibinfo{pages}{174408}.
  \DOIprefix\doi{10.1103/PhysRevB.107.174408}.
\bibitem[{Rau et~al.(2014)Rau, Lee, and Kee}]{Rau_PhysRevLett.112.077204}
\bibinfo{author}{J.~G. Rau}, \bibinfo{author}{E.~K.-H. Lee},
  \bibinfo{author}{H.-Y. Kee},
\newblock \bibinfo{title}{{Generic Spin Model for the Honeycomb Iridates beyond
  the Kitaev Limit}},
\newblock \bibinfo{journal}{Phys. Rev. Lett.} \bibinfo{volume}{112}
  (\bibinfo{year}{2014}) \bibinfo{pages}{077204}.
  \DOIprefix\doi{10.1103/PhysRevLett.112.077204}.
\bibitem[{Hayami et~al.(2016)Hayami, Kusunose, and Motome}]{hayami2016emergent}
\bibinfo{author}{S.~Hayami}, \bibinfo{author}{H.~Kusunose},
  \bibinfo{author}{Y.~Motome},
\newblock \bibinfo{title}{Emergent spin-valley-orbital physics by spontaneous
  parity breaking},
\newblock \bibinfo{journal}{J. Phys.: Condens. Matter} \bibinfo{volume}{28}
  (\bibinfo{year}{2016}) \bibinfo{pages}{395601}.
  \DOIprefix\doi{10.1088/0953-8984/28/39/395601}.
\bibitem[{Nikoli\'{c}(2021)}]{Nikolic_PhysRevB.103.155151}
\bibinfo{author}{P.~Nikoli\'{c}},
\newblock \bibinfo{title}{Dynamics of local magnetic moments induced by
  itinerant weyl electrons},
\newblock \bibinfo{journal}{Phys. Rev. B} \bibinfo{volume}{103}
  (\bibinfo{year}{2021}) \bibinfo{pages}{155151}.
  \DOIprefix\doi{10.1103/PhysRevB.103.155151}.
\bibitem[{Ruderman and Kittel(1954)}]{Ruderman}
\bibinfo{author}{M.~A. Ruderman}, \bibinfo{author}{C.~Kittel},
\newblock \bibinfo{title}{Indirect exchange coupling of nuclear magnetic
  moments by conduction electrons},
\newblock \bibinfo{journal}{Phys. Rev.} \bibinfo{volume}{96}
  (\bibinfo{year}{1954}) \bibinfo{pages}{99--102}.
  \DOIprefix\doi{10.1103/PhysRev.96.99}.
\bibitem[{Kasuya(1956)}]{Kasuya}
\bibinfo{author}{T.~Kasuya},
\newblock \bibinfo{title}{A theory of metallic ferro- and antiferromagnetism on
  {Zener's} model},
\newblock \bibinfo{journal}{Prog. Theor. Phys.} \bibinfo{volume}{16}
  (\bibinfo{year}{1956}) \bibinfo{pages}{45--57}.
  \DOIprefix\doi{10.1143/PTP.16.45}.
\bibitem[{Yosida(1957)}]{Yosida1957}
\bibinfo{author}{K.~Yosida},
\newblock \bibinfo{title}{Magnetic properties of {Cu-Mn} alloys},
\newblock \bibinfo{journal}{Phys. Rev.} \bibinfo{volume}{106}
  (\bibinfo{year}{1957}) \bibinfo{pages}{893--898}.
  \DOIprefix\doi{10.1103/PhysRev.106.893}.
\bibitem[{Hayami(2022)}]{Hayami_PhysRevB.105.224411}
\bibinfo{author}{S.~Hayami},
\newblock \bibinfo{title}{Skyrmion crystal with integer and fractional skyrmion
  numbers in a nonsymmorphic lattice structure with the screw axis},
\newblock \bibinfo{journal}{Phys. Rev. B} \bibinfo{volume}{105}
  (\bibinfo{year}{2022}) \bibinfo{pages}{224411}.
  \DOIprefix\doi{10.1103/PhysRevB.105.224411}.
\bibitem[{Kato et~al.(2021)Kato, Hayami, and Motome}]{Kato_PhysRevB.104.224405}
\bibinfo{author}{Y.~Kato}, \bibinfo{author}{S.~Hayami},
  \bibinfo{author}{Y.~Motome},
\newblock \bibinfo{title}{Spin excitation spectra in helimagnetic states:
  Proper-screw, cycloid, vortex-crystal, and hedgehog lattices},
\newblock \bibinfo{journal}{Phys. Rev. B} \bibinfo{volume}{104}
  (\bibinfo{year}{2021}) \bibinfo{pages}{224405}.
  \DOIprefix\doi{10.1103/PhysRevB.104.224405}.
\bibitem[{Hayami and Yambe(2023)}]{Hayami_PhysRevB.107.174435}
\bibinfo{author}{S.~Hayami}, \bibinfo{author}{R.~Yambe},
\newblock \bibinfo{title}{Field direction dependent skyrmion crystals in
  noncentrosymmetric cubic magnets: A comparison between the point groups
  {$(O,T)$} and ${T}_{\mathrm{d}}$},
\newblock \bibinfo{journal}{Phys. Rev. B} \bibinfo{volume}{107}
  (\bibinfo{year}{2023}) \bibinfo{pages}{174435}.
  \DOIprefix\doi{10.1103/PhysRevB.107.174435}.
\bibitem[{Hayami(2024)}]{Hayami_PhysRevB.109.054422}
\bibinfo{author}{S.~Hayami},
\newblock \bibinfo{title}{Hybrid skyrmion and antiskyrmion phases in polar
  {${C}_{4\mathrm{v}}$} systems},
\newblock \bibinfo{journal}{Phys. Rev. B} \bibinfo{volume}{109}
  (\bibinfo{year}{2024}) \bibinfo{pages}{054422}.
  \DOIprefix\doi{10.1103/PhysRevB.109.054422}.
\bibitem[{Hayami and Motome(2018)}]{Hayami_PhysRevLett.121.137202}
\bibinfo{author}{S.~Hayami}, \bibinfo{author}{Y.~Motome},
\newblock \bibinfo{title}{{N}\'eel- and {Bloch}-type magnetic vortices in
  {R}ashba metals},
\newblock \bibinfo{journal}{Phys. Rev. Lett.} \bibinfo{volume}{121}
  (\bibinfo{year}{2018}) \bibinfo{pages}{137202}.
  \DOIprefix\doi{10.1103/PhysRevLett.121.137202}.
\bibitem[{Puphal et~al.(2020)Puphal, Pomjakushin, Kanazawa, Ukleev, Gawryluk,
  Ma, Naamneh, Plumb, Keller, Cubitt, Pomjakushina, and
  White}]{Puphal_PhysRevLett.124.017202}
\bibinfo{author}{P.~Puphal}, \bibinfo{author}{V.~Pomjakushin},
  \bibinfo{author}{N.~Kanazawa}, \bibinfo{author}{V.~Ukleev},
  \bibinfo{author}{D.~J. Gawryluk}, \bibinfo{author}{J.~Ma},
  \bibinfo{author}{M.~Naamneh}, \bibinfo{author}{N.~C. Plumb},
  \bibinfo{author}{L.~Keller}, \bibinfo{author}{R.~Cubitt},
  \bibinfo{author}{E.~Pomjakushina}, \bibinfo{author}{J.~S. White},
\newblock \bibinfo{title}{{Topological Magnetic Phase in the Candidate Weyl
  Semimetal CeAlGe}},
\newblock \bibinfo{journal}{Phys. Rev. Lett.} \bibinfo{volume}{124}
  (\bibinfo{year}{2020}) \bibinfo{pages}{017202}.
  \DOIprefix\doi{10.1103/PhysRevLett.124.017202}.
\bibitem[{Hayami and Yambe(2022)}]{Hayami_PhysRevB.105.224423}
\bibinfo{author}{S.~Hayami}, \bibinfo{author}{R.~Yambe},
\newblock \bibinfo{title}{Skyrmion crystal under ${D}_{3h}$ point group: Role
  of out-of-plane {Dzyaloshinskii}-{Moriya} interaction},
\newblock \bibinfo{journal}{Phys. Rev. B} \bibinfo{volume}{105}
  (\bibinfo{year}{2022}) \bibinfo{pages}{224423}.
  \DOIprefix\doi{10.1103/PhysRevB.105.224423}.
\bibitem[{Okigami et~al.(2022)Okigami, Yambe, and
  Hayami}]{Okigami_doi:10.7566/JPSJ.91.103701}
\bibinfo{author}{K.~Okigami}, \bibinfo{author}{R.~Yambe},
  \bibinfo{author}{S.~Hayami},
\newblock \bibinfo{title}{Engineering a skyrmion crystal in
  ferromagnetic/antiferromagnetic bilayers based on magnetic frustration
  mechanism},
\newblock \bibinfo{journal}{J. Phys. Soc. Jpn.} \bibinfo{volume}{91}
  (\bibinfo{year}{2022}) \bibinfo{pages}{103701}.
  \DOIprefix\doi{10.7566/JPSJ.91.103701}.
\bibitem[{Hayami(2020)}]{hayami2020multiple}
\bibinfo{author}{S.~Hayami},
\newblock \bibinfo{title}{Multiple-{$Q$} magnetism by anisotropic
  bilinear-biquadratic interactions in momentum space},
\newblock \bibinfo{journal}{J. Magn. Magn. Mater.} \bibinfo{volume}{513}
  (\bibinfo{year}{2020}) \bibinfo{pages}{167181}.
  \DOIprefix\doi{10.1016/j.jmmm.2020.167181}.
\bibitem[{Hayami(2022)}]{Hayami_doi:10.7566/JPSJ.91.093701}
\bibinfo{author}{S.~Hayami},
\newblock \bibinfo{title}{Stability of skyrmion crystal phase in
  centrosymmetric distorted triangular-lattice antiferromagnets},
\newblock \bibinfo{journal}{J. Phys. Soc. Jpn.} \bibinfo{volume}{91}
  (\bibinfo{year}{2022}) \bibinfo{pages}{093701}.
  \DOIprefix\doi{10.7566/JPSJ.91.093701}.
\bibitem[{Hayami(2023{\natexlab{a}})}]{hayami2023orthorhombic}
\bibinfo{author}{S.~Hayami},
\newblock \bibinfo{title}{Orthorhombic distortion and rectangular skyrmion
  crystal in a centrosymmetric tetragonal host},
\newblock \bibinfo{journal}{J. Phys.: Mater.} \bibinfo{volume}{6}
  (\bibinfo{year}{2023}{\natexlab{a}}) \bibinfo{pages}{014006}.
  \DOIprefix\doi{10.1088/2515-7639/acab89}.
\bibitem[{Hayami(2023{\natexlab{b}})}]{Hayami_PhysRevB.108.094416}
\bibinfo{author}{S.~Hayami},
\newblock \bibinfo{title}{Chern insulating state with double-{$Q$} ordering
  wave vectors at the brillouin zone boundary},
\newblock \bibinfo{journal}{Phys. Rev. B} \bibinfo{volume}{108}
  (\bibinfo{year}{2023}{\natexlab{b}}) \bibinfo{pages}{094416}.
  \DOIprefix\doi{10.1103/PhysRevB.108.094416}.
\bibitem[{Hayami(2024)}]{hayami2023anisotropic}
\bibinfo{author}{S.~Hayami},
\newblock \bibinfo{title}{Anisotropic skyrmion crystal on a centrosymmetric
  square lattice under an in-plane magnetic field},
\newblock \bibinfo{journal}{J. Magn. Magn. Mater.} \bibinfo{volume}{604}
  (\bibinfo{year}{2024}) \bibinfo{pages}{172293}.
  \DOIprefix\doi{https://doi.org/10.1016/j.jmmm.2024.172293}.
\bibitem[{Hayami(2023)}]{Hayami_PhysRevB.108.094415}
\bibinfo{author}{S.~Hayami},
\newblock \bibinfo{title}{Checkerboard bubble lattice formed by octuple-period
  quadruple-{$Q$} spin density waves},
\newblock \bibinfo{journal}{Phys. Rev. B} \bibinfo{volume}{108}
  (\bibinfo{year}{2023}) \bibinfo{pages}{094415}.
  \DOIprefix\doi{10.1103/PhysRevB.108.094415}.
\bibitem[{Hayami(2022)}]{hayami2022square}
\bibinfo{author}{S.~Hayami},
\newblock \bibinfo{title}{Square skyrmion crystal in centrosymmetric systems
  with locally inversion-asymmetric layers},
\newblock \bibinfo{journal}{J. Phys.: Condens. Matter} \bibinfo{volume}{34}
  (\bibinfo{year}{2022}) \bibinfo{pages}{365802}.
  \DOIprefix\doi{10.1088/1361-648X/ac7bcb}.
\bibitem[{Hayami and Yambe(2022)}]{Hayami_PhysRevB.105.104428}
\bibinfo{author}{S.~Hayami}, \bibinfo{author}{R.~Yambe},
\newblock \bibinfo{title}{Helicity locking of a square skyrmion crystal in a
  centrosymmetric lattice system without vertical mirror symmetry},
\newblock \bibinfo{journal}{Phys. Rev. B} \bibinfo{volume}{105}
  (\bibinfo{year}{2022}) \bibinfo{pages}{104428}.
  \DOIprefix\doi{10.1103/PhysRevB.105.104428}.
\bibitem[{Hayami(2022)}]{hayami2022skyrmion}
\bibinfo{author}{S.~Hayami},
\newblock \bibinfo{title}{Skyrmion crystals in centrosymmetric triangular
  magnets under hexagonal and trigonal single-ion anisotropy},
\newblock \bibinfo{journal}{J. Magn. Magn. Mater.} \bibinfo{volume}{553}
  (\bibinfo{year}{2022}) \bibinfo{pages}{169220}.
  \DOIprefix\doi{10.1016/j.jmmm.2022.169220}.
\bibitem[{Okubo et~al.(2012)Okubo, Chung, and
  Kawamura}]{Okubo_PhysRevLett.108.017206}
\bibinfo{author}{T.~Okubo}, \bibinfo{author}{S.~Chung},
  \bibinfo{author}{H.~Kawamura},
\newblock \bibinfo{title}{Multiple-$q$ states and the skyrmion lattice of the
  triangular-lattice {Heisenberg} antiferromagnet under magnetic fields},
\newblock \bibinfo{journal}{Phys. Rev. Lett.} \bibinfo{volume}{108}
  (\bibinfo{year}{2012}) \bibinfo{pages}{017206}.
  \DOIprefix\doi{10.1103/PhysRevLett.108.017206}.
\bibitem[{Mitsumoto and Kawamura(2021)}]{Mitsumoto_PhysRevB.104.184432}
\bibinfo{author}{K.~Mitsumoto}, \bibinfo{author}{H.~Kawamura},
\newblock \bibinfo{title}{Replica symmetry breaking in the {RKKY}
  skyrmion-crystal system},
\newblock \bibinfo{journal}{Phys. Rev. B} \bibinfo{volume}{104}
  (\bibinfo{year}{2021}) \bibinfo{pages}{184432}.
  \DOIprefix\doi{10.1103/PhysRevB.104.184432}.
\bibitem[{Mitsumoto and Kawamura(2022)}]{Mitsumoto_PhysRevB.105.094427}
\bibinfo{author}{K.~Mitsumoto}, \bibinfo{author}{H.~Kawamura},
\newblock \bibinfo{title}{Skyrmion crystal in the {RKKY} system on the
  two-dimensional triangular lattice},
\newblock \bibinfo{journal}{Phys. Rev. B} \bibinfo{volume}{105}
  (\bibinfo{year}{2022}) \bibinfo{pages}{094427}.
  \DOIprefix\doi{10.1103/PhysRevB.105.094427}.
\bibitem[{Leonov and Mostovoy(2015)}]{leonov2015multiply}
\bibinfo{author}{A.~O. Leonov}, \bibinfo{author}{M.~Mostovoy},
\newblock \bibinfo{title}{Multiply periodic states and isolated skyrmions in an
  anisotropic frustrated magnet},
\newblock \bibinfo{journal}{Nat. Commun.} \bibinfo{volume}{6}
  (\bibinfo{year}{2015}) \bibinfo{pages}{8275}.
  \DOIprefix\doi{10.1038/ncomms9275}.
\bibitem[{Lin and Hayami(2016)}]{Lin_PhysRevB.93.064430}
\bibinfo{author}{S.-Z. Lin}, \bibinfo{author}{S.~Hayami},
\newblock \bibinfo{title}{Ginzburg-{Landau} theory for skyrmions in
  inversion-symmetric magnets with competing interactions},
\newblock \bibinfo{journal}{Phys. Rev. B} \bibinfo{volume}{93}
  (\bibinfo{year}{2016}) \bibinfo{pages}{064430}.
  \DOIprefix\doi{10.1103/PhysRevB.93.064430}.
\bibitem[{Hayami et~al.(2016{\natexlab{a}})Hayami, Lin, and
  Batista}]{Hayami_PhysRevB.93.184413}
\bibinfo{author}{S.~Hayami}, \bibinfo{author}{S.-Z. Lin},
  \bibinfo{author}{C.~D. Batista},
\newblock \bibinfo{title}{Bubble and skyrmion crystals in frustrated magnets
  with easy-axis anisotropy},
\newblock \bibinfo{journal}{Phys. Rev. B} \bibinfo{volume}{93}
  (\bibinfo{year}{2016}{\natexlab{a}}) \bibinfo{pages}{184413}.
  \DOIprefix\doi{10.1103/PhysRevB.93.184413}.
\bibitem[{Hayami et~al.(2016{\natexlab{b}})Hayami, Lin, Kamiya, and
  Batista}]{Hayami_PhysRevB.94.174420}
\bibinfo{author}{S.~Hayami}, \bibinfo{author}{S.-Z. Lin},
  \bibinfo{author}{Y.~Kamiya}, \bibinfo{author}{C.~D. Batista},
\newblock \bibinfo{title}{Vortices, skyrmions, and chirality waves in
  frustrated {Mott} insulators with a quenched periodic array of impurities},
\newblock \bibinfo{journal}{Phys. Rev. B} \bibinfo{volume}{94}
  (\bibinfo{year}{2016}{\natexlab{b}}) \bibinfo{pages}{174420}.
  \DOIprefix\doi{10.1103/PhysRevB.94.174420}.
\bibitem[{Hayami(2022)}]{Hayami_PhysRevB.105.184426}
\bibinfo{author}{S.~Hayami},
\newblock \bibinfo{title}{Multifarious skyrmion phases on a trilayer triangular
  lattice},
\newblock \bibinfo{journal}{Phys. Rev. B} \bibinfo{volume}{105}
  (\bibinfo{year}{2022}) \bibinfo{pages}{184426}.
  \DOIprefix\doi{10.1103/PhysRevB.105.184426}.
\bibitem[{Ozawa et~al.(2017)Ozawa, Hayami, and
  Motome}]{Ozawa_PhysRevLett.118.147205}
\bibinfo{author}{R.~Ozawa}, \bibinfo{author}{S.~Hayami},
  \bibinfo{author}{Y.~Motome},
\newblock \bibinfo{title}{Zero-field skyrmions with a high topological number
  in itinerant magnets},
\newblock \bibinfo{journal}{Phys. Rev. Lett.} \bibinfo{volume}{118}
  (\bibinfo{year}{2017}) \bibinfo{pages}{147205}.
  \DOIprefix\doi{10.1103/PhysRevLett.118.147205}.
\bibitem[{Hayami and Motome(2019)}]{Hayami_PhysRevB.99.094420}
\bibinfo{author}{S.~Hayami}, \bibinfo{author}{Y.~Motome},
\newblock \bibinfo{title}{Effect of magnetic anisotropy on skyrmions with a
  high topological number in itinerant magnets},
\newblock \bibinfo{journal}{Phys. Rev. B} \bibinfo{volume}{99}
  (\bibinfo{year}{2019}) \bibinfo{pages}{094420}.
  \DOIprefix\doi{10.1103/PhysRevB.99.094420}.
\bibitem[{Hayami(2024)}]{Hayami_PhysRevB.109.014415}
\bibinfo{author}{S.~Hayami},
\newblock \bibinfo{title}{Three-sublattice antiferro-type and ferri-type
  skyrmion crystals in magnets without the {Dzyaloshinskii-Moriya}
  interaction},
\newblock \bibinfo{journal}{Phys. Rev. B} \bibinfo{volume}{109}
  (\bibinfo{year}{2024}) \bibinfo{pages}{014415}.
  \DOIprefix\doi{10.1103/PhysRevB.109.014415}.
\bibitem[{Hayami(2021)}]{Hayami_10.1088/1367-2630/ac3683}
\bibinfo{author}{S.~Hayami},
\newblock \bibinfo{title}{Temperature-driven transition from skyrmion to bubble
  crystals in centrosymmetric itinerant magnets},
\newblock \bibinfo{journal}{New J. Phys.} \bibinfo{volume}{23}
  (\bibinfo{year}{2021}) \bibinfo{pages}{113032}.
  \DOIprefix\doi{10.1088/1367-2630/ac3683}.
\bibitem[{Utesov(2022)}]{Utesov_PhysRevB.105.054435}
\bibinfo{author}{O.~I. Utesov},
\newblock \bibinfo{title}{Mean-field description of skyrmion lattice in
  hexagonal frustrated antiferromagnets},
\newblock \bibinfo{journal}{Phys. Rev. B} \bibinfo{volume}{105}
  (\bibinfo{year}{2022}) \bibinfo{pages}{054435}.
  \DOIprefix\doi{10.1103/PhysRevB.105.054435}.
\bibitem[{Okumura et~al.(2022)Okumura, Hayami, Kato, and
  Motome}]{Okumura_doi:10.7566/JPSJ.91.093702}
\bibinfo{author}{S.~Okumura}, \bibinfo{author}{S.~Hayami},
  \bibinfo{author}{Y.~Kato}, \bibinfo{author}{Y.~Motome},
\newblock \bibinfo{title}{Magnetic hedgehog lattice in a centrosymmetric cubic
  metal},
\newblock \bibinfo{journal}{J. Phys. Soc. Jpn.} \bibinfo{volume}{91}
  (\bibinfo{year}{2022}) \bibinfo{pages}{093702}.
  \DOIprefix\doi{10.7566/JPSJ.91.093702}.
\bibitem[{Momoi et~al.(1997)Momoi, Kubo, and Niki}]{Momoi_PhysRevLett.79.2081}
\bibinfo{author}{T.~Momoi}, \bibinfo{author}{K.~Kubo},
  \bibinfo{author}{K.~Niki},
\newblock \bibinfo{title}{Possible chiral phase transition in two-dimensional
  solid ${}^{3}\mathrm{He}$},
\newblock \bibinfo{journal}{Phys. Rev. Lett.} \bibinfo{volume}{79}
  (\bibinfo{year}{1997}) \bibinfo{pages}{2081--2084}.
  \DOIprefix\doi{10.1103/PhysRevLett.79.2081}.
\bibitem[{Yoshida et~al.(2012)Yoshida, Schr\"oder, Ferriani, Serrate, Kubetzka,
  von Bergmann, Heinze, and Wiesendanger}]{Yoshida_PhysRevLett.108.087205}
\bibinfo{author}{Y.~Yoshida}, \bibinfo{author}{S.~Schr\"oder},
  \bibinfo{author}{P.~Ferriani}, \bibinfo{author}{D.~Serrate},
  \bibinfo{author}{A.~Kubetzka}, \bibinfo{author}{K.~von Bergmann},
  \bibinfo{author}{S.~Heinze}, \bibinfo{author}{R.~Wiesendanger},
\newblock \bibinfo{title}{Conical spin-spiral state in an ultrathin film driven
  by higher-order spin interactions},
\newblock \bibinfo{journal}{Phys. Rev. Lett.} \bibinfo{volume}{108}
  (\bibinfo{year}{2012}) \bibinfo{pages}{087205}.
  \DOIprefix\doi{10.1103/PhysRevLett.108.087205}.
\bibitem[{Ueland et~al.(2012)Ueland, Miclea, Kato, Ayala-Valenzuela, McDonald,
  Okazaki, Tobash, Torrez, Ronning, Movshovich, Z., Martin, and
  Thompson}]{ueland2012controllable}
\bibinfo{author}{B.~Ueland}, \bibinfo{author}{C.~Miclea},
  \bibinfo{author}{Y.~Kato}, \bibinfo{author}{O.~Ayala-Valenzuela},
  \bibinfo{author}{R.~McDonald}, \bibinfo{author}{R.~Okazaki},
  \bibinfo{author}{P.~Tobash}, \bibinfo{author}{M.~Torrez},
  \bibinfo{author}{F.~Ronning}, \bibinfo{author}{R.~Movshovich},
  \bibinfo{author}{F.~Z.}, \bibinfo{author}{E.~B.~I. Martin},
  \bibinfo{author}{J.~Thompson},
\newblock \bibinfo{title}{Controllable chirality-induced geometrical {Hall}
  effect in a frustrated highly correlated metal},
\newblock \bibinfo{journal}{Nat. Commun.} \bibinfo{volume}{3}
  (\bibinfo{year}{2012}) \bibinfo{pages}{1067}.
  \DOIprefix\doi{10.1038/ncomms2075}.
\bibitem[{Mankovsky et~al.(2020)Mankovsky, Polesya, and
  Ebert}]{Mankovsky_PhysRevB.101.174401}
\bibinfo{author}{S.~Mankovsky}, \bibinfo{author}{S.~Polesya},
  \bibinfo{author}{H.~Ebert},
\newblock \bibinfo{title}{Extension of the standard {Heisenberg} {Hamiltonian}
  to multispin exchange interactions},
\newblock \bibinfo{journal}{Phys. Rev. B} \bibinfo{volume}{101}
  (\bibinfo{year}{2020}) \bibinfo{pages}{174401}.
  \DOIprefix\doi{10.1103/PhysRevB.101.174401}.
\bibitem[{Paul et~al.(2020)Paul, Haldar, von Malottki, and
  Heinze}]{paul2020role}
\bibinfo{author}{S.~Paul}, \bibinfo{author}{S.~Haldar}, \bibinfo{author}{S.~von
  Malottki}, \bibinfo{author}{S.~Heinze},
\newblock \bibinfo{title}{Role of higher-order exchange interactions for
  skyrmion stability},
\newblock \bibinfo{journal}{Nat. Commun.} \bibinfo{volume}{11}
  (\bibinfo{year}{2020}) \bibinfo{pages}{4756}.
  \DOIprefix\doi{10.1038/s41467-020-18473-x}.
\bibitem[{Brinker et~al.(2020)Brinker, dos Santos~Dias, and
  Lounis}]{Brinker_PhysRevResearch.2.033240}
\bibinfo{author}{S.~Brinker}, \bibinfo{author}{M.~dos Santos~Dias},
  \bibinfo{author}{S.~Lounis},
\newblock \bibinfo{title}{Prospecting chiral multisite interactions in
  prototypical magnetic systems},
\newblock \bibinfo{journal}{Phys. Rev. Research} \bibinfo{volume}{2}
  (\bibinfo{year}{2020}) \bibinfo{pages}{033240}.
  \DOIprefix\doi{10.1103/PhysRevResearch.2.033240}.
\bibitem[{Lounis(2020)}]{lounis2020multiple}
\bibinfo{author}{S.~Lounis},
\newblock \bibinfo{title}{Multiple-scattering approach for multi-spin chiral
  magnetic interactions: application to the one-and two-dimensional {R}ashba
  electron gas},
\newblock \bibinfo{journal}{New J. Phys.} \bibinfo{volume}{22}
  (\bibinfo{year}{2020}) \bibinfo{pages}{103003}.
  \DOIprefix\doi{10.1088/1367-2630/abb514}.
\bibitem[{Spethmann et~al.(2020)Spethmann, Meyer, von Bergmann, Wiesendanger,
  Heinze, and Kubetzka}]{Spethmann_PhysRevLett.124.227203}
\bibinfo{author}{J.~Spethmann}, \bibinfo{author}{S.~Meyer},
  \bibinfo{author}{K.~von Bergmann}, \bibinfo{author}{R.~Wiesendanger},
  \bibinfo{author}{S.~Heinze}, \bibinfo{author}{A.~Kubetzka},
\newblock \bibinfo{title}{Discovery of magnetic single- and triple-$\mathbf{q}$
  states in $\mathrm{Mn}/\mathrm{Re}(0001)$},
\newblock \bibinfo{journal}{Phys. Rev. Lett.} \bibinfo{volume}{124}
  (\bibinfo{year}{2020}) \bibinfo{pages}{227203}.
  \DOIprefix\doi{10.1103/PhysRevLett.124.227203}.
\bibitem[{Simon et~al.(2020)Simon, Donges, Szunyogh, and
  Nowak}]{Simon_PhysRevMaterials.4.084408}
\bibinfo{author}{E.~Simon}, \bibinfo{author}{A.~Donges},
  \bibinfo{author}{L.~Szunyogh}, \bibinfo{author}{U.~Nowak},
\newblock \bibinfo{title}{{Noncollinear antiferromagnetic states in Ru-based
  Heusler compounds induced by biquadratic coupling}},
\newblock \bibinfo{journal}{Phys. Rev. Materials} \bibinfo{volume}{4}
  (\bibinfo{year}{2020}) \bibinfo{pages}{084408}.
  \DOIprefix\doi{10.1103/PhysRevMaterials.4.084408}.
\bibitem[{Mendive-Tapia et~al.(2021)Mendive-Tapia, dos Santos~Dias, Grytsiuk,
  Staunton, Bl\"ugel, and Lounis}]{Mendive-Tapia_PhysRevB.103.024410}
\bibinfo{author}{E.~Mendive-Tapia}, \bibinfo{author}{M.~dos Santos~Dias},
  \bibinfo{author}{S.~Grytsiuk}, \bibinfo{author}{J.~B. Staunton},
  \bibinfo{author}{S.~Bl\"ugel}, \bibinfo{author}{S.~Lounis},
\newblock \bibinfo{title}{{Short period magnetization texture of B20-MnGe
  explained by thermally fluctuating local moments}},
\newblock \bibinfo{journal}{Phys. Rev. B} \bibinfo{volume}{103}
  (\bibinfo{year}{2021}) \bibinfo{pages}{024410}.
  \DOIprefix\doi{10.1103/PhysRevB.103.024410}.
\bibitem[{Brinker et~al.(2019)Brinker, dos Santos~Dias, and
  Lounis}]{brinker2019chiral}
\bibinfo{author}{S.~Brinker}, \bibinfo{author}{M.~dos Santos~Dias},
  \bibinfo{author}{S.~Lounis},
\newblock \bibinfo{title}{The chiral biquadratic pair interaction},
\newblock \bibinfo{journal}{New J. Phys.} \bibinfo{volume}{21}
  (\bibinfo{year}{2019}) \bibinfo{pages}{083015}.
  \DOIprefix\doi{10.1088/1367-2630/ab35c9}.
\bibitem[{L\'aszl\'offy et~al.(2019)L\'aszl\'offy, R\'ozsa, Palot\'as, Udvardi,
  and Szunyogh}]{Laszloffy_PhysRevB.99.184430}
\bibinfo{author}{A.~L\'aszl\'offy}, \bibinfo{author}{L.~R\'ozsa},
  \bibinfo{author}{K.~Palot\'as}, \bibinfo{author}{L.~Udvardi},
  \bibinfo{author}{L.~Szunyogh},
\newblock \bibinfo{title}{{Magnetic structure of monatomic Fe chains on
  Re(0001): Emergence of chiral multispin interactions}},
\newblock \bibinfo{journal}{Phys. Rev. B} \bibinfo{volume}{99}
  (\bibinfo{year}{2019}) \bibinfo{pages}{184430}.
  \DOIprefix\doi{10.1103/PhysRevB.99.184430}.
\bibitem[{Takahashi(1977)}]{takahashi1977half}
\bibinfo{author}{M.~Takahashi},
\newblock \bibinfo{title}{Half-filled {Hubbard} model at low temperature},
\newblock \bibinfo{journal}{J. Phys. C: Solid State Phys.} \bibinfo{volume}{10}
  (\bibinfo{year}{1977}) \bibinfo{pages}{1289}.
  \DOIprefix\doi{10.1088/0022-3719/10/8/031}.
\bibitem[{Yoshimori and Inagaki(1978)}]{yoshimori1978fourth}
\bibinfo{author}{A.~Yoshimori}, \bibinfo{author}{S.~Inagaki},
\newblock \bibinfo{title}{{Fourth Order Interaction Effects on the
  Antiferromagnetic Structures. I. fcc {Hubbard} Model}},
\newblock \bibinfo{journal}{J. Phys. Soc. Jpn.} \bibinfo{volume}{44}
  (\bibinfo{year}{1978}) \bibinfo{pages}{101--107}.
  \DOIprefix\doi{https://doi.org/10.1143/JPSJ.44.101}.
\bibitem[{Bulaevskii et~al.(2008)Bulaevskii, Batista, Mostovoy, and
  Khomskii}]{Bulaevskii_PhysRevB.78.024402}
\bibinfo{author}{L.~N. Bulaevskii}, \bibinfo{author}{C.~D. Batista},
  \bibinfo{author}{M.~V. Mostovoy}, \bibinfo{author}{D.~I. Khomskii},
\newblock \bibinfo{title}{Electronic orbital currents and polarization in mott
  insulators},
\newblock \bibinfo{journal}{Phys. Rev. B} \bibinfo{volume}{78}
  (\bibinfo{year}{2008}) \bibinfo{pages}{024402}.
  \DOIprefix\doi{10.1103/PhysRevB.78.024402}.
\bibitem[{Hoffmann and Bl\"ugel(2020)}]{Hoffmann_PhysRevB.101.024418}
\bibinfo{author}{M.~Hoffmann}, \bibinfo{author}{S.~Bl\"ugel},
\newblock \bibinfo{title}{Systematic derivation of realistic spin models for
  beyond-{Heisenberg} solids},
\newblock \bibinfo{journal}{Phys. Rev. B} \bibinfo{volume}{101}
  (\bibinfo{year}{2020}) \bibinfo{pages}{024418}.
  \DOIprefix\doi{10.1103/PhysRevB.101.024418}.
\bibitem[{Li et~al.(2021)Li, Yu, Lou, Feng, Whangbo, and Xiang}]{li2021spin}
\bibinfo{author}{X.~Li}, \bibinfo{author}{H.~Yu}, \bibinfo{author}{F.~Lou},
  \bibinfo{author}{J.~Feng}, \bibinfo{author}{M.-H. Whangbo},
  \bibinfo{author}{H.~Xiang},
\newblock \bibinfo{title}{{Spin Hamiltonians in Magnets: Theories and
  Computations}},
\newblock \bibinfo{journal}{Molecules} \bibinfo{volume}{26}
  (\bibinfo{year}{2021}) \bibinfo{pages}{803}.
  \DOIprefix\doi{10.3390/molecules26040803}.
\bibitem[{Akagi et~al.(2012)Akagi, Udagawa, and
  Motome}]{Akagi_PhysRevLett.108.096401}
\bibinfo{author}{Y.~Akagi}, \bibinfo{author}{M.~Udagawa},
  \bibinfo{author}{Y.~Motome},
\newblock \bibinfo{title}{Hidden multiple-spin interactions as an origin of
  spin scalar chiral order in frustrated {Kondo} lattice models},
\newblock \bibinfo{journal}{Phys. Rev. Lett.} \bibinfo{volume}{108}
  (\bibinfo{year}{2012}) \bibinfo{pages}{096401}.
  \DOIprefix\doi{10.1103/PhysRevLett.108.096401}.
\bibitem[{Hayami and Motome(2014)}]{Hayami_PhysRevB.90.060402}
\bibinfo{author}{S.~Hayami}, \bibinfo{author}{Y.~Motome},
\newblock \bibinfo{title}{{Multiple-$Q$} instability by $(d$-${}2)$-dimensional
  connections of {F}ermi surfaces},
\newblock \bibinfo{journal}{Phys. Rev. B} \bibinfo{volume}{90}
  (\bibinfo{year}{2014}) \bibinfo{pages}{060402(R)}.
  \DOIprefix\doi{10.1103/PhysRevB.90.060402}.
\bibitem[{Ozawa et~al.(2016)Ozawa, Hayami, Barros, Chern, Motome, and
  Batista}]{Ozawa_doi:10.7566/JPSJ.85.103703}
\bibinfo{author}{R.~Ozawa}, \bibinfo{author}{S.~Hayami},
  \bibinfo{author}{K.~Barros}, \bibinfo{author}{G.-W. Chern},
  \bibinfo{author}{Y.~Motome}, \bibinfo{author}{C.~D. Batista},
\newblock \bibinfo{title}{Vortex crystals with chiral stripes in itinerant
  magnets},
\newblock \bibinfo{journal}{J. Phys. Soc. Jpn.} \bibinfo{volume}{85}
  (\bibinfo{year}{2016}) \bibinfo{pages}{103703}.
  \DOIprefix\doi{10.7566/JPSJ.85.103703}.
\bibitem[{Kato et~al.(2010)Kato, Martin, and
  Batista}]{Kato_PhysRevLett.105.266405}
\bibinfo{author}{Y.~Kato}, \bibinfo{author}{I.~Martin}, \bibinfo{author}{C.~D.
  Batista},
\newblock \bibinfo{title}{Stability of the spontaneous quantum {Hall} state in
  the triangular {Kondo}-lattice model},
\newblock \bibinfo{journal}{Phys. Rev. Lett.} \bibinfo{volume}{105}
  (\bibinfo{year}{2010}) \bibinfo{pages}{266405}.
  \DOIprefix\doi{10.1103/PhysRevLett.105.266405}.
\bibitem[{Barros and Kato(2013)}]{Barros_PhysRevB.88.235101}
\bibinfo{author}{K.~Barros}, \bibinfo{author}{Y.~Kato},
\newblock \bibinfo{title}{Efficient {Langevin} simulation of coupled classical
  fields and fermions},
\newblock \bibinfo{journal}{Phys. Rev. B} \bibinfo{volume}{88}
  (\bibinfo{year}{2013}) \bibinfo{pages}{235101}.
  \DOIprefix\doi{10.1103/PhysRevB.88.235101}.
\bibitem[{Venderbos et~al.(2012)Venderbos, Kourtis, van~den Brink, and
  Daghofer}]{Venderbos_PhysRevLett.108.126405}
\bibinfo{author}{J.~W.~F. Venderbos}, \bibinfo{author}{S.~Kourtis},
  \bibinfo{author}{J.~van~den Brink}, \bibinfo{author}{M.~Daghofer},
\newblock \bibinfo{title}{Fractional quantum-{Hall} liquid spontaneously
  generated by strongly correlated ${t}_{2g}$ electrons},
\newblock \bibinfo{journal}{Phys. Rev. Lett.} \bibinfo{volume}{108}
  (\bibinfo{year}{2012}) \bibinfo{pages}{126405}.
  \DOIprefix\doi{10.1103/PhysRevLett.108.126405}.
\bibitem[{Jiang et~al.(2015)Jiang, Zhang, Zhou, and
  Wang}]{Jiang_PhysRevLett.114.216402}
\bibinfo{author}{K.~Jiang}, \bibinfo{author}{Y.~Zhang},
  \bibinfo{author}{S.~Zhou}, \bibinfo{author}{Z.~Wang},
\newblock \bibinfo{title}{Chiral spin density wave order on the frustrated
  honeycomb and bilayer triangle lattice {Hubbard} model at half-filling},
\newblock \bibinfo{journal}{Phys. Rev. Lett.} \bibinfo{volume}{114}
  (\bibinfo{year}{2015}) \bibinfo{pages}{216402}.
  \DOIprefix\doi{10.1103/PhysRevLett.114.216402}.
\bibitem[{Venderbos(2016)}]{Venderbos_PhysRevB.93.115108}
\bibinfo{author}{J.~W.~F. Venderbos},
\newblock \bibinfo{title}{Multi-$q$ hexagonal spin density waves and
  dynamically generated spin-orbit coupling: Time-reversal invariant analog of
  the chiral spin density wave},
\newblock \bibinfo{journal}{Phys. Rev. B} \bibinfo{volume}{93}
  (\bibinfo{year}{2016}) \bibinfo{pages}{115108}.
  \DOIprefix\doi{10.1103/PhysRevB.93.115108}.
\bibitem[{Barros et~al.(2014)Barros, Venderbos, Chern, and
  Batista}]{Barros_PhysRevB.90.245119}
\bibinfo{author}{K.~Barros}, \bibinfo{author}{J.~W.~F. Venderbos},
  \bibinfo{author}{G.-W. Chern}, \bibinfo{author}{C.~D. Batista},
\newblock \bibinfo{title}{Exotic magnetic orderings in the kagome
  {Kondo}-lattice model},
\newblock \bibinfo{journal}{Phys. Rev. B} \bibinfo{volume}{90}
  (\bibinfo{year}{2014}) \bibinfo{pages}{245119}.
  \DOIprefix\doi{10.1103/PhysRevB.90.245119}.
\bibitem[{Ghosh et~al.(2016)Ghosh, O'Brien, Henley, and
  Lawler}]{Ghosh_PhysRevB.93.024401}
\bibinfo{author}{S.~Ghosh}, \bibinfo{author}{P.~O'Brien},
  \bibinfo{author}{C.~L. Henley}, \bibinfo{author}{M.~J. Lawler},
\newblock \bibinfo{title}{Phase diagram of the {Kondo} lattice model on the
  kagome lattice},
\newblock \bibinfo{journal}{Phys. Rev. B} \bibinfo{volume}{93}
  (\bibinfo{year}{2016}) \bibinfo{pages}{024401}.
  \DOIprefix\doi{10.1103/PhysRevB.93.024401}.
\bibitem[{Ozawa et~al.(2017)Ozawa, Hayami, Barros, and
  Motome}]{Ozawa_PhysRevB.96.094417}
\bibinfo{author}{R.~Ozawa}, \bibinfo{author}{S.~Hayami},
  \bibinfo{author}{K.~Barros}, \bibinfo{author}{Y.~Motome},
\newblock \bibinfo{title}{Shape of magnetic domain walls formed by coupling to
  mobile charges},
\newblock \bibinfo{journal}{Phys. Rev. B} \bibinfo{volume}{96}
  (\bibinfo{year}{2017}) \bibinfo{pages}{094417}.
  \DOIprefix\doi{10.1103/PhysRevB.96.094417}.
\bibitem[{Chern et~al.(2018)Chern, Barros, Wang, Suwa, and
  Batista}]{Chern_PhysRevB.97.035120}
\bibinfo{author}{G.-W. Chern}, \bibinfo{author}{K.~Barros},
  \bibinfo{author}{Z.~Wang}, \bibinfo{author}{H.~Suwa}, \bibinfo{author}{C.~D.
  Batista},
\newblock \bibinfo{title}{Semiclassical dynamics of spin density waves},
\newblock \bibinfo{journal}{Phys. Rev. B} \bibinfo{volume}{97}
  (\bibinfo{year}{2018}) \bibinfo{pages}{035120}.
  \DOIprefix\doi{10.1103/PhysRevB.97.035120}.
\bibitem[{Kobayashi and Hayami(2022)}]{Kobayashi_PhysRevB.106.L140406}
\bibinfo{author}{K.~Kobayashi}, \bibinfo{author}{S.~Hayami},
\newblock \bibinfo{title}{Skyrmion and vortex crystals in the {Hubbard} model},
\newblock \bibinfo{journal}{Phys. Rev. B} \bibinfo{volume}{106}
  (\bibinfo{year}{2022}) \bibinfo{pages}{L140406}.
  \DOIprefix\doi{10.1103/PhysRevB.106.L140406}.
\bibitem[{Agterberg and Yunoki(2000)}]{Agterberg_PhysRevB.62.13816}
\bibinfo{author}{D.~F. Agterberg}, \bibinfo{author}{S.~Yunoki},
\newblock \bibinfo{title}{Spin-flux phase in the {Kondo} lattice model with
  classical localized spins},
\newblock \bibinfo{journal}{Phys. Rev. B} \bibinfo{volume}{62}
  (\bibinfo{year}{2000}) \bibinfo{pages}{13816--13819}.
  \DOIprefix\doi{10.1103/PhysRevB.62.13816}.
\bibitem[{Venderbos et~al.(2012)Venderbos, Daghofer, van~den Brink, and
  Kumar}]{Venderbos_PhysRevLett.109.166405}
\bibinfo{author}{J.~W.~F. Venderbos}, \bibinfo{author}{M.~Daghofer},
  \bibinfo{author}{J.~van~den Brink}, \bibinfo{author}{S.~Kumar},
\newblock \bibinfo{title}{Switchable quantum anomalous {Hall} state in a
  strongly frustrated lattice magnet},
\newblock \bibinfo{journal}{Phys. Rev. Lett.} \bibinfo{volume}{109}
  (\bibinfo{year}{2012}) \bibinfo{pages}{166405}.
  \DOIprefix\doi{10.1103/PhysRevLett.109.166405}.
\bibitem[{Solenov et~al.(2012)Solenov, Mozyrsky, and
  Martin}]{Solenov_PhysRevLett.108.096403}
\bibinfo{author}{D.~Solenov}, \bibinfo{author}{D.~Mozyrsky},
  \bibinfo{author}{I.~Martin},
\newblock \bibinfo{title}{Chirality waves in two-dimensional magnets},
\newblock \bibinfo{journal}{Phys. Rev. Lett.} \bibinfo{volume}{108}
  (\bibinfo{year}{2012}) \bibinfo{pages}{096403}.
  \DOIprefix\doi{10.1103/PhysRevLett.108.096403}.
\bibitem[{Hayami and Motome(2015)}]{hayami_PhysRevB.91.075104}
\bibinfo{author}{S.~Hayami}, \bibinfo{author}{Y.~Motome},
\newblock \bibinfo{title}{Topological semimetal-to-insulator phase transition
  between noncollinear and noncoplanar multiple-$q$ states on a
  square-to-triangular lattice},
\newblock \bibinfo{journal}{Phys. Rev. B} \bibinfo{volume}{91}
  (\bibinfo{year}{2015}) \bibinfo{pages}{075104}.
  \DOIprefix\doi{10.1103/PhysRevB.91.075104}.
\bibitem[{Hayami et~al.(2016)Hayami, Ozawa, and
  Motome}]{Hayami_PhysRevB.94.024424}
\bibinfo{author}{S.~Hayami}, \bibinfo{author}{R.~Ozawa},
  \bibinfo{author}{Y.~Motome},
\newblock \bibinfo{title}{Engineering chiral density waves and topological band
  structures by multiple-$q$ superpositions of collinear up-up-down-down
  orders},
\newblock \bibinfo{journal}{Phys. Rev. B} \bibinfo{volume}{94}
  (\bibinfo{year}{2016}) \bibinfo{pages}{024424}.
  \DOIprefix\doi{10.1103/PhysRevB.94.024424}.
\bibitem[{Shahzad and Sengupta(2017)}]{Shahzad_PhysRevB.96.224402}
\bibinfo{author}{M.~Shahzad}, \bibinfo{author}{P.~Sengupta},
\newblock \bibinfo{title}{Noncollinear magnetic ordering in a frustrated
  magnet: Metallic regime and the role of frustration},
\newblock \bibinfo{journal}{Phys. Rev. B} \bibinfo{volume}{96}
  (\bibinfo{year}{2017}) \bibinfo{pages}{224402}.
  \DOIprefix\doi{10.1103/PhysRevB.96.224402}.
\bibitem[{Okada et~al.(2018)Okada, Kato, and Motome}]{Okada_PhysRevB.98.224406}
\bibinfo{author}{K.~N. Okada}, \bibinfo{author}{Y.~Kato},
  \bibinfo{author}{Y.~Motome},
\newblock \bibinfo{title}{Multiple-$q$ magnetic orders in
  {R}ashba-{D}resselhaus metals},
\newblock \bibinfo{journal}{Phys. Rev. B} \bibinfo{volume}{98}
  (\bibinfo{year}{2018}) \bibinfo{pages}{224406}.
  \DOIprefix\doi{10.1103/PhysRevB.98.224406}.
\bibitem[{Su et~al.(2020)Su, Hayami, and Lin}]{Su_PhysRevResearch.2.013160}
\bibinfo{author}{Y.~Su}, \bibinfo{author}{S.~Hayami}, \bibinfo{author}{S.-Z.
  Lin},
\newblock \bibinfo{title}{Dimension transcendence and anomalous charge
  transport in magnets with moving multiple-$q$ spin textures},
\newblock \bibinfo{journal}{Phys. Rev. Research} \bibinfo{volume}{2}
  (\bibinfo{year}{2020}) \bibinfo{pages}{013160}.
  \DOIprefix\doi{10.1103/PhysRevResearch.2.013160}.
\bibitem[{Marcus et~al.(2018)Marcus, Kim, Tutmaher, Rodriguez-Rivera, Birk,
  Niedermeyer, Lee, Fisk, and Broholm}]{Marcus_PhysRevLett.120.097201}
\bibinfo{author}{G.~G. Marcus}, \bibinfo{author}{D.-J. Kim},
  \bibinfo{author}{J.~A. Tutmaher}, \bibinfo{author}{J.~A. Rodriguez-Rivera},
  \bibinfo{author}{J.~O. Birk}, \bibinfo{author}{C.~Niedermeyer},
  \bibinfo{author}{H.~Lee}, \bibinfo{author}{Z.~Fisk}, \bibinfo{author}{C.~L.
  Broholm},
\newblock \bibinfo{title}{{Multi-$q$ Mesoscale Magnetism in
  ${\mathrm{CeAuSb}}_{2}$}},
\newblock \bibinfo{journal}{Phys. Rev. Lett.} \bibinfo{volume}{120}
  (\bibinfo{year}{2018}) \bibinfo{pages}{097201}.
  \DOIprefix\doi{10.1103/PhysRevLett.120.097201}.
\bibitem[{Seo et~al.(2020)Seo, Wang, Thomas, Rahn, Carmo, Ronning, Bauer, dos
  Reis, Janoschek, Thompson, Fernandes, and Rosa}]{Seo_PhysRevX.10.011035}
\bibinfo{author}{S.~Seo}, \bibinfo{author}{X.~Wang}, \bibinfo{author}{S.~M.
  Thomas}, \bibinfo{author}{M.~C. Rahn}, \bibinfo{author}{D.~Carmo},
  \bibinfo{author}{F.~Ronning}, \bibinfo{author}{E.~D. Bauer},
  \bibinfo{author}{R.~D. dos Reis}, \bibinfo{author}{M.~Janoschek},
  \bibinfo{author}{J.~D. Thompson}, \bibinfo{author}{R.~M. Fernandes},
  \bibinfo{author}{P.~F.~S. Rosa},
\newblock \bibinfo{title}{{Nematic State in ${\mathrm{CeAuSb}}_{2}$}},
\newblock \bibinfo{journal}{Phys. Rev. X} \bibinfo{volume}{10}
  (\bibinfo{year}{2020}) \bibinfo{pages}{011035}.
  \DOIprefix\doi{10.1103/PhysRevX.10.011035}.
\bibitem[{Chern(2010)}]{Chern_PhysRevLett.105.226403}
\bibinfo{author}{G.-W. Chern},
\newblock \bibinfo{title}{Noncoplanar magnetic ordering driven by itinerant
  electrons on the pyrochlore lattice},
\newblock \bibinfo{journal}{Phys. Rev. Lett.} \bibinfo{volume}{105}
  (\bibinfo{year}{2010}) \bibinfo{pages}{226403}.
  \DOIprefix\doi{10.1103/PhysRevLett.105.226403}.
\bibitem[{Hayami et~al.(2014{\natexlab{a}})Hayami, Misawa, and
  Motome}]{hayami2014charge}
\bibinfo{author}{S.~Hayami}, \bibinfo{author}{T.~Misawa},
  \bibinfo{author}{Y.~Motome},
\newblock \bibinfo{title}{{Charge Order with a Noncoplanar Triple-$Q$ Magnetic
  Order on a Cubic Lattice}},
\newblock \bibinfo{journal}{JPS Conf. Proc.} \bibinfo{volume}{3}
  (\bibinfo{year}{2014}{\natexlab{a}}) \bibinfo{pages}{016016}.
  \DOIprefix\doi{10.7566/JPSCP.3.016016}.
\bibitem[{Hayami et~al.(2014{\natexlab{b}})Hayami, Misawa, Yamaji, and
  Motome}]{Hayami_PhysRevB.89.085124}
\bibinfo{author}{S.~Hayami}, \bibinfo{author}{T.~Misawa},
  \bibinfo{author}{Y.~Yamaji}, \bibinfo{author}{Y.~Motome},
\newblock \bibinfo{title}{Three-dimensional dirac electrons on a cubic lattice
  with noncoplanar multiple-{$Q$} order},
\newblock \bibinfo{journal}{Phys. Rev. B} \bibinfo{volume}{89}
  (\bibinfo{year}{2014}{\natexlab{b}}) \bibinfo{pages}{085124}.
  \DOIprefix\doi{10.1103/PhysRevB.89.085124}.
\bibitem[{Shimizu et~al.(2021)Shimizu, Okumura, Kato, and
  Motome}]{Shimizu_PhysRevB.103.054427}
\bibinfo{author}{K.~Shimizu}, \bibinfo{author}{S.~Okumura},
  \bibinfo{author}{Y.~Kato}, \bibinfo{author}{Y.~Motome},
\newblock \bibinfo{title}{Phase transitions between helices, vortices, and
  hedgehogs driven by spatial anisotropy in chiral magnets},
\newblock \bibinfo{journal}{Phys. Rev. B} \bibinfo{volume}{103}
  (\bibinfo{year}{2021}) \bibinfo{pages}{054427}.
  \DOIprefix\doi{10.1103/PhysRevB.103.054427}.
\bibitem[{Grytsiuk et~al.(2020)Grytsiuk, Hanke, Hoffmann, Bouaziz, Gomonay,
  Bihlmayer, Lounis, Mokrousov, and Bl{\"u}gel}]{grytsiuk2020topological}
\bibinfo{author}{S.~Grytsiuk}, \bibinfo{author}{J.-P. Hanke},
  \bibinfo{author}{M.~Hoffmann}, \bibinfo{author}{J.~Bouaziz},
  \bibinfo{author}{O.~Gomonay}, \bibinfo{author}{G.~Bihlmayer},
  \bibinfo{author}{S.~Lounis}, \bibinfo{author}{Y.~Mokrousov},
  \bibinfo{author}{S.~Bl{\"u}gel},
\newblock \bibinfo{title}{Topological--chiral magnetic interactions driven by
  emergent orbital magnetism},
\newblock \bibinfo{journal}{Nat. Commun.} \bibinfo{volume}{11}
  (\bibinfo{year}{2020}) \bibinfo{pages}{511}.
  \DOIprefix\doi{10.1038/s41467-019-14030-3}.
\bibitem[{B\"omerich et~al.(2020)B\"omerich, Heinen, and
  Rosch}]{Bomerich_PhysRevB.102.100408}
\bibinfo{author}{T.~B\"omerich}, \bibinfo{author}{L.~Heinen},
  \bibinfo{author}{A.~Rosch},
\newblock \bibinfo{title}{Skyrmion and tetarton lattices in twisted bilayer
  graphene},
\newblock \bibinfo{journal}{Phys. Rev. B} \bibinfo{volume}{102}
  (\bibinfo{year}{2020}) \bibinfo{pages}{100408}.
  \DOIprefix\doi{10.1103/PhysRevB.102.100408}.
\bibitem[{Binz et~al.(2006)Binz, Vishwanath, and
  Aji}]{Binz_PhysRevLett.96.207202}
\bibinfo{author}{B.~Binz}, \bibinfo{author}{A.~Vishwanath},
  \bibinfo{author}{V.~Aji},
\newblock \bibinfo{title}{Theory of the helical spin crystal: A candidate for
  the partially ordered state of {MnSi}},
\newblock \bibinfo{journal}{Phys. Rev. Lett.} \bibinfo{volume}{96}
  (\bibinfo{year}{2006}) \bibinfo{pages}{207202}.
  \DOIprefix\doi{10.1103/PhysRevLett.96.207202}.
\bibitem[{Binz and Vishwanath(2006)}]{Binz_PhysRevB.74.214408}
\bibinfo{author}{B.~Binz}, \bibinfo{author}{A.~Vishwanath},
\newblock \bibinfo{title}{Theory of helical spin crystals: Phases, textures,
  and properties},
\newblock \bibinfo{journal}{Phys. Rev. B} \bibinfo{volume}{74}
  (\bibinfo{year}{2006}) \bibinfo{pages}{214408}.
  \DOIprefix\doi{10.1103/PhysRevB.74.214408}.
\bibitem[{Park and Han(2011)}]{Park_PhysRevB.83.184406}
\bibinfo{author}{J.-H. Park}, \bibinfo{author}{J.~H. Han},
\newblock \bibinfo{title}{Zero-temperature phases for chiral magnets in three
  dimensions},
\newblock \bibinfo{journal}{Phys. Rev. B} \bibinfo{volume}{83}
  (\bibinfo{year}{2011}) \bibinfo{pages}{184406}.
  \DOIprefix\doi{10.1103/PhysRevB.83.184406}.
\bibitem[{Becker et~al.(2015)Becker, Hermanns, Bauer, Garst, and
  Trebst}]{Michael_PhysRevB.91.155135}
\bibinfo{author}{M.~Becker}, \bibinfo{author}{M.~Hermanns},
  \bibinfo{author}{B.~Bauer}, \bibinfo{author}{M.~Garst},
  \bibinfo{author}{S.~Trebst},
\newblock \bibinfo{title}{Spin-orbit physics of $j=\frac{1}{2}$ {Mott}
  insulators on the triangular lattice},
\newblock \bibinfo{journal}{Phys. Rev. B} \bibinfo{volume}{91}
  (\bibinfo{year}{2015}) \bibinfo{pages}{155135}.
  \DOIprefix\doi{10.1103/PhysRevB.91.155135}.
\bibitem[{Lee and Kim(2015)}]{Lee_PhysRevB.91.064407}
\bibinfo{author}{E.~K.-H. Lee}, \bibinfo{author}{Y.~B. Kim},
\newblock \bibinfo{title}{Theory of magnetic phase diagrams in hyperhoneycomb
  and harmonic-honeycomb iridates},
\newblock \bibinfo{journal}{Phys. Rev. B} \bibinfo{volume}{91}
  (\bibinfo{year}{2015}) \bibinfo{pages}{064407}.
  \DOIprefix\doi{10.1103/PhysRevB.91.064407}.
\bibitem[{Janssen et~al.(2016)Janssen, Andrade, and
  Vojta}]{Lukas_PhysRevLett.117.277202}
\bibinfo{author}{L.~Janssen}, \bibinfo{author}{E.~C. Andrade},
  \bibinfo{author}{M.~Vojta},
\newblock \bibinfo{title}{Honeycomb-lattice {Heisenberg}-{Kitaev} model in a
  magnetic field: Spin canting, metamagnetism, and vortex crystals},
\newblock \bibinfo{journal}{Phys. Rev. Lett.} \bibinfo{volume}{117}
  (\bibinfo{year}{2016}) \bibinfo{pages}{277202}.
  \DOIprefix\doi{10.1103/PhysRevLett.117.277202}.
\bibitem[{Rousochatzakis et~al.(2016)Rousochatzakis, R\"ossler, van~den Brink,
  and Daghofer}]{Rousochatzakis2016}
\bibinfo{author}{I.~Rousochatzakis}, \bibinfo{author}{U.~K. R\"ossler},
  \bibinfo{author}{J.~van~den Brink}, \bibinfo{author}{M.~Daghofer},
\newblock \bibinfo{title}{{Kitaev} anisotropy induces mesoscopic ${Z}_{2}$
  vortex crystals in frustrated hexagonal antiferromagnets},
\newblock \bibinfo{journal}{Phys. Rev. B} \bibinfo{volume}{93}
  (\bibinfo{year}{2016}) \bibinfo{pages}{104417}.
  \DOIprefix\doi{10.1103/PhysRevB.93.104417}.
\bibitem[{Yao and Dong(2016)}]{yao2016topological}
\bibinfo{author}{X.~Yao}, \bibinfo{author}{S.~Dong},
\newblock \bibinfo{title}{Topological triple-vortex lattice stabilized by mixed
  frustration in expanded honeycomb {Kitaev}-{Heisenberg} model},
\newblock \bibinfo{journal}{Sci. Rep.} \bibinfo{volume}{6}
  (\bibinfo{year}{2016}) \bibinfo{pages}{26750}.
  \DOIprefix\doi{10.1038/srep26750}.
\bibitem[{Chern et~al.(2017)Chern, Sizyuk, Price, and
  Perkins}]{Chern_PhysRevB.95.144427}
\bibinfo{author}{G.-W. Chern}, \bibinfo{author}{Y.~Sizyuk},
  \bibinfo{author}{C.~Price}, \bibinfo{author}{N.~B. Perkins},
\newblock \bibinfo{title}{Kitaev-{Heisenberg} model in a magnetic field:
  Order-by-disorder and commensurate-incommensurate transitions},
\newblock \bibinfo{journal}{Phys. Rev. B} \bibinfo{volume}{95}
  (\bibinfo{year}{2017}) \bibinfo{pages}{144427}.
  \DOIprefix\doi{10.1103/PhysRevB.95.144427}.
\bibitem[{Maksimov et~al.(2019)Maksimov, Zhu, White, and
  Chernyshev}]{Maksimov_PhysRevX.9.021017}
\bibinfo{author}{P.~A. Maksimov}, \bibinfo{author}{Z.~Zhu},
  \bibinfo{author}{S.~R. White}, \bibinfo{author}{A.~L. Chernyshev},
\newblock \bibinfo{title}{Anisotropic-exchange magnets on a triangular lattice:
  Spin waves, accidental degeneracies, and dual spin liquids},
\newblock \bibinfo{journal}{Phys. Rev. X} \bibinfo{volume}{9}
  (\bibinfo{year}{2019}) \bibinfo{pages}{021017}.
  \DOIprefix\doi{10.1103/PhysRevX.9.021017}.
\bibitem[{Amoroso et~al.(2020)Amoroso, Barone, and
  Picozzi}]{amoroso2020spontaneous}
\bibinfo{author}{D.~Amoroso}, \bibinfo{author}{P.~Barone},
  \bibinfo{author}{S.~Picozzi},
\newblock \bibinfo{title}{Spontaneous skyrmionic lattice from anisotropic
  symmetric exchange in a {Ni}-halide monolayer},
\newblock \bibinfo{journal}{Nat. Commun.} \bibinfo{volume}{11}
  (\bibinfo{year}{2020}) \bibinfo{pages}{5784}.
  \DOIprefix\doi{10.1038/s41467-020-19535-w}.
\bibitem[{Lin et~al.(1973)Lin, Grundy, and Giess}]{lin1973bubble}
\bibinfo{author}{Y.~Lin}, \bibinfo{author}{P.~Grundy},
  \bibinfo{author}{E.~Giess},
\newblock \bibinfo{title}{Bubble domains in magnetostatically coupled garnet
  films},
\newblock \bibinfo{journal}{Appl. Phys. Lett.} \bibinfo{volume}{23}
  (\bibinfo{year}{1973}) \bibinfo{pages}{485--487}.
  \DOIprefix\doi{10.1063/1.1654968}.
\bibitem[{Malozemoff and Slonczewski(1979)}]{malozemoff1979magnetic}
\bibinfo{author}{A.~Malozemoff}, \bibinfo{author}{J.~Slonczewski},
\newblock \bibinfo{title}{Magnetic domain walls in bubble materials academic},
\newblock \bibinfo{journal}{New York}  (\bibinfo{year}{1979})
  \bibinfo{pages}{382}.
\bibitem[{Garel and Doniach(1982)}]{Garel_PhysRevB.26.325}
\bibinfo{author}{T.~Garel}, \bibinfo{author}{S.~Doniach},
\newblock \bibinfo{title}{Phase transitions with spontaneous modulation-the
  dipolar ising ferromagnet},
\newblock \bibinfo{journal}{Phys. Rev. B} \bibinfo{volume}{26}
  (\bibinfo{year}{1982}) \bibinfo{pages}{325--329}.
  \DOIprefix\doi{10.1103/PhysRevB.26.325}.
\bibitem[{Takao(1983)}]{takao1983study}
\bibinfo{author}{S.~Takao},
\newblock \bibinfo{title}{A study of magnetization distribution of submicron
  bubbles in sputtered {Ho-Co} thin films},
\newblock \bibinfo{journal}{J. Magn. Magn. Mater.} \bibinfo{volume}{31}
  (\bibinfo{year}{1983}) \bibinfo{pages}{1009--1010}.
  \DOIprefix\doi{10.1016/0304-8853(83)90772-2}.
\bibitem[{Ezawa(2010)}]{Ezawa_PhysRevLett.105.197202}
\bibinfo{author}{M.~Ezawa},
\newblock \bibinfo{title}{Giant skyrmions stabilized by dipole-dipole
  interactions in thin ferromagnetic films},
\newblock \bibinfo{journal}{Phys. Rev. Lett.} \bibinfo{volume}{105}
  (\bibinfo{year}{2010}) \bibinfo{pages}{197202}.
  \DOIprefix\doi{10.1103/PhysRevLett.105.197202}.
\bibitem[{Utesov(2021)}]{Utesov_PhysRevB.103.064414}
\bibinfo{author}{O.~I. Utesov},
\newblock \bibinfo{title}{Thermodynamically stable skyrmion lattice in a
  tetragonal frustrated antiferromagnet with dipolar interaction},
\newblock \bibinfo{journal}{Phys. Rev. B} \bibinfo{volume}{103}
  (\bibinfo{year}{2021}) \bibinfo{pages}{064414}.
  \DOIprefix\doi{10.1103/PhysRevB.103.064414}.
\bibitem[{Lin(2024)}]{lin2024skyrmion}
\bibinfo{author}{S.-Z. Lin},
\newblock \bibinfo{title}{Skyrmion lattice in centrosymmetric magnets with
  local {Dzyaloshinsky-Moriya} interaction},
\newblock \bibinfo{journal}{Mater. Today Quantum} \bibinfo{volume}{2}
  (\bibinfo{year}{2024}) \bibinfo{pages}{100006}.
\bibitem[{Bogdanov et~al.(2002)Bogdanov, R\"o\ss{}ler, Wolf, and
  M\"uller}]{Bogdanov_PhysRevB.66.214410}
\bibinfo{author}{A.~N. Bogdanov}, \bibinfo{author}{U.~K. R\"o\ss{}ler},
  \bibinfo{author}{M.~Wolf}, \bibinfo{author}{K.-H. M\"uller},
\newblock \bibinfo{title}{Magnetic structures and reorientation transitions in
  noncentrosymmetric uniaxial antiferromagnets},
\newblock \bibinfo{journal}{Phys. Rev. B} \bibinfo{volume}{66}
  (\bibinfo{year}{2002}) \bibinfo{pages}{214410}.
  \DOIprefix\doi{10.1103/PhysRevB.66.214410}.
\bibitem[{Buhl et~al.(2017)Buhl, Freimuth, Bl{\"u}gel, and
  Mokrousov}]{buhl2017topological}
\bibinfo{author}{P.~M. Buhl}, \bibinfo{author}{F.~Freimuth},
  \bibinfo{author}{S.~Bl{\"u}gel}, \bibinfo{author}{Y.~Mokrousov},
\newblock \bibinfo{title}{Topological spin {Hall} effect in antiferromagnetic
  skyrmions},
\newblock \bibinfo{journal}{Phys. Status Solidi RRL} \bibinfo{volume}{11}
  (\bibinfo{year}{2017}) \bibinfo{pages}{1700007}.
  \DOIprefix\doi{10.1002/pssr.201700007}.
\bibitem[{G\"obel et~al.(2017)G\"obel, Mook, Henk, and
  Mertig}]{Gobel_PhysRevB.96.060406}
\bibinfo{author}{B.~G\"obel}, \bibinfo{author}{A.~Mook},
  \bibinfo{author}{J.~Henk}, \bibinfo{author}{I.~Mertig},
\newblock \bibinfo{title}{Antiferromagnetic skyrmion crystals: Generation,
  topological {Hall}, and topological spin {Hall} effect},
\newblock \bibinfo{journal}{Phys. Rev. B} \bibinfo{volume}{96}
  (\bibinfo{year}{2017}) \bibinfo{pages}{060406}.
  \DOIprefix\doi{10.1103/PhysRevB.96.060406}.
\bibitem[{Akosa et~al.(2018)Akosa, Tretiakov, Tatara, and
  Manchon}]{Akosa_PhysRevLett.121.097204}
\bibinfo{author}{C.~A. Akosa}, \bibinfo{author}{O.~A. Tretiakov},
  \bibinfo{author}{G.~Tatara}, \bibinfo{author}{A.~Manchon},
\newblock \bibinfo{title}{Theory of the topological spin {Hall} effect in
  antiferromagnetic skyrmions: Impact on current-induced motion},
\newblock \bibinfo{journal}{Phys. Rev. Lett.} \bibinfo{volume}{121}
  (\bibinfo{year}{2018}) \bibinfo{pages}{097204}.
  \DOIprefix\doi{10.1103/PhysRevLett.121.097204}.
\bibitem[{Rosales et~al.(2015)Rosales, Cabra, and
  Pujol}]{Rosales_PhysRevB.92.214439}
\bibinfo{author}{H.~D. Rosales}, \bibinfo{author}{D.~C. Cabra},
  \bibinfo{author}{P.~Pujol},
\newblock \bibinfo{title}{Three-sublattice skyrmion crystal in the
  antiferromagnetic triangular lattice},
\newblock \bibinfo{journal}{Phys. Rev. B} \bibinfo{volume}{92}
  (\bibinfo{year}{2015}) \bibinfo{pages}{214439}.
  \DOIprefix\doi{10.1103/PhysRevB.92.214439}.
\bibitem[{D\'{\i}az et~al.(2019)D\'{\i}az, Klinovaja, and
  Loss}]{Diaz_hysRevLett.122.187203}
\bibinfo{author}{S.~A. D\'{\i}az}, \bibinfo{author}{J.~Klinovaja},
  \bibinfo{author}{D.~Loss},
\newblock \bibinfo{title}{Topological magnons and edge states in
  antiferromagnetic skyrmion crystals},
\newblock \bibinfo{journal}{Phys. Rev. Lett.} \bibinfo{volume}{122}
  (\bibinfo{year}{2019}) \bibinfo{pages}{187203}.
  \DOIprefix\doi{10.1103/PhysRevLett.122.187203}.
\bibitem[{Osorio et~al.(2017)Osorio, Rosales, Sturla, and
  Cabra}]{Osorio_PhysRevB.96.024404}
\bibinfo{author}{S.~A. Osorio}, \bibinfo{author}{H.~D. Rosales},
  \bibinfo{author}{M.~B. Sturla}, \bibinfo{author}{D.~C. Cabra},
\newblock \bibinfo{title}{Composite spin crystal phase in antiferromagnetic
  chiral magnets},
\newblock \bibinfo{journal}{Phys. Rev. B} \bibinfo{volume}{96}
  (\bibinfo{year}{2017}) \bibinfo{pages}{024404}.
  \DOIprefix\doi{10.1103/PhysRevB.96.024404}.
\bibitem[{Liu et~al.(2020)Liu, dos Santos~Dias, and
  Lounis}]{liu2020theoretical}
\bibinfo{author}{Z.~Liu}, \bibinfo{author}{M.~dos Santos~Dias},
  \bibinfo{author}{S.~Lounis},
\newblock \bibinfo{title}{Theoretical investigation of antiferromagnetic
  skyrmions in a triangular monolayer},
\newblock \bibinfo{journal}{J. Phys.: Condens. Matter} \bibinfo{volume}{32}
  (\bibinfo{year}{2020}) \bibinfo{pages}{425801}.
  \DOIprefix\doi{10.1088/1361-648X/ab96ef}.
\bibitem[{Tom\'e and Rosales(2021)}]{Tome_PhysRevB.103.L020403}
\bibinfo{author}{M.~Tom\'e}, \bibinfo{author}{H.~D. Rosales},
\newblock \bibinfo{title}{Topological phase transition driven by magnetic field
  and topological {Hall} effect in an antiferromagnetic skyrmion lattice},
\newblock \bibinfo{journal}{Phys. Rev. B} \bibinfo{volume}{103}
  (\bibinfo{year}{2021}) \bibinfo{pages}{L020403}.
  \DOIprefix\doi{10.1103/PhysRevB.103.L020403}.
\bibitem[{Mukherjee et~al.(2021{\natexlab{a}})Mukherjee, Kathyat, and
  Kumar}]{mukherjee2021antiferromagnetic}
\bibinfo{author}{A.~Mukherjee}, \bibinfo{author}{D.~S. Kathyat},
  \bibinfo{author}{S.~Kumar},
\newblock \bibinfo{title}{Antiferromagnetic skyrmion crystals in the {R}ashba
  {H}und’s insulator on triangular lattice},
\newblock \bibinfo{journal}{Sci. Rep.} \bibinfo{volume}{11}
  (\bibinfo{year}{2021}{\natexlab{a}}) \bibinfo{pages}{9566}.
  \DOIprefix\doi{10.1038/s41598-021-88556-2}.
\bibitem[{Mukherjee et~al.(2021{\natexlab{b}})Mukherjee, Kathyat, and
  Kumar}]{Mukherjee_PhysRevB.103.134424}
\bibinfo{author}{A.~Mukherjee}, \bibinfo{author}{D.~S. Kathyat},
  \bibinfo{author}{S.~Kumar},
\newblock \bibinfo{title}{Antiferromagnetic skyrmions and skyrmion density wave
  in a {R}ashba-coupled hund insulator},
\newblock \bibinfo{journal}{Phys. Rev. B} \bibinfo{volume}{103}
  (\bibinfo{year}{2021}{\natexlab{b}}) \bibinfo{pages}{134424}.
  \DOIprefix\doi{10.1103/PhysRevB.103.134424}.
\bibitem[{Wang et~al.(2021)Wang, Su, Lin, and
  Batista}]{Wang_PhysRevB.103.104408}
\bibinfo{author}{Z.~Wang}, \bibinfo{author}{Y.~Su}, \bibinfo{author}{S.-Z.
  Lin}, \bibinfo{author}{C.~D. Batista},
\newblock \bibinfo{title}{Meron, skyrmion, and vortex crystals in
  centrosymmetric tetragonal magnets},
\newblock \bibinfo{journal}{Phys. Rev. B} \bibinfo{volume}{103}
  (\bibinfo{year}{2021}) \bibinfo{pages}{104408}.
  \DOIprefix\doi{10.1103/PhysRevB.103.104408}.
\bibitem[{Hayami and Yambe(2020)}]{Hayami_doi:10.7566/JPSJ.89.103702}
\bibinfo{author}{S.~Hayami}, \bibinfo{author}{R.~Yambe},
\newblock \bibinfo{title}{Degeneracy lifting of {N}\'eel, {Bloch}, and
  anti-skyrmion crystals in centrosymmetric tetragonal systems},
\newblock \bibinfo{journal}{J. Phys. Soc. Jpn.} \bibinfo{volume}{89}
  (\bibinfo{year}{2020}) \bibinfo{pages}{103702}.
  \DOIprefix\doi{10.7566/JPSJ.89.103702}.
\bibitem[{Hayami(2024)}]{Hayami_PhysRevB.109.184419}
\bibinfo{author}{S.~Hayami},
\newblock \bibinfo{title}{Field-induced transformation between triangular and
  square skyrmion crystals in a tetragonal polar magnet},
\newblock \bibinfo{journal}{Phys. Rev. B} \bibinfo{volume}{109}
  (\bibinfo{year}{2024}) \bibinfo{pages}{184419}.
  \DOIprefix\doi{10.1103/PhysRevB.109.184419}.
\bibitem[{Okubo et~al.(2011)Okubo, Nguyen, and
  Kawamura}]{Okubo_PhysRevB.84.144432}
\bibinfo{author}{T.~Okubo}, \bibinfo{author}{T.~H. Nguyen},
  \bibinfo{author}{H.~Kawamura},
\newblock \bibinfo{title}{Cubic and noncubic multiple-$q$ states in the
  {Heisenberg} antiferromagnet on the pyrochlore lattice},
\newblock \bibinfo{journal}{Phys. Rev. B} \bibinfo{volume}{84}
  (\bibinfo{year}{2011}) \bibinfo{pages}{144432}.
  \DOIprefix\doi{10.1103/PhysRevB.84.144432}.
\bibitem[{Rosales et~al.(2013)Rosales, Cabra, Lamas, Pujol, and
  Zhitomirsky}]{Rosales_PhysRevB.87.104402}
\bibinfo{author}{H.~D. Rosales}, \bibinfo{author}{D.~C. Cabra},
  \bibinfo{author}{C.~A. Lamas}, \bibinfo{author}{P.~Pujol},
  \bibinfo{author}{M.~E. Zhitomirsky},
\newblock \bibinfo{title}{Broken discrete symmetries in a frustrated honeycomb
  antiferromagnet},
\newblock \bibinfo{journal}{Phys. Rev. B} \bibinfo{volume}{87}
  (\bibinfo{year}{2013}) \bibinfo{pages}{104402}.
  \DOIprefix\doi{10.1103/PhysRevB.87.104402}.
\bibitem[{Kato and Motome(2022)}]{Kato_PhysRevB.105.174413}
\bibinfo{author}{Y.~Kato}, \bibinfo{author}{Y.~Motome},
\newblock \bibinfo{title}{Magnetic field--temperature phase diagrams for
  multiple-$q$ magnetic ordering: Exact steepest descent approach to long-range
  interacting spin systems},
\newblock \bibinfo{journal}{Phys. Rev. B} \bibinfo{volume}{105}
  (\bibinfo{year}{2022}) \bibinfo{pages}{174413}.
  \DOIprefix\doi{10.1103/PhysRevB.105.174413}.
\bibitem[{Kato and Motome(2023)}]{Kato_PhysRevB.107.094437}
\bibinfo{author}{Y.~Kato}, \bibinfo{author}{Y.~Motome},
\newblock \bibinfo{title}{Hidden topological transitions in emergent magnetic
  monopole lattices},
\newblock \bibinfo{journal}{Phys. Rev. B} \bibinfo{volume}{107}
  (\bibinfo{year}{2023}) \bibinfo{pages}{094437}.
  \DOIprefix\doi{10.1103/PhysRevB.107.094437}.
\bibitem[{Kamiya and Batista(2014)}]{Kamiya_PhysRevX.4.011023}
\bibinfo{author}{Y.~Kamiya}, \bibinfo{author}{C.~D. Batista},
\newblock \bibinfo{title}{Magnetic vortex crystals in frustrated mott
  insulator},
\newblock \bibinfo{journal}{Phys. Rev. X} \bibinfo{volume}{4}
  (\bibinfo{year}{2014}) \bibinfo{pages}{011023}.
  \DOIprefix\doi{10.1103/PhysRevX.4.011023}.
\bibitem[{Wang et~al.(2015)Wang, Kamiya, Nevidomskyy, and
  Batista}]{Wang_PhysRevLett.115.107201}
\bibinfo{author}{Z.~Wang}, \bibinfo{author}{Y.~Kamiya}, \bibinfo{author}{A.~H.
  Nevidomskyy}, \bibinfo{author}{C.~D. Batista},
\newblock \bibinfo{title}{Three-dimensional crystallization of vortex strings
  in frustrated quantum magnets},
\newblock \bibinfo{journal}{Phys. Rev. Lett.} \bibinfo{volume}{115}
  (\bibinfo{year}{2015}) \bibinfo{pages}{107201}.
  \DOIprefix\doi{10.1103/PhysRevLett.115.107201}.
\bibitem[{Marmorini and Momoi(2014)}]{Marmorini2014}
\bibinfo{author}{G.~Marmorini}, \bibinfo{author}{T.~Momoi},
\newblock \bibinfo{title}{Magnon condensation with finite degeneracy on the
  triangular lattice},
\newblock \bibinfo{journal}{Phys. Rev. B} \bibinfo{volume}{89}
  (\bibinfo{year}{2014}) \bibinfo{pages}{134425}.
  \DOIprefix\doi{10.1103/PhysRevB.89.134425}.
\bibitem[{Ueda et~al.(2016)Ueda, Akagi, and Shannon}]{Ueda_PhysRevA.93.021606}
\bibinfo{author}{H.~T. Ueda}, \bibinfo{author}{Y.~Akagi},
  \bibinfo{author}{N.~Shannon},
\newblock \bibinfo{title}{Quantum solitons with emergent interactions in a
  model of cold atoms on the triangular lattice},
\newblock \bibinfo{journal}{Phys. Rev. A} \bibinfo{volume}{93}
  (\bibinfo{year}{2016}) \bibinfo{pages}{021606(R)}.
  \DOIprefix\doi{10.1103/PhysRevA.93.021606}.
\bibitem[{Lohani et~al.(2019)Lohani, Hickey, Masell, and
  Rosch}]{Lohani_PhysRevX.9.041063}
\bibinfo{author}{V.~Lohani}, \bibinfo{author}{C.~Hickey},
  \bibinfo{author}{J.~Masell}, \bibinfo{author}{A.~Rosch},
\newblock \bibinfo{title}{Quantum skyrmions in frustrated ferromagnets},
\newblock \bibinfo{journal}{Phys. Rev. X} \bibinfo{volume}{9}
  (\bibinfo{year}{2019}) \bibinfo{pages}{041063}.
  \DOIprefix\doi{10.1103/PhysRevX.9.041063}.
\bibitem[{Sotnikov et~al.(2021)Sotnikov, Mazurenko, Colbois, Mila, Katsnelson,
  and Stepanov}]{Sotnikov_PhysRevB.103.L060404}
\bibinfo{author}{O.~M. Sotnikov}, \bibinfo{author}{V.~V. Mazurenko},
  \bibinfo{author}{J.~Colbois}, \bibinfo{author}{F.~Mila},
  \bibinfo{author}{M.~I. Katsnelson}, \bibinfo{author}{E.~A. Stepanov},
\newblock \bibinfo{title}{Probing the topology of the quantum analog of a
  classical skyrmion},
\newblock \bibinfo{journal}{Phys. Rev. B} \bibinfo{volume}{103}
  (\bibinfo{year}{2021}) \bibinfo{pages}{L060404}.
  \DOIprefix\doi{10.1103/PhysRevB.103.L060404}.
\bibitem[{Siegl et~al.(2022)Siegl, Vedmedenko, Stier, Thorwart, and
  Posske}]{Siegl_PhysRevResearch.4.023111}
\bibinfo{author}{P.~Siegl}, \bibinfo{author}{E.~Y. Vedmedenko},
  \bibinfo{author}{M.~Stier}, \bibinfo{author}{M.~Thorwart},
  \bibinfo{author}{T.~Posske},
\newblock \bibinfo{title}{Controlled creation of quantum skyrmions},
\newblock \bibinfo{journal}{Phys. Rev. Res.} \bibinfo{volume}{4}
  (\bibinfo{year}{2022}) \bibinfo{pages}{023111}.
  \DOIprefix\doi{10.1103/PhysRevResearch.4.023111}.
\bibitem[{Haller et~al.(2022)Haller, Groenendijk, Habibi, Michels, and
  Schmidt}]{Haller_PhysRevResearch.4.043113}
\bibinfo{author}{A.~Haller}, \bibinfo{author}{S.~Groenendijk},
  \bibinfo{author}{A.~Habibi}, \bibinfo{author}{A.~Michels},
  \bibinfo{author}{T.~L. Schmidt},
\newblock \bibinfo{title}{Quantum skyrmion lattices in {Heisenberg}
  ferromagnets},
\newblock \bibinfo{journal}{Phys. Rev. Res.} \bibinfo{volume}{4}
  (\bibinfo{year}{2022}) \bibinfo{pages}{043113}.
  \DOIprefix\doi{10.1103/PhysRevResearch.4.043113}.
\bibitem[{Wollny et~al.(2011)Wollny, Fritz, and
  Vojta}]{Wollny_PhysRevLett.107.137204}
\bibinfo{author}{A.~Wollny}, \bibinfo{author}{L.~Fritz},
  \bibinfo{author}{M.~Vojta},
\newblock \bibinfo{title}{Fractional impurity moments in two-dimensional
  noncollinear magnets},
\newblock \bibinfo{journal}{Phys. Rev. Lett.} \bibinfo{volume}{107}
  (\bibinfo{year}{2011}) \bibinfo{pages}{137204}.
  \DOIprefix\doi{10.1103/PhysRevLett.107.137204}.
\bibitem[{Sen et~al.(2012)Sen, Damle, and Moessner}]{Sen_PhysRevB.86.205134}
\bibinfo{author}{A.~Sen}, \bibinfo{author}{K.~Damle},
  \bibinfo{author}{R.~Moessner},
\newblock \bibinfo{title}{Vacancy-induced spin textures and their interactions
  in a classical spin liquid},
\newblock \bibinfo{journal}{Phys. Rev. B} \bibinfo{volume}{86}
  (\bibinfo{year}{2012}) \bibinfo{pages}{205134}.
  \DOIprefix\doi{10.1103/PhysRevB.86.205134}.
\bibitem[{Maryasin and Zhitomirsky(2013)}]{Maryasin_PhysRevLett.111.247201}
\bibinfo{author}{V.~S. Maryasin}, \bibinfo{author}{M.~E. Zhitomirsky},
\newblock \bibinfo{title}{Triangular antiferromagnet with nonmagnetic
  impurities},
\newblock \bibinfo{journal}{Phys. Rev. Lett.} \bibinfo{volume}{111}
  (\bibinfo{year}{2013}) \bibinfo{pages}{247201}.
  \DOIprefix\doi{10.1103/PhysRevLett.111.247201}.
\bibitem[{Maryasin and Zhitomirsky(2015)}]{maryasin2015collective}
\bibinfo{author}{V.~Maryasin}, \bibinfo{author}{M.~Zhitomirsky},
\newblock \bibinfo{title}{Collective impurity effects in the {Heisenberg}
  triangular antiferromagnet},
\newblock \bibinfo{journal}{J. Phys.: Conf. Ser.} \bibinfo{volume}{592}
  (\bibinfo{year}{2015}) \bibinfo{pages}{012112}.
  \DOIprefix\doi{10.1088/1742-6596/592/1/012112}.
\bibitem[{Lin et~al.(2016)Lin, Hayami, and
  Batista}]{Lin_PhysRevLett.116.187202}
\bibinfo{author}{S.-Z. Lin}, \bibinfo{author}{S.~Hayami},
  \bibinfo{author}{C.~D. Batista},
\newblock \bibinfo{title}{Magnetic vortex induced by nonmagnetic impurity in
  frustrated magnets},
\newblock \bibinfo{journal}{Phys. Rev. Lett.} \bibinfo{volume}{116}
  (\bibinfo{year}{2016}) \bibinfo{pages}{187202}.
  \DOIprefix\doi{10.1103/PhysRevLett.116.187202}.
\bibitem[{Lin and Batista(2018)}]{Lin_PhysRevLett.120.077202}
\bibinfo{author}{S.-Z. Lin}, \bibinfo{author}{C.~D. Batista},
\newblock \bibinfo{title}{Face centered cubic and hexagonal close packed
  skyrmion crystals in centrosymmetric magnets},
\newblock \bibinfo{journal}{Phys. Rev. Lett.} \bibinfo{volume}{120}
  (\bibinfo{year}{2018}) \bibinfo{pages}{077202}.
  \DOIprefix\doi{10.1103/PhysRevLett.120.077202}.
\bibitem[{Hayami(2023)}]{hayami2023multiple}
\bibinfo{author}{S.~Hayami},
\newblock \bibinfo{title}{Multiple-$q$ instability under fourth-order inplane
  single-ion anisotropy},
\newblock \bibinfo{journal}{J. Magn. Magn. Mater.} \bibinfo{volume}{570}
  (\bibinfo{year}{2023}) \bibinfo{pages}{170507}.
  \DOIprefix\doi{https://doi.org/10.1016/j.jmmm.2023.170507}.
\bibitem[{Mochizuki et~al.(2018)Mochizuki, Ihara, Ohe, and
  Takeuchi}]{mochizuki2018highly}
\bibinfo{author}{M.~Mochizuki}, \bibinfo{author}{K.~Ihara},
  \bibinfo{author}{J.-i. Ohe}, \bibinfo{author}{A.~Takeuchi},
\newblock \bibinfo{title}{Highly efficient induction of spin polarization by
  circularly polarized electromagnetic waves in the {Rashba} spin-orbit
  systems},
\newblock \bibinfo{journal}{Appl. Phys. Lett.} \bibinfo{volume}{112}
  (\bibinfo{year}{2018}). \DOIprefix\doi{https://doi.org/10.1063/1.5022262}.
\bibitem[{Miyake and Mochizuki(2020)}]{Miyake_PhysRevB.101.094419}
\bibinfo{author}{M.~Miyake}, \bibinfo{author}{M.~Mochizuki},
\newblock \bibinfo{title}{Creation of nanometric magnetic skyrmions by global
  application of circularly polarized microwave magnetic field},
\newblock \bibinfo{journal}{Phys. Rev. B} \bibinfo{volume}{101}
  (\bibinfo{year}{2020}) \bibinfo{pages}{094419}.
  \DOIprefix\doi{10.1103/PhysRevB.101.094419}.
\bibitem[{Eto and Mochizuki(2021)}]{Eto_PhysRevB.104.104425}
\bibinfo{author}{R.~Eto}, \bibinfo{author}{M.~Mochizuki},
\newblock \bibinfo{title}{Dynamical switching of magnetic topology in
  microwave-driven itinerant magnet},
\newblock \bibinfo{journal}{Phys. Rev. B} \bibinfo{volume}{104}
  (\bibinfo{year}{2021}) \bibinfo{pages}{104425}.
  \DOIprefix\doi{10.1103/PhysRevB.104.104425}.
\bibitem[{Matsumoto et~al.(2017)Matsumoto, Chimata, and
  Koga}]{matsumoto2017symmetry}
\bibinfo{author}{M.~Matsumoto}, \bibinfo{author}{K.~Chimata},
  \bibinfo{author}{M.~Koga},
\newblock \bibinfo{title}{Symmetry analysis of spin-dependent electric dipole
  and its application to magnetoelectric effects},
\newblock \bibinfo{journal}{J. Phys. Soc. Jpn.} \bibinfo{volume}{86}
  (\bibinfo{year}{2017}) \bibinfo{pages}{034704}.
  \DOIprefix\doi{10.7566/JPSJ.86.034704}.
\bibitem[{Yambe and Hayami(2023)}]{Yambe_PhysRevB.108.064420}
\bibinfo{author}{R.~Yambe}, \bibinfo{author}{S.~Hayami},
\newblock \bibinfo{title}{Symmetry analysis of light-induced magnetic
  interactions via {Floquet} engineering},
\newblock \bibinfo{journal}{Phys. Rev. B} \bibinfo{volume}{108}
  (\bibinfo{year}{2023}) \bibinfo{pages}{064420}.
  \DOIprefix\doi{10.1103/PhysRevB.108.064420}.
\bibitem[{Yambe and Hayami(2024{\natexlab{a}})}]{Yambe_PhysRevB.109.064428}
\bibinfo{author}{R.~Yambe}, \bibinfo{author}{S.~Hayami},
\newblock \bibinfo{title}{Scalar spin chirality induced by a circularly
  polarized electric field in a classical kagome magnet},
\newblock \bibinfo{journal}{Phys. Rev. B} \bibinfo{volume}{109}
  (\bibinfo{year}{2024}{\natexlab{a}}) \bibinfo{pages}{064428}.
  \DOIprefix\doi{10.1103/PhysRevB.109.064428}.
\bibitem[{Yambe and Hayami(2024{\natexlab{b}})}]{yambe2024dynamical}
\bibinfo{author}{R.~Yambe}, \bibinfo{author}{S.~Hayami},
\newblock \bibinfo{title}{Dynamical generation of skyrmion and bimeron crystals
  by a circularly polarized electric field in frustrated magnets},
\newblock \bibinfo{journal}{Phys. Rev. B} \bibinfo{volume}{110}
  (\bibinfo{year}{2024}{\natexlab{b}}) \bibinfo{pages}{014428}.
  \DOIprefix\doi{10.1103/PhysRevB.110.014428}.
\bibitem[{Dzyaloshinskii(1964)}]{dzyaloshinskii1964theory}
\bibinfo{author}{I.~Dzyaloshinskii},
\newblock \bibinfo{title}{Theory of helicoidal structures in antiferromagnets.
  i. nonmetals},
\newblock \bibinfo{journal}{Sov. Phys. JETP} \bibinfo{volume}{19}
  (\bibinfo{year}{1964}) \bibinfo{pages}{960--971}.
\bibitem[{Kataoka and Nakanishi(1981)}]{kataoka1981helical}
\bibinfo{author}{M.~Kataoka}, \bibinfo{author}{O.~Nakanishi},
\newblock \bibinfo{title}{Helical spin density wave due to antisymmetric
  exchange interaction},
\newblock \bibinfo{journal}{J. Phys. Soc. Jpn.} \bibinfo{volume}{50}
  (\bibinfo{year}{1981}) \bibinfo{pages}{3888--3896}.
  \DOIprefix\doi{10.1143/JPSJ.50.3888}.
\bibitem[{Berg and L\''{u}scher(1981)}]{BERG1981412}
\bibinfo{author}{B.~Berg}, \bibinfo{author}{M.~L\''{u}scher},
\newblock \bibinfo{title}{Definition and statistical distributions of a
  topological number in the lattice o(3) $\sigma$-model},
\newblock \bibinfo{journal}{Nucl. Phys. B} \bibinfo{volume}{190}
  (\bibinfo{year}{1981}) \bibinfo{pages}{412--424}.
  \DOIprefix\doi{https://doi.org/10.1016/0550-3213(81)90568-X}.
\bibitem[{Hayami and Yambe(2021)}]{Hayami_PhysRevResearch.3.043158}
\bibinfo{author}{S.~Hayami}, \bibinfo{author}{R.~Yambe},
\newblock \bibinfo{title}{Locking of skyrmion cores on a centrosymmetric
  discrete lattice: Onsite versus offsite},
\newblock \bibinfo{journal}{Phys. Rev. Research} \bibinfo{volume}{3}
  (\bibinfo{year}{2021}) \bibinfo{pages}{043158}.
  \DOIprefix\doi{10.1103/PhysRevResearch.3.043158}.
\bibitem[{Yang et~al.(2016)Yang, Liu, and Han}]{Yang2016}
\bibinfo{author}{S.-G. Yang}, \bibinfo{author}{Y.-H. Liu},
  \bibinfo{author}{J.~H. Han},
\newblock \bibinfo{title}{Formation of a topological monopole lattice and its
  dynamics in three-dimensional chiral magnets},
\newblock \bibinfo{journal}{Phys. Rev. B} \bibinfo{volume}{94}
  (\bibinfo{year}{2016}) \bibinfo{pages}{054420}.
  \DOIprefix\doi{10.1103/PhysRevB.94.054420}.
\bibitem[{Aoyama and Kawamura(2021)}]{Aoyama_PhysRevB.103.014406}
\bibinfo{author}{K.~Aoyama}, \bibinfo{author}{H.~Kawamura},
\newblock \bibinfo{title}{{Hedgehog-lattice spin texture in classical
  Heisenberg antiferromagnets on the breathing pyrochlore lattice}},
\newblock \bibinfo{journal}{Phys. Rev. B} \bibinfo{volume}{103}
  (\bibinfo{year}{2021}) \bibinfo{pages}{014406}.
  \DOIprefix\doi{10.1103/PhysRevB.103.014406}.
\bibitem[{Aoyama and Kawamura(2022)}]{Aoyama_PhysRevB.106.064412}
\bibinfo{author}{K.~Aoyama}, \bibinfo{author}{H.~Kawamura},
\newblock \bibinfo{title}{{Hedgehog lattice and field-induced chirality in
  breathing-pyrochlore Heisenberg antiferromagnets}},
\newblock \bibinfo{journal}{Phys. Rev. B} \bibinfo{volume}{106}
  (\bibinfo{year}{2022}) \bibinfo{pages}{064412}.
  \DOIprefix\doi{10.1103/PhysRevB.106.064412}.
\bibitem[{Eto and Mochizuki(2024)}]{Eto_PhysRevLett.132.226705}
\bibinfo{author}{R.~Eto}, \bibinfo{author}{M.~Mochizuki},
\newblock \bibinfo{title}{{Theory of Collective Excitations in the
  Quadruple-$Q$ Magnetic Hedgehog Lattices}},
\newblock \bibinfo{journal}{Phys. Rev. Lett.} \bibinfo{volume}{132}
  (\bibinfo{year}{2024}) \bibinfo{pages}{226705}.
  \DOIprefix\doi{10.1103/PhysRevLett.132.226705}.
\bibitem[{Brey et~al.(1996)Brey, Fertig, C{\^o}t{\'e}, and
  MacDonald}]{brey1996skyrme}
\bibinfo{author}{L.~Brey}, \bibinfo{author}{H.~Fertig},
  \bibinfo{author}{R.~C{\^o}t{\'e}}, \bibinfo{author}{A.~MacDonald},
\newblock \bibinfo{title}{Skyrme and meron crystals in quantum {Hall}
  ferromagnets},
\newblock \bibinfo{journal}{Phys. Scr.} \bibinfo{volume}{1996}
  (\bibinfo{year}{1996}) \bibinfo{pages}{154}.
  \DOIprefix\doi{10.1088/0031-8949/1996/T66/027}.
\bibitem[{Ezawa(2011)}]{Ezawa_PhysRevB.83.100408}
\bibinfo{author}{M.~Ezawa},
\newblock \bibinfo{title}{Compact merons and skyrmions in thin chiral magnetic
  films},
\newblock \bibinfo{journal}{Phys. Rev. B} \bibinfo{volume}{83}
  (\bibinfo{year}{2011}) \bibinfo{pages}{100408(R)}.
  \DOIprefix\doi{10.1103/PhysRevB.83.100408}.
\bibitem[{Lin et~al.(2015)Lin, Saxena, and Batista}]{Lin_PhysRevB.91.224407}
\bibinfo{author}{S.-Z. Lin}, \bibinfo{author}{A.~Saxena},
  \bibinfo{author}{C.~D. Batista},
\newblock \bibinfo{title}{Skyrmion fractionalization and merons in chiral
  magnets with easy-plane anisotropy},
\newblock \bibinfo{journal}{Phys. Rev. B} \bibinfo{volume}{91}
  (\bibinfo{year}{2015}) \bibinfo{pages}{224407}.
  \DOIprefix\doi{10.1103/PhysRevB.91.224407}.
\bibitem[{Tan et~al.(2016)Tan, Li, Scholl, Arenholz, Young, Li, Hwang, and
  Qiu}]{Tan_PhysRevB.94.014433}
\bibinfo{author}{A.~Tan}, \bibinfo{author}{J.~Li}, \bibinfo{author}{A.~Scholl},
  \bibinfo{author}{E.~Arenholz}, \bibinfo{author}{A.~T. Young},
  \bibinfo{author}{Q.~Li}, \bibinfo{author}{C.~Hwang}, \bibinfo{author}{Z.~Q.
  Qiu},
\newblock \bibinfo{title}{Topology of spin meron pairs in coupled
  {Ni/Fe/Co/Cu(001)} disks},
\newblock \bibinfo{journal}{Phys. Rev. B} \bibinfo{volume}{94}
  (\bibinfo{year}{2016}) \bibinfo{pages}{014433}.
  \DOIprefix\doi{10.1103/PhysRevB.94.014433}.
\bibitem[{Hayami et~al.(2018)Hayami, Ozawa, and
  Motome}]{Hayami_PhysRevB.98.019903}
\bibinfo{author}{S.~Hayami}, \bibinfo{author}{R.~Ozawa},
  \bibinfo{author}{Y.~Motome},
\newblock \bibinfo{title}{{Erratum: Effective bilinear-biquadratic model for
  noncoplanar ordering in itinerant magnets [Phys. Rev. B 95, 224424 (2017)]}},
\newblock \bibinfo{journal}{Phys. Rev. B} \bibinfo{volume}{98}
  (\bibinfo{year}{2018}) \bibinfo{pages}{019903}.
  \DOIprefix\doi{10.1103/PhysRevB.98.019903}.
\bibitem[{Mukherjee et~al.(2022)Mukherjee, Kathyat, and
  Kumar}]{Mukherjee_PhysRevB.105.075102}
\bibinfo{author}{A.~Mukherjee}, \bibinfo{author}{D.~S. Kathyat},
  \bibinfo{author}{S.~Kumar},
\newblock \bibinfo{title}{Engineering antiferromagnetic skyrmions and
  antiskyrmions at metallic interfaces},
\newblock \bibinfo{journal}{Phys. Rev. B} \bibinfo{volume}{105}
  (\bibinfo{year}{2022}) \bibinfo{pages}{075102}.
  \DOIprefix\doi{10.1103/PhysRevB.105.075102}.
\bibitem[{Hayami and Hattori(2023)}]{hayami2023multipleJPSJ}
\bibinfo{author}{S.~Hayami}, \bibinfo{author}{K.~Hattori},
\newblock \bibinfo{title}{Multiple-$q$ dipole-quadrupole instability in spin-1
  triangular-lattice systems},
\newblock \bibinfo{journal}{J. Phys. Soc. Jpn.} \bibinfo{volume}{92}
  (\bibinfo{year}{2023}) \bibinfo{pages}{124709}.
  \DOIprefix\doi{10.7566/JPSJ.92.124709}.
\bibitem[{Ishitobi and Hattori(2021)}]{Ishitobi_PhysRevB.104.L241110}
\bibinfo{author}{T.~Ishitobi}, \bibinfo{author}{K.~Hattori},
\newblock \bibinfo{title}{Triple-$q$ quadrupole-octupole order scenario for
  {${\mathrm{PrV}}_{2}{\mathrm{Al}}_{20}$}},
\newblock \bibinfo{journal}{Phys. Rev. B} \bibinfo{volume}{104}
  (\bibinfo{year}{2021}) \bibinfo{pages}{L241110}.
  \DOIprefix\doi{10.1103/PhysRevB.104.L241110}.
\bibitem[{Ishitobi and Hattori(2023)}]{Ishitobi_PhysRevB.107.104413}
\bibinfo{author}{T.~Ishitobi}, \bibinfo{author}{K.~Hattori},
\newblock \bibinfo{title}{Triple-$\mathcal{Q}$ partial magnetic orders induced
  by quadrupolar interactions: Triforce order scenario for
  {${\mathrm{UNi}}_{4}\mathrm{B}$}},
\newblock \bibinfo{journal}{Phys. Rev. B} \bibinfo{volume}{107}
  (\bibinfo{year}{2023}) \bibinfo{pages}{104413}.
  \DOIprefix\doi{10.1103/PhysRevB.107.104413}.
\bibitem[{Zhang and Lin(2024)}]{zhang2024multipolar}
\bibinfo{author}{H.~Zhang}, \bibinfo{author}{S.-Z. Lin},
\newblock \bibinfo{title}{{Multipolar Skyrmion Crystals in Non-Kramers Doublet
  Systems}},
\newblock \bibinfo{journal}{arXiv:2404.14404}  (\bibinfo{year}{2024}).
\bibitem[{Sharma et~al.(2023)Sharma, Wang, and Batista}]{sharma2023machine}
\bibinfo{author}{V.~Sharma}, \bibinfo{author}{Z.~Wang}, \bibinfo{author}{C.~D.
  Batista},
\newblock \bibinfo{title}{Machine learning assisted derivation of minimal
  low-energy models for metallic magnets},
\newblock \bibinfo{journal}{npj Comput. Mater.} \bibinfo{volume}{9}
  (\bibinfo{year}{2023}) \bibinfo{pages}{192}.
  \DOIprefix\doi{https://doi.org/10.1038/s41524-023-01137-x}.
\bibitem[{Okigami and Hayami(2024)}]{okigami2024exploring}
\bibinfo{author}{K.~Okigami}, \bibinfo{author}{S.~Hayami},
\newblock \bibinfo{title}{Exploring topological spin order by inverse
  hamiltonian design: A new stabilization mechanism for square skyrmion
  crystals},
\newblock \bibinfo{journal}{arXiv:2407.01159}  (\bibinfo{year}{2024}).

\end{thebibliography}

\end{document}